\documentclass[journal]{IEEEtran}
\usepackage{graphicx}
\usepackage{amssymb}
\usepackage{epstopdf}
\usepackage{color}
\usepackage{subfigure}
\usepackage[cmex10]{amsmath}
\usepackage{cite}
\usepackage{url,hyperref}
\usepackage[normalem]{ulem} 
\newcommand{\Xeqref}[1]{\eqref{#1}}

\newcommand{\Xsecref}[1]{\secref{#1}}

\newcommand{\Xtabref}[1]{\tabref{#1}}

\newcommand{\Xfoot}[1]{footnote~\ref{#1}}

{\bf}{\it}
{\em}

\DeclareMathAlphabet{\mathsfbf}{OT1}{cmss}{bx}{n}
\DeclareSymbolFont{bbold}{U}{bbold}{m}{n}
\DeclareSymbolFontAlphabet{\mathbbold}{bbold}
\renewcommand{\hat}{\widehat}
\renewcommand{\tilde}{\widetilde}
\renewcommand{\bar}{\overline}
\newcommand{\Dirac}{\mathbbold{1}}

\newcommand{\non}{\nonumber}

\newcommand{\ba}[1]{\begin{array}{#1}}
\newcommand{\ea}{\end{array}}
\newcommand{\beq}{\begin{equation}}
\newcommand{\eeq}{\end{equation}}
\newcommand{\beqar}{\begin{eqnarray}}
\newcommand{\eeqar}{\end{eqnarray}}
\newcommand{\beqars}{\begin{eqnarray*}}
\newcommand{\eeqars}{\end{eqnarray*}}
\newcommand{\bmat}{\left[\ba{cccccc}}
\newcommand{\emat}{\ea\right]}

 \newcommand{\defn}{\triangleq}
 
 \newcommand{\uvec}[1]{\ensuremath{\boldsymbol{\underline{#1}}}}

 \newcommand{\hvec}[1]{\ensuremath{\boldsymbol{\Hat{#1}}}}
    
 \renewcommand{\vec}[1]{\ensuremath{\boldsymbol{#1}}}

 \newcommand{\norm}[1]{\ensuremath{\| #1 \|}}
 \newcommand{\mc}[1]{\ensuremath{\mathcal{#1}}}

 \newcommand{\Real}{{\mathbb{R}}}
 
 \newcommand{\Int}{{\mathbb{Z}}}

 \newcommand{\tran}{^\textsf{T}}
 \newcommand{\herm}{^\textsf{H}}

\newcommand{\Nor}{\mathcal{N}}
 \newcommand{\df}{\mathsf{df}}

 \DeclareMathOperator{\E}{E}
 \DeclareMathOperator{\var}{var}

 \DeclareMathOperator*{\argmax}{arg\, max}
 \DeclareMathOperator*{\argmin}{arg\, min}

 \DeclareMathOperator{\med}{median}
 \renewcommand{\eqref}[1]{(\ref{eq:#1})}
 
 \newcommand{\Figref}[1]{Figure~\ref{fig:#1}}
 \newcommand{\figref}[1]{Fig.~\ref{fig:#1}}
 \newcommand{\tabref}[1]{Table~\ref{tab:#1}}
 \newcommand{\secref}[1]{Sec.~\ref{sec:#1}}
 \newcommand{\Secref}[1]{Section~\ref{sec:#1}}
 \newcommand{\appref}[1]{Appendix~\ref{app:#1}}

 \newcommand{\textb}[1]{\textcolor{blue}{#1}}

\usepackage{psfrag}

\newcommand{\putFrag}[4]{\begin{figure}[htbp]
                            \begin{center}
                            #4
                            \includegraphics[width=#3in]{figures/#1.eps}
                            \end{center}
			    \caption{#2}
			    \label{fig:#1}
                          \end{figure} }

 \newcommand{\putTable}[3]{\begin{table}[tp]
  			    \centering
		            #3
     			    \caption{#2}
     			    \label{tab:#1}
			    \vspace{-5mm}
			  \end{table} }

 \newcommand{\giv}{\,|\,}
 \newcommand{\biggiv}{\,\big|\,}
 \newcommand{\Biggiv}{\,\Big|\,}
 \newcommand{\bigggiv}{\,\bigg|\,}

 \newcommand{\fxnvar}[3]{\Delta_{#1{\scriptscriptstyle \rightarrow} #2}^{#3}}
 \newcommand{\varfxn}[3]{\Delta_{#1{\scriptscriptstyle \leftarrow} #2}^{#3}}
 \newcommand{\X}{\textsf{x}}
 \newcommand{\uX}{\underline{\textsf{x}}}
 \newcommand{\A}{\textsf{a}}
 \newcommand{\uA}{\underline{\textsf{a}}}
 \newcommand{\Y}{\textsf{y}}
 \newcommand{\uY}{\underline{\textsf{y}}}
 \newcommand{\Z}{\textsf{z}}
 \newcommand{\uZ}{\underline{\textsf{z}}}
 
 \newcommand{\p}{\textsf{p}}
 \newcommand{\R}{\textsf{r}}
 \newcommand{\Q}{\textsf{q}}
 \newcommand{\vX}{\textsf{\textbf{\textit{X}}}}
 
 \newcommand{\vA}{\textsf{\textbf{\textit{A}}}}
 \newcommand{\vY}{\textsf{\textbf{\textit{Y}}}}
 \newcommand{\vvY}{\textsf{\textbf{\textit{y}}}}
 \newcommand{\vZ}{\textsf{\textbf{\textit{Z}}}}
 \newcommand{\vvZ}{\textsf{\textbf{\textit{z}}}}

 \newcommand{\vTheta}{\mathsfbf{\Theta}}
 \newcommand{\const}{\text{\sf const}}

 \newcommand{\Ord}{O}

\graphicspath{{./}} 

\renewcommand{\textb}[1]{#1}

\begin{document}
\setlength{\arraycolsep}{0.4mm}
\title{Bilinear Generalized Approximate Message Passing}

\author{Jason~T.~Parker, 
        Philip~Schniter, 
        and~Volkan~Cevher
\thanks{J. Parker is with the Sensors Directorate, Air Force Research Laboratory, Dayton, OH 45433, USA e-mail: \textb{jason.parker.13@us.af.mil}.  His work on this project has been supported by \textb{AFOSR Lab Task 11RY02COR}.}
\thanks{P. Schniter is with the Dept. ECE, The Ohio State University, 2015 Neil Ave., Columbus OH 43210, e-mail: schniter@ece.osu.edu, phone 614.247.6488, fax 614.292.7596. His work on this project has been supported by NSF grants IIP-0968910, CCF-1018368, CCF-1218754, and by DARPA/ONR grant N66001-10-1-4090.}
\thanks{V. Cevher is with \'{E}cole Polytechnique F\'{e}d\'{e}rale de Lausanne. Email: volkan.cevher@epfl.ch. His work is supported in part by the European Commission under the grants MIRG-268398 and ERC Future Proof, and by the Swiss Science Foundation under the grants SNF 200021-132548, SNF 200021-146750 and SNF CRSII2-147633.}
\thanks{Portions of this work were presented at the 2011 Workshop on Signal Processing with Adaptive Sparse Structured Representations \cite{Schniter:SPARS:11}, the 2012 Information Theory and Applications Workshop \cite{Schniter:ITA:12}, and at the 2012 Asilomar Conference on Signals, Systems, and Computers \cite{Parker:Asil:12}.} 
}

\markboth{\today}{}

\maketitle

\begin{abstract}
We extend the generalized approximate message passing (G-AMP) approach, originally proposed for high-dimensional generalized-linear regression in the context of compressive sensing, to the generalized-bilinear case, which enables its application to matrix completion, robust PCA, dictionary learning, and related matrix-factorization problems.
In the first part of the paper, we derive our Bilinear G-AMP (BiG-AMP) algorithm as an approximation of the sum-product belief propagation algorithm in the high-dimensional limit, where central-limit theorem arguments and Taylor-series approximations apply, and under the assumption of statistically independent matrix entries with known priors. 
In addition, we propose an adaptive damping mechanism that aids convergence under finite problem sizes, an expectation-maximization (EM)-based method to automatically tune the parameters of the assumed priors, and two rank-selection strategies.
In the second part of the paper, we discuss the specializations of EM-BiG-AMP to the problems of matrix completion, robust PCA, and dictionary learning, and present the results of an extensive empirical study comparing EM-BiG-AMP to state-of-the-art algorithms on each problem.
Our numerical results, using both synthetic and real-world datasets, demonstrate that EM-BiG-AMP yields excellent reconstruction accuracy (often best in class) while maintaining competitive runtimes and avoiding the need to tune algorithmic parameters. 
\end{abstract}

\IEEEpeerreviewmaketitle

\section{Introduction}	\label{sec:intro}

In this work, we present a new algorithmic framework for the following \emph{generalized bilinear} inference problem: estimate the matrices $\vec{A} \!=\! [a_{mn}] \!\in\! \Real^{M \times N}$ and $\vec{X} \!=\! [x_{nl}] \!\in\! \Real^{N \times L}$ from a matrix observation $\vec{Y}\in\Real^{M \times \textb{L}}$ that is statistically coupled to their product, $\vec{Z}\!\defn\!\vec{AX}$.
In doing so, we treat $\vec{A}$ and $\vec{X}$ as realizations of random matrices $\vA$ and $\vX$ with known separable pdfs (or pmfs in the case of discrete models), i.e.,
\begin{align}
p_{\vA}(\vec{A}) &= \prod_m \prod_n p_{\A_{mn}}(a_{mn})\label{eq:pA}\\
p_{\vX}(\vec{X}) &= \prod_n \prod_l p_{\X_{nl}}(x_{nl})\label{eq:pX},
\end{align}
and we likewise assume that the likelihood function of $\vec{Z}$ is known and separable, i.e.,
\begin{align}
p_{\vY|\vZ}(\vec{Y}\giv\vec{Z}) &= \prod_m \prod_l p_{\Y_{ml}|\Z_{ml}}(y_{ml}\giv z_{ml}).\label{eq:pYgivZ}
\end{align}

Recently, various special cases of this problem have gained the intense interest of the research community, e.g.,
\begin{enumerate}
\item
\emph{Matrix Completion}:
In this problem, one observes a few (possibly noise-corrupted) entries of a low-rank matrix and the goal is to infer the missing entries.
In our framework, $\vec{Z}\!=\!\vec{AX}$ would represent the complete low-rank matrix (with tall $\vec{A}$ and wide $\vec{X}$) and $p_{\Y_{ml}|\Z_{ml}}$ the observation mechanism, which would be (partially) informative about $\Z_{ml}$ at the observed entries $(m,l)\in\Omega$ and non-informative at the missing entries $(m,l)\notin\Omega$.
\item
\emph{Robust PCA}:
Here, the objective is to recover a low-rank matrix (or its principal components) observed in the presence of noise and sparse outliers.
In our framework, $\vec{Z}\!=\!\vec{AX}$ could again represent the low-rank matrix, and $p_{\Y_{ml}|\Z_{ml}}$ the noise-and-outlier-corrupted observation mechanism. 
Alternatively, $\vec{X}$ could also capture the outliers, as described in the sequel.
\item
\emph{Dictionary Learning}:
Here, the objective is to learn a dictionary $\vec{A}$ for which there exists a sparse data representation $\vec{X}$ such that $\vec{AX}$ closely matches the observed data $\vec{Y}$.
In our framework, $\{p_{\X_{nl}}\}$ would be chosen to induce sparsity, $\vec{Z}=\vec{AX}$ would represent the noiseless observations, and $\{p_{\Y_{ml}|\Z_{ml}}\}$ would model the (possibly noisy) observation mechanism.
\end{enumerate}

While a plethora of approaches to these problems have been proposed based on 
optimization techniques
(e.g., {\cite{jtp_Cai2010,jtp_Ma2011,jtp_Zhou2010,jtp_Recht2010,jtp_Lin2010,jtp_Tao2011,jtp_Zhou2011,Ghasemi:ICASSP:11,jtp_Wen2012a,Kyrillidis:12,jtp_Marjanovic2012}}),
greedy methods 
(e.g., {\cite{jtp_Haldar2009,jtp_Balzano2010,jtp_Keshavan2010,jtp_Dai2010,jtp_He2011}}),
Bayesian sampling methods 
(e.g., {\cite{jtp_Salakhutdinov2008,jtp_Ding2011}}),
variational methods
(e.g., {\cite{jtp_Lim2007,Babacan:TSP:12,jtp_Tipping1999,jtp_Leger2011,jtp_Wang2012}}),
and discrete message passing (e.g., \cite{jtp_Kim2010}),
ours is based on the \emph{Approximate Message Passing} (AMP) framework, an instance of loopy belief propagation (LBP) \cite{Frey:NIPS:97} that was recently developed to tackle \emph{linear} \cite{Donoho:PNAS:09,Donoho:ITW:10a,Montanari:Chap:12} and \emph{generalized linear} \cite{Rangan:ISIT:11} inference problems encountered in the context of compressive sensing (CS). 
In the generalized-linear CS problem, one estimates $\vec{x}\in\Real^{N}$ from observations 
$\vec{y}\in\Real^M$ that are statistically coupled to the transform outputs $\vec{z}=\vec{Ax}$ through a separable likelihood function $p_{\vvY|\vvZ}(\vec{y}|\vec{z})$, where in this case the transform $\vec{A}$ is \emph{fixed and known}.

In the context of CS, the AMP framework yields algorithms with remarkable properties:
i) solution trajectories that, in the large-system limit (i.e., as $M,N\rightarrow\infty$ with $M/N$ fixed, under iid sub-Gaussian $\vec{A}$) are governed by a state-evolution whose fixed points---when unique---yield the true posterior means \cite{Bayati:TIT:11,Javanmard:IMA:13} and 
ii) a low implementation complexity (i.e., dominated by one multiplication with $\vec{A}$ and $\vec{A}\tran$ per iteration, and relatively few iterations) \cite{Montanari:Chap:12}.
Thus, a natural question is whether the AMP framework can be successfully applied to the generalized \emph{bilinear} problem described earlier.

In this manuscript, we propose an AMP-based approach to generalized bilinear inference that we henceforth refer to as \emph{Bilinear Generalized AMP} (BiG-AMP), and we uncover special cases under which the general approach can be simplified.
In addition, we propose
an adaptive damping \textb{\cite{Rangan:ISIT:14}} mechanism,
an expectation-maximization (EM)-based \cite{Dempster:JRSS:77} method of tuning the parameters of \textb{$p_{\A_{mn}}$, $p_{\X_{nl}}$, and $p_{\Y_{ml}|\Z_{ml}}$} (in case they are unknown),
and methods to select the rank $N$ (in case it is unknown).
\textb{In the case that $p_{\A_{mn}}$, $p_{\X_{nl}}$, and/or $p_{\Y_{ml}|\Z_{ml}}$ are completely unknown, they can be modeled as Gaussian-mixtures with mean/variance/weight parameters learned via EM \cite{Vila:TSP:13}.}
Finally, we present a detailed numerical investigation of BiG-AMP applied to matrix completion, robust PCA, and dictionary learning.
Our empirical results show that BiG-AMP yields an excellent combination of estimation accuracy
and runtime when compared to existing state-of-the-art algorithms for each application.

Although the AMP methodology \textb{is itself restricted to} separable known pdfs \eqref{pA}-\eqref{pYgivZ}, \textb{the results of Part II suggest that this limitation is not an issue for many practical problems of interest.  
However, in problems where the separability assumption is too constraining, it can be relaxed}
through the use of hidden (coupling) variables, as originally proposed in the context of ``turbo-AMP'' \cite{Schniter:CISS:10} and applied to BiG-AMP in \cite{Vila:SPIE:13}.
Due to space limitations, however, this approach will not be discussed here.
Finally, although we focus on real-valued random variables, all of the methodology described in this work can be easily extended to circularly symmetric complex-valued random variables.

We now discuss related work.
One possibility of applying AMP methods to matrix completion was suggested by Montanari in \cite[Sec.~9.7.3]{Montanari:Chap:12} but the approach described there differs from BiG-AMP in that it was i) constructed from a factor graph with \emph{vector-valued} variables and ii) restricted to the (incomplete) additive white Gaussian noise (AWGN) observation model.
Moreover, no concrete algorithm nor performance evaluation was reported.
Since we first reported on BiG-AMP in \cite{Schniter:SPARS:11,Schniter:ITA:12}, Rangan and Fletcher~\cite{jtp_Rangan2012a} proposed an AMP-based approach for the estimation of \emph{rank-one} matrices from AWGN-corrupted observations, and showed that it can be characterized by a state evolution in the large-system limit.
More recently, Krzakala, M{\'e}zard, and Zdeborov{\'a}~\cite{jtp_Krzakala2013} proposed an AMP-based approach to blind calibration and dictionary learning in AWGN that bears similarity to a special case of BiG-AMP, and derived a state-evolution using the cavity method. 
Their method, however, was not numerically successful in solving dictionary learning problems \cite{jtp_Krzakala2013}.
The BiG-AMP algorithm that we derive here is a generalization of those in \cite{jtp_Rangan2012a,jtp_Krzakala2013} in that it handles \emph{generalized} bilinear observations rather than AWGN-corrupted ones.
Moreover, our work is the first to detail adaptive damping, parameter tuning, and rank-selection mechanisms \textb{for AMP based bilinear inference}, 
and it is the first to present an in-depth numerical investigation involving both synthetic and real-world datasets.
An application/extension of the BiG-AMP algorithm described here to hyperspectral unmixing (an instance of non-negative matrix factorization) was recently proposed in \cite{Vila:SPIE:13}.

The remainder of the document is organized as follows.
\Secref{derivation} derives the BiG-AMP algorithm, and \secref{simple} presents several special-case simplifications of BiG-AMP.
\Secref{coststep} describes the adaptive damping mechanism, and \secref{learning} the EM-based tuning of prior parameters and selection of rank $N$.
Application-specific issues and numerical results demonstrating the efficacy of our approach for matrix completion, robust PCA, and dictionary learning, are discussed in Sections~\ref{sec:MC}--\ref{sec:DL}, respectively, and
concluding remarks are offered in \secref{conc}.

\emph{Notation}:
Throughout, we use san-serif font (e.g., $\X$) for random variables and serif font (e.g., $x$) otherwise.
We use boldface capital letters (e.g., $\vX$ and $\vec{X}$) for matrices, boldface small letters (e.g., $\textsf{\textit{\textbf{x}}}$ and $\vec{x}$) for vectors, and non-bold small letters (e.g., $\X$ and $x$) for scalars.
We then use $p_{\X}(x)$ to denote the pdf of random quantity $\X$, and $\Nor(x;\hat{x},\nu^x)$ to denote the Gaussian pdf for a scalar random variable with mean $\hat{x}$ and variance $\nu^x$.
Also, we use $\E\{\X\}$ and $\var\{\X\}$ to denote mean and variance of $\X$, respectively, 
and $D(p_1\|p_2)$ for the Kullback-Leibler (KL) divergence between pdfs $p_1$ and $p_2$.
For a matrix $\vec{X}$, we use 
$x_{nl}=[\vec{X}]_{nl}$ to denote the entry in the $n^{th}$ row and $l^{th}$ column,
$\|\vec{X}\|_F$ to denote the Frobenius norm, and
$\vec{X}\tran$ to denote transpose.
Similarly, we use $x_n$ to denote the $n^{th}$ entry in vector $\vec{x}$ and
$\norm{\vec{x}}_p=(\sum_n |x_n|^p)^{1/p}$ to denote the $\ell_p$ norm. 

\section{Bilinear Generalized AMP}
\label{sec:derivation}

%


\subsection{Problem Formulation}

For the statistical model \eqref{pA}-\eqref{pYgivZ}, the posterior distribution is
\begin{align}
\lefteqn{p_{\vX,\vA|\vY}(\vec{X},\vec{A}\giv\vec{Y}) }\nonumber\\
&= p_{\vY|\vX,\vA}(\vec{Y}\giv \vec{X},\vec{A}) 
   \,p_{\vX}(\vec{X}) 
   \,p_{\vA}(\vec{A})/p_{\vY}(\vec{Y)} 
\label{eq:Bayes}\\
&\propto p_{\vY|\vZ}(\vec{Y}\giv \vec{AX}) 
 	\,p_{\vX}(\vec{X})
	\,p_{\vA}(\vec{A}) \label{eq:scaling}\\
&= \bigg[ \prod_m \prod_l p_{\Y_{ml}|\Z_{ml}}\Big(y_{ml}\Biggiv \sum_{k} a_{mk} x_{kl} \Big)\bigg]\nonumber\\
&\quad\times \bigg[ \prod_n \prod_l p_{\X_{nl}}(x_{nl}) \bigg]  \bigg[\prod_m \prod_n p_{\A_{mn}}(a_{mn})\bigg],	\label{eq:post}
\end{align}
where \eqref{Bayes} employs Bayes' rule and $\propto$ denotes equality up to a constant scale factor.

The posterior distribution can be represented with a factor graph, as depicted in \figref{fg_bigamp2}. 
There, the factors of $p_{\vX,\vA|\vY}$ from \eqref{post} are represented by ``factor nodes'' that appear as black boxes, and the random variables are represented by ``variable nodes'' that appear as white circles. 
Each variable node is connected to every factor node in which that variable appears. 
The observed data $\{y_{ml}\}$ are treated as parameters of the $p_{\Y_{ml}|\Z_{ml}}$ factor nodes in the middle of the graph, and not as random variables. 
The structure of \figref{fg_bigamp2} becomes intuitive when recalling that $\vZ=\vA\vX$ implies $\Z_{ml} = \sum_{n=1}^N \A_{mn}\X_{nl}$.

\putFrag{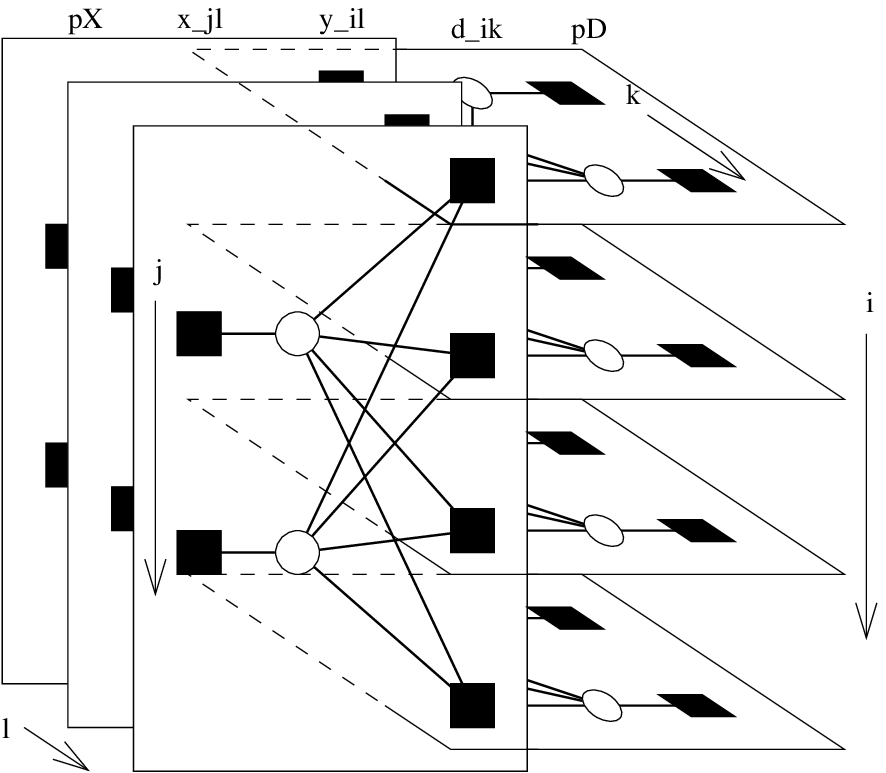}{The factor graph for generalized bilinear inference for (toy-sized) problem dimensions $M=4$, $L=3$, and $N=2$.}{2.0}
   {\psfrag{l}[][Bl][0.9]{$l$}
    \psfrag{k}[][Bl][0.9]{$k$}
    \psfrag{i}[B][Bl][0.9]{$m$}
    \psfrag{j}[B][Bl][0.9]{$n$}
    \psfrag{x_jl}[b][Bl][0.7]{$\X_{nl}$}
    \psfrag{y_il}[b][Bl][0.7]{$p_{\Y_{ml}|\Z_{ml}}$~}
    \psfrag{d_ik}[b][Bl][0.7]{~~$\A_{mk}$}
    \psfrag{pX}[b][Bl][0.7]{$p_{\X_{nl}}$}
    \psfrag{pD}[b][Bl][0.7]{$p_{\A_{mk}}$}
    }

\subsection{Loopy Belief Propagation}			\label{sec:LBP}

In this work, we aim to compute minimum mean-squared error (MMSE) estimates of $\vec{X}$ and $\vec{A}$, i.e., the means\footnote{Another worthwhile objective could be to compute the joint MAP estimate $\arg\max_{\vec{X},\vec{A}}p_{\vX,\vA|\vY}(\vec{X},\vec{A}\giv\vec{Y})$; we leave this to future work.} of the marginal posteriors $p_{\X_{nl}|\vY}(\cdot\giv\vec{Y})$ and $p_{\A_{mn}|\vY}(\cdot\giv\vec{Y})$, for all pairs $(n,l)$ and $(m,n)$.
Although exact computation of these quantities is generally prohibitive, they can be efficiently approximated using loopy belief propagation (LBP) \cite{Frey:NIPS:97}.

In LBP, beliefs about the random variables (in the form of pdfs or log pdfs) are propagated among the nodes of the factor graph until they converge.
The standard way to compute these beliefs, known as the \emph{sum-product algorithm} (SPA) \cite{Pearl:Book:88,Kschischang:TIT:01}, stipulates that the belief emitted by a variable node along a given edge of the graph is computed as the product of the incoming beliefs from all other edges, whereas the belief emitted by a factor node along a given edge is computed as the integral of the product of the factor associated with that node and the incoming beliefs on all other edges.
The product of all beliefs impinging on a given variable node yields the posterior pdf for that variable.
In cases where the factor graph has no loops, exact marginal posteriors result from two (i.e., forward and backward) passes of the SPA \cite{Pearl:Book:88,Kschischang:TIT:01}.
For loopy factor graphs, exact inference is in general NP hard \cite{Cooper:AI:90} and so LBP does not guarantee correct posteriors.
That said, LBP has shown state-of-the-art performance in many applications,
such as 
inference on Markov random fields \cite{Freeman:IJCV:00},
turbo decoding \cite{McEliece:JSAC:98},
LDPC decoding \cite{MacKay:Book:03},
multiuser detection \cite{Boutros:TIT:02}, 
and compressive sensing \cite{Donoho:PNAS:09,Donoho:ITW:10a,Rangan:ISIT:11,Bayati:TIT:11,Javanmard:IMA:13}.

In high-dimensional inference problems, exact implementation of the SPA is impractical, motivating approximations of the SPA.
A notable example is the \emph{generalized approximate message passing} (GAMP) algorithm, developed in \cite{Rangan:ISIT:11} to solve the generalized CS problem, which exploits the ``blessings of dimensionality'' that arise when $\vec{A}$ is a sufficiently large and dense and which was rigorously analyzed in \cite{Javanmard:IMA:13}.
In the sequel, we derive an algorithm for the generalized bilinear inference BiG-AMP algorithm that employs GAMP-like approximations to the SPA on the factor graph in \figref{fg_bigamp2}.
As we shall see, the approximations are primarily based on central-limit-theorem (CLT) and Taylor-series arguments.

\subsection{Sum-product Algorithm}	\label{sec:SPA}
In our formulation of the SPA, messages take the form of log-pdfs with arbitrary constant offsets. 
For example, the iteration-$t$ (where $t\in\Int$) message $\fxnvar{m}{nl}{\X}(t,.)$ can be converted to the pdf $\frac{1}{C}\exp(\fxnvar{m}{nl}{\X}(t,.))$, where the choice of scale factor $C=\int_{x_{nl}} \exp(\fxnvar{m}{nl}{\X}(t,x_{nl}))$ ensures that the pdf integrates to one.
Four types of message will be used, as specified in \tabref{logpdf}.
We also find it convenient to express the (iteration-$t$ SPA-approximated) posterior pdfs $p_{\X_{nl}|\vY}(t,.\giv\vec{Y})$ and $p_{\A_{mn}|\vY}(t,.\giv\vec{Y})$ in the log domain as $\Delta_{nl}^{\X}(t,.)$ and $\Delta_{mn}^{\A}(t,.)$, respectively, again with arbitrary constant offsets.

\putTable{logpdf}{SPA message definitions at iteration $t\in\Int$.}
{\begin{tabular}{|r|l|} \hline
  $\fxnvar{m}{nl}{\X}(t,.)$ 
  & SPA message from node $p_{\Y_{ml}|\Z_{ml}}$ to node $\X_{nl}$ \\
  $\varfxn{m}{nl}{\X}(t,.)$ 
  & SPA message from node $\X_{nl}$ to node $p_{\Y_{ml}|\Z_{ml}}$\\
  $\fxnvar{l}{mn}{\A}(t,.)$ 
  & SPA message from node $p_{\Y_{ml}|\Z_{ml}}$ to node $\A_{mn}$\\
  $\varfxn{l}{mn}{\A}(t,.)$ 
  & SPA message from node $\A_{mn}$ to node $p_{\Y_{ml}|\Z_{ml}}$\\
  $\Delta_{nl}^{\X}(t,.)$ 
  & SPA-approximated log posterior pdf of $\X_{nl}$\\
  $\Delta_{mn}^{\A}(t,.)$ 
  & SPA-approximated log posterior pdf of $\A_{mn}$\\\hline
 \end{tabular}
}

Applying the SPA to the factor graph in \figref{fg_bigamp2}, we arrive at the following update rules for the four messages in \tabref{logpdf}. 
\begin{align}
\lefteqn{ \fxnvar{m}{nl}{\X}(t,x_{nl}) }\nonumber\\[-2mm]
&=\log \int_{\{a_{mk}\}_{k=1}^N ,\{x_{rl}\}_{r \ne n}} 
p_{\Y_{ml}|\Z_{ml}}\bigg(y_{ml}\bigggiv\sum_{k=1}^N a_{mk} x_{kl}\bigg)
\nonumber\\&\quad\times 
\prod_{r \ne n}\exp\Big(\varfxn{m}{rl}{\X}(t,x_{rl})\Big) \prod_{k=1}^N \exp\Big(\varfxn{l}{mk}{\A}(t,a_{mk})\Big) 
\nonumber\\&\quad 
+\const 
\label{eq:zTox}\\
\lefteqn{ \varfxn{m}{nl}{\X}(t\!+\!1,x_{nl}) }\nonumber\\
&= \log p_{\X_{nl}}(x_{nl}) + \sum_{k\ne m} \fxnvar{k}{nl}{\X}(t,x_{nl})
+ \const 
\label{eq:xToz}\\
\lefteqn{ \fxnvar{l}{mn}{\A}(t,a_{mn}) }\nonumber\\[-2mm]
&= \log \int_{\{a_{mr}\}_{r \ne n} ,\{x_{kl}\}_{k=1}^N} 
p_{\Y_{ml}|\Z_{ml}}\Big(y_{ml}\bigggiv\sum_{k=1}^N a_{mk} x_{kl}\Big)
\nonumber\\&\quad\times
\prod_{k=1}^N \exp\Big(\varfxn{m}{kl}{\X}(t,x_{kl})\Big) \prod_{r \ne n} \exp\Big(\varfxn{l}{mr}{\A}(t,a_{mr})\Big)
\nonumber\\&\quad
+\const
\label{eq:zToA}\\
\lefteqn{ \varfxn{l}{mn}{\A}(t\!+\!1,a_{mn}) }\nonumber\\
&= \log p_{\A_{mn}}(a_{mn}) + \sum_{k\ne l} \fxnvar{k}{mn}{\A}(t,a_{mn})
+ \const ,
\label{eq:AToz}
\end{align}
where $\const$ is an arbitrary constant (w.r.t $x_{nl}$ in \eqref{zTox} and \eqref{xToz}, and w.r.t $a_{mn}$ in \eqref{zToA} and \eqref{AToz}).
In the sequel, we denote the mean and variance of the pdf $\frac{1}{C}\exp(\varfxn{m}{nl}{\X}(t,.))$ by $\hat{x}_{m,nl}(t)$ and $\nu^x_{m,nl}(t)$, respectively, and we denote the mean and variance of $\frac{1}{C}\exp(\varfxn{l}{mn}{\A}(t,.))$ by $\hat{a}_{l,mn}(t)$ and $\nu^a_{l,mn}(t)$.
For the log-posteriors, the SPA implies 
\begin{align}
\lefteqn{ \Delta_{nl}^{\X}(t\!+\!1,x_{nl}) }\nonumber\\[-3mm]
&= \log p_{\X_{nl}}(x_{nl}) + \sum_{m=1}^M \fxnvar{m}{nl}{\X}(t,x_{nl}) 
+ \const 
\label{eq:Dx}\\
\lefteqn{ \Delta_{mn}^{\A}(t\!+\!1,a_{mn}) }\nonumber\\[-3mm]
&= \log p_{\A_{mn}}(a_{mn}) + \sum_{l=1}^L \fxnvar{l}{mn}{\A}(t,a_{mn})
+ \const,
\label{eq:Da}
\end{align}
and we denote the mean and variance of $\frac{1}{C}\exp(\Delta_{nl}^{\X}(t,.))$ by $\hat{x}_{nl}(t)$ and $\nu^x_{nl}(t)$, and the mean and variance of $\frac{1}{C}\exp(\Delta_{mn}^{\A}(t,.))$ by $\hat{a}_{mn}(t)$ and $\nu^a_{mn}(t)$.

\subsection{Approximated Factor-to-Variable Messages}	\label{sec:fTov}
We now apply AMP approximations to the SPA updates \eqref{zTox}-\eqref{Da}.
As we shall see, the approximations are based primarily on central-limit-theorem (CLT) and Taylor-series arguments \textb{that become exact} in the large-system limit, where $M,L,N \to \infty$ with fixed ratios $M/N$ and $L/N$. 
\textb{
(Due to the use of finite $M,L,N$ in practice, we still regard them as approximations.)
In particular, our derivation will neglect terms that vanish relative to others as $N\rightarrow\infty$, which requires that we establish certain scaling conventions. 
First}, we assume w.l.o.g\footnote{Other scalings on $\E\{\Z_{ml}^2\}$, $\E\{\X_{nl}^2\}$, and $\E\{\A_{mn}^2\}$ could be used as long as they are consistent with the relationship $\Z_{ml}=\sum_{n=1}^N \A_{mn}\X_{nl}$.}  
that $\E\{\Z_{ml}^2\}$ and $\E\{\X_{nl}^2\}$ scale as $\Ord(1)$, 
\textb{i.e., that the magnitudes of these elements do not change as $N\rightarrow\infty$. 
In this case, the relationship $\Z_{ml}=\sum_{n=1}^N \A_{mn}\X_{nl}$ implies that $\E\{\A_{mn}^2\}$ must scale} as $\Ord(1/N)$. 
These scalings are assumed to hold for random variables $\Z_{ml}$, $\A_{mn}$, and $\X_{ml}$ distributed according to the prior pdfs, according to the pdfs corresponding to the SPA messages \eqref{zTox}-\eqref{AToz}, and according to the pdfs corresponding to the SPA posterior approximations \eqref{Dx}-\eqref{Da}.
These assumptions lead straightforwardly to the scalings of 
$\hat{z}_{ml}(t)$, $\nu^z_{ml}(t)$, $\hat{x}_{m,nl}(t)$, $\nu^x_{m,nl}(t)$, $\hat{x}_{nl}(t)$, $\nu^x_{nl}(t)$, $\hat{a}_{l,mn}(t)$, $\nu^a_{l,mn}(t)$, $\hat{a}_{mn}(t)$, and $\nu^a_{mn}(t)$ specified in \tabref{termOrders}.
Furthermore, because $\fxnvar{m}{nl}{\X}(t,\cdot)$ and $\Delta_{nl}^{\X}(t,\cdot)$ differ by only one term out of $M$, it is reasonable to assume \cite{Montanari:Chap:12,Rangan:ISIT:11} that the corresponding 
difference in means $\hat{x}_{m,nl}(t) - \hat{x}_{nl}(t)$ and variances $\nu^x_{m,nl}(t) - \nu^x_{nl}(t)$ are both $\Ord(1/\sqrt{N})$, which then implies that $\hat{x}^2_{m,nl}(t) - \hat{x}^2_{nl}(t)$ is also $\Ord(1/\sqrt{N})$. 
Similarly, because $\fxnvar{l}{mn}{\A}(t,\cdot)$ and $\Delta_{mn}^{\A}(t,\cdot)$ differ by only one term out of $N$, where $\hat{a}_{l,mn}(t)$ and $\hat{a}_{mn}(t)$ are $\Ord(1/\sqrt{N})$, it is reasonable to assume that $\hat{a}_{l,mn}(t) - \hat{a}_{mn}(t)$ is $\Ord(1/N)$ and that both $\nu^a_{l,mn}(t) - \nu^a_{mn}(t)$ and
$\hat{a}^2_{l,mn}(t) - \hat{a}^2_{mn}(t)$ are $\Ord(1/N^{3/2})$.
The remaining entries in \tabref{termOrders} will be explained below.

\putTable{termOrders}{BiG-AMP variable scalings in the large-system limit.}
{\begin{tabular}{|@{\;}c@{\;}|@{\;}c@{~}||@{~}c@{\;}|@{\;}c@{~}||@{~}c@{\;}|@{\;}c@{\;}||} \hline
$\hat{z}_{ml}(t)$ & $\Ord(1)$ 
	&  $\nu^z_{ml}(t)$ & $\Ord(1)$
	& $\hat{x}_{m,nl}(t) - \hat{x}_{nl}(t)$
	& $\Ord(\frac{1}{\sqrt{N}})$
	\\[0.5mm]
$\hat{x}_{m,nl}(t)$ & $\Ord(1)$ 
	&  $\nu^x_{m,nl}(t)$& $\Ord(1)$
        & $\hat{x}^2_{m,nl}(t) - \hat{x}^2_{nl}(t)$
	& $\Ord(\frac{1}{\sqrt{N}})$
	\\
$\hat{x}_{nl}(t)$ & $\Ord(1)$ 
	&  $\nu^x_{nl}(t)$& $\Ord(1)$
	& $\nu^x_{m,nl}(t) - \nu^x_{nl}(t)$
	& $\Ord(\frac{1}{\sqrt{N}})$
	\\
$\hat{a}_{l,mn}(t)$ & $\Ord(\frac{1}{\sqrt{N}})$ 
	&  $\nu^a_{l,mn}(t)$ & $\Ord(\frac{1}{N})$
	& $\hat{a}_{l,mn}(t) - \hat{a}_{mn}(t)$
	& $\Ord(\frac{1}{N})$
	\\
$\hat{a}_{mn}(t)$ & $\Ord(\frac{1}{\sqrt{N}})$ 
	&  $\nu^a_{mn}(t)$ & $\Ord(\frac{1}{N})$
	& $\hat{a}^2_{l,mn}(t) - \hat{a}^2_{mn}(t)$
	& $\Ord(\frac{1}{N^{3/2}})$
	\\
$\hat{p}_{ml}(t)$ & $\Ord(1)$ 
	&  $\nu^p_{ml}(t)$ & $\Ord(1)$
	& $\nu^a_{l,mn}(t) - \nu^a_{mn}(t)$
	& $\Ord(\frac{1}{N^{3/2}})$
	\\
$\hat{r}_{m,nl}(t)$ & $\Ord(1)$ 
	&  $\nu^r_{m,nl}(t)$& $\Ord(1)$
	& $\hat{r}_{m,nl}(t) - \hat{r}_{nl}(t)$
	& $\Ord(\frac{1}{\sqrt{N}})$ 
	\\
$\hat{r}_{nl}(t)$ & $\Ord(1)$ 
	&  $\nu^r_{nl}(t)$& $\Ord(1)$
	& $\nu^r_{m,nl}(t) - \nu^r_{nl}(t)$
	& $\Ord(\frac{1}{N})$
	\\
$\hat{q}_{l,mn}(t)$ & $\Ord(\frac{1}{\sqrt{N}})$ 
	&  $\nu^q_{l,mn}(t)$ & $\Ord(\frac{1}{N})$
	& $\hat{q}_{l,mn}(t) - \hat{q}_{mn}(t)$
	& $\Ord(\frac{1}{N})$
	\\
$\hat{q}_{mn}(t)$ & $\Ord(\frac{1}{\sqrt{N}})$ 
	&  $\nu^q_{mn}(t)$ & $\Ord(\frac{1}{N})$
	& $\nu^q_{l,mn}(t) - \nu^q_{mn}(t)$
	& $\Ord(\frac{1}{N^2})$
	\\
$\hat{s}_{ml}(t)$ & $\Ord(1)$ 
	&  $\nu^s_{ml}(t)$ & $\Ord(1)$
	&&
	\\[0.5mm]
\hline
 \end{tabular}
}

We start by approximating the message $\fxnvar{m}{nl}{\X}(t,.)$.
Expanding \eqref{zTox}, we find 
\begin{align}
\lefteqn{ \fxnvar{m}{nl}{\X}(t,x_{nl}) }\nonumber\\[-6mm]
&=\log \int_{\{a_{mk}\}_{k=1}^N ,\{x_{rl}\}_{r \ne n}} \hspace{-6mm} p_{\Y_{ml}|\Z_{ml}}\Big(y_{ml}\Biggiv\overbrace{ a_{mn}x_{nl} +\!\!\!\sum_{k=1\neq n}^N \!\!\! a_{mk} x_{kl}}^{\displaystyle z_{ml}}\Big)
\non\\[-1mm]&\quad\times 
\prod_{r \ne n} \exp\Big(\varfxn{m}{rl}{\X}(t,x_{rl})\Big) \prod_{k=1}^N \exp\Big(\varfxn{l}{mk}{\A}(t,a_{mk})\Big)
\non\\&\quad
+ \const . 
\label{eq:mlTonl}
\end{align}
For large $N$, the CLT motivates the treatment of $\Z_{ml}$, the random variable associated with the $z_{ml}$ identified in \eqref{mlTonl}, conditioned on $\X_{nl}=x_{nl}$, as Gaussian and thus completely characterized by a (conditional) mean and variance. 
Defining the zero-mean r.v.s $\tilde{\A}_{l,mn} \defn \A_{mn} - \hat{a}_{l,mn}(t)$ and $\tilde{\X}_{m,nl} = \X_{nl} - \hat{x}_{m,nl}(t)$, where $\A_{mn}\sim\frac{1}{C}\exp(\varfxn{l}{mn}{\A}(t,\cdot))$ and $\X_{nl}\sim\frac{1}{C}\exp(\varfxn{m}{nl}{\X}(t,\cdot))$, we can write
\begin{eqnarray}
\Z_{ml} 
&=& \big(\hat{a}_{l,mn}(t) + \tilde{\A}_{l,mn}\big)\X_{nl} +
\sum_{k \ne n} \big( \hat{a}_{l,mk}(t)\,\hat{x}_{m,kl}(t)
\nonumber\\[-1mm]&&\mbox{}
+ \hat{a}_{l,mk}(t)\tilde{\X}_{m,kl} 
+ \tilde{\A}_{l,mk}\hat{x}_{m,kl}(t) + \tilde{\A}_{l,mk}\tilde{\X}_{m,kl} \big) 
\qquad
\label{eq:zgivx}
\end{eqnarray} 
after which it is straightforward to see that
\begin{align}
\E\{\Z_{ml}\giv \X_{nl}=x_{nl}\} 
&= \hat{a}_{l,mn}(t)x_{nl} + \hat{p}_{n,ml}(t) \label{eq:Ez}\\
\var\{\Z_{ml} \giv \X_{nl}=x_{nl}\} 
&= \nu^a_{l,mn}(t) x_{nl}^2 + \nu^p_{n,ml}(t) \label{eq:Varz}
\end{align}
for
\begin{align}
\hat{p}_{n,ml}(t) 
&\defn \sum_{k \ne n} \hat{a}_{l,mk}(t) \hat{x}_{m,kl}(t) \\
\nu^p_{n,ml}(t)
&\defn \sum_{k \ne n} \big( \hat{a}_{l,mk}^2(t)\nu^x_{m,kl}(t) + \nu^a_{l,mk}(t) \hat{x}_{m,kl}^2(t) 
\nonumber\\&\quad
+ \nu^a_{l,mk}(t)\nu^x_{m,kl}(t) \big) .
\end{align}
With this conditional-Gaussian approximation, \eqref{mlTonl} becomes
\begin{align}
&\fxnvar{m}{nl}{\X}(t,x_{nl}) \approx \const + \log \int_{z_{ml}}p_{\Y_{ml}|\Z_{ml}}(y_{ml}\giv z_{ml})
\\&\quad\times 
\Nor\big(z_{ml}; \hat{a}_{l,mn}(t)x_{nl} \!+\! \hat{p}_{n,ml}(t),\nu^a_{l,mn}(t) x_{nl}^2 \!+\! \nu^p_{n,ml}(t) \big)\nonumber\\
&= H_{ml}\Big(\hat{a}_{l,mn}(t)x_{nl} + \hat{p}_{n,ml}(t),
\nonumber\\&\qquad
\nu^a_{l,mn}(t) x_{nl}^2 + \nu^p_{n,ml}(t); y_{ml}\Big) 
+ \const 
\label{eq:mTOnXl}
\end{align}
in terms of the function
\begin{align}
H_{ml}\big(\textb{\hat{q}},\nu^q; y\big) &\defn \log \int_z p_{\Y_{ml}|\Z_{ml}}(y \giv z) \,\Nor(z;\textb{\hat{q}},\nu^q). \label{eq:H}
\end{align}

Unlike the original SPA message \eqref{zTox}, the approximation \eqref{mTOnXl} requires only a single integration. 
Still, additional simplifications are possible.
First, notice that $\hat{p}_{n,ml}(t)$ and $\nu^p_{n,ml}(t)$ differ from the corresponding $n$-invariant quantities
\begin{align}
\hat{p}_{ml}(t) 
&\defn \sum_{k=1}^N \hat{a}_{l,mk}(t) \hat{x}_{m,kl}(t)\label{eq:pml}\\
\nu^p_{ml}(t) 
&\defn \sum_{k=1}^N \big( \hat{a}_{l,mk}^2(t)\nu^x_{m,kl}(t) + \nu^a_{l,mk}(t) \hat{x}^2_{m,kl}(t) 
\nonumber\\&\quad
+ \nu^a_{l,mk}(t)\nu^x_{m,kl}(t) \big)
\label{eq:nupml}
\end{align}
by one term.
In the sequel, we will assume that $\hat{p}_{ml}(t)$ and $\nu^p_{ml}(t)$ are $\Ord(1)$ since these quantities can be recognized as the mean and variance, respectively, of an estimate of $\Z_{ml}$, which is $\Ord(1)$.
Writing the $H_{ml}$ term in \eqref{mTOnXl} using \eqref{pml}-\eqref{nupml},
\begin{align}
\lefteqn{H_{ml}\Big(\hat{a}_{l,mn}(t)x_{nl} + \hat{p}_{n,ml}(t),\nu^a_{l,mn}(t) x_{nl}^2 + \nu^p_{n,ml}(t); y_{ml}\Big) }\non\\
&= H_{ml}\Big(\hat{a}_{l,mn}(t)\big(x_{nl} - \hat{x}_{m,nl}(t)\big) + \hat{p}_{ml}(t),\nonumber\\
&\qquad \nu^a_{l,mn}(t) \big(x_{nl}^2 - \hat{x}^2_{m,nl}(t)\big) - \hat{a}_{l,mn}^2(t)\nu^x_{m,nl}(t)
\nonumber\\& \qquad -
\nu^a_{l,mn}(t)\nu^x_{m,nl}(t) + \nu^p_{ml}(t); y_{ml}\Big)\\
&= H_{ml}\Big(\hat{a}_{l,mn}(t)\big(x_{nl} - \hat{x}_{nl}(t)\big) + \hat{p}_{ml}(t) + \Ord(1/N),\nonumber\\
&\qquad \nu^a_{l,mn}(t) \big(x_{nl}^2 - \hat{x}^2_{nl}(t)\big) + \nu^p_{ml}(t) + \Ord(1/N); y_{ml}\Big) \label{eq:dropDest_step}
\end{align}
where in \eqref{dropDest_step} we used the facts that $\hat{a}_{l,mn}(t)(\hat{x}_{nl}(t) - \hat{x}_{m,nl}(t))$ and $\nu^a_{l,mn}(t)(\hat{x}_{m,nl}^2(t) - \hat{x}_{nl}^2(t)))-\hat{a}_{l,mn}^2(t)\nu^x_{m,nl}(t)
- \nu^a_{l,mn}(t)\nu^x_{m,nl}(t)$ are both $\Ord(1/N)$.

Rewriting \eqref{mTOnXl} using a Taylor series expansion in $x_{nl}$ about the point $\hat{x}_{nl}(t)$, we get
\begin{align}
\lefteqn{ \fxnvar{m}{nl}{\X}(t,x_{nl}) 
\approx \const }
\nonumber\\&\quad
+ H_{ml}\big( \hat{p}_{ml}(t) + \Ord(1/N),  \nu^p_{ml}(t) + \Ord(1/N); y_{ml} \big)\non
\\&\quad+ 
\hat{a}_{l,mn}(t)\big(x_{nl} - \hat{x}_{nl}(t)\big) 
\non\\&\quad\quad\times
 H'_{ml}\big( \hat{p}_{ml}(t) + \Ord(1/N),  \nu^p_{ml}(t) + \Ord(1/N); y_{ml} \big)
\non\\&\quad+ 
 2\nu^a_{l,mn}(t)\hat{x}_{nl}(t)\big(x_{nl} - \hat{x}_{nl}(t)\big)
\non\\&\quad\quad\times
 \dot{H}_{ml}\big( \hat{p}_{ml}(t) + \Ord(1/N),  \nu^p_{ml}(t) + \Ord(1/N); y_{ml} \big)
 \non\\&\quad\textb{+
 \nu^a_{l,mn}(t)\big(x_{nl} - \hat{x}_{nl}(t)\big)^2}
\non\\&\quad\quad\textb{\times
  \dot{H}_{ml}\big( \hat{p}_{ml}(t) + \Ord(1/N),  \nu^p_{ml}(t) + \Ord(1/N); y_{ml} \big)}
\non\\&\quad+ 
 \frac{1}{2} \hat{a}^2_{l,mn}(t)\big(x_{nl} - \hat{x}_{nl}(t)\big)^2
\non\\&\quad\quad\times
  H''_{ml}\big( \hat{p}_{ml}(t) + \Ord(1/N),  \nu^p_{ml}(t) + \Ord(1/N); y_{ml} \big)
  \non\\&\quad + \textb{\Ord(1/N^{3/2})},
\label{eq:dropDestTay} 
\end{align}
where $H_{ml}'$ and $H_{mn}''$ are the first two derivatives of $H_{mn}$ w.r.t its first argument and $\dot{H}_{ml}$ is the first derivative w.r.t its second argument. 
\textb{Note that, in \eqref{dropDestTay} and elsewhere, the higher-order terms in the Taylor's expansion are written solely in terms of their scaling dependence on $N$, which is what will eventually allow us to neglect these terms (in the large-system limit).} 

We now approximate \eqref{dropDestTay} by dropping terms that vanish, relative to the second-to-last term in \eqref{dropDestTay}, as $N\rightarrow\infty$. 
Since this second-to-last term is $\Ord(1/N)$ due to the scalings of $\hat{a}^2_{l,mn}(t)$, $\hat{p}_{ml}(t)$, and $\nu^p_{ml}(t)$, we drop terms that are of order $\Ord(1/N^{3/2})$, such as the final term.
We also replace $\nu^a_{l,mn}(t)$ with $\nu^a_{mn}(t)$, and $\hat{a}_{l,mn}^2(t)$ with \textb{$\hat{a}_{mn}^2(t)$}, since in both cases the difference is $\Ord(1/N^{3/2})$. 
\textb{Finally, we drop the $\Ord(1/N)$ terms inside the $H_{ml}$ derivatives, which can be justified by taking a Taylor series expansion of these derivatives with respect to the $\Ord(1/N)$ perturbations and verifying that the higher-order terms in this latter expansion are $\Ord(1/N^{3/2})$.}
All of these approximations are analogous to those made in previous AMP derivations, e.g., \textb{\cite{Donoho:ITW:10a}},  \cite{Montanari:Chap:12}, and \cite{Rangan:ISIT:11}.

Applying these approximations to \eqref{dropDestTay} and absorbing $x_{nl}$-invariant terms into the $\const$ term, we obtain
\begin{align}
{\fxnvar{m}{nl}{\X}(t,x_{nl})} 
&\approx \big[ \hat{s}_{ml}(t)\hat{a}_{l,mn}(t) + \nu^s_{ml}(t)\hat{a}^2_{mn}(t)\hat{x}_{nl}(t) \big]
\nonumber\\&\quad
\times  x_{nl} - \tfrac{1}{2}\big[\nu^s_{ml}(t)\hat{a}_{mn}^2(t) \textb{- \nu^a_{mn}(t)}
\nonumber\\&\quad
\textb{\times ( \hat{s}^2_{ml}(t) - \nu^s_{ml}(t)   )} \big]  x_{nl}^2 + \const,
\label{eq:XmTon}
\end{align}
where we used the relationship 
\begin{eqnarray}
\dot{H}_{ml}\big(\textb{\hat{q}},\nu^q; y\big)
&=& \frac{1}{2}\left[\left(H'_{ml}\big(\textb{\hat{q}},\nu^q; y\big)\right)^2 +  H''_{ml}\big(\textb{\hat{q}},\nu^q; y\big) \right] \quad
\end{eqnarray}
and defined
\begin{align}
\hat{s}_{ml}(t) 
&\defn H'_{ml}\big( \hat{p}_{ml}(t),  \nu^p_{ml}(t); y_{ml} \big)\label{eq:shat}\\
\nu^s_{ml}(t) 
&\defn -H''_{ml}\big( \hat{p}_{ml}(t),  \nu^p_{ml}(t); y_{ml} \big).\label{eq:nushat}
\end{align}
Note that \eqref{XmTon} is essentially a Gaussian approximation to the pdf $\frac{1}{C}\exp(\fxnvar{m}{nl}{\X}(t,.))$. 

\textb{We show in \appref{Hderivatives} that}
\begin{align}
\hat{s}_{ml}(t) 
&= \frac{1}{\nu^p_{ml}(t)} \big( \hat{z}_{ml}(t) - \hat{p}_{ml}(t)  \big) 
\label{eq:s} \\
\nu^s_{ml}(t) 
&= \textb{\frac{1}{\nu^p_{ml}(t)} \left( 1 - \frac{\nu^z_{ml}(t)}{\nu^p_{ml}(t)}\right)} ,
\label{eq:nus} 
\end{align}
for the conditional mean and variance
\begin{align}
\hat{z}_{ml}(t) &\defn \E\{\Z_{ml}\giv \p_{ml}\!=\!\hat{p}_{ml}(t);\nu^p_{ml}(t)\} \label{eq:zhat}\\
\nu^z_{ml}(t) &\defn \var\{\Z_{ml}\giv \p_{ml}\!=\!\hat{p}_{ml}(t);\nu^p_{ml}(t)\}, \label{eq:zvar}
\end{align}
computed according to the (conditional) pdf 
\begin{align}
\lefteqn{ p_{\Z_{ml}|\p_{ml}}\big(z_{ml}\giv\hat{p}_{ml}(t);\nu^p_{ml}(t)\big) }\non\\
&\defn 
\textstyle \frac{1}{C} \, p_{\Y_{ml}|\Z_{ml}}(y_{ml} \giv z_{ml}) \, \Nor\big(z_{ml};\hat{p}_{ml}(t),\nu^p_{ml}(t)\big) ,
\label{eq:pZgivYP}
\end{align}
where here $C=\int_{z} p_{\Y_{ml}|\Z_{ml}}(y_{ml} \giv z) \Nor\big(z;\hat{p}_{ml}(t),\nu^p_{ml}(t)\big)$.
In fact,
\eqref{pZgivYP} is BiG-AMP's iteration-$t$ approximation to the true marginal posterior $p_{\Z_{ml}|\vY}(\cdot|\vec{Y})$. 
We note that \eqref{pZgivYP} can also be interpreted as the (exact) posterior pdf for $\Z_{ml}$ given the likelihood $p_{\Y_{ml}|\Z_{ml}}(y_{ml}|\cdot)$ from \eqref{pYgivZ} and the prior $\Z_{ml}\sim \Nor\big(\hat{p}_{ml}(t),\nu^p_{ml}(t)\big)$ that is implicitly assumed by iteration-$t$ BiG-AMP.

Since $\vZ\tran = \vX\tran\vA\tran$, the derivation of the BiG-AMP approximation of $\fxnvar{l}{mn}{\A}(t,.)$ closely follows the derivation for $\fxnvar{m}{nl}{\X}(t,.)$. 
In particular, it starts with (similar to \eqref{mlTonl})
\begin{align}
\lefteqn{ \fxnvar{l}{mn}{\A}(t,a_{mn}) }\nonumber\\[-4mm]
&=\log \int_{\{a_{mk}\}_{k\ne n} ,\{x_{rl}\}_{r =1}^N} \hspace{-5mm}p_{\Y_{ml}|\Z_{ml}}\Big(y_{ml}\Biggiv\overbrace{ a_{mn}x_{nl} +\sum_{k\ne n} a_{mk} x_{kl}}^{\displaystyle z_{ml}}\Big)
\non\\[-1mm]&\quad\times 
\prod_{r =1}^N \exp\Big(\varfxn{m}{rl}{\X}(t,x_{rl})\Big) \prod_{k\ne n} \exp\Big(\varfxn{l}{mk}{\A}(t,a_{mk})\Big)
\nonumber\\&\quad
+ \const, \label{eq:mlTomn}
\end{align}
where again the CLT motivates the treatment of $\Z_{ml}$, conditioned on $\A_{mn}=a_{mn}$, as Gaussian. 
Eventually we arrive at the Taylor-series approximation (similar to \eqref{XmTon})
\begin{align}
\fxnvar{l}{mn}{\A}(t,a_{mn}) 
&\approx \Big[ \hat{s}_{ml}(t)\hat{x}_{m,nl}(t) + \nu^s_{ml}(t)\hat{x}^2_{nl}(t)\hat{a}_{mn}(t)\Big] 
\non\\&\quad
\times  a_{mn} 
- \tfrac{1}{2}\big[ 
 \nu^s_{ml}(t)\hat{x}_{nl}^2(t) \textb{- \nu^x_{nl}(t)}
\non\\&\quad
\textb{\times ( \hat{s}^2_{ml}(t) - \nu^s_{ml}(t)   )} \big]  a_{mn}^2
\non\\&\quad
+ \const.
\label{eq:AlTon}
\end{align}

\subsection{Approximated Variable-to-Factor Messages}	\label{sec:vTof}
We now turn to approximating the messages flowing from the variable nodes to the factor nodes. 
Starting with \eqref{xToz} and plugging in \eqref{XmTon} we obtain
\begin{eqnarray}
\lefteqn{ 
\varfxn{m}{nl}{\X}(t\!+\!1,x_{nl}) 
}\nonumber\\ 
&\approx& \const + \log p_{\X_{nl}}(x_{nl}) 
+\sum_{k \ne m} \Bigg( \Big[ \hat{s}_{kl}(t)\hat{a}_{l,kn}(t) 
\nonumber\\&&
+ \nu^s_{kl}(t)\hat{a}^2_{kn}(t)\hat{x}_{nl}(t) \Big] x_{nl} 
-\tfrac{1}{2} \Big[\nu^s_{kl}(t)\hat{a}_{kn}^2(t)
\nonumber\\&&
\textb{- \nu^a_{kn}(t) \big(\hat{s}^2_{kl}(t) - \nu^s_{kl}(t)  \big)} \Big]  x_{nl}^2 \Bigg)\\
&=& \const + \log p_{\X_{nl}}(x_{nl}) - \frac{1}{2\nu^r_{m,nl}(t)} \big( x_{nl} - \hat{r}_{m,nl}(t) \big)^2 \quad \\
&=& \const + \log\Big( p_{\X_{nl}}(x_{nl})\Nor\big(x_{nl};\hat{r}_{m,nl}(t),\nu^r_{m,nl}(t)\big) \Big)
\label{eq:xTozFinal}
\end{eqnarray}
for
\begin{align}
\nu^r_{m,nl}(t) &\defn \bigg(\sum_{k \ne m} \hat{a}^2_{kn}(t)\nu^s_{kl}(t) 
	\textb{- \nu^a_{kn}(t) \big(\hat{s}^2_{kl}(t) - \nu^s_{kl}(t)  \big)} \bigg)^{-1}\label{eq:nurM}\\
\hat{r}_{m,nl}(t) &\defn \hat{x}_{nl}(t) \Big(1 + \nu^r_{m,nl}(t) \sum_{k \ne m}    \nu^a_{kn}(t) [\hat{s}^2_{kl}(t) - \nu^s_{kl}(t) ]    \Big)
\non\\&\quad
+ \nu^r_{m,nl}(t) \sum_{k \ne m} \hat{a}_{l,kn}(t) \hat{s}_{kl}(t). \label{eq:rhatM}
\end{align}
Since $\hat{a}^2_{mn}(t)$ \textb{and $\nu^a_{mn}(t)$ are} $\Ord(1/N)$, \textb{and recalling $\hat{s}^2_{ml}(t)$ and} $\nu^s_{ml}(t)$ \textb{are} $\Ord(1)$, we take $\nu^r_{m,nl}(t)$ to be $\Ord(1)$. 
Meanwhile, since $\hat{r}_{m,nl}(t)$ is an estimate of $\X_{nl}$, we reason that it is $\Ord(1)$.

The mean and variance of the pdf associated with the $\varfxn{m}{nl}{\X}(t\!+\!1,.)$ approximation in \eqref{xTozFinal} are
\begin{align}
\lefteqn{ 
\hat{x}_{m,nl}(t\!+\!1)
}\nonumber\\ 
&\defn \underbrace{\frac{1}{C} \int_x x\, p_{\X_{nl}}(x) \Nor\big(x;\hat{r}_{m,nl}(t), \nu^r_{m,nl}(t)\big)}_{\displaystyle \defn g_{\X_{nl}}(\hat{r}_{m,nl}(t),\nu^r_{m,nl}(t)) } \label{eq:xmnl}\\
\lefteqn{ 
\nu^x_{m,nl}(t\!+\!1) 
}\nonumber\\
&\defn \underbrace{\frac{1}{C} \int_x \big|x - \hat{x}_{m,nl}(t\!+\!1)\big|^2
p_{\X_{nl}}(x) \Nor\big(x;\hat{r}_{m,nl}(t), \nu^r_{m,nl}(t)\big) }_{ \displaystyle \nu^r_{m,nl}(t) \, g'_{\X_{nl}}(\hat{r}_{m,nl}(t),\nu^r_{m,nl}(t))}
\nonumber\\[-5mm]&\label{eq:nuxmnl}
\end{align}
where here $C = \int_x p_{\X_{nl}}(x) \Nor\big(x;\hat{r}_{m,nl}(t), \nu^r_{m,nl}(t)\big)$ and $g'_{\X_nl}$ denotes the derivative of $g_{\X_{nl}}$ with respect to the first argument. 
The fact that \eqref{xmnl} and \eqref{nuxmnl} are related through a derivative was shown in \cite{Rangan:ISIT:11}.

We now derive approximations of $\hat{x}_{m,nl}(t)$ and $\nu^x_{m,nl}(t)$ that avoid the dependence on the destination node $m$.
For this, we introduce $m$-invariant versions of $\hat{r}_{m,nl}(t)$ and $\nu^r_{m,nl}(t)$:
\begin{align}
\nu^r_{nl}(t) 
&\defn \bigg(\sum_{m=1}^M \hat{a}^2_{mn}(t)\nu^s_{ml}(t) 
	\textb{- \nu^a_{mn}(t) \big(\hat{s}^2_{ml}(t) - \nu^s_{ml}(t)}  \big)\bigg)^{-1}
\label{eq:nur}\\
\hat{r}_{nl}(t) 
&\defn \hat{x}_{nl}(t) \Big(1 + \nu^r_{nl}(t) \sum^M_{m=1}    \nu^a_{mn}(t) [\hat{s}^2_{ml}(t) - \nu^s_{ml}(t) ]    \Big)
\non\\[-1mm]&\quad
+ \nu^r_{nl}(t) \sum_{m = 1}^M \hat{a}_{l,mn}(t) \hat{s}_{ml}(t) .
\label{eq:rhat}
\end{align}
Comparing \eqref{nur}-\eqref{rhat} with \eqref{nurM}-\eqref{rhatM} and applying previously established scalings \textb{from} \tabref{termOrders} reveals that $\nu^r_{m,nl}(t) - \nu^r_{nl}(t)$ is $\Ord(1/N)$ and that $\hat{r}_{m,nl}(t) = \hat{r}_{nl}(t) \textb{-} \nu^r_{nl}(t)\hat{a}_{mn}(t)\hat{s}_{ml}(t) + \Ord(1/N)$, \textb{so that} \eqref{xmnl} implies
\begin{eqnarray}
\lefteqn{ 
\hat{x}_{m,nl}(t\!+\!1) 
}\non\\
&=& g_{\X_{nl}}\big(\hat{r}_{nl}(t) - \nu^r_{nl}(t)\hat{a}_{mn}(t) \hat{s}_{ml}(t) + \Ord(1/N),
\non\\&&\qquad \nu^r_{nl}(t) + \Ord(1/N) \big) \label{eq:hatx_approx1}\\
&=& g_{\X_{nl}}\big(\hat{r}_{nl}(t) - \nu^r_{nl}(t)\hat{a}_{mn}(t) \hat{s}_{ml}(t), \nu^r_{nl}(t)\big) + \Ord(1/N) \qquad \label{eq:hatx_approx2}\\
&=& g_{\X_{nl}}\big(\hat{r}_{nl}(t),\nu^r_{nl}(t)\big) 
\label{eq:taylorx}\\&&\quad
-\nu^r_{nl}(t)\hat{a}_{mn}(t) \hat{s}_{ml}(t) \, g'_{\X_{nl}}\big(\hat{r}_{nl}(t),\nu^r_{nl}(t)\big) + \Ord(1/N)\non\\
&\approx& \hat{x}_{nl}(t\!+\!1) - \hat{a}_{mn}(t) \hat{s}_{ml}(t) \nu^x_{nl}(t\!+\!1). \label{eq:xmnlforp}
\end{eqnarray}
Above, \eqref{hatx_approx2} follows from taking Taylor series expansions around each of the $\Ord(1/N)$ perturbations in \eqref{hatx_approx1}; 
\eqref{taylorx} \textb{follows} from a Taylor series expansion in the first argument of \eqref{hatx_approx2} about the point $\hat{r}_{nl}(t)$; and 
\eqref{xmnlforp} follows by neglecting the $\Ord(1/N)$ term (which vanishes relative to the others in the large-system limit) and applying the definitions 
\begin{align}
\hat{x}_{nl}(t\!+\!1) 
&\defn g_{\X_{nl}}\big(\hat{r}_{nl}(t),\nu^r_{nl}(t)\big)\label{eq:xhatDefined}\\
\nu^x_{nl}(t\!+\!1) 
&\defn \nu^r_{nl}(t)g'_{\X_{nl}}\big(\hat{r}_{nl}(t),\nu^r_{nl}(t)\big), \label{eq:nuxDefined}
\end{align}
which match \eqref{xmnl}-\eqref{nuxmnl} sans the $m$ dependence. 
Note that \eqref{xmnlforp} confirms that the difference $\hat{x}_{m,nl}(t)-\hat{x}_{nl}(t)$ is $\Ord(1/\sqrt{N})$, as was assumed at the start of the BiG-AMP derivation.
Likewise, taking Taylor series expansions of $g'_{\X_{nl}}$ in \eqref{nuxmnl} about the point $\hat{r}_{nl}(t)$ in the first argument and about the point $\nu^r_{nl}(t)$ in the second argument and then comparing the result with \eqref{nuxDefined} confirms that $\nu^x_{m,nl}(t) - \nu^x_{nl}(t)$ is $\Ord(1/\sqrt{N})$. 

We then repeat the above procedure to derive an approximation to $\varfxn{l}{mn}{\A}(t\!+\!1,.)$ analogous to \eqref{xTozFinal}, whose corresponding mean is then further approximated as
\begin{align}
\hat{a}_{l,mn}(t\!+\!1) 
&\approx \hat{a}_{mn}(t\!+\!1) - \hat{x}_{nl}(t) \hat{s}_{ml}(t) \nu^a_{mn}(t\!+\!1), \label{eq:almnforp}
\end{align}
for
\begin{align}
\hat{a}_{mn}(t\!+\!1) 
&\defn g_{\A_{mn}}\big(\hat{q}_{mn}(t) ,\nu^q_{mn}(t) \big)\label{eq:ahatDefined}\\
\nu^a_{mn}(t\!+\!1) 
&\defn \nu^q_{mn}(t) g'_{\A_{mn}}\big(\hat{q}_{mn}(t) ,\nu^q_{mn}(t) \big)\label{eq:nuaDefined} \\
g_{\A_{mn}}(\hat{q},\nu^q) 
&\defn \frac{ \int_a a\, p_{\A_{mn}}(a) \Nor(a;\hat{q},\nu^q) }
	{ \int_a p_{\A_{mn}}(a) \Nor(a;\hat{q},\nu^q) } 
\end{align}
where
\begin{align}
\nu^q_{mn}(t) 
&\defn \bigg(\sum_{l = 1}^L \hat{x}_{nl}^2(t) \nu^s_{ml}(t) 
	\textb{- \nu^x_{nl}(t) \big( \hat{s}^2_{ml}(t) - \nu^s_{ml}(t)  \big)} \bigg)^{-1} \label{eq:nuq}  \\
\hat{q}_{mn}(t) 
&\defn \hat{a}_{mn}(t) \Big(1 + \nu^q_{mn}(t) \sum_{l=1}^L \nu^x_{nl}(t) [ \hat{s}^2_{ml}(t) - \nu^s_{ml}(t)   ]         \Big)
\non\\&\quad
 + \nu^q_{mn}(t) \sum_{l = 1}^L \hat{x}_{m,nl}(t) \hat{s}_{ml}(t).\label{eq:what}
\end{align}
Arguments analogous to the discussion following \eqref{rhatM} justify the remaining scalings in \tabref{termOrders}.

\subsection{Closing the Loop} 		\label{sec:loop}
The penultimate step in the derivation of BiG-AMP is to approximate earlier steps that use $\hat{a}_{l,mn}(t)$ and $\hat{x}_{m,nl}(t)$ in place of $\hat{a}_{mn}(t)$ and $\hat{x}_{nl}(t)$.
For this, we start by plugging \eqref{xmnlforp} and \eqref{almnforp} into \eqref{pml}, which yields\footnote{Recall that the error of the approximation in \eqref{xmnlforp} is $\Ord(1/N)$ and the error in \eqref{almnforp} is $\Ord(1/N^{3/2})$.}
\begin{align}
\lefteqn{
\hat{p}_{ml}(t) \, = \, \Ord(1/\sqrt{N}) + 
\overbrace{\sum_{n=1}^N \hat{a}_{mn}(t) \hat{x}_{nl}(t) }^{\displaystyle \defn \bar{p}_{ml}(t)}  - \hat{s}_{ml}(t\!-\!1)
}\nonumber
\\&\quad\times
\sum_{n=1}^N \Big(\nu^a_{mn}(t) \hat{x}_{nl}(t)\hat{x}_{nl}(t\!-\!1) + \hat{a}_{mn}(t)\hat{a}_{mn}(t\!-\!1)  \nu^x_{nl}(t)  \Big) 
\non\\&\quad
+\hat{s}^2_{ml}(t\!-\!1) \sum_{n=1}^N \hat{a}_{mn}(t\!-\!1)\nu^a_{mn}(t) \nu^x_{nl}(t) \hat{x}_{nl}(t\!-\!1) \label{eq:preOnsager}\\
&\approx  \bar{p}_{ml}(t)
- \hat{s}_{ml}(t\!-\!1) \underbrace{ \sum_{n=1}^N \Big( \nu^a_{mn}(t) \hat{x}_{nl}^2(t) + \hat{a}_{mn}^2(t)\nu^x_{nl}(t) \Big) }_{
\displaystyle \defn \bar{\nu}^p_{ml}(t)}, \nonumber\\[-5mm]&\label{eq:Onsager}
\end{align}
where, for \eqref{Onsager}, we used $\hat{a}_{mn}^2(t)$ in place of $\hat{a}_{mn}(t)\hat{a}_{mn}(t\!-\!1)$, used $\hat{x}_{nl}^2(t)$ in place of $\hat{x}_{nl}(t)\hat{x}_{nl}(t\!-\!1)$, and neglected terms that are $\Ord(1/\sqrt{N})$, since they vanish relative to the remaining $\Ord(1)$ terms in the large-system limit.

Next we plug \eqref{xmnlforp}, \eqref{almnforp}, $\nu^x_{m,nl}(t) = \nu^x_{nl}(t) + \Ord(1/\sqrt{N})$, and $\nu^a_{l,mn}(t) = \nu^a_{mn}(t) + \Ord(1/N^{3/2})$ into \eqref{nupml}, giving
\begin{align}
\nu^p_{ml}(t)
&= \bar{\nu}^p_{ml}(t) + \sum_{n=1}^N \nu^a_{mn}(t)\nu^x_{nl}(t)
\label{eq:muOnsager1} \\&\quad
-2\hat{s}_{ml}(t\!-\!1) 
\sum_{n=1}^N \Big( \nu^a_{mn}(t)\hat{a}_{mn}(t)   \hat{x}_{nl}(t\!-\!1)\nu^x_{nl}(t) 
\non\\&\qquad
+ \nu^a_{mn}(t)\hat{a}_{mn}(t\!-\!1)   \hat{x}_{nl}(t)\nu^x_{nl}(t) \Big) 
\non\\&\quad
+ \hat{s}_{ml}(t\!-\!1)^2 
\sum_{n=1}^N \Big( (\nu^a_{mn}(t))^2\hat{x}_{nl}^2(t\!-\!1)\nu^x_{nl}(t) 
\non\\&\qquad
+ \nu^a_{mn}(t)\hat{a}_{mn}^2(t\!-\!1)(\nu^x_{nl}(t))^2 \Big) + \Ord(1/\sqrt{N}) \non\\
&\approx \bar{\nu}^p_{ml}(t) + \sum_{n=1}^N \nu^a_{mn}(t)\nu^x_{nl}(t),
\label{eq:muOnsager}
\end{align}
where \eqref{muOnsager} retains only the $\Ord(1)$ terms from \eqref{muOnsager1}.

Similarly, we plug \eqref{almnforp} into \eqref{rhat} and \eqref{xmnlforp} into \eqref{what} to obtain
\begin{align}
\hat{r}_{nl}(t) 
&\approx \hat{x}_{nl}(t) \left(1 - \frac{\sum_{m=1}^M \nu^a_{mn}(t)\nu^s_{ml}(t)}{\sum_{m=1}^M \hat{a}^2_{mn}(t)\nu^s_{ml}(t)}   \right)
\non\\&\quad
+ \nu^r_{nl}(t) \sum_{m = 1}^M \hat{a}_{mn}(t) \hat{s}_{ml}(t)\\
\hat{q}_{mn}(t) 
&\approx \hat{a}_{mn}(t) \left(1 - \frac{\sum_{l=1}^L \nu^x_{nl}(t)\nu^s_{ml}(t)}{\sum_{l=1}^L \hat{x}^2_{nl}(t)\nu^s_{ml}(t)}   \right)
\non\\&\quad
+ \nu^q_{mn}(t) \sum_{l = 1}^L \hat{x}_{nl}(t) \hat{s}_{ml}(t),
\end{align}
where the approximations involve the use of $\hat{s}^2_{ml}(t)$ in place of $\hat{s}_{ml}(t)\hat{s}_{ml}(t-1)$, of $\hat{a}_{mn}(t)$ in place of $\hat{a}_{mn}(t-1)$, of $\hat{x}_{nl}(t)$ in place of $\hat{x}_{nl}(t-1)$, and the dropping of terms that vanish in the large-system limit. 
\textb{Finally, we make the approximations
\begin{align}
\nu^r_{nl}(t) 
&\approx \bigg(\sum_{m=1}^M \hat{a}^2_{mn}(t)\nu^s_{ml}(t) \bigg)^{-1}
\label{eq:nur2}\\
\nu^q_{mn}(t) 
&\approx \bigg(\sum_{l = 1}^L \hat{x}_{nl}^2(t) \nu^s_{ml}(t)  \bigg)^{-1},
\label{eq:nuq2}
\end{align}
by neglecting the $\hat{s}^2_{ml}(t) - \nu^s_{ml}(t)$ terms in \eqref{nur} and \eqref{nuq}, as explained in \appref{s2nus}.}
\subsection{Approximated Posteriors}		\label{sec:post}

The final step in the BiG-AMP derivation is to approximate the SPA posterior log-pdfs in \eqref{Dx} and \eqref{Da}.
Plugging \eqref{XmTon} and \eqref{AlTon} into those expressions, we get
\begin{align}
\lefteqn{ \Delta^{\X}_{nl}(t\!+\!1,x_{nl}) }\non\\
&\approx \const + \log\Big( p_{\X_{nl}}\!(x_{nl})\,\Nor\big(x_{nl};\hat{r}_{nl}(t),\nu^r_{nl}(t)\big) \Big) \label{eq:DxFinal} \\
\lefteqn{ \Delta^{\A}_{mn}(t\!+\!1,a_{mn}) }\non\\
&\approx \const + \log\Big( p_{\A_{mn}}\!(a_{mn})\,\Nor\big(a_{mn};\hat{q}_{mn}(t),\nu^q_{mn}(t)\big) \Big) \label{eq:DaFinal}
\end{align}
using steps similar to \eqref{xTozFinal}.
The associated pdfs are
\begin{align}
\lefteqn{ p_{\X_{nl}|\R_{nl}}\big(x_{nl}\giv\hat{r}_{nl}(t);\nu^r_{nl}(t)\big) }\non\\
&\defn {\textstyle \frac{1}{C_x}}\, p_{\X_{nl}}(x_{nl}) \, \Nor\big(x_{nl};\hat{r}_{nl}(t),\nu^r_{nl}(t)\big)
	\label{eq:pXgivR}\\
\lefteqn{ p_{\A_{mn}|\Q_{mn}}\big(a_{mn}\giv \hat{q}_{mn}(t);\nu^q_{mn}(t)\big) }\non\\
&\defn {\textstyle \frac{1}{C_a}}\, p_{\A_{mn}}(a_{mn}) \, \Nor\big(a_{mn};\hat{q}_{mn}(t),\nu^q_{mn}(t)\big)
	\label{eq:pAgivQ}
\end{align}
for $C_x\defn \int_{x} p_{\X_{nl}}(x) \Nor\big(x;\hat{r}_{nl}(t),\nu^r_{nl}(t)\big)$ and $C_a\defn \int_{a} p_{\A_{mn}}(a) \Nor\big(a;\hat{q}_{mn}(t),\nu^q_{mn}(t)\big)$, which are iteration-$t$ BiG-AMP's approximations to the true marginal posteriors $p_{\X_{nl}|\vY}(x_{nl}\giv\vec{Y})$ and $p_{\A_{mn}|\vY}(a_{mn}\giv\vec{Y})$, respectively. 

Note that $\hat{x}_{nl}(t\!+\!1)$ and $\nu^x_{nl}(t\!+\!1)$ from \eqref{xhatDefined}-\eqref{nuxDefined} are the mean and variance, respectively, of the posterior pdf in \eqref{pXgivR}.
Note also that \eqref{pXgivR} can be interpreted as the (exact) posterior pdf of $\X_{nl}$ given the observation $\R_{nl}=\hat{r}_{nl}(t)$ under the prior model $\X_{nl}\sim p_{\X_{nl}}$ and the likelihood model $p_{\R_{nl}|\X_{nl}}(\hat{r}_{nl}(t)\giv x_{nl}; \nu^r_{nl}(t))=\Nor\big(\hat{r}_{nl}(t);x_{nl},\nu^r_{nl}(t)\big)$ implicitly assumed by iteration-$t$ BiG-AMP.
Analogous statements can be made about the posterior pdf of $\A_{mn}$ in \eqref{pAgivQ}.

This completes the derivation of BiG-AMP.

\subsection{Algorithm Summary}		\label{sec:summary}

The BiG-AMP algorithm derived in Sections~\ref{sec:SPA}~to~\ref{sec:post} is summarized in \tabref{bigAmp}. 
There, we have included a maximum number of iterations, $T_\textrm{max}$ and a stopping condition (R17) based on the (normalized) change in the residual and a user-defined parameter $\tau_\textrm{BiG-AMP}$. 
We have also written the algorithm in a more general form that allows the use of complex-valued quantities [note the complex conjugates in (R10) and (R12)], in which case $\Nor$ in (D1)-(D3) would be circular complex Gaussian.
For ease of interpretation, \tabref{bigAmp} does not include the important damping modifications that will be detailed in \secref{damp}.
Suggestions for the initializations in (I2) will be given in the sequel.

\putTable{bigAmp}{The BiG-AMP Algorithm}{\scriptsize
\begin{equation*}
\begin{array}{|lrcl@{}r|}\hline
  \multicolumn{2}{|l}{\textsf{definitions:}}&&&\\[-1mm]
  &p_{\Z_{ml}|\p_{ml}}(z|\hat{p};\nu^p)
   &\defn& \frac{p_{\Y_{ml}|\Z_{ml}}(y_{ml}|z) \,\mc{N}(z;\hat{p},\nu^p)}
	{\int_{z'} p_{\Y_{ml}|\Z_{ml}}(y_{ml}|z') \,\mc{N}(z';\hat{p},\nu^p)} &\text{(D1)}\\
  &p_{\X_{nl}|\R_{nl}}(x|\hat{r};\nu^r)
   &\defn& \frac{p_{\X_{nl}}\!(x) \,\mc{N}(x;\hat{r},\nu^r)}
        {\int_{x'}p_{\X_{nl}}\!(x') \,\mc{N}(x';\hat{r},\nu^r)}&\text{(D2)}\\
  &p_{\A_{mn}|\Q_{mn}}(a|\hat{q};\nu^q)
    &\defn& \frac{p_{\A_{mn}}\!(a) \,\mc{N}(a;\hat{q},\nu^q)}
         {\int_{a'}p_{\A_{mn}}\!(a') \,\mc{N}(a';\hat{q},\nu^q)}&\text{(D3)}\\
  \multicolumn{2}{|l}{\textsf{initialization:}}&&&\\
  &\forall m,l:
   \hat{s}_{ml}(0) &=& 0 & \text{(I1)}\\
  &\forall m,n,l: \textsf{choose~} &
  \multicolumn{2}{l}{\nu^x_{nl}(1), \hat{x}_{nl}(1), \nu^a_{mn}(1), \hat{a}_{mn}(1)} &\text{(I2)}\\
  \multicolumn{2}{|l}{\textsf{for $t=1,\dots T_\textrm{max}$}}&&&\\
  &\forall m,l:
   \bar{\nu}^p_{ml}(t)
   &=& \textstyle \sum_{n=1}^{N} |\hat{a}_{mn}(t)|^2 \nu^x_{nl}(t) + \nu^a_{mn}(t) |\hat{x}_{nl}(t)|^2 \!\!& \text{(R1)}\\[0.5mm]
  &\forall m,l:
   \bar{p}_{ml}(t)
   &=& \textstyle \sum_{n=1}^{N} \hat{a}_{mn}(t) \hat{x}_{nl}(t) & \text{(R2)}\\
    &\forall m,l:
   \nu^p_{ml}(t)
   &=& \textstyle \bar{\nu}^p_{ml}(t) + \sum_{n=1}^{N} \nu_{mn}^a(t) \nu^x_{nl}(t) & \text{(R3)}\\
  &\forall m,l:
   \hat{p}_{ml}(t) &=&
   \textstyle \bar{p}_{ml}(t) - \hat{s}_{ml}(t\!-\!1)\bar{\nu}^p_{ml}(t)& \text{(R4)}\\[0.5mm]
  &\forall m,l:
   \nu^z_{ml}(t) &=&
   \textstyle \var\{\Z_{ml}\giv\p_{ml}\!=\!\hat{p}_{ml}(t);\nu^p_{ml}(t)\} & \text{(R5)}\\[0.5mm]
  &\forall m,l:
    \hat{z}_{ml}(t) &=&
   \textstyle  \E\{\Z_{ml}\giv\p_{ml}\!=\!\hat{p}_{ml}(t);\nu^p_{ml}(t)\} & \text{(R6)}\\
  &\forall m,l:
   \nu^s_{ml}(t) &=&
   \textstyle {\textb{(1 -  \nu^z_{ml}(t)/\nu^p_{ml}(t))/\nu^p_{ml}(t)}}  & \text{(R7)}\\
  &\forall m,l:
   \hat{s}_{ml}(t) &=&
   	\textstyle ( \hat{z}_{ml}(t) - \hat{p}_{ml}(t))/\nu^p_{ml}(t) & \text{(R8)}\\[1mm]
 &\forall n,l:
   \nu^r_{nl}(t)
   &=& \textstyle \big(\sum_{m=1}^{M} |\hat{a}_{mn}(t)|^2 \nu^s_{ml}(t)
	\big)^{-1} & \text{(R9)}\\
  &\forall n,l:
   \hat{r}_{nl}(t)
   &=& \textstyle \hat{x}_{nl}(t) ( 1 - \nu^r_{nl}(t) \sum_{m=1}^{M} \nu^a_{mn}(t) \nu^s_{ml}(t)   ) &\\
   &&&+ \nu^r_{nl}(t) \sum_{m=1}^{M} \hat{a}_{mn}^*(t)
	\hat{s}_{ml}(t)  & \text{(R10)}\\
  &\forall m,n:
   \nu^q_{mn}(t)
   &=& \textstyle \big(\sum_{l=1}^{L} |\hat{x}_{nl}(t)|^2 \nu^s_{ml}(t)
	\big)^{-1} & \text{(R11)}\\
  &\forall m,n:
   \hat{q}_{mn}(t)
   &=& \textstyle \hat{a}_{mn}(t)(1 - \nu^q_{mn}(t) \sum_{l=1}^{L} \nu^x_{nl}(t) \nu^s_{ml}(t)    ) &\\
   &&&+ \nu^q_{mn}(t) \sum_{l=1}^{L} \hat{x}_{nl}^*(t)
	\hat{s}_{ml}(t)  & \text{(R12)}\\[1mm]
  &\forall n,l:
	\nu^x_{nl}(t\!+\!1) &=&
		 \var\{\X_{nl}\giv\R_{nl}\!=\!\hat{r}_{nl}(t); \nu^r_{nl}(t)\} & \text{(R13)}\\
  &\forall n,l:
    \hat{x}_{nl}(t\!+\!1) &=&
    	 \E\{\X_{nl}\giv \R_{nl}\!=\!\hat{r}_{nl}(t); \nu^r_{nl}(t)\} & \text{(R14)}\\
  &\forall m,n:
	\nu^a_{mn}(t\!+\!1) &=&
		 \var\{\A_{mn}\giv\Q_{mn}\!=\!\hat{q}_{mn}(t); \nu^q_{mn}(t)\}& \text{(R15)}\\
  &\forall m,n:
	\hat{a}_{mn}(t\!+\!1) &=&
		 \E\{\A_{mn}\giv\Q_{mn}\!=\!\hat{q}_{mn}(t); \nu^q_{mn}(t)\} & \text{(R16)}\\[1mm]
   \multicolumn{4}{|c}{\textsf{if $\sum_{m,l} |\bar{p}_{ml}(t) - \bar{p}_{ml}(t\!-\!1)|^2 \le \tau_\textrm{BiG-AMP} \sum_{m,l} |\bar{p}_{ml}(t)|^2$, \textb{\textsf{stop}}}}&\text{(R17)}\\
    \multicolumn{2}{|l}{\textsf{end}}&&&\\\hline
\end{array}
\end{equation*}
}

We note that BiG-AMP avoids the use of SVD or QR decompositions, lending itself to simple and potentially parallel implementations. 
Its complexity order is dominated\footnote{The computations in steps (R4)-(R8) are $\Ord(ML)$, while the remainder of the algorithm is $\Ord(MN + NL)$.  Thus, as $N$ grows, the matrix multiplies dominate the complexity.} by ten matrix multiplications per iteration [in steps (R1)-(R3) and (R9)-(R12)], each requiring $MNL$ multiplications, although simplifications will be discussed in \secref{simple}. 

The steps in \tabref{bigAmp} can be interpreted as follows.
(R1)-(R2) compute a ``plug-in'' estimate $\vec{\bar{P}}$ of the matrix product $\vZ\!=\!\vA\vX$ and a corresponding set of element-wise variances $\{\bar{\nu}^p_{ml}\}$.
(R3)-(R4) then apply ``Onsager'' correction (see \cite{Montanari:Chap:12} and \cite{Rangan:ISIT:11} for discussions in the contexts of AMP and GAMP, respectively) to obtain the corresponding quantities $\vec{\hat{P}}$ and $\{\nu^p_{ml}\}$.
Using these quantities, (R5)-(R6) compute the (approximate) marginal posterior means $\vec{\hat{Z}}$ and variances $\{\nu^z_{ml}\}$ of $\vZ$.
Steps (R7)-(R8) then use these posterior moments to compute the scaled residual $\vec{\hat{S}}$ and a set of inverse-residual-variances $\{\nu^s_{ml}\}$.
This interpretation becomes clear in the case of AWGN observations with noise variance $\nu^w$, 
where
\begin{align}
p_{\Y_{ml}|\Z_{ml}}(y_{ml}\giv z_{ml}) 
&= \Nor(y_{ml};z_{ml},\nu^w).	\label{eq:AWGN}
\end{align}
and hence
\begin{align}
   \nu^s_{ml} 
   = \frac{1}{\nu^p_{ml}+\nu^w}              \label{eq:nushat_AWGN}
   \text{~~and~~} \hat{s}_{ml} 
   = \frac{y_{ml}-\hat{p}_{ml}}{\nu^p_{ml}+\nu^w}. 
\end{align}
Steps (R9)-(R10) then use the residual terms $\hvec{S}$ and $\{\nu^s_{ml}\}$ to compute $\hvec{R}$ and $\{\nu^r_{nl}\}$, where $\hat{r}_{nl}$ can be interpreted as a $\nu^r_{nl}$-variance-AWGN corrupted observation of the true $\X_{nl}$.
Similarly, (R11)-(R12) compute $\hvec{Q}$ and $\{\nu^q_{mn}\}$, where $\hat{q}_{mn}$ can be interpreted as a $\nu^q_{mn}$-variance-AWGN corrupted observation of the true $\A_{mn}$.
Finally, (R13)-(R14) merge these AWGN-corrupted observations with the priors $\{p_{\X_{nl}}\}$ to produce the posterior means $\hvec{X}$ and variances $\{\nu^{x}_{nl}\}$; (R15)-(R16) do the same for the $\A_{mn}$ quantities.

The BiG-AMP algorithm in \tabref{bigAmp} is a direct (although non-trivial) extension of the GAMP algorithm for compressive sensing \cite{Rangan:ISIT:11}, which estimates $\vX$ assuming perfectly known $\vA$, and even stronger similarities to the $\vA$-uncertain GAMP from \cite{Parker:ASIL:11}, which estimates $\vX$ assuming knowledge of the marginal means and variances of unknown random $\vA$, but which makes no attempt to estimate $\vA$ itself.
In \secref{PIAWGN}, a simplified version of BiG-AMP will be developed that is similar to the Bayesian-AMP algorithm \cite{Donoho:ITW:10a} for compressive sensing.

\section{BiG-AMP Simplifications}	\label{sec:simple} 
We now describe simplifications of the BiG-AMP algorithm from \tabref{bigAmp} that result from additional approximations and from the use of specific priors $p_{\Y_{ml}|\Z_{ml}}$, $p_{\X_{nl}}$, and $p_{\A_{mn}}$ that arise in practical applications of interest. 

\subsection{Scalar Variances}		\label{sec:scalar}
The BiG-AMP algorithm in \tabref{bigAmp} stores and processes a number of element-wise variance terms whose values vary across the elements (e.g., $\nu^x_{nl}$ can vary across $n$ and $l$).
The use of scalar variances (i.e., uniform across $m,n,l$) significantly reduces the memory and complexity of the algorithm.

To derive scalar-variance BiG-AMP, we first assume
$\forall n,l: \nu^x_{nl}(t) \approx 
\nu^x(t) \defn \frac{1}{NL} \sum_{n=1}^N \sum_{l=1}^L \nu^x_{nl}(t)$ and
$\forall m,n: \nu^a_{mn}(t) \approx 
\nu^a(t) \defn \frac{1}{MN} \sum_{m=1}^M \sum_{n=1}^N \nu^a_{mn}(t)$, so from (R1) 
\begin{align}
\bar{\nu}_{ml}^p(t) 
&\approx \nu^x(t) \sum_{n=1}^N |\hat{a}_{mn}(t)|^2 + \nu^a(t) \sum_{n=1}^N|\hat{x}_{nl}(t)|^2\\
&\approx \frac{\norm{\vec{\hat{A}}(t)}_F^2}{M} \nu^x(t) + \frac{\norm{\vec{\hat{X}}(t)}_F^2}{L} \nu^a(t) 
\defn \bar{\nu}^p(t).	\label{eq:nup_bar}
\end{align}
Note that using \eqref{nup_bar} in place of (R1) avoids two matrix multiplies.
Plugging these approximations into (R3) gives
\begin{align}
\nu^p_{ml}(t)
&\approx \bar{\nu}^p(t) + N\nu^a(t)\nu^x(t) 
\defn \nu^p(t) 	\label{eq:nupScalar}
\end{align}
which, when used in place of (R3), avoids another matrix multiply.
Even with the above scalar-variance approximations, $\{\nu^s_{ml}(t)\}$ from (R5) are not guaranteed to be equal (except in special cases like AWGN $p_{\Y_{ml}|\Z_{ml}}$). 
Still, they can be approximated as such using
$\nu^s(t) \defn \frac{1}{ML} \sum_{m=1}^M\sum_{l=1}^L \nu^s_{ml}(t)$,
in which case 
\begin{align}
\nu^r_{nl}(t)
&\approx \frac{1}{\nu^s(t) \sum_{m=1}^M |\hat{a}_{mn}(t)|^2 } 
\approx \frac{N}{\nu^s(t)\norm{\vec{\hat{A}}(t)}_F^2} \defn \nu^r(t)\label{eq:nurScalar} \\
\nu^q_{mn}(t)
&\approx \frac{1}{\nu^s(t) \sum_{l=1}^L |\hat{x}_{nl}(t)|^2 } 
\approx \frac{N}{\nu^s(t)\norm{\vec{\hat{X}}(t)}_F^2} \defn \nu^q(t)\label{eq:nuqScalar} .
\end{align}
Using \eqref{nurScalar} in place of (R9) and \eqref{nuqScalar} in place of (R11) avoids two matrix multiplies and $NL\!+\!MN\!-\!2$ scalar divisions, and furthermore allows (R10) and (R12) to be implemented as 
\begin{align}
\hat{r}_{nl}(t) 
&= \hat{x}_{nl}(t) \bigg(1 - \frac{MN \nu^a(t)}{\norm{\vec{\hat{A}}(t)}_F^2}  \bigg)
+ \nu^r(t) \sum_{m = 1}^M \hat{a}_{mn}(t) \hat{s}_{ml}(t) \label{eq:hatrScalar}\\
\hat{q}_{mn}(t) 
&= \hat{a}_{mn}(t) \bigg(1 - \frac{NL\nu^x(t)}{\norm{\vec{\hat{X}}(t)}_F^2}   \bigg)
+ \nu^q(t) \sum_{l = 1}^L \hat{x}_{nl}(t) \hat{s}_{ml}(t) \label{eq:hatqScalar},
\end{align}
saving two more matrix multiplies, and leaving a total of only three matrix multiplies per iteration.

\subsection{Possibly Incomplete AWGN Observations}	\label{sec:PIAWGN}
We now consider a particular observation model wherein the elements of $\vec{Z}\!=\!\vec{AX}$ are AWGN-corrupted at a subset of indices $\Omega \subset (1\dots M) \!\times\! (1\dots L)$ and unobserved at the remaining indices, noting that the standard AWGN model \eqref{AWGN} is the special case where $|\Omega|\!=\!ML$.
This ``possibly incomplete AWGN'' (PIAWGN) model arises in a number of important applications, such as matrix completion and dictionary learning.

We can state the PIAWGN model probabilistically as
\begin{align}
p_{\Y_{ml}|\Z_{ml}}(y_{ml}\giv z_{ml})
&= \begin{cases}
  \Nor(y_{ml};z_{ml},\nu^w) & (m,l) \in \Omega \\
  \Dirac_{y_{ml}} & (m,l) \notin \Omega ,
\end{cases}\label{eq:PIAWGN}
\end{align}
where $\nu^w$ is the noise variance on the non-missing observations and $\Dirac_{y}$ denotes a point mass at $y\!=\!0$. 
Thus, at the observed entries $(m,l)\in\Omega$, the quantities $\hat{s}_{ml}$ and $\nu^s_{ml}$ calculated using the AWGN expressions \eqref{nushat_AWGN}, while at the ``missing'' entries $(m,l)\notin\Omega$, where $y_{ml}$ is invariant to $z_{ml}$, we have 
$\E\{\Z_{ml}\giv \p_{ml}\!=\!\hat{p}_{ml};\nu^p_{ml}\} \!=\! \hat{p}_{ml}$ and 
$\var\{\Z_{ml}\giv \p_{ml}\!=\!\hat{p}_{ml};\nu^p_{ml}\} \!=\! \nu^p_{ml}$, 
so that $\hat{s}_{ml} \!=\!  0$ and
$\nu^s_{ml} \!=\! 0$.		
This is expected, given that $\nu^s$ can be interpreted as an inverse residual variance and $\hat{s}$ as a $\nu^s$-scaled residual.
In summary, the PIAWGN model yields
\begin{align}
\hat{s}_{ml}(t)  &= 
  \begin{cases}
  \frac{y_{ml}-\hat{p}_{ml}(t)}{\nu^p_{ml}(t)+\nu^w} & (m,l)\notin\Omega\\
  0 & (m,l)\notin\Omega\\
  \end{cases} \label{eq:hats_piawgn} \\
\nu^s_{ml}(t)  &=
  \begin{cases}
  \frac{1}{\nu^p_{ml}(t)+\nu^w} & (m,l)\notin\Omega\\
  0 & (m,l)\notin\Omega\\
  \end{cases} . \label{eq:nus_piawgn}
\end{align}

When the PIAWGN model is combined with the scalar-variance approximations from \secref{scalar}, BiG-AMP simplifies considerably.
To see this, we start by using $\nu^p(t)$ from \eqref{nupScalar} in place of $\nu^p_{ml}(t)$ in \eqref{hats_piawgn}-\eqref{nus_piawgn}, resulting in
\begin{align}
\vec{\hat{S}}(t) 
&= P_\Omega \bigg( \frac{\vec{Y} - \vec{\hat{P}}(t)}{\nu^p(t) + \nu^w} \bigg) 	\label{eq:shatOmega}\\
\nu^s(t) 
&= \frac{\delta}{\nu^w + \nu^p(t)} , \label{eq:nusOmega}
\end{align}
where $\delta\defn\frac{|\Omega|}{ML}$ denotes the fraction of observed entries 
and $P_\Omega:\Real^{M\times L}\rightarrow\Real^{M\times L}$ is the projection operator defined by
\begin{align}
\big[ P_\Omega(\vec{Z}) \big]_{ml} &\defn
\begin{cases}
z_{ml} &(m,l) \in \Omega\\
0 &(m,l) \notin \Omega 
\end{cases}.	\label{eq:proj}
\end{align}
We can then write (R10) and (R12) as
\begin{align}
\vec{\hat{R}}(t) 
&= \vec{\hat{X}}(t)\bigg(1 - \frac{MN \nu^a(t)}{\norm{\vec{\hat{A}}(t)}_F^2}  \bigg) 
+ \frac{N}{\delta \norm{\vec{\hat{A}}(t)}_F^2} \vec{\hat{A}\herm}(t) \hvec{V}(t)
\\
\vec{\hat{Q}}(t) 
&= \vec{\hat{A}}(t) \left(1 - \frac{NL \nu^x(t)}{\norm{\vec{\hat{X}}(t)}_F^2}  \right)
+ \frac{N}{\delta \norm{\vec{\hat{X}}(t)}_F^2} \vec{\hat{V}}(t) \vec{\hat{X}\herm}(t) 
\end{align}
using \eqref{hatrScalar}-\eqref{hatqScalar} and \eqref{shatOmega}-\eqref{proj} with
\begin{align}
\vec{\hat{V}}(t) 
&\defn P_\Omega\big(\vec{Y} - \vec{\hat{P}}(t)\big) \\
&= P_\Omega\big(\vec{Y} - \vec{\bar{P}}(t)\big) + \bar{\nu}^p(t) \vec{\hat{S}}(t\!-\!1)\\
&= P_\Omega\big(\vec{Y} - \vec{\bar{P}}(t)\big)  + \frac{\bar{\nu}^p(t)}{\nu^p(t\!-\!1) + \nu^w} \vec{\hat{V}}(t\!-\!1), 
\end{align}
since $P_\Omega$ is a projection operator, and using (R4) and \eqref{shatOmega}.

Scalar-variance BiG-AMP under PIAWGN observations is summarized in \tabref{bigAmpPI}.
Note that the residual matrix $\vec{\hat{U}}(t)\defn P_{\Omega}(\vec{Y}-\vec{\hat{A}}(t)\vec{\hat{X}}(t))$ needs to be computed and stored only at the observed entries $(m,l)\in\Omega$, leading to significant savings\footnote{Similar computational savings also occur with incomplete non-Gaussian observations.} when the observations are highly incomplete (i.e., $|\Omega|\ll ML$).
The same is true for the Onsager-corrected residual, $\vec{\hat{V}}(t)$. 
Thus, the algorithm in \tabref{bigAmpPI} involves only three (partial) matrix multiplies [in steps (R3p), (R8p), and (R10p), respectively], each of which can be computed using only $N|\Omega|$ scalar multiplies.

We note that Krzakala, M{\'e}zard, and Zdeborov{\'a} recently proposed an AMP-based approach to blind calibration and dictionary learning \cite{jtp_Krzakala2013} that bears close similarity\footnote{\label{Krzakala}The approach in \cite{jtp_Krzakala2013} does not compute (or use) $\nu^p(t)$ as given in lines (R4p)-(R5p) of \tabref{bigAmpPI}, but rather uses an empirical average of the squared Onsager-corrected residual in place of our $\nu^p(t)+\nu^w$ throughout their algorithm.} to BiG-AMP under the special case of AWGN-corrupted observations (i.e., $|\Omega|=ML$) and scalar variances. 
Their derivation differs significantly from that in \secref{derivation} due to the many simplifications offered by this special case.

\putTable{bigAmpPI}{Scalar-variance BiG-AMP with PIAWGN $p_{\Y|\Z}$}{\scriptsize
\begin{equation*}
\begin{array}{|lrcl@{}r|}\hline
  \multicolumn{2}{|l}{\textsf{initialization:}}&&&\\
  &\vec{\hat{V}}(0) &=& \vec{0} & \text{(I1p)}\\
  \multicolumn{2}{|r}{\textsf{choose}} & \multicolumn{2}{l}{\nu^x(1), \vec{\hat{X}}(1), \nu^a(1), \vec{\hat{A}}(1) } &\text{(I2p)}\\
  \multicolumn{3}{|l}{\textsf{for $t=1,\dots T_\textrm{max}$}}&&\\
  &G_a(t)
   &=& \frac{N}{\delta \|\vec{\hat{A}}(t)\|_F^2} & \text{(R1p)}\\
  &G_x(t)
   &=& \frac{N}{\delta \|\vec{\hat{X}}(t)\|_F^2} & \text{(R2p)}\\
  &\vec{\hat{U}}(t)
   &=& P_{\Omega}\big(\vec{Y}-\vec{\hat{A}}(t)\vec{\hat{X}}(t)\big) & \text{(R3p)}\\
  &\bar{\nu}^p(t)
   &=& \big(\frac{\nu^x(t)}{M G_a(t)} + \frac{\nu^a(t)}{L G_x(t)} \big)
   	\frac{N}{\delta} & \text{(R4p)}\\[0.5mm]
  &\nu^p(t)
   &=& \bar{\nu}^p(t) + N \nu^a(t) \nu^x(t) & \text{(R5p)}\\
  &\vec{\hat{V}}(t) &=& \vec{\hat{U}}(t)
	 + \frac{\bar{\nu}^p(t)}{\nu^p(t\!-\!1)+\nu^w}\vec{\hat{V}}(t\!-\!1)
	 & \text{(R6p)}\\[1mm]
  &\nu^r(t)
   &=& G_a(t)\big(\nu^p(t)+\nu^w\big) & \text{(R7p)}\\
  &\vec{\hat{R}}(t)
   &=& (1 - M\delta\nu^a(t)G_a(t)) \vec{\hat{X}}(t)+G_a(t)\vec{\hat{A}\herm}(t)\vec{\hat{V}}(t)
	& \text{(R8p)}\\
  &\nu^q(t)
   &=& G_x(t)\big(\nu^p(t)+\nu^w\big) & \text{(R9p)}\\
  &\vec{\hat{Q}}(t)
   &=&(1 - L\delta\nu^x(t)G_x(t)) \vec{\hat{A}}(t)+G_x(t)\vec{\hat{V}}(t)\vec{\hat{X}\herm}(t)
	& \text{(R10p)}\\[1mm]
  &\nu^x(t\!+\!\!1) &=&
	 \frac{1}{NL} \sum_{n=1}^N\sum_{l=1}^L \var\{\X_{nl}\giv \vec{Y}; \hat{r}_{nl}(t), \nu^r(t)\}& \text{(R11p)}\\[0.5mm]
  &\forall n,l:
    \hat{x}_{nl}(t\!+\!\!1) &=&
    	 \E\{\X_{nl}\giv \vec{Y}; \hat{r}_{nl}(t), \nu^r(t)\} & \text{(R12p)}\\[0.5mm]
  &\nu^a(t\!+\!\!1) &=&
	\frac{1}{MN} \!\!\sum_{m=1}^M \!\sum_{n=1}^N  \!\var\{\A_{mn}|\vec{Y}; \hat{q}_{mn}(t), \nu^q(t)\} & \text{(R13p)}\\[0.5mm]
  &\forall m,n:
	\hat{a}_{mn}(t\!+\!\!1) &=&
		  \E\{\A_{mn}\giv \vec{Y}; \hat{q}_{mn}(t), \nu^q(t)\} & \text{(R14p)}\\[1mm]
   \multicolumn{4}{|c}{\textsf{if $\|\vec{\hat{U}}(t) - \vec{\hat{U}}(t\!-\!1)\|_F^2 \le \tau_\textrm{BiG-AMP} \|\vec{\hat{U}}(t)\|_F^2$, \textb{\textsf{stop}}}}&\text{(R15p)}\\
    \multicolumn{2}{|l}{\textsf{end}}&&&\\\hline
\end{array}
\end{equation*}
}

\subsection{Zero-mean iid Gaussian Priors on $\vA$ and $\vX$} 	\label{sec:gaussian}
In this section we will investigate the simplifications that result in the case that both $p_{\A_{mn}}$ and $p_{\X_{nl}}$ are zero-mean iid Gaussian, i.e.,
\begin{align}
p_{\X_{nl}}(x) &= \Nor(x; 0, \nu^x_0) ~\forall n,l \label{eq:x_gauss}\\
p_{\A_{mn}}(a) &= \Nor(a; 0, \nu^a_0) ~\forall m,n \label{eq:a_gauss} ,
\end{align}
which, as will be discussed later, is appropriate for matrix completion.
In this case, straightforward calculations reveal that
$\E\{\X_{nl}\giv \R_{nl}\!=\!\hat{r}_{nl}; \nu^r_{nl}\} =\hat{r}_{nl}\nu^x_0/(\nu^r_{nl}+\nu^x_0)$ and $\var\{\X_{nl}\giv \R_{nl}\!=\!\hat{r}_{nl}; \nu^r_{nl}\})=\nu^x_0 \nu^r_{nl}/(\nu^r_{nl}+\nu^x_0)$ 
and, similarly, that
$\E\{\A_{mn}\giv \Q_{mn}\!=\!\hat{q}_{mn}; \nu^q_{mn}\}  =\hat{q}_{mn}\nu^a_0/(\nu^q_{mn}+\nu^a_0)$ and $\var\{\A_{mn}\giv\Q_{mn}\!=\!\hat{q}_{mn}, \nu^q_{mn}\}=\nu^a_0 \nu^q_{mn}/(\nu^q_{mn}+\nu^a_0)$.
Combining these iid Gaussian simplifications with the scalar-variance simplifications from \secref{scalar} yields an algorithm
whose computational cost is dominated by three matrix multiplies per iteration, each with a cost of $MNL$ scalar multiplies. 
The precise number of multiplies it consumes depends on the assumed likelihood model that determines steps (R7g)-(R8g).

Additionally incorporating the PIAWGN observations from \secref{PIAWGN} reduces the cost of the three matrix multiplies to only $N|\Omega|$ scalar multiplies each, and yields the ``BiG-AMP-Lite'' algorithm summarized in \tabref{bigAmpL}, consuming $(3N + 5)|\Omega| + 3(MN + NL) + 29$ multiplies per iteration.

\putTable{bigAmpL}{BiG-AMP-Lite: Scalar-variance, PIAWGN, Gaussian $p_{\X}$ and $p_{\A}$}{\scriptsize
\begin{equation*}
\begin{array}{|lrcl@{}r|}\hline
  \multicolumn{2}{|l}{\textsf{initialization:}}&&&\\
  &\vec{\hat{V}}(0) &=& \vec{0} & \text{(I1i)}\\
  \multicolumn{2}{|r}{\textsf{choose}} & \multicolumn{2}{l}{\nu^x(1), \vec{\hat{X}}(1), \nu^a(1), \vec{\hat{A}}(1)} &\text{(I2i)}\\
  \multicolumn{4}{|l}{\textsf{for $t=1,\dots T_\textrm{max}$}}&\\
  &G_a(t)
   &=& \frac{N}{\delta \|\vec{\hat{A}}(t)\|_F^2} & \text{(R1i)}\\
  &G_x(t)
   &=& \frac{N}{\delta \|\vec{\hat{X}}(t)\|_F^2} & \text{(R2i)}\\
  &\vec{\hat{U}}(t)
   &=& P_{\Omega}\big(\vec{Y}-\vec{\hat{A}}(t)\vec{\hat{X}}(t)\big) & \text{(R3i)}\\
  &\bar{\nu}^p(t)
   &=& \big(\frac{\nu^x(t)}{M G_a(t)} + \frac{\nu^a(t)}{L G_x(t)} \big)
   	\frac{N}{\delta} & \text{(R4i)}\\[0.5mm]
  &\nu^p(t)
   &=& \bar{\nu}^p(t) + N \nu^a(t) \nu^x(t) & \text{(R5i)}\\
  &\vec{\hat{V}}(t) &=& \vec{\hat{U}}(t)
	 + \frac{\bar{\nu}^p(t)}{\nu^p(t\!-\!1)+\nu^w}\vec{\hat{V}}(t\!-\!1)
	 & \text{(R6i)}\\[1mm]
 &\nu^r(t)
  &=& G_a(t)\big(\nu^p(t)+\nu^w\big) & \text{(R7i)}\\
 &\nu^q(t)
  &=& G_x(t)\big(\nu^p(t)+\nu^w\big) & \text{(R8i)}\\
  &\nu^x(t\!+\!1) 
  	&=& \big(\frac{1}{\nu^r(t)}+\frac{1}{\nu^x_0}\big)^{-1} & \text{(R9i)}\\
  &\vec{\hat{X}}(t\!+\!1) &=&
    	 \frac{\nu^x(t+1)}{\nu^r(t)} 
	\big((1 - M\delta\nu^a(t)G_a(t))\vec{\hat{X}}(t)    
							&\\&&&\mbox{}
	+G_a(t)\vec{\hat{A}\herm}(t)\vec{\hat{V}}(t)\big) 
	~& \text{(R10i)}\\[0.5mm]
  &\nu^a(t\!+\!1) 
  	&=& \big(\frac{1}{\nu^q(t)}+\frac{1}{\nu^a_0}\big)^{-1} & \text{(R11i)}\\
  &\vec{\hat{A}}(t\!+\!1) &=&
    	 \frac{\nu^a(t+1)}{\nu^q(t)} 
	\big((1 - L\delta\nu^x(t)G_x(t))\vec{\hat{A}}(t)
							&\\&&&\mbox{}
	+G_x(t)\vec{\hat{V}}(t)\vec{\hat{X}\herm}(t)\big) 
	~& \text{(R12i)}\\[0.5mm]
  \multicolumn{4}{|c}{\textsf{if $\|\vec{\hat{U}}(t) - \vec{\hat{U}}(t\!-\!1)\|_F^2 \le \tau_\textrm{BiG-AMP} \|\vec{\hat{U}}(t)\|_F^2$, \textb{\textsf{stop}}}}&\text{(R13i)}\\
    \multicolumn{2}{|l}{\textsf{end}}&&&\\\hline
\end{array}
\end{equation*}
}
\section{Adaptive Damping}    \label{sec:coststep}
The approximations made in the BiG-AMP derivation presented in \secref{derivation} were well-justified in the large system limit, i.e., the case where $M,N,L\rightarrow\infty$ with fixed $\frac{M}{N}$ and $\frac{L}{N}$.
In practical applications, however, these dimensions (especially $N$) are finite, and hence the algorithm presented in \secref{derivation} may diverge. 
\textb{In case of compressive sensing, the use of ``damping'' with GAMP yields provable convergence guarantees with arbitrary matrices \cite{Rangan:ISIT:14}.
Here, we propose to incorporate damping into BiG-AMP.}
Moreover, we propose to \emph{adapt} the damping of these variables to ensure that a particular cost criterion decreases monotonically (or near-monotonically), as described in the sequel.
\textb{The specific damping strategy that we adopt is similar to that described in \cite{Schniter:ALL:12} and coded in \cite{GAMPmatlab}.}

\subsection{Damping}			\label{sec:damp}
In BiG-AMP, the iteration-$t$ damping factor $\beta(t)\in(0,1]$ is used to slow the evolution of certain variables, namely $\bar{\nu}^p_{ml}$, $\nu^p_{ml}$, $\nu^s_{ml}$, $\hat{s}_{ml}$, $\hat{x}_{nl}$, and $\hat{a}_{mn}$. 
To do this, steps (R1), (R3), (R7), and (R8) in \tabref{bigAmp} are replaced with
\begin{align}
\bar{\nu}^p_{ml}(t)
&=  \beta(t) \bigg(\sum_{n=1}^{N} |\hat{a}_{mn}(t)|^2 \nu^x_{nl}(t) + \nu^a_{mn}(t) |\hat{x}_{nl}(t)|^2\bigg)
\non\\&\quad
+(1 - \beta(t))\bar{\nu}^p_{ml}(t-1) \\
\nu^p_{ml}(t)
&= \beta(t)  \bigg( \bar{\nu}^p_{ml}(t) + \sum_{n=1}^{N} \nu_{mn}^a(t) \nu^x_{nl}(t)\bigg)
\non\\&\quad
+(1 - \beta(t)) \nu^p_{ml}(t-1)\\
\nu^s_{ml}(t) 
&= \beta(t)\big(( \nu^z_{ml}(t)/\nu^p_{ml}(t)  - 1)/\nu^p_{ml}(t) \big)
\non\\&\quad
+(1-\beta(t)) \nu^s_{ml}(t\!-\!1)\\
\hat{s}_{ml}(t) 
&= \beta(t) \big( \hat{z}_{ml}(t) - \hat{p}_{ml}(t))/\nu^p_{ml}(t)\big)
\non\\&\quad 
+(1-\beta(t)) \hat{s}_{ml}(t\!-\!1) ,
\end{align}
and the following are inserted between (R8) and (R9):
\begin{align}
\bar{x}_{nl}(t\!+\!1) 
&= \beta(t)\hat{x}_{nl}(t\!+\!1) + (1-\beta(t))\bar{x}_{nl}(t)\\
\bar{a}_{mn}(t\!+\!1) 
&= \beta(t)\hat{a}_{mn}(t\!+\!1) + (1-\beta(t))\bar{a}_{mn}(t).
\end{align}
The newly defined state variables $\bar{x}_{nl}(t)$ and $\bar{a}_{mn}(t)$ are then used in place of $\hat{x}_{nl}(t)$ and $\hat{a}_{mn}(t)$ in steps (R9)-(R12) [but not (R1)-(R2)] of \tabref{bigAmp}.
A similar approach can be used for the algorithm in \tabref{bigAmpPI} (with the damping applied to $\vec{\hat{V}}(t)$ instead of $\vec{\hat{S}}(t)$) and those in 
\tabref{bigAmpL}.
Notice that, when $\beta(t)\!=\!1$, the damping has no effect, whereas 
when $\beta(t)\!=\!0$, all quantities become frozen in $t$. 

\subsection{Adaptive Damping}			\label{sec:adapt}
The idea behind adaptive damping is to monitor a chosen cost criterion $J(t)$ and decrease $\beta(t)$ when the cost has not decreased sufficiently\footnote{%
The following adaptation procedure is borrowed from GAMPmatlab\cite{GAMPmatlab}, where it has been established to work well in the context of GAMP-based compressive sensing.
When the current cost $J(t)$ is not smaller than the largest cost in the most recent \texttt{stepWindow} iterations, then the ``step'' is deemed unsuccessful, the damping factor $\beta(t)$ is reduced by the factor \texttt{stepDec}, and the step is attempted again.
These attempts continue until either the cost criterion decreases or the damping factor reaches \texttt{stepMin}, at which point the step is considered successful, or the iteration count exceeds $T_{\max}$ or the damping factor reaches \texttt{stepTol}, at which point the algorithm terminates. 
When a step is deemed successful, the damping factor is increased by the factor \texttt{stepInc}, up to the allowed maximum value \texttt{stepMax}. 
} relative to $\{J(\tau)\}_{\tau=t-1-T}^{t-1}$ for some ``step window'' $T\geq 0$.
This mechanism allows the cost criterion to increase over short intervals of $T$ iterations and in this sense is similar to the procedure used by SpaRSA~\cite{Wright:TSP:09}. 
We now describe how the cost criterion $J(t)$ is constructed,
\textb{building on ideas in \cite{Rangan:ISIT:13}.}

Notice that, for fixed observations $\vec{Y}$, the joint posterior pdf solves the (trivial) KL-divergence minimization problem 
\begin{align}
p_{\vX,\vA|\vY} &= \argmin_{b_{\vX,\vA}} D(b_{\vX,\vA}\|p_{\vX,\vA|\vY}).\label{eq:originalProblem}
\end{align}
The factorized form \eqref{scaling} of the posterior allows us to write 
\begin{eqnarray}
\lefteqn{ D(b_{\vX,\vA}\|p_{\vX,\vA|\vY}) - \log p_{\vY}(\vec{Y})} \non\\
&=& \int_{\vec{A},\vec{X}} \!\!\!b_{\vX,\vA}(\vec{A},\vec{X}) 
\log\frac{b_{\vX,\vA}(\vec{A},\vec{X})}
		{p_{\vY|\vZ}(\vec{Y}\giv \vec{AX}) 
		\,p_{\vX}(\vec{X})
		\,p_{\vA}(\vec{A})}  \label{eq:varStep1} \\
&=& D(b_{\vX,\vA}\|p_{\vA}p_{\vX}) 
   	- \int_{\vec{A},\vec{X}} \!\!\! b_{\vX,\vA}(\vec{A},\vec{X}) 
	\log p_{\vY|\vZ}(\vec{Y}\giv \vec{AX}) 
\qquad \label{eq:KLdiv}
\end{eqnarray}
Equations \eqref{originalProblem} and \eqref{KLdiv} then imply that
\begin{eqnarray}
p_{\vX,\vA|\vY} 
&=& \argmin_{b_{\vX,\vA}} J(b_{\vX,\vA}) \label{eq:varProblem} \\
J(b_{\vX,\vA})
&\defn& D(b_{\vX,\vA}\|p_{\vA}p_{\vX}) - \E_{b_{\vX,\vA}}\big\{ \log p_{\vY|\vZ}(\vec{Y}\giv \vA\vX) \big\}.
\qquad \label{eq:exactCost}
\end{eqnarray}

To judge whether a given time-$t$ BiG-AMP approximation ``$b_{\vX,\vA}(t)$'' of the joint posterior $p_{\vX,\vA|\vY}$ is better than the previous approximation $b_{\vX,\vA}(t\!-\!1)$, one could in principle plug the posterior approximation expressions \eqref{pXgivR}-\eqref{pAgivQ} into \eqref{exactCost} and then check whether $J(b_{\vX,\vA}(t))<J(b_{\vX,\vA}(t\!-\!1))$.
But, since the expectation in \eqref{exactCost} is difficult to evaluate, we approximate the cost \eqref{exactCost} by using, in place of $\vA\vX$, an independent Gaussian matrix\footnote{The GAMP work \cite{Rangan:ISIT:13} uses a similar approximation.} whose component means and variances are matched to those of $\vA\vX$.
Taking the joint BiG-AMP posterior approximation $b_{\vX,\vA}(t)$ to be the product of the marginals from \eqref{pXgivR}-\eqref{pAgivQ}, the resulting component means and variances are 
\begin{eqnarray}
\E_{b_{\vX,\vA}(t)}\{[\vA\vX]_{ml}\} 
&=& \sum_n \E_{b_{\vX,\vA}(t)}\{\A_{mn}\}\E_{b_{\vX,\vA}(t)}\{\X_{nl}\} \\
&=& \sum_n \hat{a}_{mn}(t)\hat{x}_{nl}(t) ~= \bar{p}_{ml}(t)\\
\var_{b_{\vX,\vA}(t)}\{[\vA\vX]_{ml}\}
&=& \sum_{n} \hat{a}_{mn}^2(t)\nu^x_{nl}(t) + \nu^a_{mn}(t)\hat{x}_{nl}^2(t) \qquad
\non\\&&\mbox{}
+ \nu^a_{mn}(t)\nu^x_{nl}(t) \\
&=& \nu^p_{ml}(t) .
\end{eqnarray}
In this way, the approximate iteration-$t$ cost becomes 
\begin{align}
\hat{J}(t) &= 
\sum_{n,l} D\Big( p_{\X_{nl}|\R_{nl}}\big(\cdot\biggiv\hat{r}_{nl}(t);\nu_{nl}^r(t)\big) \Big\|\, p_{\X_{nl}}(\cdot) \Big) 		\label{eq:cost} 
\\&\quad+
\sum_{m,n} D\Big( p_{\A_{mn}|\Q_{mn}}\big(\cdot\biggiv\hat{q}_{mn}(t);\nu_{mn}^q(t)\big) \Big\|\, p_{\A_{mn}}(\cdot) \Big) 
\non\\&\quad-
\sum_{m,l} \E_{\Z_{ml}\sim\mathcal{N}(\bar{p}_{ml}(t);\nu^p_{ml}(t))}\big\{ \log p_{\Y_{ml}|\Z_{ml}}(y_{ml} \giv \Z_{ml}) \big\}.	\non
\end{align}
Intuitively, the first term in \eqref{cost} penalizes the deviation between the (BiG-AMP approximated) posterior and the assumed prior on $\vX$, 
the second penalizes the deviation between the (BiG-AMP approximated) posterior and the assumed prior on $\vA$, and 
the third term rewards highly likely estimates $\vZ$.

\section{Parameter Tuning and Rank Selection} \label{sec:learning}	

\subsection{Parameter Tuning via Expectation Maximization} \label{sec:EM}

Recall that BiG-AMP requires the specification of priors $p_{\vX}(\vec{X})=\prod_{n,l}p_{\X_{nl}}(x_{nl})$, $p_{\vA}(\vec{A})=\prod_{m,n}p_{\A_{mn}}(a_{mn})$, and $p_{\vY|\vZ}(\vec{Y}|\vec{Z})=\prod_{m,l} p_{\Y_{ml}|\Z_{ml}}(y_{ml}|z_{ml})$. 
In practice, although one may know appropriate families for these distributions, the exact parameters that govern them are generally unknown.
For example, one may have good reason to believe apriori that the observations are AWGN corrupted, justifying the choice $p_{\Y_{ml}|\Z_{ml}}(y_{ml}|z_{ml})=\mc{N}(y_{ml};z_{ml},\nu^w)$, but the noise variance $\nu^w$ may be unknown.
In this section, we outline a methodology that takes a given set of BiG-AMP parameterized priors $\{p_{\X_{nl}}(\cdot;\vec{\theta}),p_{\A_{mn}}(\cdot;\vec{\theta}),p_{\Y_{ml}|\Z_{ml}}(y_{ml}|\cdot;\vec{\theta})\}_{\forall m,n,l}$ and tunes the parameter vector $\vec{\theta}$ using an expectation-maximization (EM) \cite{Dempster:JRSS:77} based approach, with the goal of maximizing the likelihood, i.e., finding $\hvec{\theta}\defn \argmax_{\vec{\theta}}p_{\vY}(\vec{Y};\vec{\theta})$. 
The approach presented here can be considered as a generalization of the GAMP-based work \cite{Vila:TSP:13} to BiG-AMP.

Taking $\vX$, $\vA$, and $\vZ$ to be the hidden variables, the EM recursion can be written as \cite{Dempster:JRSS:77}
\begin{align}
\vec{\hat{\theta}}^{k+1} 
&= \argmax_{\vec{\theta}} \E\Big\{\log p_{\vX,\vA,\vZ,\vY}(\vX,\vA,\vZ,\vY;\vec{\theta})\Biggiv \vec{Y}; \hvec{\theta}^k\Big\} \nonumber \\
&= \argmax_{\vec{\theta}} \bigg\{
	\sum_{n,l} \E\Big\{\log p_{\X_{nl}}(\X_{nl};\vec{\theta}) \Biggiv \vec{Y}; \vec{\hat{\theta}}^k \Big\}	
	\label{eq:EM}
	\\&\quad
	+ \sum_{m,n} \E\Big\{\log p_{\A_{mn}}(\A_{mn};\vec{\theta}) \Biggiv \vec{Y}; \vec{\hat{\theta}}^k\Big\}
	\nonumber\\&\quad
	+ \sum_{m,l} \E\Big\{\log p_{\Y_{ml}|\Z_{ml}}(y_{ml}\giv \Z_{ml};\vec{\theta}) \Biggiv \vec{Y}; \vec{\hat{\theta}}^k \Big\}
	\bigg\}	\nonumber
\end{align}
where for \eqref{EM} we used the fact
$p_{\vX,\vA,\vZ,\vY}(\vX,\vA,\vZ,\vY;\vec{\theta})
= p_{\vX}(\vX;\vec{\theta}) p_{\vA}(\vA;\vec{\theta})
p_{\vY|\vZ}(\vY|\vZ;\vec{\theta}) \,\Dirac_{\vZ-\vA\vX}$
and the factorizability of $p_{\vX}$, $p_{\vA}$, and $p_{\vY|\vZ}$. 
As can be seen from \eqref{EM}, knowledge of the marginal posteriors $\{p_{\X_{nl}|\vY}, p_{\A_{mn}|\vY}, p_{\Z_{ml}|\vY}\}_{\forall m,n,l}$ is sufficient to compute the EM update.
Since the exact marginal posteriors are unknown, we employ BiG-AMP's approximations from \eqref{pXgivR}, \eqref{pAgivQ}, and \eqref{pZgivYP} for approximate EM.
In addition, we adopt the ``incremental'' update strategy from \cite{Neal:Jordan:99}, where the maximization over $\vec{\theta}$ is performed one element at a time while holding the others fixed.

As a concrete example, consider updating the noise variance $\nu^w$ under the PIAWGN model \eqref{PIAWGN}.  
Equation \eqref{EM} suggests
\begin{align}
(\nu^w)^{k+1} 
&= \argmax_{\nu^w} \sum_{(m,l) \in \Omega} \int_{z_{ml}}
p_{\Z_{ml}|\vY}(z_{ml}|\vec{Y})
\non\\&\quad\times 
\log \mc{N}(y_{ml};z_{ml},\nu^w) ,	\label{eq:nuw}
\end{align}
where the true marginal posterior 
$p_{\Z_{ml}|\vY}(\cdot|\vec{Y})$
is replaced with the most recent BiG-AMP approximation
$p_{\Z_{ml}|\p_{ml}}(\cdot|\hat{p}_{ml}(T_{\max});\nu^p_{ml}(T_{\max}),\vec{\hat{\theta}}^k)$, where ``most recent'' is with respect to both EM and BiG-AMP iterations.
Zeroing the derivative of the sum in \eqref{nuw} with respect to $\nu^w$, 
\begin{align}
(\nu^w)^{k+1} &= \frac{1}{|\Omega|} \sum_{(m,l) \in \Omega} \big(y_{ml}-\hat{z}_{ml}(T_{\max})\big)^2 + \nu^z_{ml}(T_{\max}),	\label{eq:nuw2}
\end{align}
where $\hat{z}_{ml}(t)$ and $\nu^z_{ml}(t)$ are the BiG-AMP approximated posterior mean and variance from \eqref{zhat}-\eqref{zvar}. 

The overall procedure can be summarized as follows.
From a suitable initialization $\hvec{\theta}^0$, BiG-AMP is run using the priors $\{p_{\X_{nl}}(\cdot;\hvec{\theta}^0),p_{\A_{mn}}(\cdot;\hvec{\theta}^0),p_{\Y_{ml}|\Z_{ml}}(y_{ml}|\cdot;\hvec{\theta}^0)\}_{\forall m,n,l}$ and iterated to completion, yielding approximate marginal posteriors on $\{\X_{nl},\A_{mn},\Z_{ml}\}_{\forall m,n,l}$.
These posteriors are used in \eqref{EM} to update the parameters $\vec{\theta}$ one element at a time, yielding $\hvec{\theta}^1$.
BiG-AMP is then run using the priors $\{p_{\X_{nl}}(\cdot;\hvec{\theta}^1),p_{\A_{mn}}(\cdot;\hvec{\theta}^1),p_{\Y_{ml}|\Z_{ml}}(y_{ml}|\cdot;\hvec{\theta}^1)\}_{\forall m,n,l}$, and so on.
A detailed discussion in the context of GAMP, along with explicit update equations for the parameters of Bernoulli-Gaussian-mixture pdfs, can be found in \cite{Vila:TSP:13}.

\subsection{Rank Selection}   \label{sec:modelOrder}
BiG-AMP and EM-BiG-AMP, as described up to this point, require the specification of the rank $N$, i.e., the number of columns in $\vA$ (and rows in $\vX$) in the matrix factorization $\vZ=\vA\vX$.
Since, in many applications, the best choice of $N$ is difficult to specify in advance, we now describe two procedures to estimate $N$ from the data $\vec{Y}$,
\textb{building on well-known rank-selection procedures}.

\subsubsection{Penalized log-likelihood maximization}  \label{sec:AIC}
Consider a set of possible models $\{\mathcal{H}_N\}_{N=1}^{\overline{N}}$ for the observation $\vec{Y}$ where, under $\mathcal{H}_N$, EM-BiG-AMP estimates $\vTheta_N = \{\vA_N, \vX_N, \vec{\theta}\}$. 
Here, the subscripts on $\vA_N$ and $\vX_N$ indicate the restriction to $N$ columns and rows, $\vec{\theta}$ refers to the vector of parameters defined in \secref{EM}, and the subscript on $\vTheta_N$ indicates the dependence of the overall number of parameters in $\vTheta_N$ with the rank $N$.
Because the selection rule $\hat{N}=\argmax_{N} p_{\vY}(\vec{Y};\mc{H}_N)$ is typically intractable, \textb{several well-known rules of the form} 
\begin{equation}
\hat{N}=\argmax_{N=1,\dots,\overline{N}} \, 2\log p_{\vY|\vTheta_N}(\vec{Y}\giv\hvec{\Theta}_N)-\eta(N) \label{eq:MOS}
\end{equation}
have been developed, such as the Bayesian Information Criterion (BIC) and Akaike's Information Criterion (AIC) \cite{jtp_Stoica2004}.
In \eqref{MOS}, $\hvec{\Theta}_N$ is the ML estimate of $\vTheta_N$ under $\vec{Y}$, and $\eta(\cdot)$ is a penalty function that depends on the \emph{effective} number of scalar parameters $N_{\text{eff}}$ estimated under model $\mc{H}_N$ (which depends on $N$) and possibly on the number of scalar parameters $|\Omega|$ that make up the observation $\vec{Y}$.

Applying this methodology to EM-BiG-AMP, where
$p_{\vY|\vTheta_N}(\vec{Y}\giv\vec{\Theta}_N) 
= p_{\vY|\vZ}(\vec{Y}\giv \vec{A}_N\vec{X}_N; \vec{\theta})$,
we obtain the rank-selection rule 
\begin{equation}
\hat{N} = \argmax_{N=1,\dots,\overline{N}} \, 2\log p_{\vY|\vZ}(\vec{Y}\giv \vec{\hat{A}}_N\vec{\hat{X}}_N; \vec{\hat{\theta}}) - \eta(N). \label{eq:orderCriteria}
\end{equation} 
Since $N_{\text{eff}}$ 
depends on the application (e.g., matrix completion, robust PCA, dictionary learning), detailed descriptions of $\eta(\cdot)$ are postponed to \textb{\cite{jtp_Parker2013a}}.

To perform the maximization over $N$ in \eqref{orderCriteria}, we start with a small hypothesis $N_1$ and run EM-BiG-AMP to completion, generating the (approximate) MMSE estimates $\hvec{A}_{N_1},\hvec{X}_{N_1}$ and ML estimate $\hvec{\theta}$, which are then used to evaluate\footnote{Since we compute approximate MMSE estimates rather than ML estimates, we are in fact evaluating a lower bound on the penalized log-likelihood.} the penalized log-likelihood in \eqref{orderCriteria}.
The $N$ hypothesis is then increased by a fixed value (i.e., $N_2=N_1+\texttt{rankStep}$), initializations of $(\vec{A}_{N_2},\vec{X}_{N_2},\vec{\theta})$ are chosen based on the previously computed $(\hvec{A}_{N_1},\hvec{X}_{N_1},\hvec{\theta})$, and EM-BiG-AMP is run to completion, yielding estimates $(\hvec{A}_{N_2},\hvec{X}_{N_2},\hvec{\theta})$ with which the penalized likelihood is again evaluated.
This process continues until either the value of the penalized log-likelihood decreases, in which case $\hat{N}$ is set at the previous (i.e., maximizing) hypothesis of $N$, or the maximum-allowed rank $\overline{N}$ is reached.

\subsubsection{Rank contraction} 	\label{sec:contraction}
We now describe an alternative rank-selection procedure that is appropriate when $\vZ$ has a ``cliff'' in its singular value profile and \textb{which is reminiscent of that used in LMaFit~\cite{jtp_Wen2012a}}. 
In this approach, EM-BiG-AMP is initially configured to use the maximum-allowed rank, i.e., $N=\overline{N}$. 
After the first EM iteration, the singular values $\{\sigma_n\}$ of the estimate $\hvec{X}$ and the corresponding pairwise ratios $R_n = \sigma_{n}/\sigma_{n+1}$ are computed,\footnote{In some cases the singular values of $\hvec{A}$ could be used instead.} from which a candidate rank estimate $\hat{N} = \argmax_n R_n$ is identified, corresponding to the largest gap in successive singular values. 
However, this candidate is accepted only if this maximizing ratio exceeds the average ratio by the \textb{user-specified} parameter $\tau_{\text{MOS}}$ (e.g., $\tau_{\text{MOS}}=5$), i.e., if
\begin{align}
R_{\hat{N}} >
\frac{\tau_{\text{MOS}}}{\overline{N}-2}\sum_{i \ne \hat{N}} R_i, \label{eq:contract}
\end{align}
and if $\hat{N}/\overline{N}$ is sufficiently small. \textb{Increasing $\tau_{\text{MOS}}$ makes the approach less prone to selecting an erroneous rank during the first few iterations, but making the value too large prevents the algorithm from detecting small gaps between the singular values.}
If $\hat{N}$ is accepted, then the matrices $\vA$ and $\vX$ are pruned to size $\hat{N}$ and EM-BiG-AMP is run to convergence.
If not, EM-BiG-AMP is run for one more iteration, after which a new candidate $\hat{N}$ is identified and checked for acceptance, and so on.

In many cases, a rank candidate is accepted after a small number of iterations,
 and thus only a few SVDs need be computed. 
This procedure has the advantage of running EM-BiG-AMP to convergence only once, rather than several times under different hypothesized ranks. 
However, when the singular values of $\vZ$ decay smoothly, this procedure can mis-estimate the rank, as discussed in \cite{jtp_Wen2012a}.

\section{Matrix Completion}             \label{sec:MC}

In this and the next two sections,
we detail the application of BiG-AMP to the problems of matrix completion (MC), robust principle components analysis (RPCA), and dictionary learning (DL), respectively.
\textb{For each application, we discuss the BiG-AMP's choice of matrix representation, priors, likelihood,} initialization, adaptive damping, EM-driven parameter learning, and rank-selection.
\textb{Also, for each application, we provide} an extensive empirical study comparing BiG-AMP to state-of-the-art solvers on both synthetic and real-world datasets.
These results demonstrate that BiG-AMP yields excellent reconstruction performance (often best in class) while maintaining competitive runtimes.
\textb{For each application of BiG-AMP discussed in the sequel, we recommend numerical settings for necessary parameter values, as well as initialization strategies when appropriate.
Although we cannot guarantee that our recommendations are universally optimal, they worked well for the range of problems considered in this paper, and we conjecture that they offer a useful starting point for further experimentation.
Nevertheless, modifications may be appropriate when applying BiG-AMP outside the range of problems considered here.}
\textb{Our BiG-AMP Matlab code can be found as part of the GAMPmatlab package at {\small \texttt{\url{https://sourceforge.net/projects/gampmatlab/}}}, including examples of BiG-AMP applied to the MC, RPCA, and DL problems.}

%
%


\subsection{Problem setup}
In matrix completion (MC) \cite{Candes:PROC:10}, one seeks to recover a rank-$N \ll \min(M,L)$ matrix $\vZ\in\Real^{M\times L}$ after observing a fraction $\delta = \frac{|\Omega|} {ML}$ of its (possibly noise-corrupted) entries, where $\Omega$ denotes the set of observations. 
 
BiG-AMP approaches the MC problem by modeling the complete matrix $\vZ$ as the product $\vZ=\vA\vX$ of random matrices $\vA\in\Real^{M\times N}$ and $\vX\in\Real^{N\times L}$ with priors of the decoupled form in \textb{\eqref{pA}-\eqref{pX}}, where $\vZ$ is probabilistically related to the observed matrix $\vec{Y}$ through a likelihood $p_{\vY|\vZ}(\vec{Y}\giv\vZ)$ of the decoupled form in \textb{\eqref{pYgivZ}}. 
To finally perform MC, BiG-AMP infers \textb{$\vA$ and $\vX$} from $\vY$ under the above model. 
\textb{The corresponding estimates $\hvec{A}$ and $\hvec{X}$ can then be multiplied to yield an estimate $\hvec{Z}=\hvec{A}\hvec{X}$ of the noiseless complete matrix $\vZ$.}

As \textb{in several existing} Bayesian approaches to matrix completion (e.g., 
\textb{\cite{Tipping:JRSSb:99,jtp_Lim2007,Salakhutdinov:NIPS:08,Lawrence:ICML:09}}), we choose Gaussian priors for the factors $\vA$ and $\vX$.
\textb{Although EM-BiG-AMP readily supports the use of priors with row- and/or column-dependent parameters, we focus on simple iid priors of the form}
\begin{align}
p_{\A_{mn}}(a) &= \Nor(a; 0,1) ~\forall m,n \label{eq:a_gauss2}\\
p_{\X_{nl}}(x) &= \Nor(x; \hat{x}_0, \nu^x_0) ~\forall n,l \label{eq:x_gauss2},
\end{align}
\textb{
where the mean and variance in \eqref{x_gauss2} can be tuned using EM-BiG-AMP, as described in the sequel, and
where the variance in \eqref{a_gauss2} is fixed to avoid a scaling ambiguity between $\vA$ and $\vX$.}
\textb{\Secref{MCexamples} demonstrates that this simple approach is effective in attacking several MC problems of interest.}
Assuming the observation noise to be additive and Gaussian, we then choose the PIAWGN model from \Xeqref{PIAWGN} for the likelihood $p_{\vY|\vZ}$ \textb{given by}
\textb{\begin{align}
p_{\Y_{ml}|\Z_{ml}}(y_{ml}\giv z_{ml})
&= \begin{cases}
  \Nor(y_{ml};z_{ml},\nu^w) & (m,l) \in \Omega \\
  \Dirac_{y_{ml}} & (m,l) \notin \Omega.
\end{cases} \label{eq:PIAWGNPart2}
\end{align}}%
Note that, by using \eqref{a_gauss2}-\eqref{x_gauss2} with $\hat{x}_0=0$ and the scalar-variance approximation from \Xsecref{scalar}, the BiG-AMP algorithm from \Xtabref{bigAmp} reduces to the simpler BiG-AMP-Lite algorithm from \Xtabref{bigAmpL} with $\nu^a_0=1$.

\subsection{Initialization} \label{sec:MCinit}
In most cases we advocate initializing the BiG-AMP quantities $\hvec{X}(1)$ and $\hvec{A}(1)$ using random draws from the priors $p_{\vX}$ and $p_{\vA}$, although setting either $\hvec{X}(1)$ or $\hvec{A}(1)$ at zero also seems to perform well in the MC application.
Although it is also possible to use SVD-based initializations of $\hvec{X}(1)$ and $\hvec{A}(1)$ (i.e., for SVD $\vec{Y} = \vec{U}\vec{\Sigma}\vec{D}^T$, set $\hvec{A}(1) = \vec{U}\vec{\Sigma}^{1/2}$ and $\hvec{X}(1) = \vec{\Sigma}^{1/2}\vec{D}^T$) as done in LMaFit~\cite{jtp_Wen2012a} and VSBL~\cite{Babacan:TSP:12}, experiments suggest that the extra computation required is rarely worthwhile for BiG-AMP. 

As for the initializations $\nu_{nl}^x(1)$ and $\nu_{mn}^a(1)$, we advocate setting them at $10$ times the prior variances in \eqref{a_gauss2}-\eqref{x_gauss2}, which has the effect of weighting the measurements $\vec{Y}$ more than the priors $p_{\vX}, p_{\vA}$ during the first few iterations.

\subsection{Adaptive damping}
For the assumed likelihood \textb{\eqref{PIAWGNPart2}} and priors \eqref{a_gauss2}-\eqref{x_gauss2}, the adaptive-damping cost criterion $\hat{J}(t)$ described in \Xsecref{adapt} reduces to 
\begin{align}
\hat{J}(t) &= 
\frac{1}{2}\sum_{n,l} \bigg[ \log \frac{\nu^x_0}{\nu^x(t)} + \bigg(\frac{\nu^x(t)}{\nu^x_0}-1\bigg) + \frac{(\hat{x}_{nl}(t) - \hat{x}_0)^2}{\nu^x_0}  \bigg] 
\non\\&\quad
+\frac{1}{2}\sum_{m,n} \bigg[ \log \frac{1}{\nu^a(t)} + \bigg(\nu^a(t) -1\bigg) + \hat{a}_{mn}^2(t) \bigg] 
\non\\&\quad
+\frac{1}{\nu^w}\bigg( \frac{1}{2}\sum_{(m,l) \in \Omega} \left(y_{ml} - \bar{p}_{ml}(t)\right)^2 + \nu^p(t) \bigg)
\non\\&\quad 
+ |\Omega| \log \sqrt{2\pi\nu^w} \label{eq:LiteCost} .
\end{align}
To derive \eqref{LiteCost}, one can start with the first term in \Xeqref{cost} and leverage the Gaussianity of the approximated posterior on $\X_{nl}$:
\begin{align}
&\sum_{n,l} D\Big( p_{\X_{nl}|\R_{nl}}\big(\cdot\biggiv\hat{r}_{nl}(t);\nu_{nl}^r(t)\big) \Big\|\, p_{\X_{nl}}(\cdot) \Big) \\
  &= \sum_{n,l} \int_{x_{nl}} \mc{N}(x_{nl};\hat{x}_{nl}(t),\nu_{nl}^x(t))
  	\log \frac{\mc{N}(x_{nl}; \hat{x}_{nl}(t),\nu_{nl}^x(t))}{\mc{N}(x_{nl};{\hat{x}_0},\nu^x_0)} \non ,
\end{align}
which then directly yields the first term in \eqref{LiteCost}.
The second term in \eqref{LiteCost} follows using a similar procedure,
and the third and fourth terms follow directly from the PIAWGN model. 

In the noise free setting (i.e., $\nu^w \rightarrow 0$), the third term in \eqref{LiteCost} dominates, omitting the need to compute the other terms.

\subsection{EM-BiG-AMP} \label{sec:MCEM}
For the likelihood \textb{\eqref{PIAWGNPart2}} and priors \eqref{a_gauss2}-\eqref{x_gauss2}, the distributional parameters $\vec{\theta} = [\nu^w, \hat{x}_0, \nu_0^x]^T$ can be tuned using the EM approach from \Xsecref{EM}.%
\footnote{For the first EM iteration, we recommend initializing BiG-AMP using $\nu_{nl}^x(1)=\nu^x_0$, $\hat{x}_{nl}(1)=\hat{x}_0$, $\nu_{mn}^a(1)=1$, and $\hat{a}_{mn}(1)$ drawn randomly from $p_{\A_{mn}}$. After the first iteration, we recommend warm-starting BiG-AMP using the values from the previous EM iteration.}
To initialize $\vec{\theta}$ for EM-BiG-AMP, we adapt the procedure outlined in \cite{Vila:TSP:13} to our matrix-completion problem, giving the EM initializations $\hat{x}_0=0$ and 
\begin{align}
\nu^w &= \frac{\norm{P_\Omega(\vec{Y})}_F^2}{(\textrm{SNR}^0 + 1) |\Omega|}\\
\nu^x_0 &= \frac{1}{N}\bigg[ \frac{\norm{P_\Omega(\vec{Y})}_F^2}{|\Omega|} - \nu^w\bigg],
\end{align}
where $\textrm{SNR}^0$ is an initial estimate of the signal-to-noise ratio that, in the absence of other knowledge, can be set at $100$.

\subsection{Rank selection}
\label{sec:MCrankLearn}
For MC rank-selection under the penalized log-likelihood strategy \Xeqref{orderCriteria}, we recommend using the \emph{small sample corrected AIC} (AICc) \cite{jtp_Stoica2004} penalty
$\eta(N) = 2\frac{|\Omega|}{|\Omega| - N_{\text{eff}} - 1} N_{\text{eff}}$.
For the MC problem, $N_{\text{eff}}=\df+3$, where $\df\defn N(M\!+\!L\!-\!N)$ counts the degrees-of-freedom in a rank-$N$ real-valued $M\times L$ matrix \cite{Candes:PROC:10} and the three additional parameters come from $\vec{\theta}$.
Based on the PIAWGN likelihood \textb{\eqref{PIAWGNPart2}} and the standard form of the ML estimate of $\nu^w$ (see, e.g., \cite[eq.\ (7)]{jtp_Stoica2004}), the update rule \Xeqref{orderCriteria} becomes 
\begin{align}
\hat{N}
&= \argmax_{N=1,\dots,\bar{N}} \bigg[
\textb{-} |\Omega| \log \bigg(\frac{1}{|\Omega|} \sum_{(m,l) \in \Omega} \big(y_{ml}-\hat{z}_{ml}(t)\big)^2 \bigg) 
\non\\&\qquad
\textb{-} 2\frac{|\Omega|(N(M+L-N)+3)}{|\Omega| - N(M+L-N)-4} \bigg].
\end{align}
We note that a similar rule (but based on BIC rather than AICc) was used for rank-selection in \cite{jtp_Marjanovic2012}.

MC rank selection can also be performed using the rank contraction scheme described in \Xsecref{contraction}. We recommend choosing the maximum rank $\overline{N}$ to be the largest value such that $\overline{N}(M+L-\overline{N}) < |\Omega|$ and setting $\tau_{\text{MOS}} = 1.5$. Since the first EM iteration runs BiG-AMP with the large value $N = \overline{N}$, we suggest limiting the number of allowed BiG-AMP iterations during this first EM iteration to $\texttt{nitFirstEM} = 50$. In many cases, the rank learning procedure will correctly reduce the rank after these first few iterations, reducing the added computational cost of the rank selection procedure.


\subsection{Matrix Completion Experiments} \label{sec:MCexamples}
We now present the results of experiments used to ascertain the performance of BiG-AMP relative to existing state-of-the-art algorithms for matrix completion. 
For these experiments, we considered
IALM~\cite{jtp_Lin2010}, a nuclear-norm based convex-optimization method;
LMaFit~\cite{jtp_Wen2012a}, a non-convex optimization-based approach using non-linear successive over-relaxation;
GROUSE~\cite{jtp_Balzano2010}, which performs gradient descent on the Grassmanian manifold; 
Matrix-ALPS~\cite{Kyrillidis:12}, a greedy hard-thresholding approach; and 
VSBL~\cite{Babacan:TSP:12}, a variational Bayes approach.
In general, we configured BiG-AMP as described in \secref{MC}\footnote{%
Unless otherwise noted, we used the BiG-AMP parameters 
$T_{\max} = 1500$ (see \Xsecref{summary} for descriptions) and the adaptive damping parameters $\texttt{stepInc} = 1.1$, $\texttt{stepDec} = 0.5$, $\texttt{stepMin} = 0.05$, $\texttt{stepMax} = 0.5$, $\texttt{stepWindow} = 1$, and $\beta(1) = \texttt{stepMin}$. 
(See \Xsecref{adapt} for descriptions). 
} and made our best attempt to configure the competing algorithms for maximum performance. 
That said, the different experiments that we ran required somewhat different parameter settings, as we detail in the sequel.

\subsubsection{Low-rank matrices}
\label{sec:phaseMC}
We first investigate recovery of rank-$N$ matrices $\vec{Z}\in\Real^{M\times L}$ from noiseless incomplete observations $\{z_{ml}\}_{(m,l)\in\Omega}$ with indices $\Omega$ chosen uniformly at random.
To do this, we evaluated the normalized mean square error (NMSE) $\frac{\norm{\vec{Z} - \hvec{Z}}_F^2}{\norm{\vec{Z}}_F^2}$ of the estimate $\hvec{Z}$ returned by the various algorithms under test, examining $10$ realizations of $(\vec{Z},\Omega)$ at each problem size $(M,L,N)$. 
Here, each realization of $\vec{Z}$ was constructed as $\vec{Z} = \vec{AX}$ for $\vec{A}$ and $\vec{X}$ with iid $\Nor(0,1)$ elements.\footnote{\textb{%
We chose the i.i.d Gaussian construction due to its frequent appearance in the matrix-completion literature.
Similar performance was observed when the low-rank factors $\vec{A}$ and $\vec{X}$ were generated in other ways, such as from the left and right singular vectors of an i.i.d Gaussian matrix.}}
All algorithms were forced\footnote{This restriction always improved the performance of the tested algorithms.} to use the true rank $N$, 
run under default settings with very minor modifications,\footnote{GROUSE was run with $\texttt{maxCycles}=600$ and $\texttt{step\_size}=0.5$, where the latter was chosen as a good compromise between phase-transition performance and runtime. VSBL was run under $\texttt{MAXITER} = 2000$ and fixed $\beta = 10^9$; adaptive selection of $\beta$ was found to produce a significant degradation in the observed phase transition.  LMaFit was run from the same random initialization as BiG-AMP and permitted at most $\texttt{maxit}=6000$ iterations.  IALM was allowed at most $2000$ iterations.  A maximum runtime of one hour per realization was enforced for all algorithms.} 
and terminated when the normalized change in either $\vec{\hat{Z}}$ or $P_\Omega(\vec{\hat{Z}})$ across iterations fell below the tolerance value of $10^{-8}$. 

\begin{figure*}[htb]
\centering
\psfrag{N (rank)}[][][0.6]{\sf rank $N$}
\psfrag{delta}[][b][0.6]{~~~\sf sampling ratio $\delta$}
 \begin{tabular}{cccc}
\includegraphics[width=1.6in]{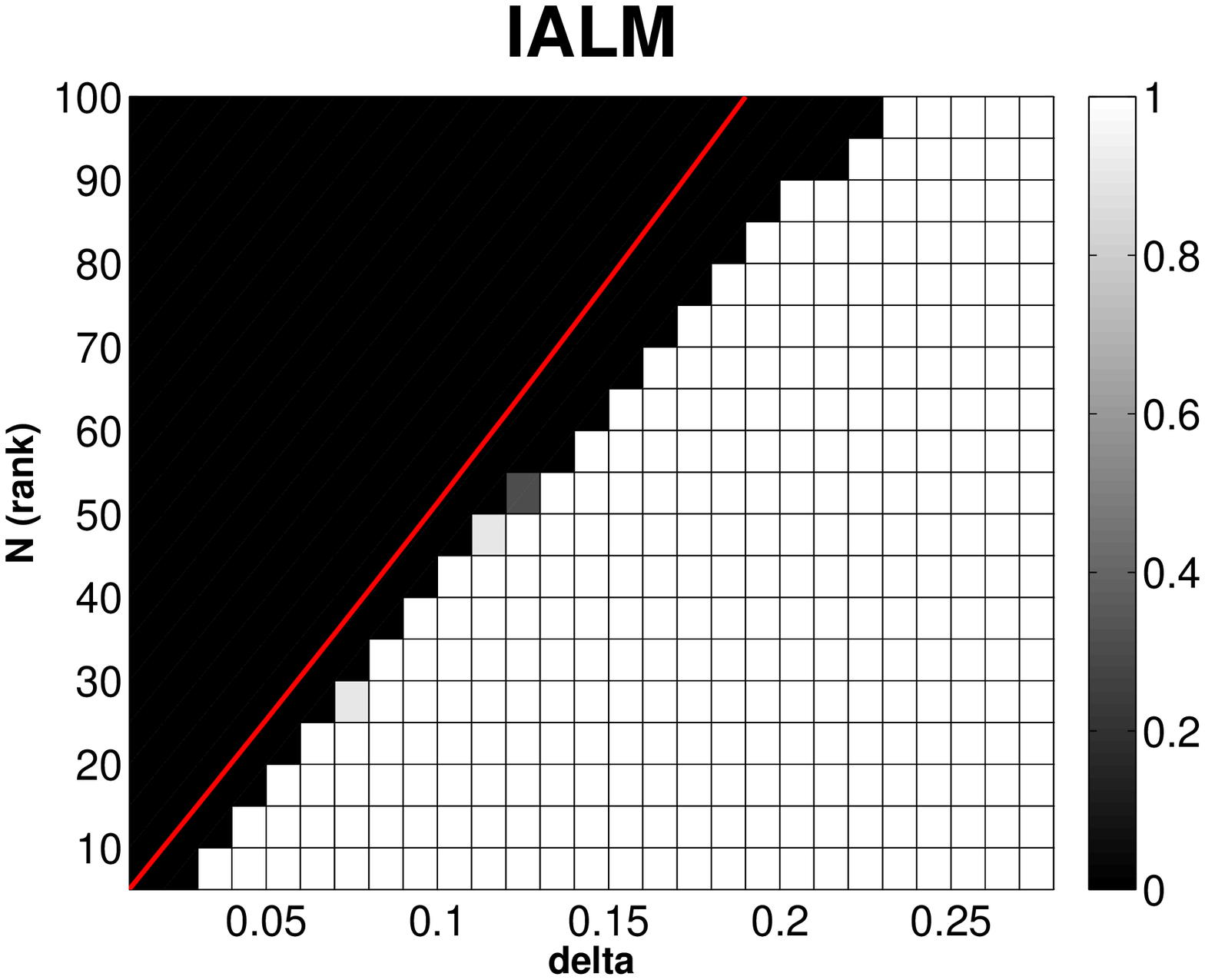}&
\includegraphics[width=1.6in]{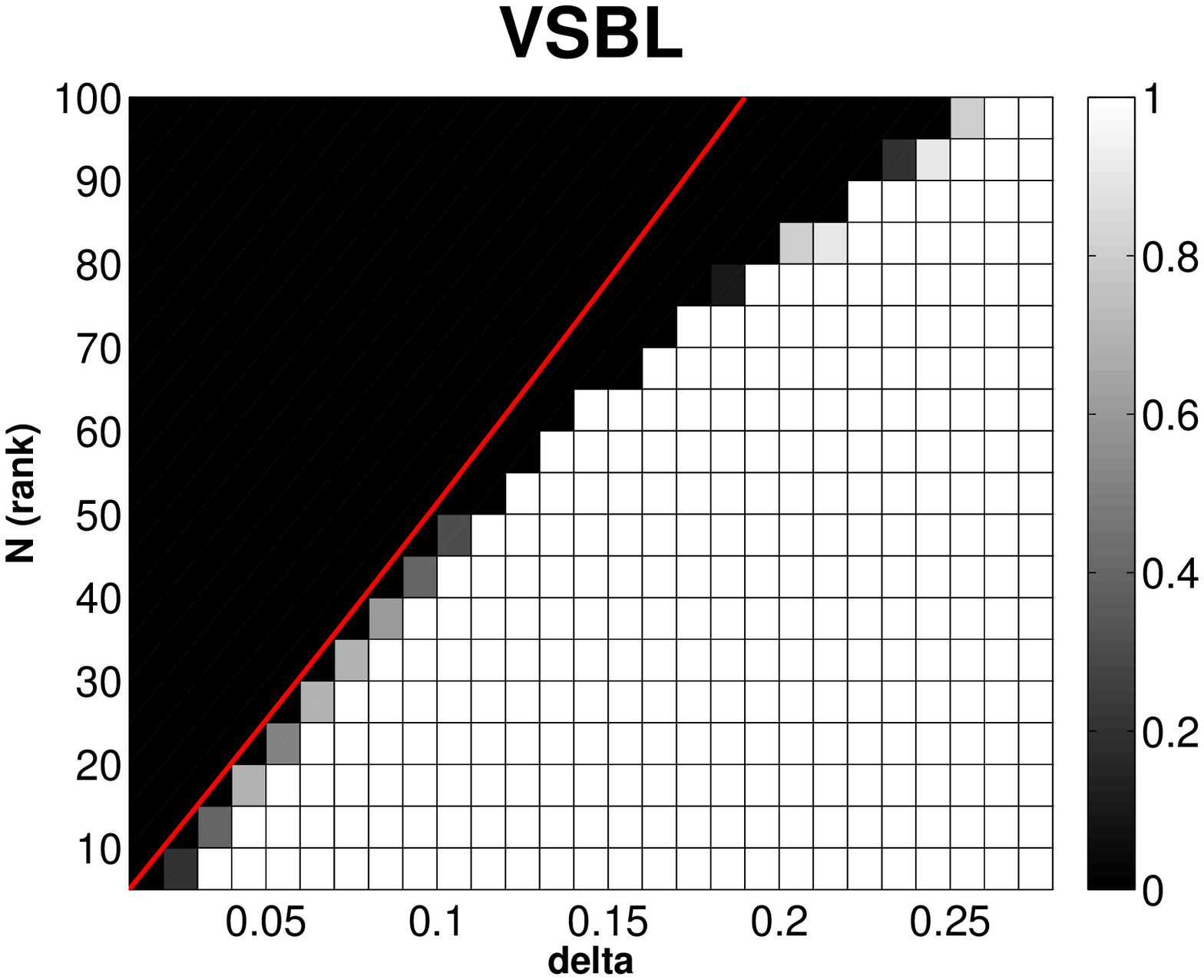}&
\includegraphics[width=1.6in]{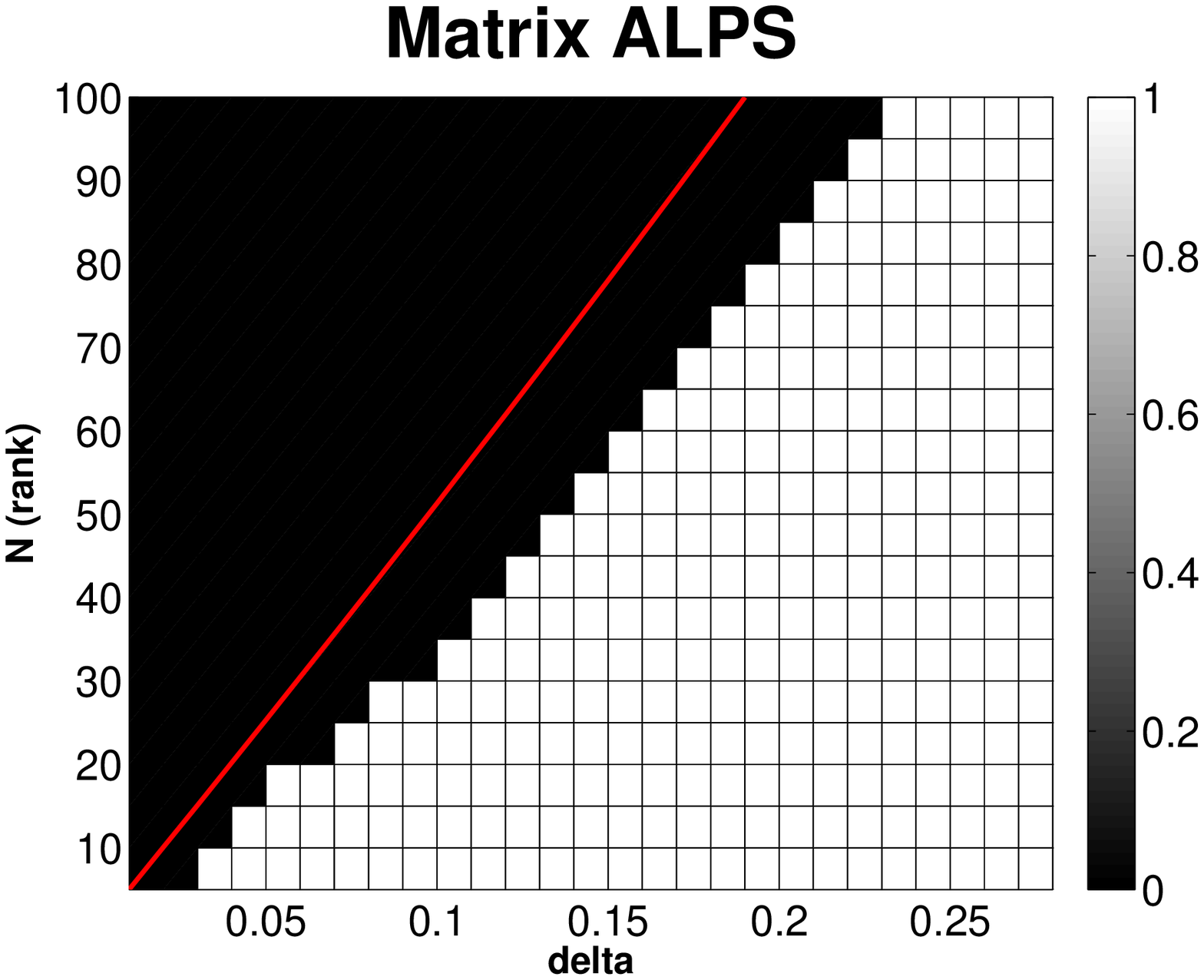}&
\includegraphics[width=1.6in]{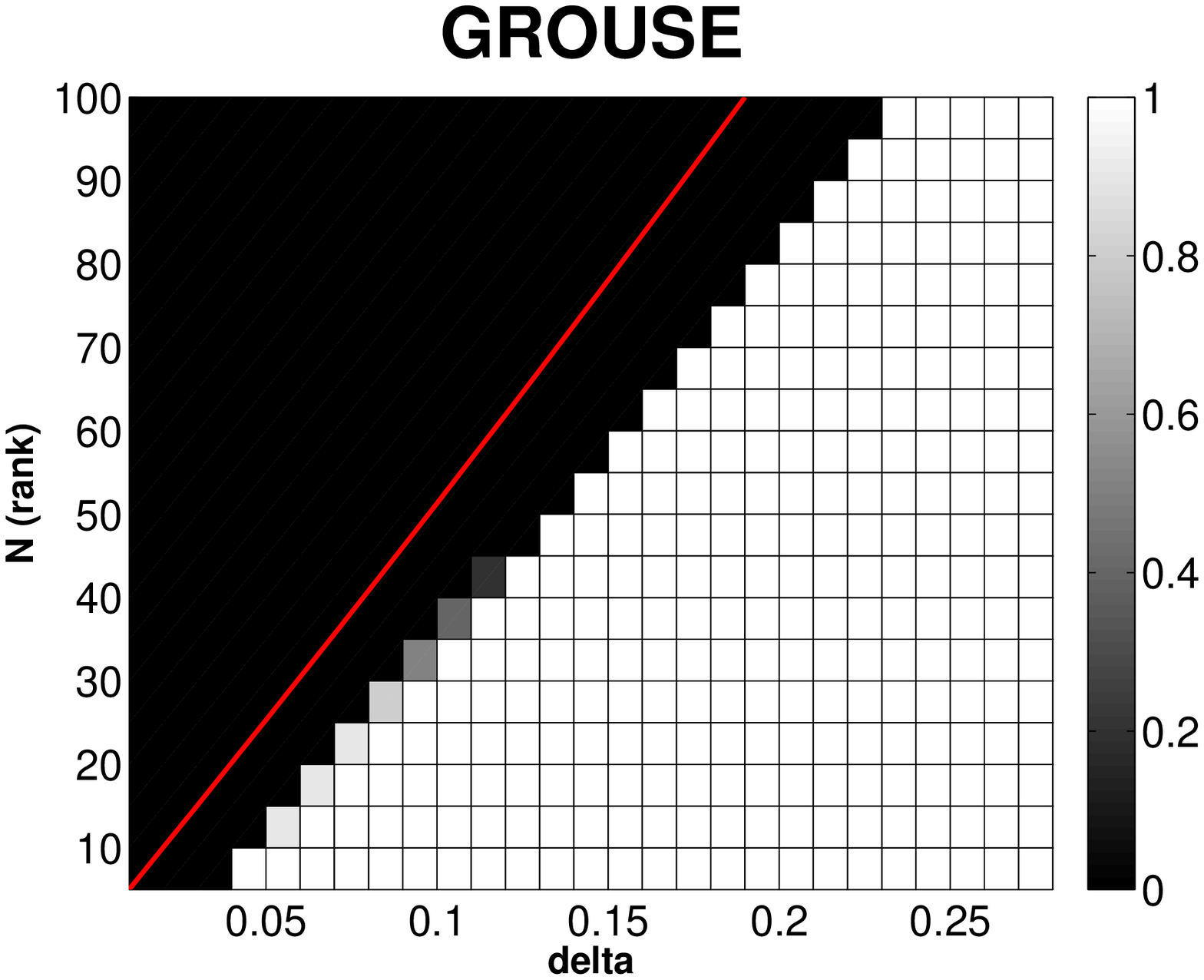}\\
\includegraphics[width=1.6in]{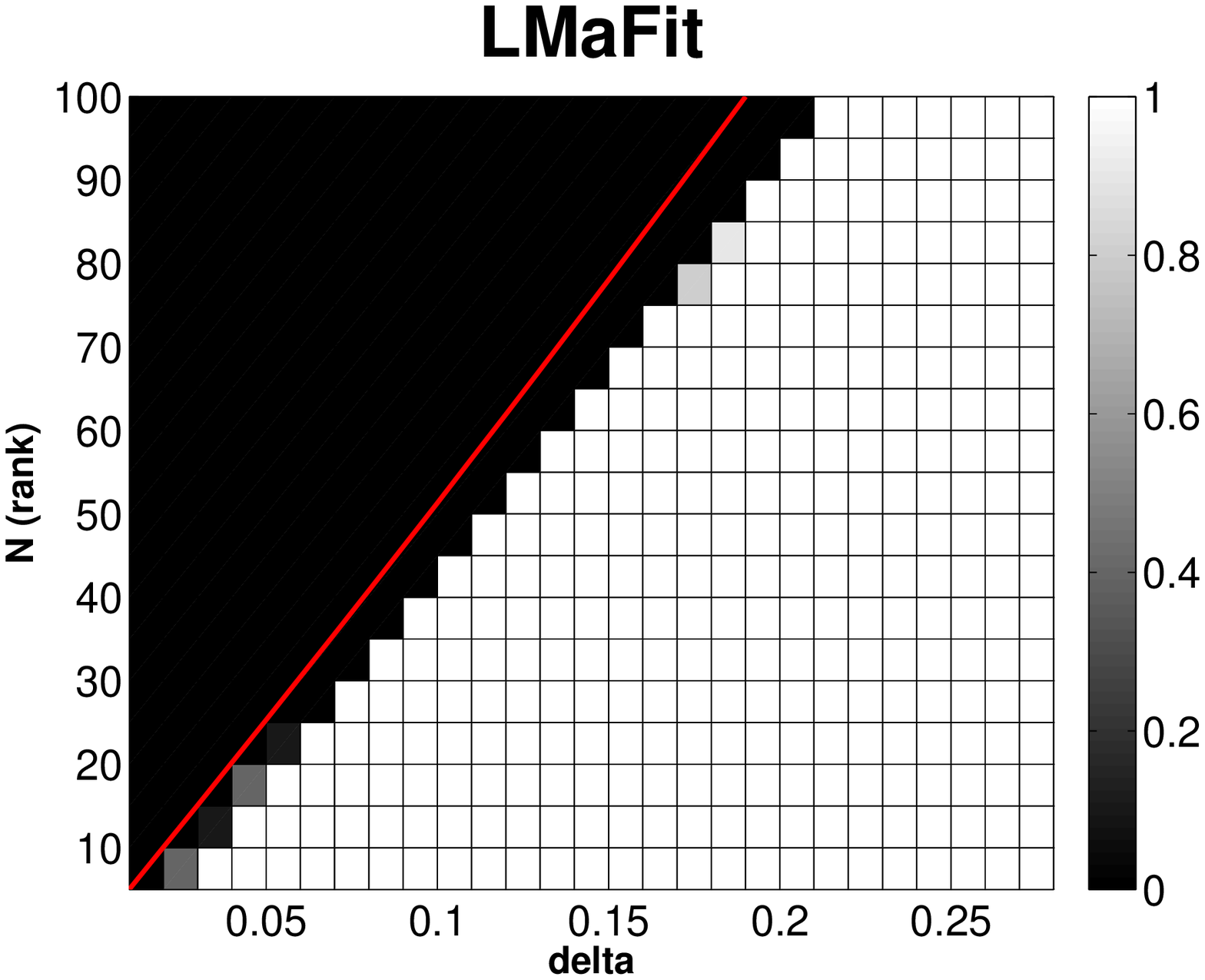}&
\includegraphics[width=1.6in]{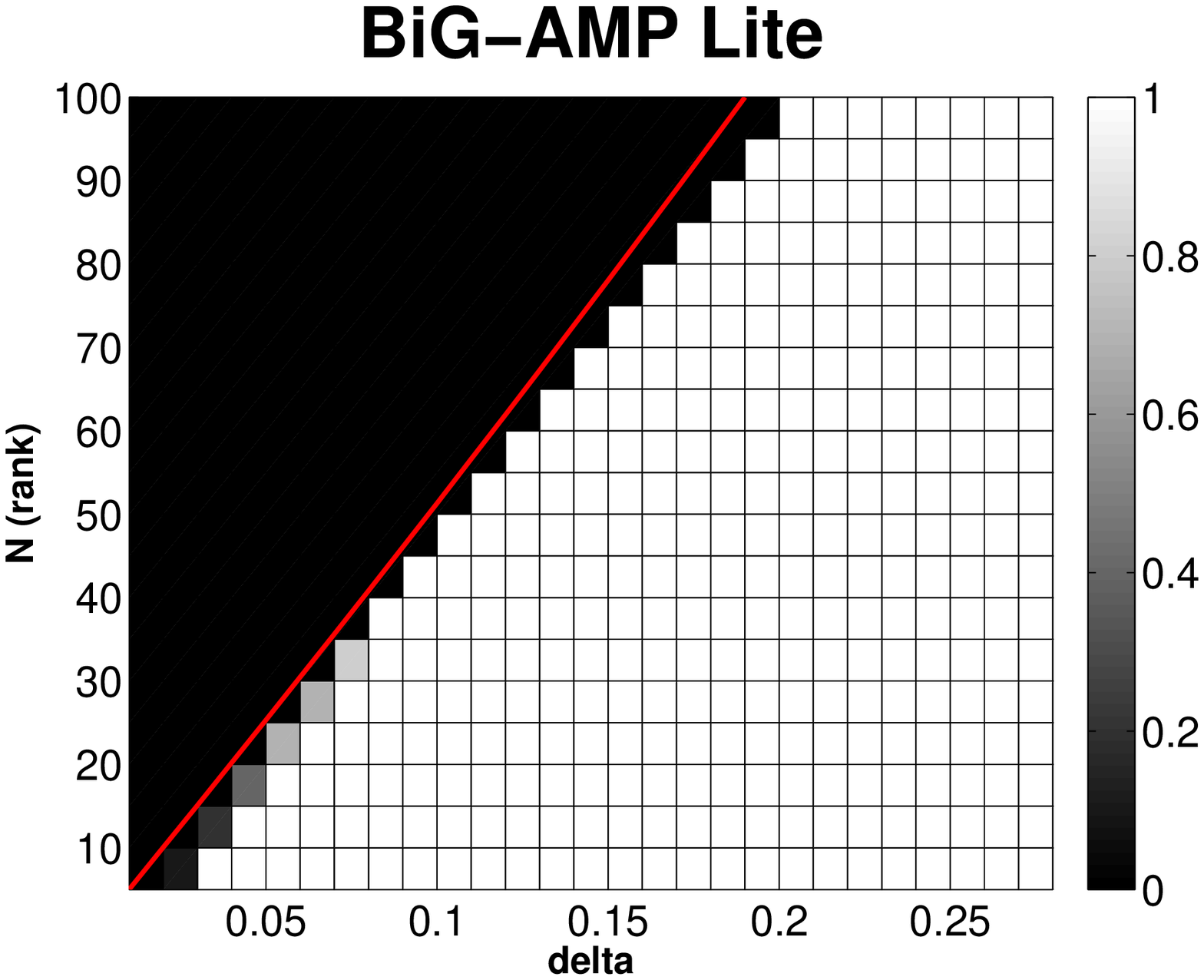}&
\includegraphics[width=1.6in]{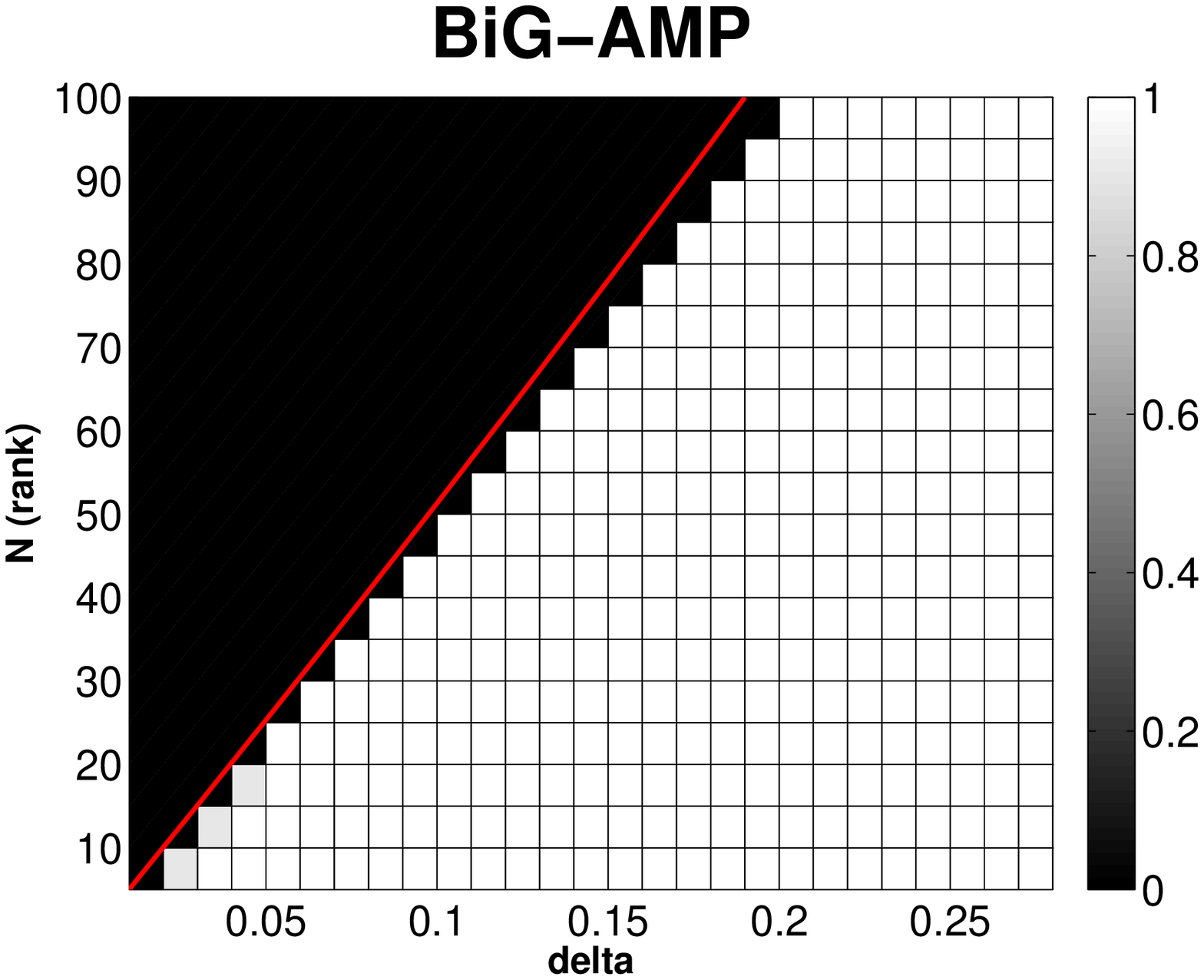}&
\includegraphics[width=1.6in]{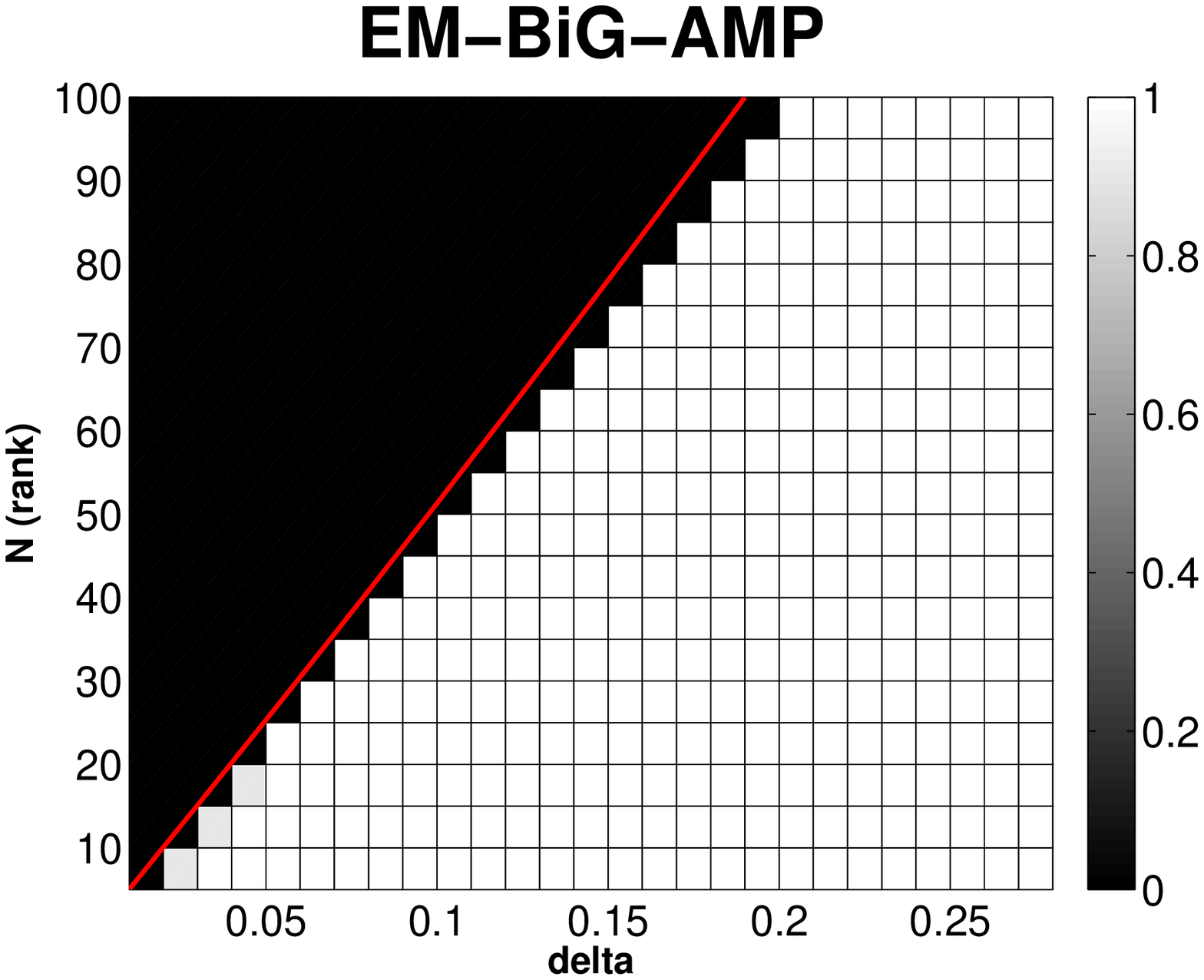}
\end{tabular}
\caption{Empirical success rates for noiseless completion of an $M\times L$ matrix sampled uniformly at random, as a function of sampling ratio $\delta=\frac{|\Omega|}{ML}$ and rank $N$.  Here, ``success'' is defined as NMSE~$<-100$~dB, success rates were computed from $10$ random realizations, and $M=L=1000$. Points above the red curve are infeasible, as described in the text.}
\label{fig:phaseMCFull}
\end{figure*}

Defining ``successful'' matrix completion as NMSE~$<-100$~dB, \figref{phaseMCFull} shows the success rate of each algorithm over a grid of sampling ratios $\delta\defn \frac{|\Omega|}{ML}$ and ranks $N$.
As a reference, the solid line superimposed on each subplot delineates the problem feasibility boundary, i.e., the values of $(\delta,N)$ yielding $|\Omega|=\df$, where $\df=N(M+L-N)$ is the degrees-of-freedom in a rank-$N$ real-valued $M\times L$ matrix; successful recovery above this line is impossible by any method.

\Figref{phaseMCFull} shows that each algorithm exhibits a sharp phase-transition separating near-certain success from near-certain failure.  
There we see that BiG-AMP yields the best PTC. 
Moreover, BiG-AMP's PTC is \emph{near optimal} in the sense of coming very close to the feasibility boundary for all tested $\delta$ and $N$.
In addition, \figref{phaseMCFull} shows that BiG-AMP-Lite yields the second-best PTC, which matches that of BiG-AMP except under very low sampling rates (e.g., $\delta<0.03$).
Recall that the only difference between the two algorithms is that BiG-AMP-Lite uses the scalar-variance simplification from \Xsecref{scalar}. 

\begin{figure*}[htb]
\psfrag{N}[][][0.8]{\sf rank $N$}
\psfrag{delta = 0.05}[][][0.8]{\sf sampling ratio $\delta = 0.05$}
\psfrag{delta = 0.1}[][][0.8]{\sf sampling ratio $\delta = 0.1$}
\psfrag{delta = 0.2}[][][0.8]{\sf sampling ratio $\delta = 0.2$}
\psfrag{delta = 0.6}[][][0.8]{\sf sampling ratio $\delta = 0.6$}
\psfrag{Z NMSE (dB)}[B][B][0.7]{\sf NMSE (dB)}
\psfrag{time (seconds)}[B][B][0.7]{\sf runtime (sec)}
\centering
 \begin{tabular}{ccc}
\includegraphics[width=2.2in]{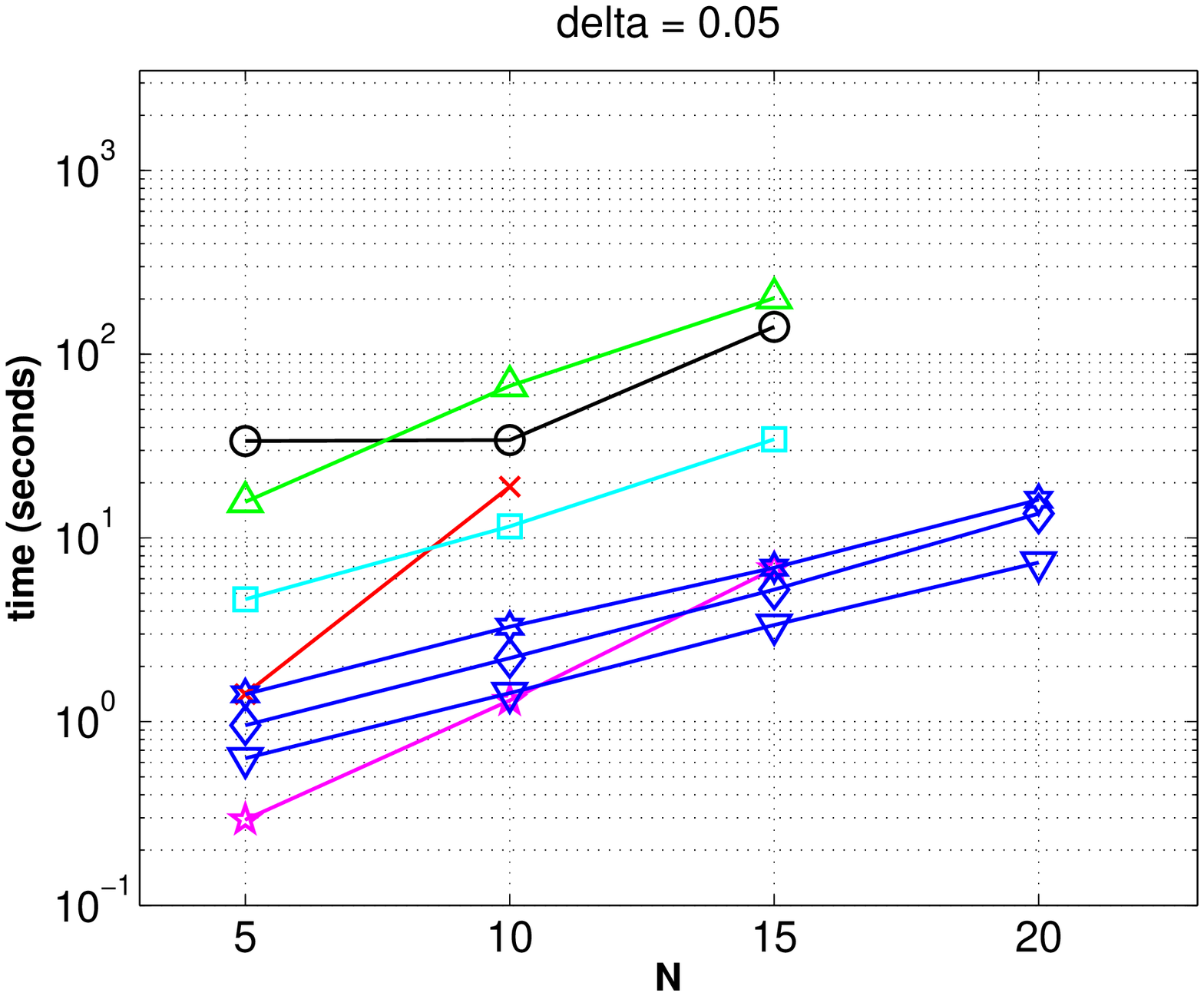}&
\includegraphics[width=2.2in]{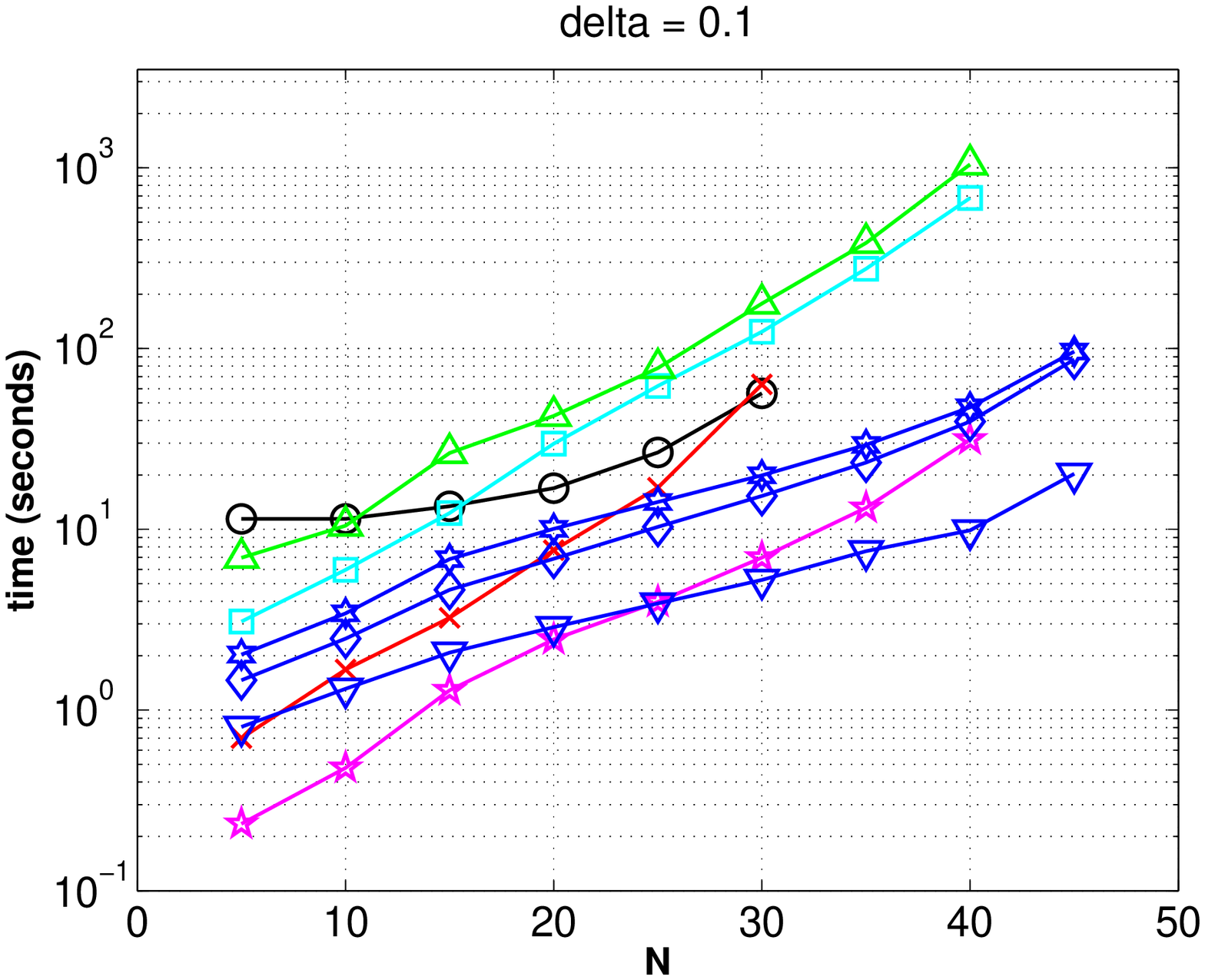}&
\includegraphics[width=2.2in]{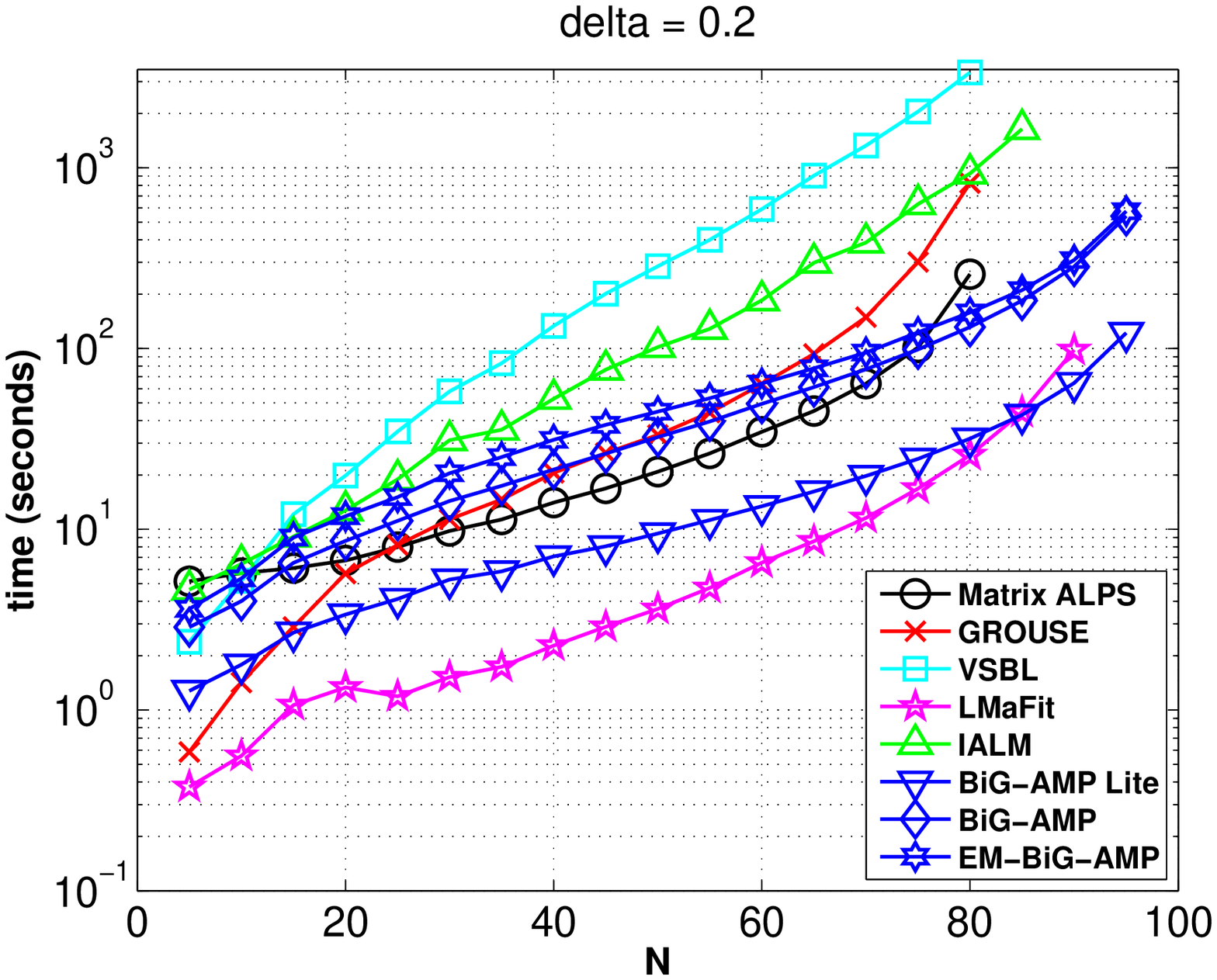}
\end{tabular}
 \caption{Runtime to NMSE$=-100$~dB for noiseless completion of an $M\times L$ matrix sampled uniformly at random, versus rank $N$, at $M=L=1000$ and several sampling ratios $\delta=\frac{|\Omega|}{ML}$. 
 All results represent median performance over $10$ trials. \textb{Missing values indicate that the algorithm did not achieve the required NMSE before termination and correspond to the black regions in \figref{phaseMCFull}.}}
 \label{fig:PhaseCutsMC}
\end{figure*}

\Figref{PhaseCutsMC} plots median runtime\footnote{The reported runtimes do not include the computations used for initialization nor those used for runtime evaluation.} to NMSE $=-100$~dB versus rank $N$ for several sampling ratios $\delta$, uncovering orders-of-magnitude differences among algorithms.
For most values of $\delta$ and $N$, LMaFit was the fastest algorithm and BiG-AMP-Lite was the second fastest, although BiG-AMP-Lite was faster than LMaFit at small $\delta$ and relatively large $N$, while BiG-AMP-Lite was slower than GROUSE at large $\delta$ and very small $N$.
In all cases, BiG-AMP-Lite was faster than IALM and VSBL, with several orders-of-magnitude difference at high rank. 
Meanwhile, EM-BiG-AMP was about $3$ to $5$ times slower than BiG-AMP-Lite.
Although none of the algorithm implementations were fully optimized, we believe that the reported runtimes are insightful, especially with regard to the scaling of runtime with rank $N$. 

\subsubsection{Approximately low-rank matrices} \label{sec:MCrankVary}
Next we evaluate the performance of recovering approximately low-rank matrices by repeating an experiment from the LMaFit paper \cite{jtp_Wen2012a}. 
For this, we constructed the complete matrix as $\vec{Z} = \vec{U\Sigma V}\tran\in\Real^{500\times 500}$, where $\vec{U}, \vec{V}$ were orthogonal matrices (built by orthogonalizing iid $\Nor(0,1)$ matrices using \textb{MATLAB's} \texttt{orth} command) and $\vec{\Sigma}$ was a positive diagonal matrix containing the singular values of $\vec{Z}$. 
Two versions of $\vec{\Sigma}$ were considered: one with exponentially decaying singular values $[\vec{\Sigma}]_{m,m} = e^{-0.3 m}$, and one with the power-law decay $[\vec{\Sigma}]_{m,m} = m^{-3}$.

\begin{figure}[htb]
\centering
\psfrag{Z NMSE (dB)}[][B][0.5]{\sf NMSE (dB)}
\psfrag{Estimated Rank}[B][][0.5]{\sf Estimated rank}
\psfrag{delta}[B][][0.5]{\sf $\delta$}
\psfrag{Power Law Decay}[B][][0.5]{\sf Power Law Decay}
\psfrag{Exponential Decay}[B][][0.5]{\sf Exponential Decay}
 \begin{tabular}{@{\hspace{-2mm}}c@{\hspace{-3.5mm}}c@{}}
\includegraphics[width=1.9in]{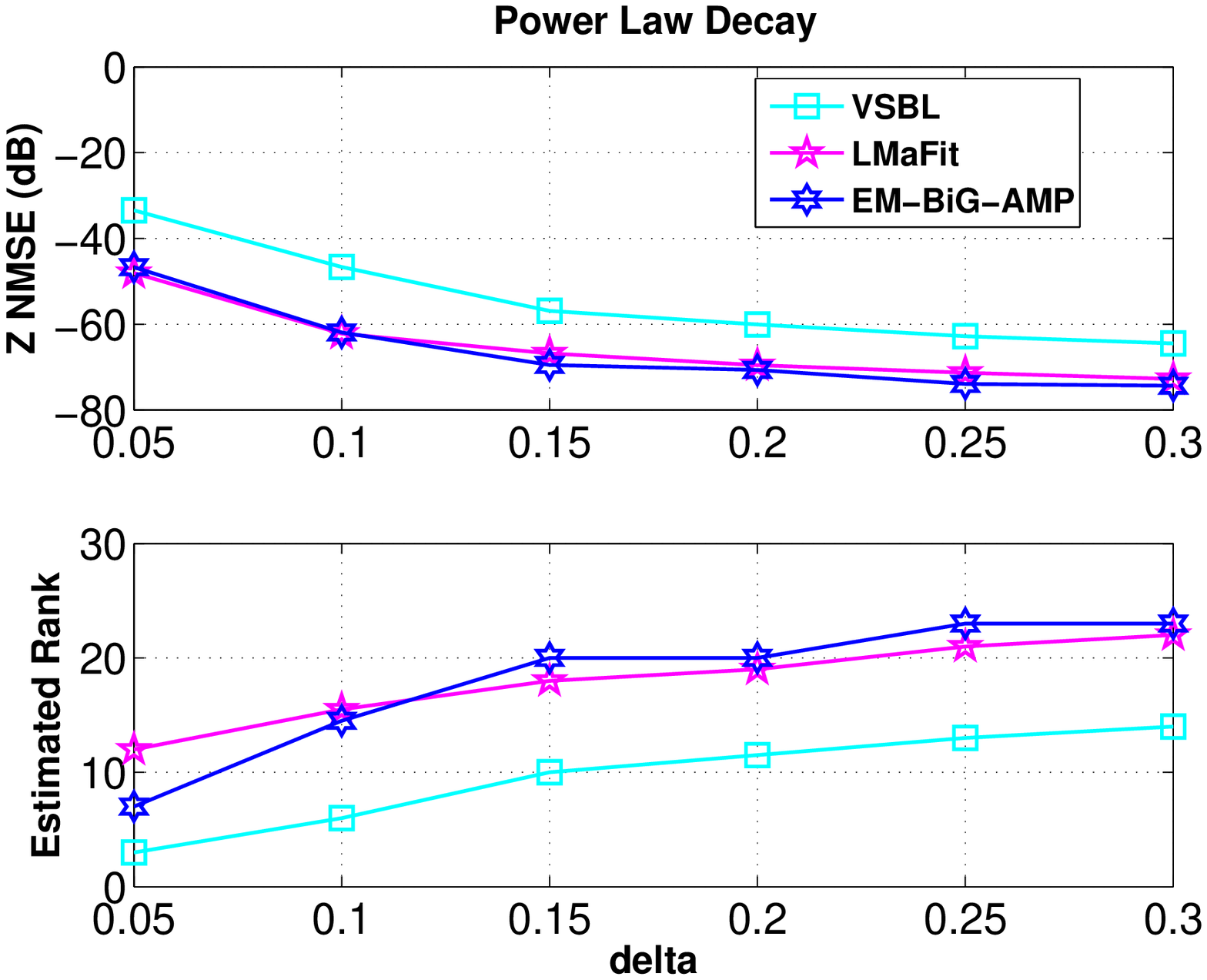}&
\includegraphics[width=1.9in]{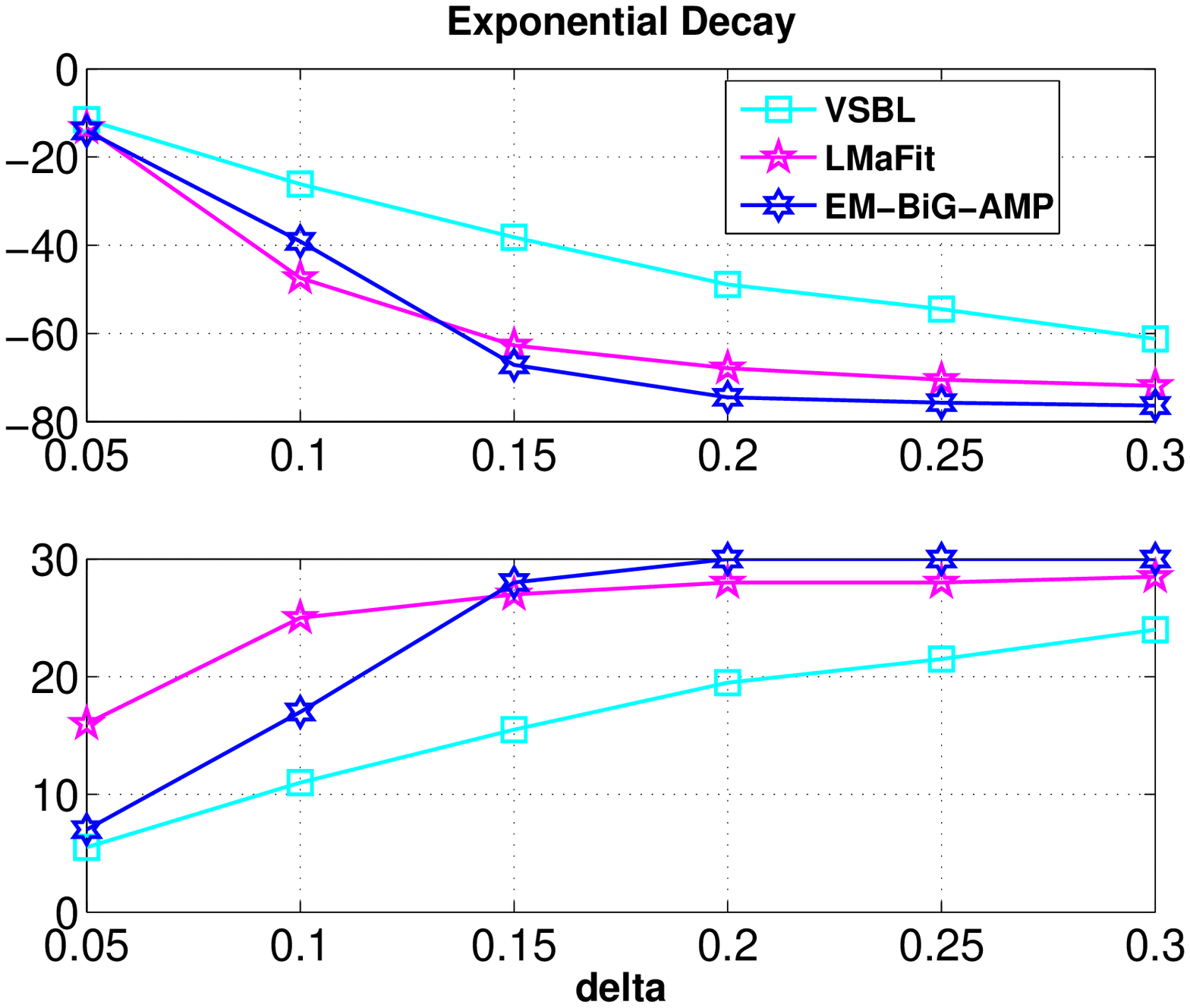}
\end{tabular}
 \caption{NMSE (top) and estimated rank (bottom) in noiseless completion of an $M\times L$ matrix sampled uniformly at random, versus sampling ratio $\delta=\frac{|\Omega|}{ML}$.  The complete matrices were approximately low-rank, with power-law (left) and exponential (right) singular-value decays and $M=L=500$. All results represent median performance over $10$ random trials.}
 \label{fig:fullRank}
\end{figure}

As in \cite{jtp_Wen2012a}, we first tried to recover $\vec{Z}$ from the noiseless incomplete observations $\{z_{ml}\}_{(m,l)\in\Omega}$, with $\Omega$ chosen uniformly at random.
\Figref{fullRank} shows the performance of several algorithms that are able to learn the underlying rank: 
LMaFit,\footnote{LMaFit was run under the settings provided in their source code for this example.} 
VSBL,\footnote{VSBL was allowed at most $100$ iterations and run with \texttt{DIMRED\_THR} $=10^3$, \texttt{UPDATE\_BETA} $=1$, and tolerance $=10^{-8}$.}
and EM-BiG-AMP under the penalized log-likelihood rank selection strategy from \Xsecref{AIC}.\footnote{Rank-selection rule \Xeqref{orderCriteria} was used with up to $5$ EM iterations for each rank hypothesis $N$, a minimum of $30$ and maximum of $100$ BiG-AMP iterations for each EM iteration, and a BiG-AMP tolerance of $10^{-8}$.}
All three algorithms were allowed a maximum rank of $\bar{N}=30$. 
The figure shows that the NMSE performance of BiG-AMP and LMaFit are similar, although BiG-AMP tends to find solutions with lower rank but comparable NMSE at low sampling ratios $\delta$. 
For this noiseless experiment, VSBL consistently estimates ranks that are too low, leading to inferior NMSEs. 

\begin{figure}[htb]
\psfrag{Z NMSE (dB)}[][B][0.5]{\sf NMSE (dB)}
\psfrag{Estimated Rank}[B][][0.5]{\sf Estimated rank}
\psfrag{delta}[B][][0.5]{\sf $\delta$}
\psfrag{SNR (dB)}[B][B][0.5]{\sf SNR (dB)}
\psfrag{Power Law Decay}[B][][0.5]{\sf Power Law Decay}
\psfrag{Exponential Decay}[B][][0.5]{\sf Exponential Decay}
\centering
 \begin{tabular}{@{\hspace{-2mm}}c@{\hspace{-3.5mm}}c@{}}
\includegraphics[width=1.9in]{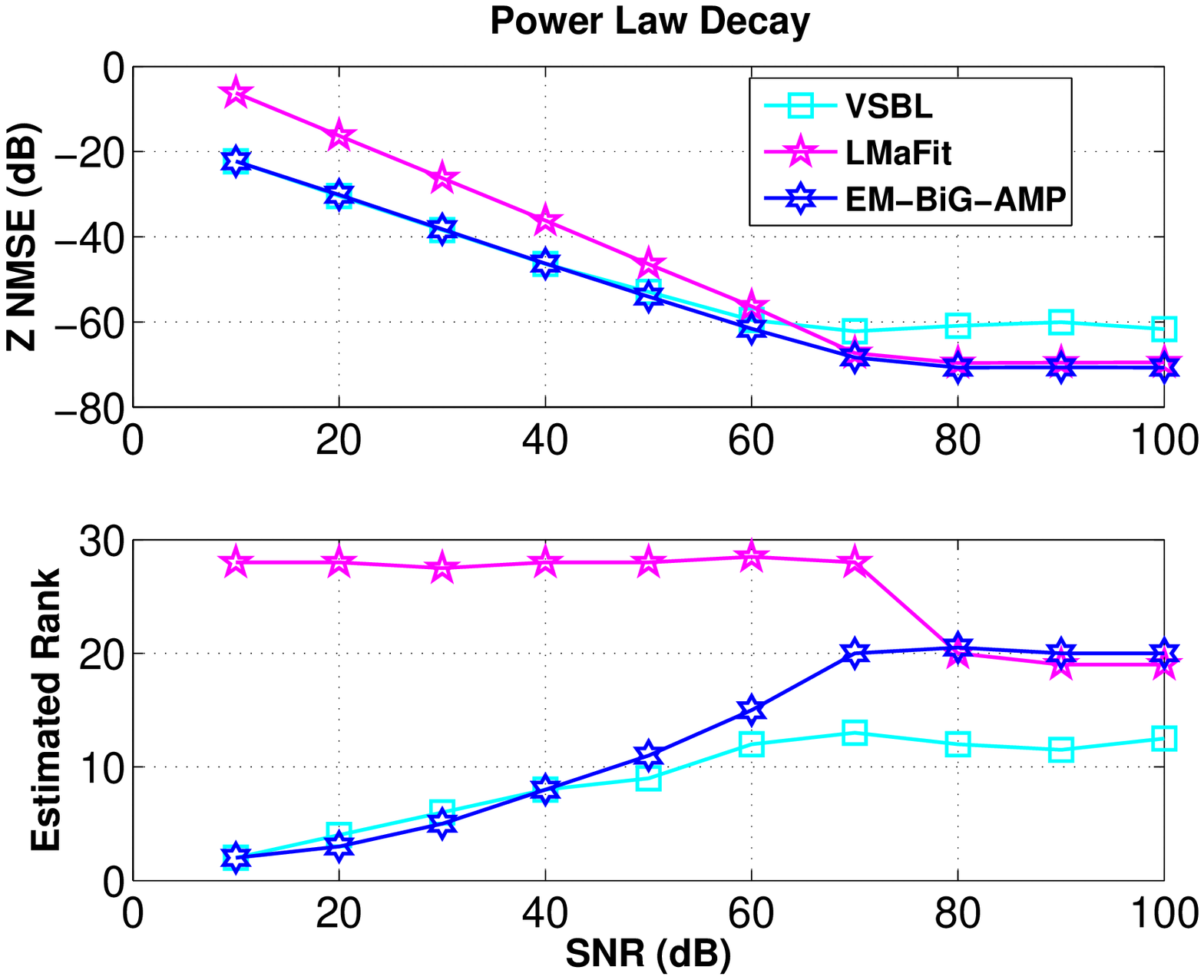}&
\includegraphics[width=1.9in]{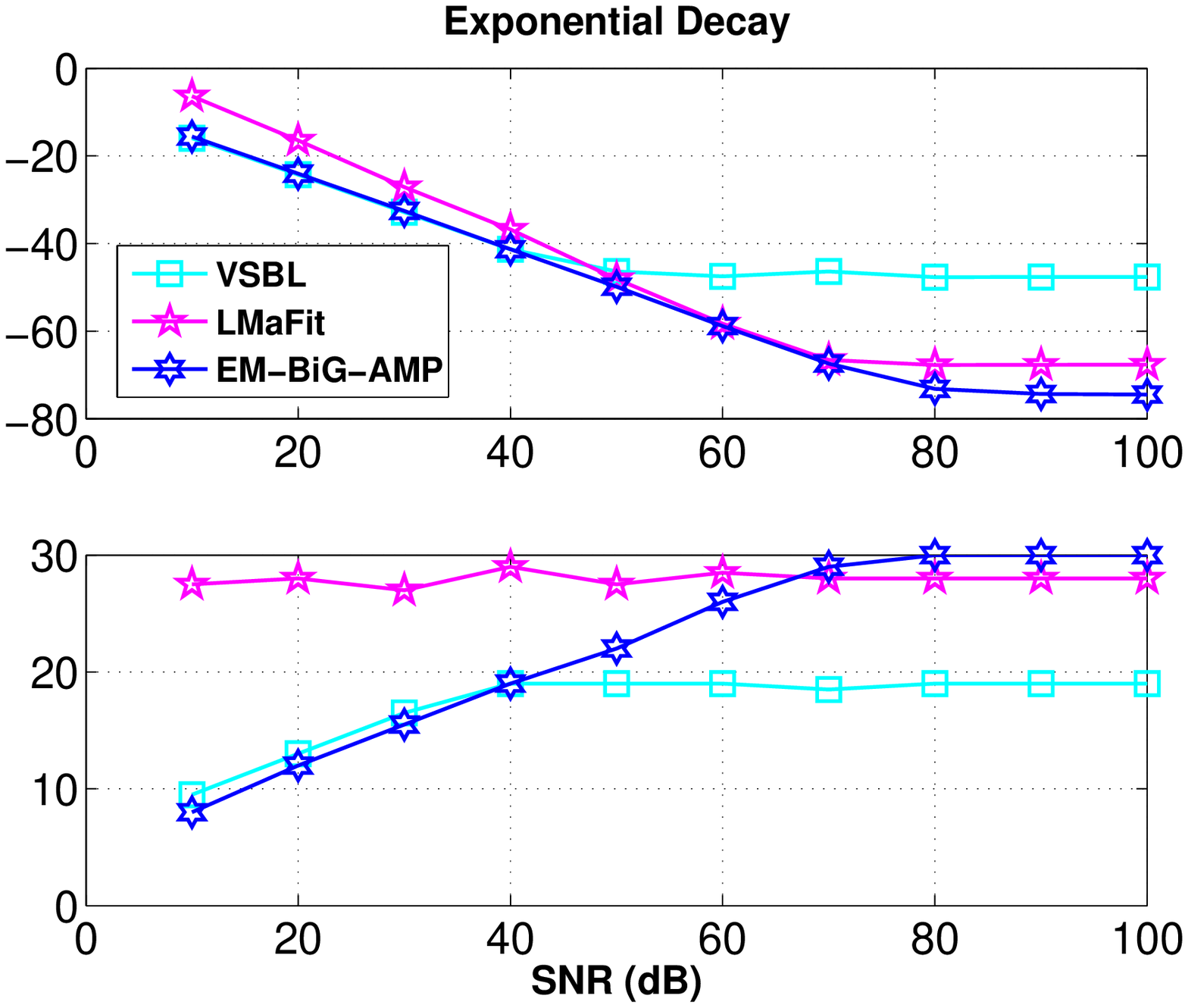}
\end{tabular}
 \caption{NMSE (top) and estimated rank (bottom), versus SNR, in noisy completion of an $500\times 500$ matrix sampled uniformly at random at rate $\delta=0.2$.  The complete matrices were approximately low-rank, with power-law (left) and exponential (right) singular-value decays.}
 \label{fig:fullRank2}
\end{figure}

Next, we examined \emph{noisy} matrix completion by constructing the matrix $\vec{Z}=\vec{U\Sigma V}\tran$ as above but then corrupting the measurements with AWGN.
\Figref{fullRank2} shows NMSE and estimated rank versus the measurement signal-to-noise ratio (SNR)
$\sum_{(m,l)\in\Omega}|z_{ml}|^2/\sum_{(m,l)\in\Omega}|y_{ml}-z_{ml}|^2$ at a sampling rate of $\delta=0.2$. 
There we see that, for SNRs~$<50$~dB, EM-BiG-AMP and VSBL offer similarly good NMSE performance and nearly identical rank estimates, whereas LMaFit overestimates the rank and thus performs worse in NMSE.
Meanwhile, for SNRs~$>50$~dB, EM-BiG-AMP and LMaFit offer similarly good NMSE performance and nearly identical rank estimates, whereas VSBL underestimates the rank and thus performs worse in NMSE.
Thus, in these examples, EM-BiG-AMP is the only algorithm to successfully estimate the rank across the full SNR range.

\subsubsection{Image completion}

\begin{figure*}[htb]
\centering
 \begin{tabular}{ccc}
\includegraphics[width=2.2in]{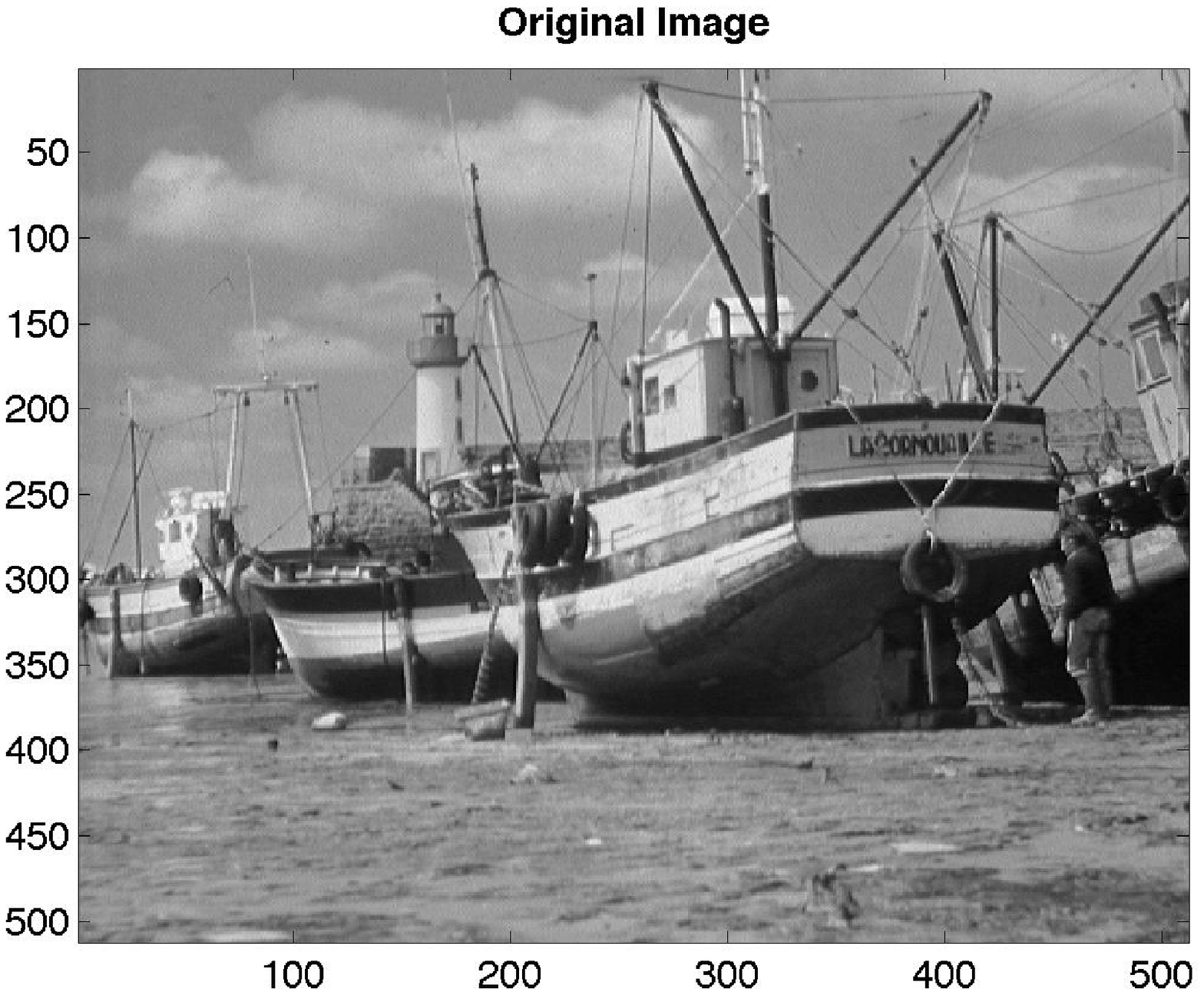}&
\includegraphics[width=2.2in]{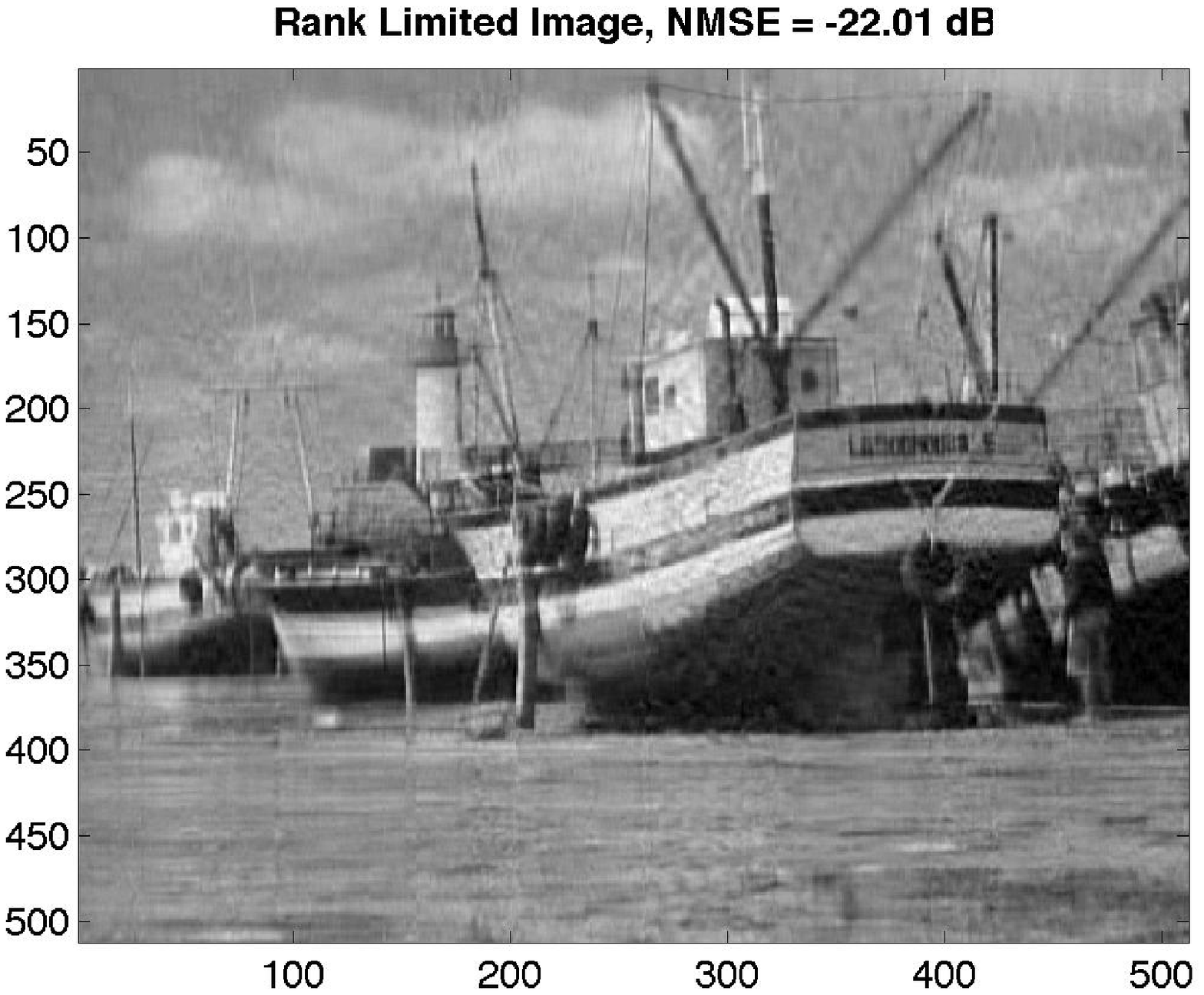}&
\includegraphics[width=2.2in]{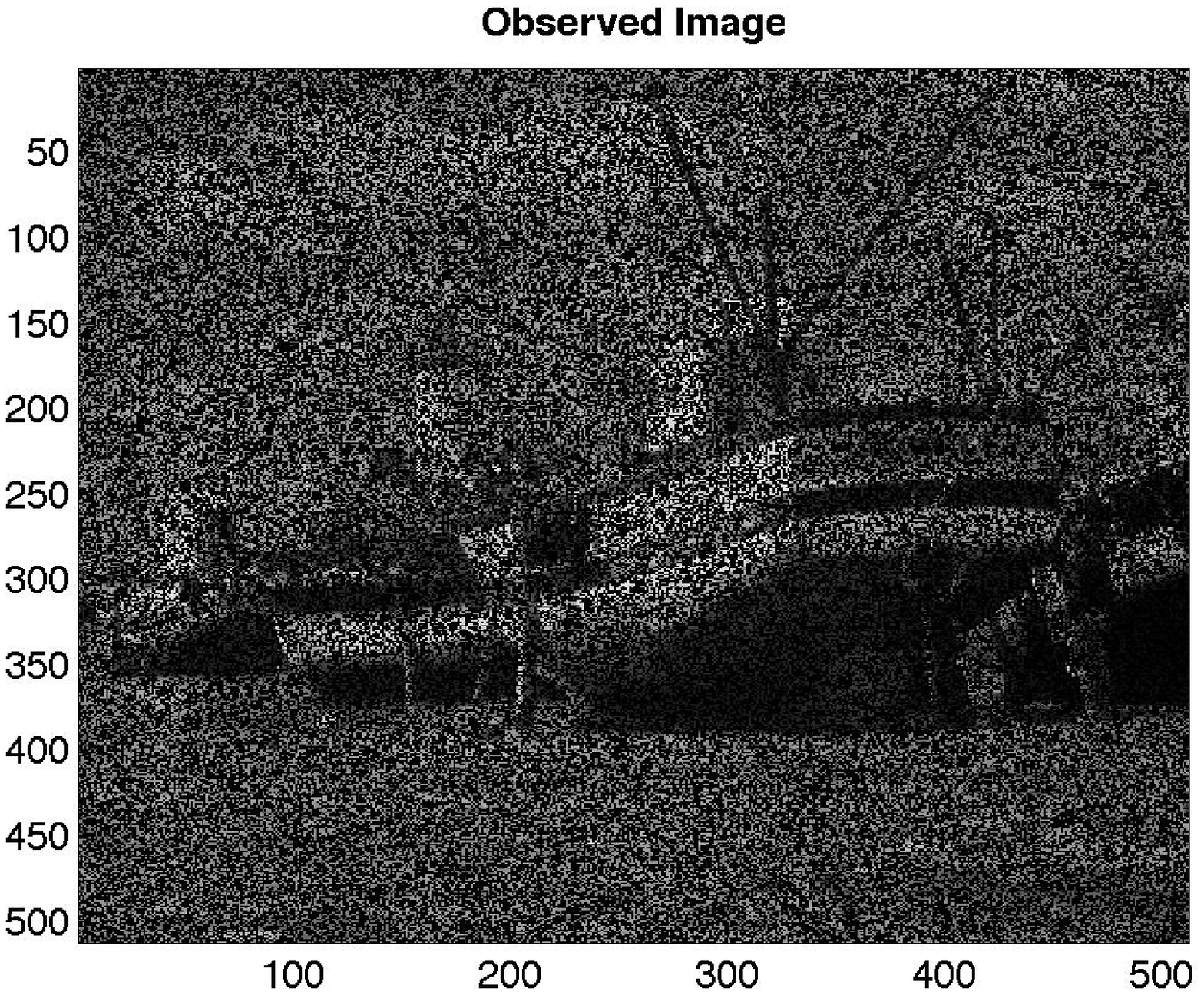}\\
\includegraphics[width=2.2in]{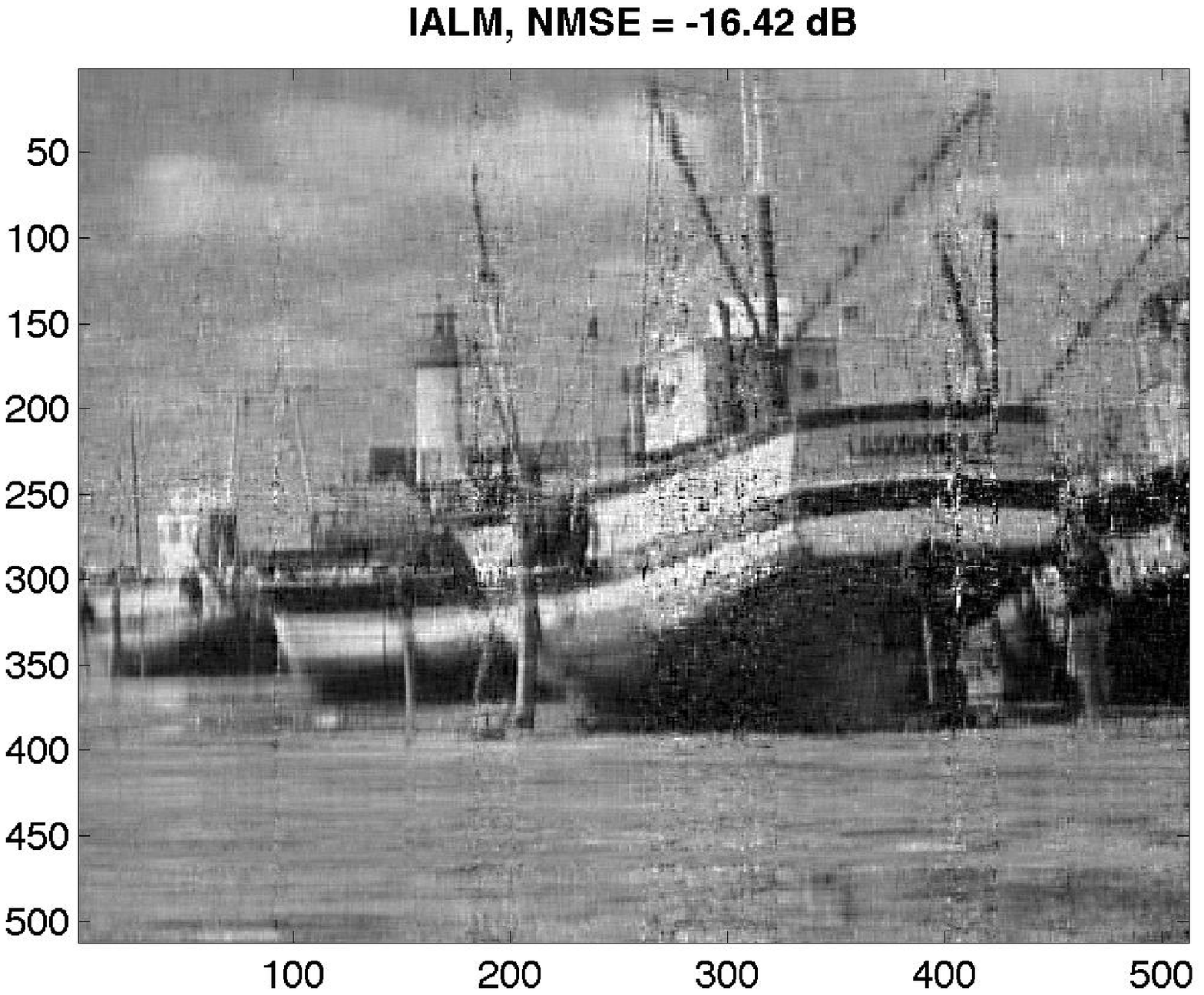}&
\includegraphics[width=2.2in]{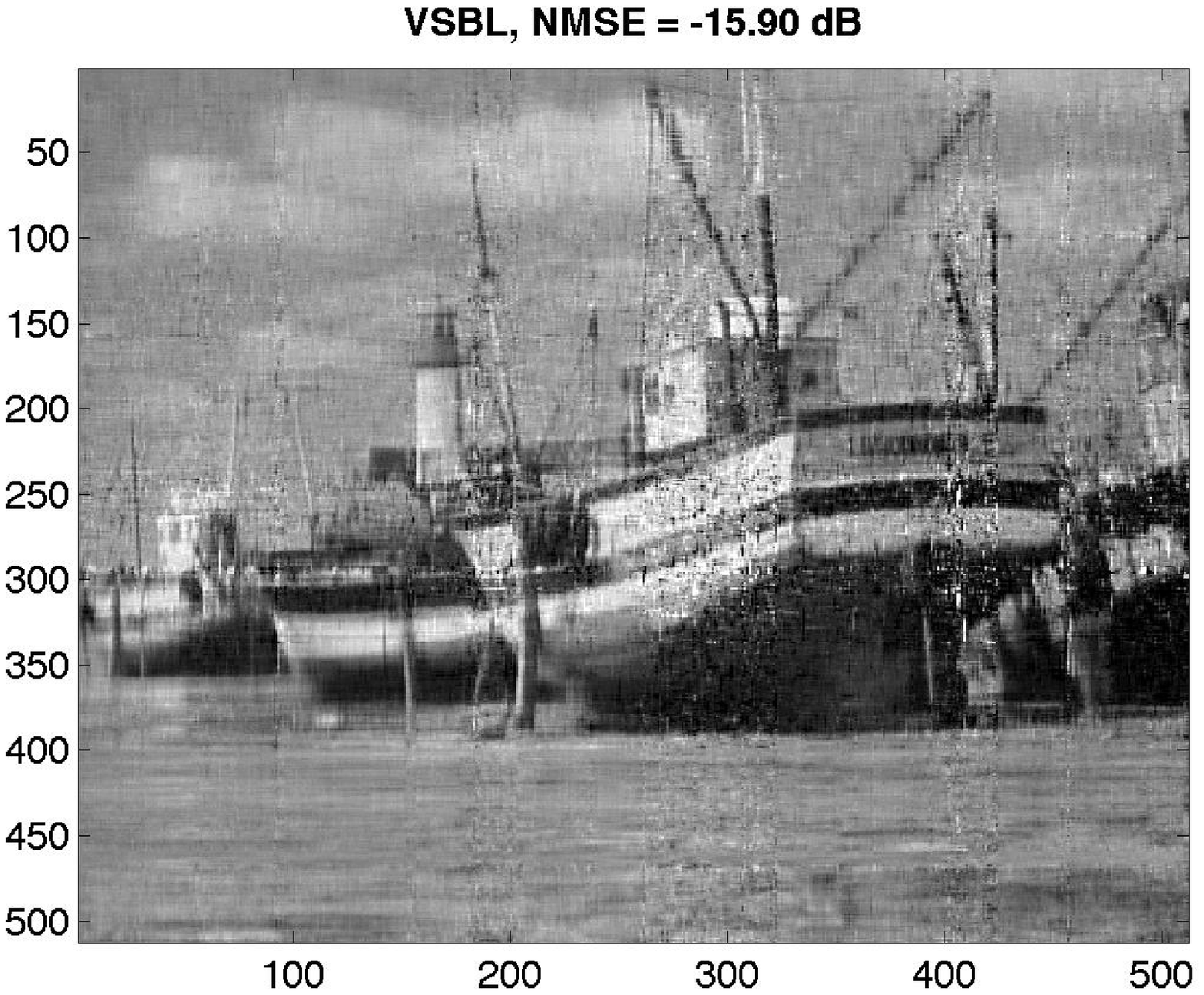}&
\includegraphics[width=2.2in]{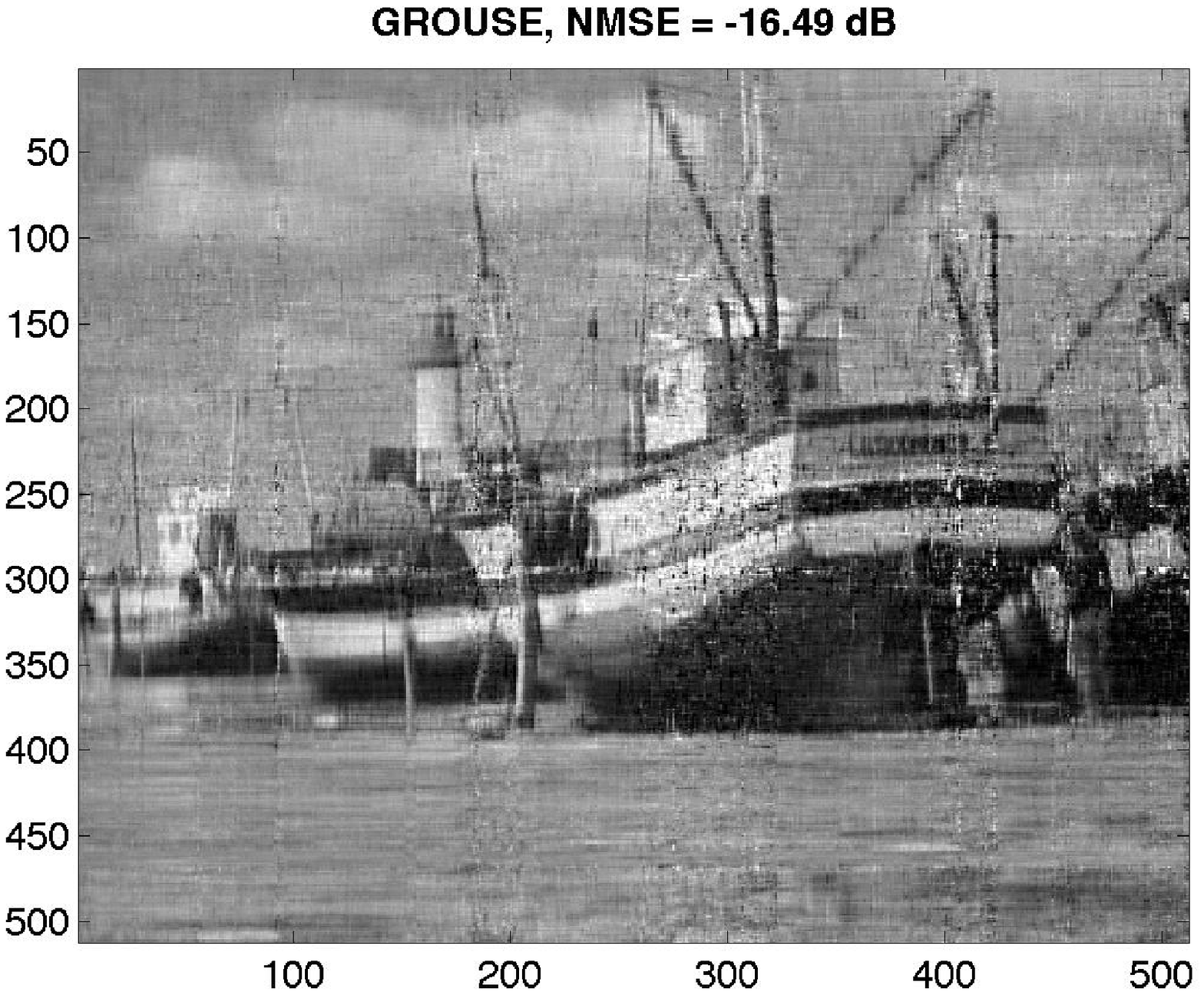}\\
\includegraphics[width=2.2in]{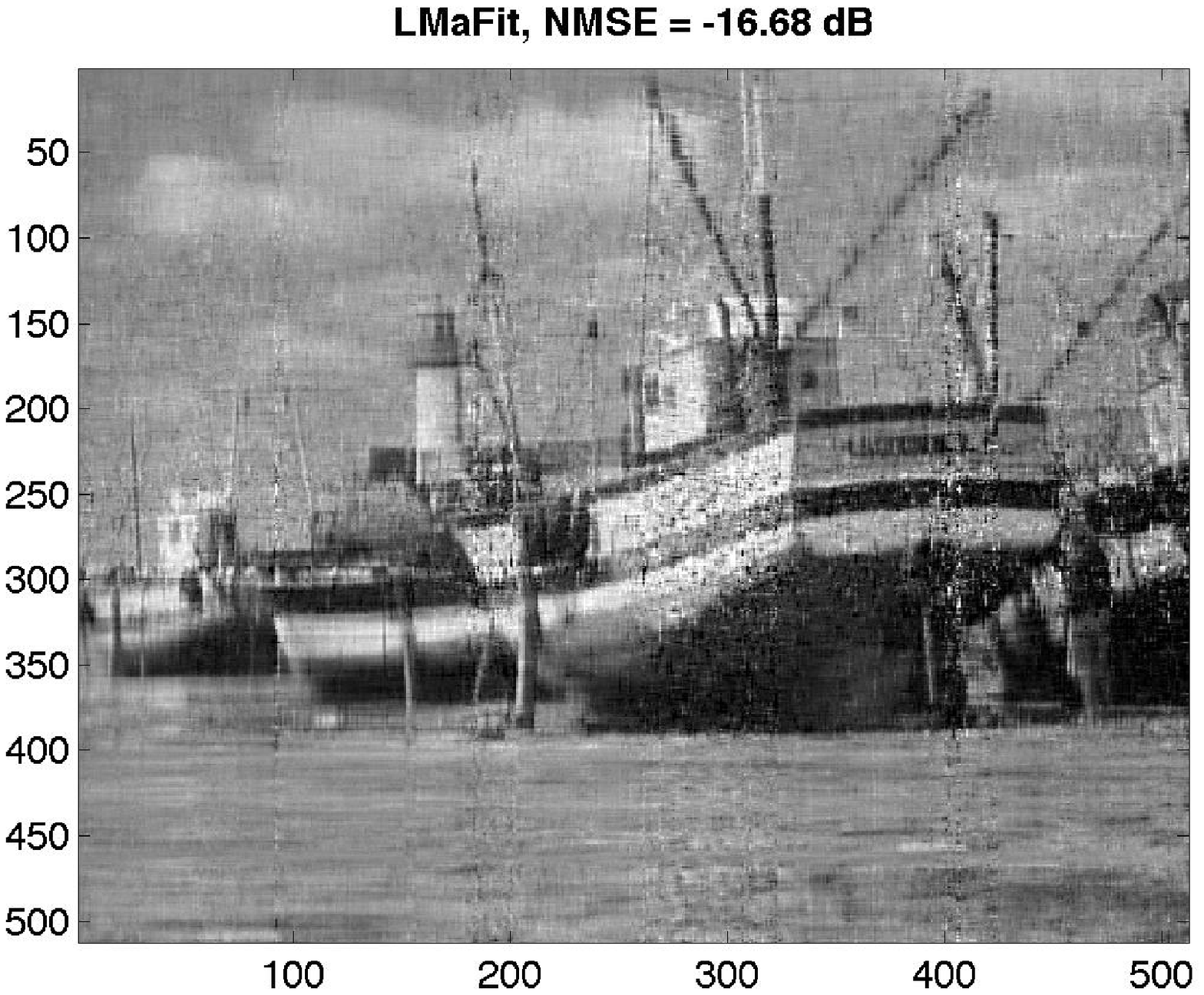}&
\includegraphics[width=2.2in]{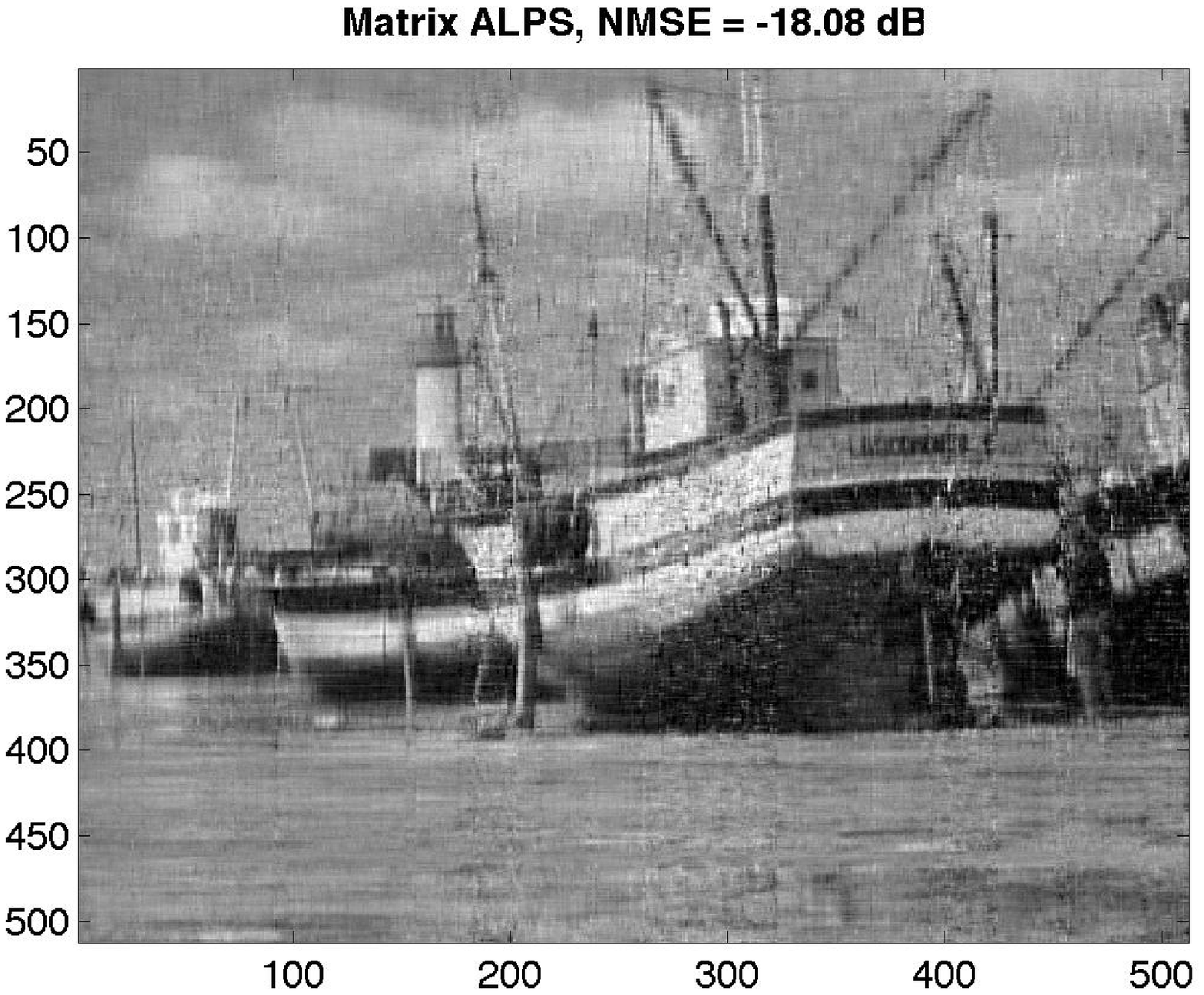}&
\includegraphics[width=2.2in]{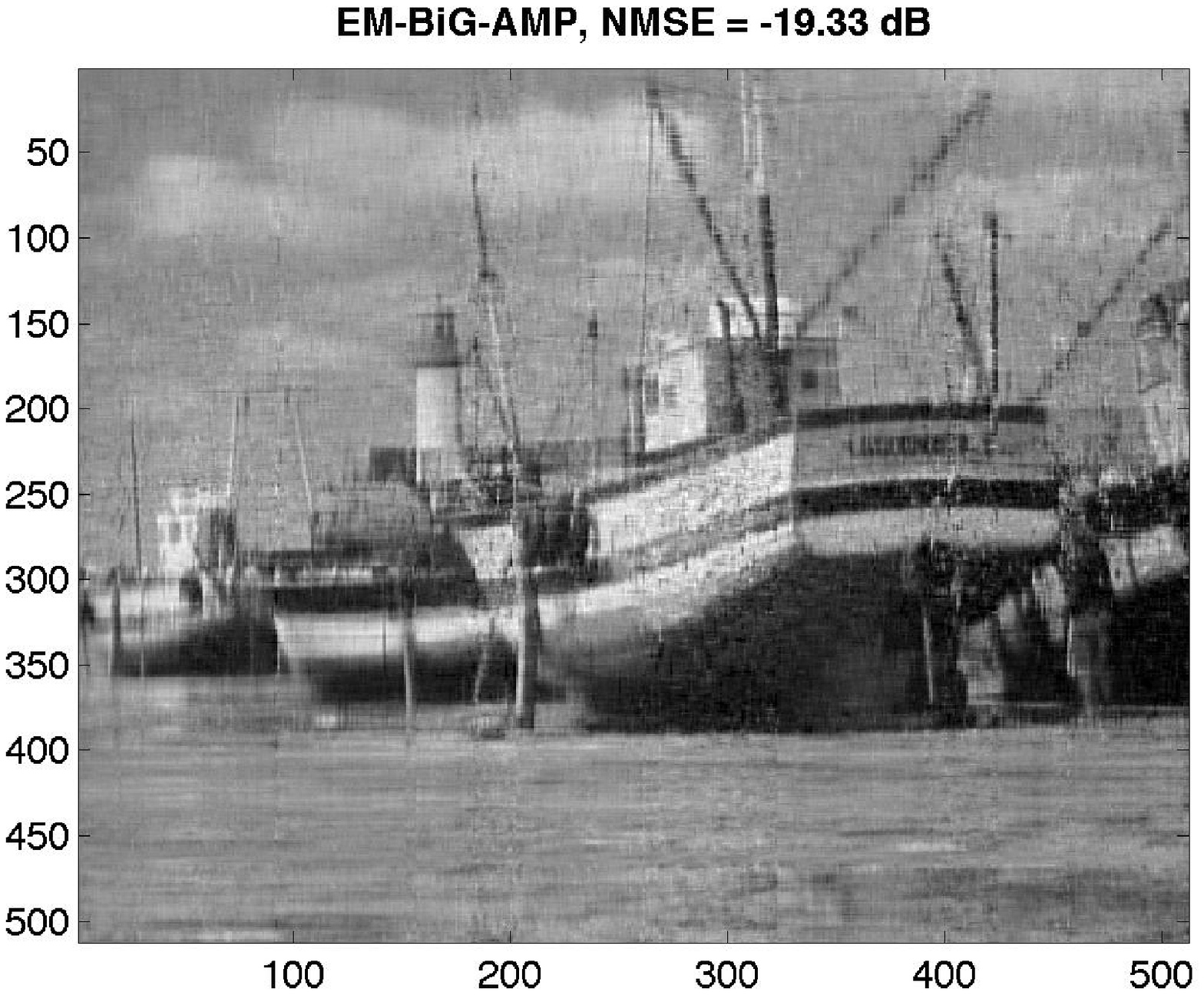}
\end{tabular}
\caption{For the image completion experiment, the complete image is shown on the top left, its best rank-$40$ approximation is shown on the top middle, and the observed image with $35\%$ of the pixels observed is shown on the top right.  The other panes show various algorithms' image reconstructions from $35\%$ of the complete-image pixels (selected uniformly at random) as well as the mean NMSE over $10$ trials.}
 \label{fig:boat}
\end{figure*}

We now compare the performance of several matrix-completion algorithms for the task of reconstructing an image from a subset of its pixels.
For this, we repeated the experiment in the Matrix-ALPS paper~\cite{Kyrillidis:12}, where the $512\times 512$ boat image was reconstructed from $35\%$ of its pixels sampled uniformly at random.
\Figref{boat} shows the complete (full-rank) image, 
the images reconstructed by several matrix-completion algorithms\footnote{All algorithms were run with a convergence tolerance of $10^{-4}$.  VSBL was run with $\beta$ hand-tuned to maximize performance, as the adaptive version did not converge on this example. GROUSE was run with $\texttt{maxCycles} = 600$ and $\texttt{step\_size} = 0.1$. Matrix-ALPS II with QR was run under default parameters and $300$ allowed iterations. Other settings are similar to earlier experiments.}
under a fixed rank of $N=40$, 
and the NMSE-minimizing rank-$40$ approximation of the complete image, computed using an SVD. 
In all cases, the sample mean of the observations was subtracted prior to processing and then added back to the estimated images, since this approach generally improved performance.
\Figref{boat} also lists the median reconstruction NMSE over $10$ sampling-index realizations $\Omega$.
From these results, it is apparent that EM-BiG-AMP provides the best NMSE, which is only $3$~dB from that of the NMSE-optimal rank-$40$ approximation.

\subsubsection{Collaborative Filtering}
In our final experiment, we investigate the performance of several matrix-completion algorithms on the task of collaborative filtering.
For this, we repeated an experiment from the VSBL paper \cite{Babacan:TSP:12} that used the MovieLens 100k dataset, which contains ratings $\{z_{ml}\}_{(m,l)\in\mc{R}}$, where $|\mc{R}|=100\,000$ and $z_{ml}\in\{1,2,3,4,5\}$, from $M=943$ users about $L=1682$ movies. 
The algorithms were provided with a randomly chosen training subset $\{z_{ml}\}_{(m,l)\in\Omega}$ of the ratings (i.e., $\Omega\subset\mc{R}$) from which they estimated the unseen ratings $\{\hat{z}_{ml}\}_{(m,l)\in\mc{R}\setminus\Omega}$.
Performance was then assessed by computing the Normalized Mean Absolute Error (NMAE) 
\begin{align}
\textrm{NMAE} &= \frac{1}{4|\mc{R}\setminus\Omega|}\sum_{(m,l) \in \mc{R}\setminus\Omega} |z_{ml} - \hat{z}_{ml}| ,	\label{eq:NMAE}
\end{align} 
where the $4$ in the denominator of \eqref{NMAE} reflects the difference between the largest and smallest user ratings (i.e., $5$ and $1$).
When constructing $\Omega$, we used a fixed percentage of the ratings given by each user and made sure that at least one rating was provided for every movie in the training set. 

\begin{figure}[htb]
\psfrag{NMAE}[B][B][0.6]{\sf NMAE}
\psfrag{Estimated Rank}[B][B][0.6]{\sf Estimated rank}
\psfrag{Percentage used for Training}[B][B][0.6]{\sf Training fraction $|\Omega|/|\mc{R}|$}
\psfrag{VSBL}[l][l][0.47]{\sf VSBL}
\psfrag{LMaFit}[l][l][0.47]{\sf LMaFit}
\psfrag{EM-BiG-AMP}[l][l][0.47]{\sf EM-BiG-AMP assuming PIAWGN}
\psfrag{EM-BiG-AMP Laplacian Noise}[l][l][0.47]{\sf EM-BiG-AMP assuming PIAWLN}
\centering
 \begin{tabular}{c}
\includegraphics[width=\columnwidth]{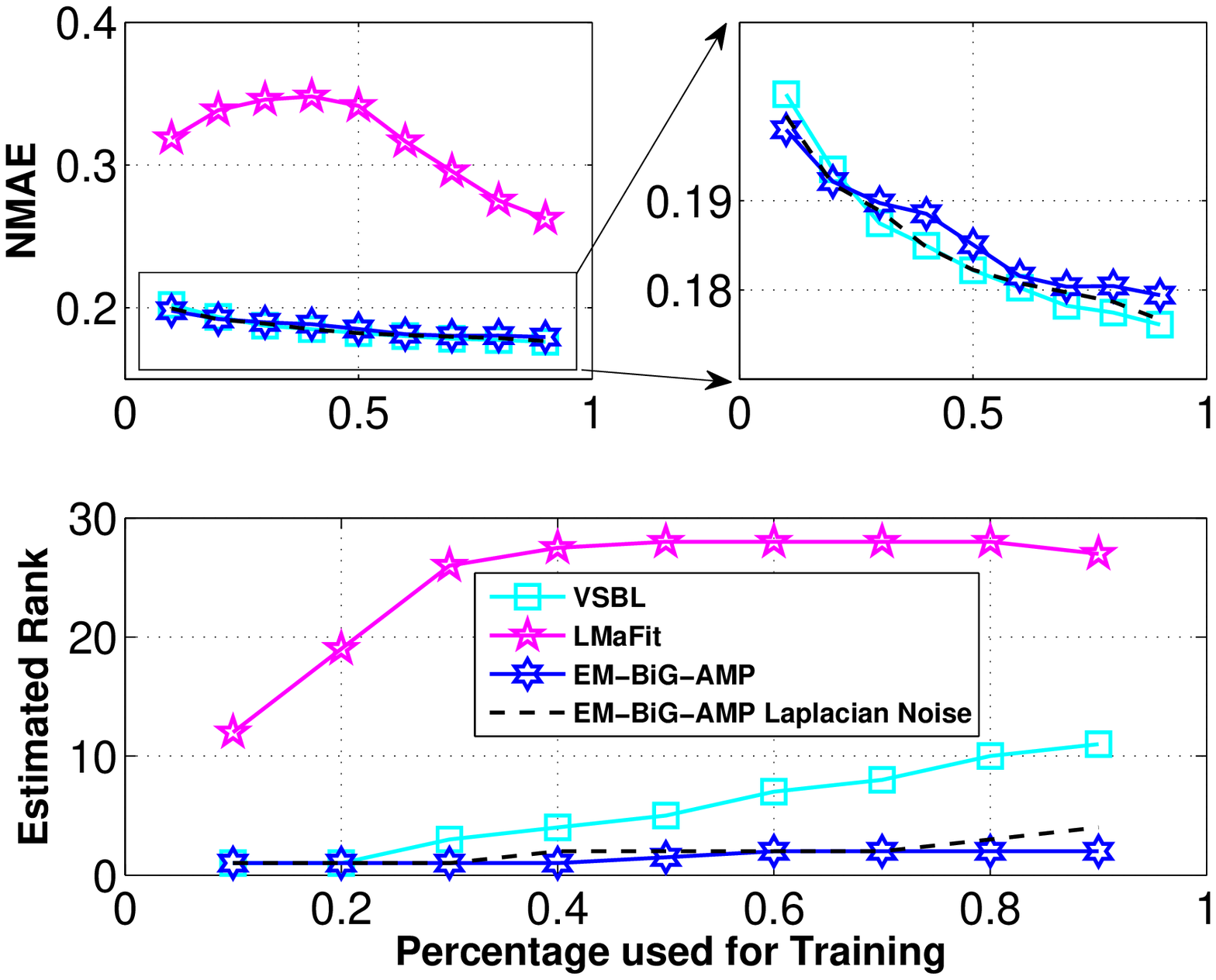}
\end{tabular}
 \caption{Median NMAE (top) and estimated rank (bottom) for movie-rating prediction versus fraction of training data $|\Omega|/|\mc{R}|$ over $10$ trials for the 100k MovieLens data set. }
 \label{fig:mlens}
\end{figure}

\Figref{mlens} reports the NMAE and estimated rank $\hat{N}$ for EM-BiG-AMP under the PIAWGN model \textb{\eqref{PIAWGNPart2}}, LMaFit, and VSBL,\footnote{VSBL was was allowed at most $100$ iterations and was run with \texttt{DIMRED\_THR}$=10^3$ and \texttt{UPDATE\_BETA}$=1$. Both VSBL and EM-BiG-AMP used a tolerance of $10^{-8}$. LMaFit was configured as for the MovieLens experiment in \cite{jtp_Wen2012a}.  Each algorithm was allowed a maximum rank of $\overline{N}=30$.} 
all of which include mechanisms for rank estimation.
\Figref{mlens} shows that, under the PIAWGN model, EM-BiG-AMP yields NMAEs that are very close to those of VSBL\footnote{The NMAE values reported for VSBL in \figref{mlens} are slightly inferior to those reported in \cite{Babacan:TSP:12}.  We attribute the discrepancy to differences in experimental setup, such as the construction of $\Omega$.  
}
but slightly inferior at larger training fractions, whereas LMaFit returns NMAEs that are substantially worse all training fractions.\footnote{The NMAE results presented here differ markedly from those in the MovieLens experiment in \cite{jtp_Wen2012a} because, in the latter paper, the entire set of ratings was used for both training \emph{and testing}, with the (trivial) result that high-rank models (e.g., $\hat{N}=94$) yield nearly zero test error.}
\Figref{mlens} also shows that LMaFit's estimated rank is much higher than those of VSBL and EM-BiG-AMP, suggesting that its poor NMAE performance is the result of overfitting.
(Recall that similar behavior was seen for noisy matrix completion in \figref{fullRank2}.)
In addition, \figref{mlens} shows that, as the training fraction increases, EM-BiG-AMP's estimated rank remains very low (i.e., $\leq 2$) while that of VSBL steady increases (to $>10$).
This prompts the question: is VSBL's excellent NMAE the result of accurate rank estimation or the use of a heavy-tailed (i.e., student's t) noise prior?

To investigate the latter question, we ran BiG-AMP under
\begin{align}
p_{\Y_{ml}|\Z_{ml}}(y_{ml}\giv z_{ml}) 
&= \begin{cases}
   \frac{\lambda}{2} \exp\big(-\lambda |y_{ml} - z_{ml}|\big) & (m,l)\in\Omega\\
   \Dirac_{y_{ml}} & (m,l)\notin\Omega,
   \end{cases}
\end{align}
i.e., a possibly incomplete additive white Laplacian noise (PIAWLN) model,
and used the EM-based approach from \Xsecref{EM} to learn the rate parameter $\lambda$. 
\Figref{mlens} shows that, under the PIAWLN model, EM-BiG-AMP essentially matches the NMAE performance of VSBL and even improves on it at very low training fractions.
Meanwhile, its estimated rank $\hat{N}$ remains low for all training fractions, suggesting that the use of a heavy-tailed noise model was the key to achieving low NMAE in this experiment. 
Fortunately, the generality and modularity of BiG-AMP made this an easy task.

\subsubsection{Summary}

In summary, the known-rank synthetic-data results above showed the EM-BiG-AMP methods yielding phase-transition curves superior to all other algorithms under test. 
In addition, they showed BiG-AMP-Lite to be the second fastest algorithm (behind LMaFit) for most combinations of sampling ratio $\delta$ and rank $N$, although it was the fastest for small $\delta$ and relatively high $N$.
Also, they showed EM-BiG-AMP was about $3$ to $5$ times slower than BiG-AMP-Lite but still much faster than IALM and VSBL at high ranks.
Meanwhile, the unknown-rank synthetic-data results above showed EM-BiG-AMP yielding excellent NMSE performance in both noiseless and noisy scenarios.
For example, in the noisy experiment, EM-BiG-AMP uniformly outperformed its competitors (LMaFit and VSBL).

In the image completion experiment, EM-BiG-AMP again outperformed all competitors, beating the second best algorithm (Matrix ALPS) by more than $1$~dB and the third best algorithm (LMaFit) by more than $2.5$~dB.
Finally, in the collaborative filtering experiment, EM-BiG-AMP (with the PIAWLN likelihood model) matched the best competitor (VSBL) in terms of NMAE, and significantly outperformed the second best (LMaFit).


\section{Robust PCA}
\label{sec:RPCA}
\subsection{Problem Setup}
\label{sec:RPCAsetup}
In robust principal components analysis (RPCA)~\cite{jtp_Candes2011}, one seeks to estimate a low-rank matrix observed in the presence of noise and large outliers. 
The data model for RPCA can be written as 
\begin{align}
\vec{Y} &= \vec{AX} + \vec{E} + \vec{W},\label{eq:RPCA}
\end{align}
where $\vec{Z}=\vec{AX}$---the product of tall $\vec{A}$ and wide $\vec{X}$---is the low-rank matrix of interest, $\vec{E}$ is a \emph{sparse} outlier matrix, and $\vec{W}$ is a dense noise matrix.
We now suggest two ways of applying BiG-AMP to the RPCA problem, both of which treat 
the elements of $\vec{A}$ as iid $\Nor(0,\nu_0^a)$ \textb{similar to \eqref{a_gauss2}}, 
the elements of $\vec{X}$ as iid $\Nor(0,\nu_0^x)$ \textb{similar to \eqref{x_gauss2}},
the \emph{non-zero} elements of $\vec{E}$ as iid $\Nor(0,\nu_1)$,
and the elements of $\vec{W}$ as iid $\Nor(0,\nu_0)$,
\textb{with $\nu_1\gg\nu_0$.} 

In the first approach, $\vec{E}+\vec{W}$ is treated as additive noise on $\vec{Z}$, leading to the likelihood model
\begin{align}
&p_{\Y_{ml}|\Z_{ml}}(y_{ml}\giv z_{ml})\non\\&= (1 - \lambda) \Nor(y_{ml}; z_{ml},\nu_0) + \lambda \Nor(y_{ml};z_{ml},\nu_0+\nu_1),\label{eq:heavyTail}
\end{align}
where $\lambda \in [0,1]$ models outlier density.

In the second approach, $\vec{W}$ is treated as additive noise but $\vec{E}$ is treated as an additional estimand.
In this case, by multiplying both sides of \eqref{RPCA} by any (known) unitary matrix $\vec{Q} \in \mathbb{R}^{M \times M}$, we can write 
\begin{align}
\underbrace{\vec{QY}}_{\defn \uvec{Y}} &= \underbrace{\bmat \vec{QA} & ~\vec{Q} \emat}_{\defn \uvec{A}} \underbrace{\bmat \vec{X} \\ \vec{E} \emat}_{\defn \uvec{X}} \mbox{} + \underbrace{\vec{QW}}_{\defn \uvec{W}}, \label{eq:augmented}
\end{align}
and apply BiG-AMP to the ``augmented'' model $\uvec{Y}=\uvec{A}\uvec{X}+\uvec{W}$.
Here, $\uvec{W}$ remains iid $\Nor(0,\nu_0)$, thus giving the likelihood
\begin{align}
p_{\uY_{ml}|\uZ_{ml}}(\underline{y}_{ml}\giv \underline{z}_{ml}) &=  \Nor(\underline{y}_{ml}; \underline{z}_{ml},\nu_0).
\end{align}
Meanwhile, \textb{we choose the following separable} priors on $\uvec{A}$ and $\uvec{X}$:
\begin{align}
p_{\uA_{mn}}(\underline{a}_{mn})
&= \begin{cases}
  \mc{N}(\underline{a}_{mn};0,\nu^a_0) & n\leq \textb{N} \\
  \mc{N}(\underline{a}_{mn};q_{mn},0) & n> \textb{N} 
  \end{cases}\\
p_{\uX_{nl}}(\underline{x}_{nl})
&= \begin{cases}
  \mc{N}(\underline{x}_{nl};0,\nu^x_0) & n\leq \textb{N} \\
  (1 - \lambda) \Dirac_{\underline{x}_{nl}} + \lambda \Nor(\underline{x}_{nl};0,\nu_1)
  	& n>\textb{N} .
  \end{cases}
\end{align}
\textb{Essentially, the first $N$ columns of $\uvec{A}$ and first $N$ rows of $\uvec{X}$ model the factors of the low-rank matrix $\vec{AX}$, and thus their elements are assigned iid Gaussian priors, similar to \eqref{a_gauss2}-\eqref{x_gauss2} in the case of matrix completion.
Meanwhile, the last $M$ rows in $\uvec{X}$ are used to represent the sparse outlier matrix $\vec{E}$, and thus their elements are assigned a Bernoulli-Gaussian prior.
Finally, the last $M$ columns of $\uvec{A}$ are used to represent the designed matrix $\vec{Q}$, and thus their elements are assigned zero-variance priors.
Since we find that BiG-AMP is numerically more stable when $\vec{Q}$ is chosen as a dense matrix, we set it equal to the singular-vector matrix of an iid $\mc{N}(0,1)$ matrix.} 
\textb{After running BiG-AMP,} we can recover an estimate of $\vec{A}$ by left multiplying the estimate of $\uvec{A}$ by $\vec{Q}\herm$.

\subsection{Initialization}
We recommend initializing $\hat{\underline{a}}_{mn}(1)$ using a random draw from its prior \textb{and} initializing \textb{$\hat{\underline{x}}_{nl}(1)$ at the mean of its prior, i.e.,} $\hat{\underline{x}}_{nl}(1)=0$. 
\textb{The latter tends to perform better than initializing $\hat{\underline{x}}_{nl}(1)$ randomly, because it allows the measurements $\vec{Y}$ to determine the initial locations of the outliers in $\vec{E}$.
As in \secref{MCinit}, we suggest initializing} $\nu^{\underline{a}}_{mn}(1)$ and $\nu^{\underline{x}}_{nl}(1)$ at $10$ times the variance of their respective priors \textb{to emphasize the role of the measurements during the first few iterations}.

\subsection{EM-BiG-AMP}
The EM approach from \Xsecref{EM} can be straightforwardly applied to BiG-AMP for RPCA: after fixing $\nu_0^a=1$, EM can be used to tune the remaining distributional parameters, $\vec{\theta} = [\nu_0, \nu_1, \nu_0^x, \lambda]^T$.
To avoid \textb{initializing} $\nu_0$ and $\nu^x_0$ \textb{with overly large values} in the presence of large outliers $e_{nl}$, we suggest the following \textb{procedure}.
First, define the set $\Gamma \defn \big\{(m,l) : \textb{|y_{ml}|} \le \med\{|y_{ml}|\}\big\}$ and its complement $\Gamma^c$. 
Then initialize
\begin{align}
\nu_0 &= \frac{\frac{1}{|\Gamma|}\sum_{(m,l)\in\Gamma} |y_{ml}|^2}{\textrm{SNR}^0 + 1}\\
\nu^x_0 &= \frac{1}{N}\textrm{SNR}^0 \nu_0 \\
\nu_1 &= \frac{1}{|\Gamma^c|}\sum_{(m,l)\in\Gamma^c} |y_{ml}|^2 ,
\end{align}
where, \textb{as in \secref{MCEM}}, we suggest setting $\textrm{SNR}^0 = 100$ in the absence of prior knowledge.
\textb{This approach uses the median to avoid including outliers among the samples used to estimate the variances of the dense-noise and low-rank components.}
Under these rules, the initialization $\lambda = 0.1$ was found to work well for most problems.

\subsection{Rank Selection}
\label{sec:RPCArankLearn}
In many applications of RPCA, such as video separation, the singular-value profile of $\vec{AX}$ exhibits a sharp cutoff, in which case it is recommended to perform rank-selection using the contraction strategy from \Xsecref{contraction}.

\subsection{Avoiding Local Minima}
\label{sec:localMinima}
Sometimes, when $N$ is very small, BiG-AMP may converge to a local solution that mistakes entire rows or columns of $\vec{AX}$ for outliers. 
Fortunately, this situation is easy to remedy \textb{with a simple heuristic procedure}: 
the posterior probability that $y_{ml}$ is outlier-corrupted can be computed for each $(m,l)$ at convergence, and if any of the row-wise sums exceeds $0.8M$ or any of the column-wise sums exceeds $0.8L$, then BiG-AMP is restarted from a new random initialization.
Experimentally, we found that one or two of such restarts is generally sufficient to avoid local minima.


\subsection{Robust PCA Experiments}   \label{sec:RPCAexamples}
In this section, we present a numerical study of the two BiG-AMP formulations of RPCA proposed in \secref{RPCA},
including a comparison to the state-of-the-art IALM~\cite{jtp_Lin2010}, LMaFit~\cite{jtp_Wen2012a}, GRASTA~\cite{jtp_He2011}, and VSBL~\cite{Babacan:TSP:12} algorithms. 
In the sequel, we use ``BiG-AMP-1'' when referring to the formulation that treats the outliers as noise, and ``BiG-AMP-2'' when referring to the formulation that explicitly estimates the outliers.

\subsubsection{Phase Transition Behavior}	\label{sec:RPCAptc}
We first study the behavior of the proposed BiG-AMP algorithms for RPCA on noise-free synthetic problems. 
For this, we generated problem realizations of the form $\vec{Y}=\vec{Z}+\vec{E}$, where the low-rank component $\vec{Z}=\vec{AX}$ was generated from $\vec{A}$ and $\vec{X}$ with iid $\Nor(0,1)$ entries, and where the sparse corruption matrix $\vec{E}$ had a fraction $\delta$ of non-zero entries that were located uniformly at random with amplitudes drawn iid uniform on $[-10,10]$.
The dimensions of $\vec{Y}\in\Real^{M\times L}$ were fixed at $M = L = 200$, the rank $N$ (of $\vec{Z}$) was varied from $10$ to $90$, and the outlier fraction $\delta$ was varied from $0.05$ to $0.45$.
\textb{We note that, under these settings, the outlier magnitudes are on the same order as the magnitudes of $\vec{Z}$, which is the most challenging case: much larger outliers would be easier to detect, after which the corrupted elements of $\vec{Y}$ could be safely treated as incomplete, whereas much smaller outliers could be treated like AWGN.}

All algorithms under test were run to a convergence tolerance of $10^{-8}$ and forced to use the true rank $N$. 
GRASTA, LMaFit, and VSBL were run under their recommended settings.\footnote{For LMaFit, however, we increased the maximum number of allowed iterations, since this improved its performance.}  
Two versions of IALM were tested: ``IALM-1,'' which uses the universal penalty parameter $\lambda_{\texttt{ALM}}=\frac{1}{\sqrt{M}}$, and ``IALM-2,'' which tries $50$ hypotheses of $\lambda_{\texttt{ALM}}$, logarithmically spaced from $\frac{1}{10\sqrt{M}}$ to $\frac{10}{\sqrt{M}}$ and uses an oracle to choose the MSE-minimizing hypothesis.
\textb{BiG-AMP-1 and BiG-AMP-2 were given perfect knowledge of the mean and variance of the entries of $\vec{A}$, $\vec{X}$, and $\vec{E}$ (although their Bernoulli-Gaussian model of $\vec{E}$ did not match the data generation process) as well as the outlier density $\lambda$, while EM-BiG-AMP-2 learned all model parameters from the data. }
BiG-AMP-1 was run under a fixed damping of $\beta=0.25$, while BiG-AMP-2 was run under adaptive damping with $\texttt{stepMin} = 0.05$ and $\texttt{stepMax} = 0.5$. 
Both variants used a maximum of $5$ restarts to avoid local minima.

\begin{figure*}[htb]
\centering
\psfrag{N (rank)}[B][][0.8]{$N$}
\psfrag{lambda}[][][0.8]{$\delta$}
\begin{tabular}{cccc}
\includegraphics[width=1.6in]{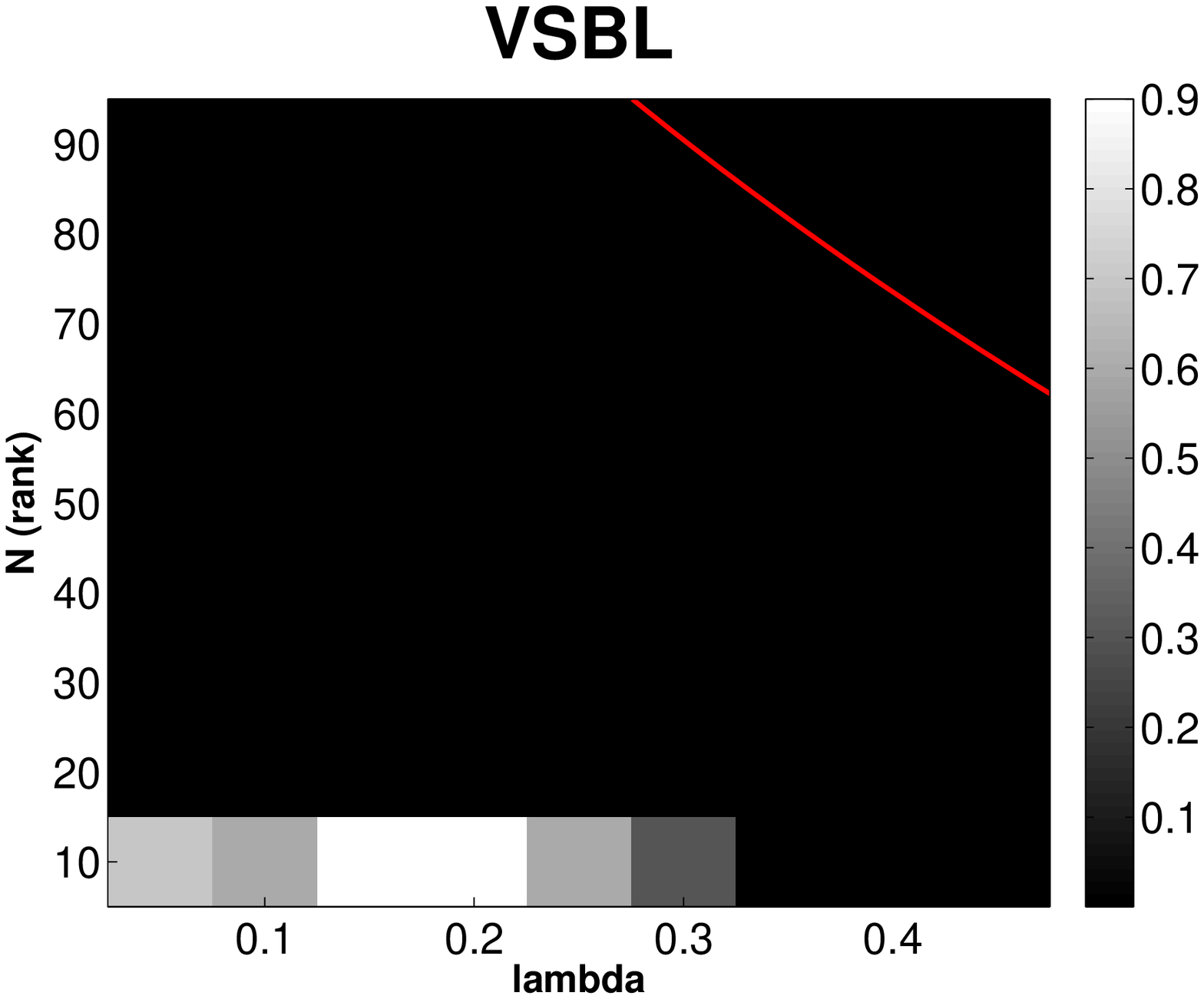}&
\includegraphics[width=1.6in]{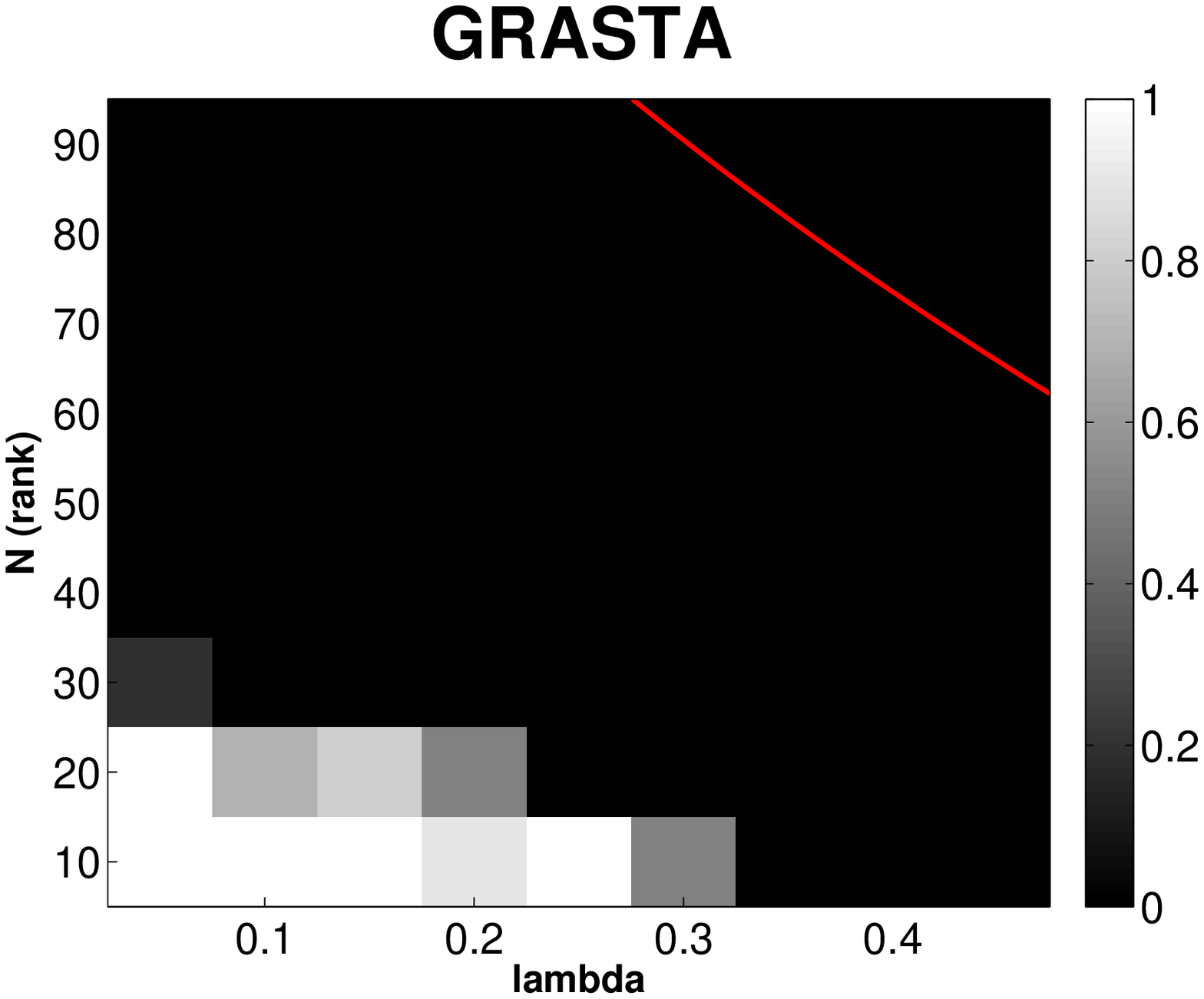}&
\includegraphics[width=1.6in]{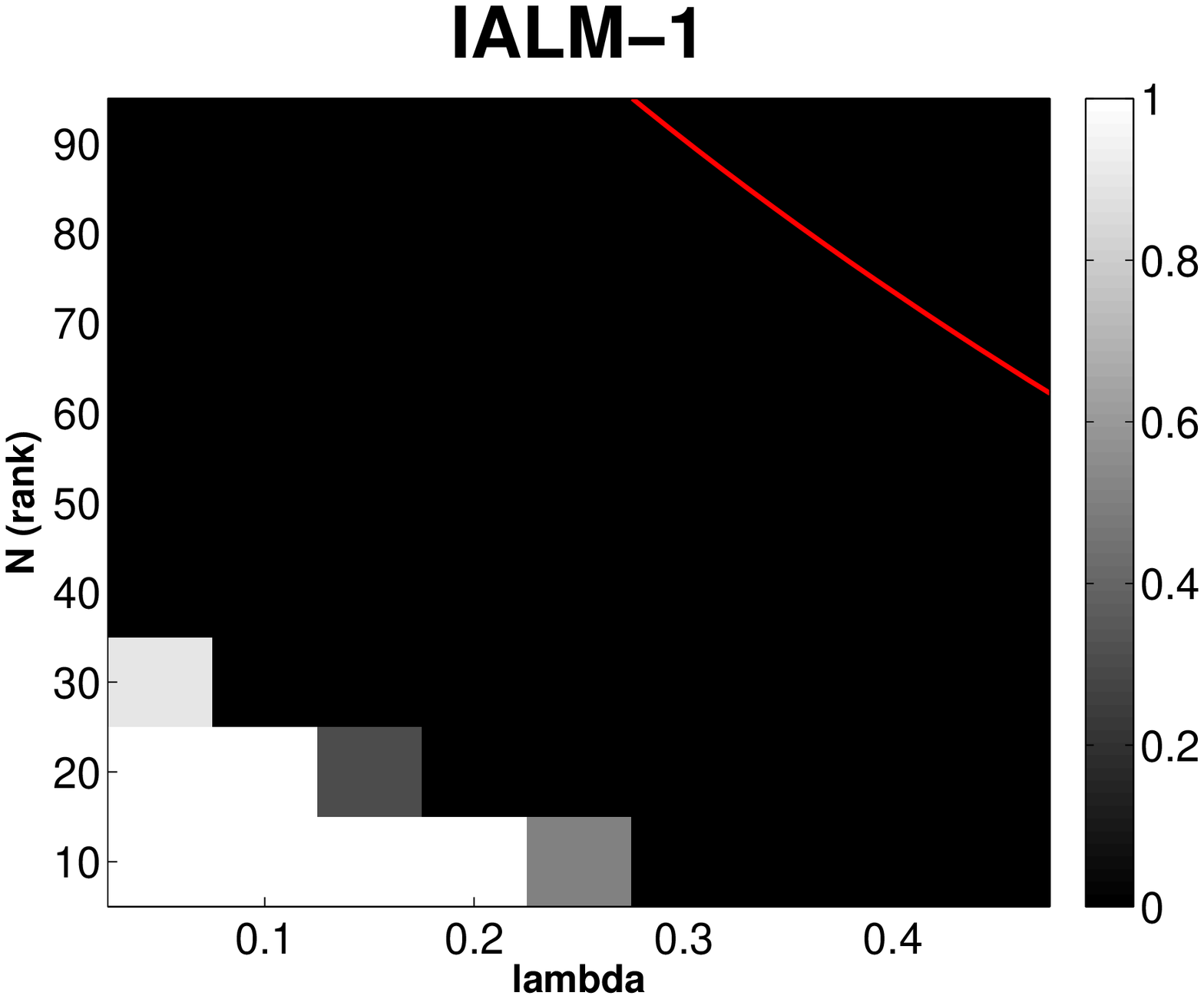}&
\includegraphics[width=1.6in]{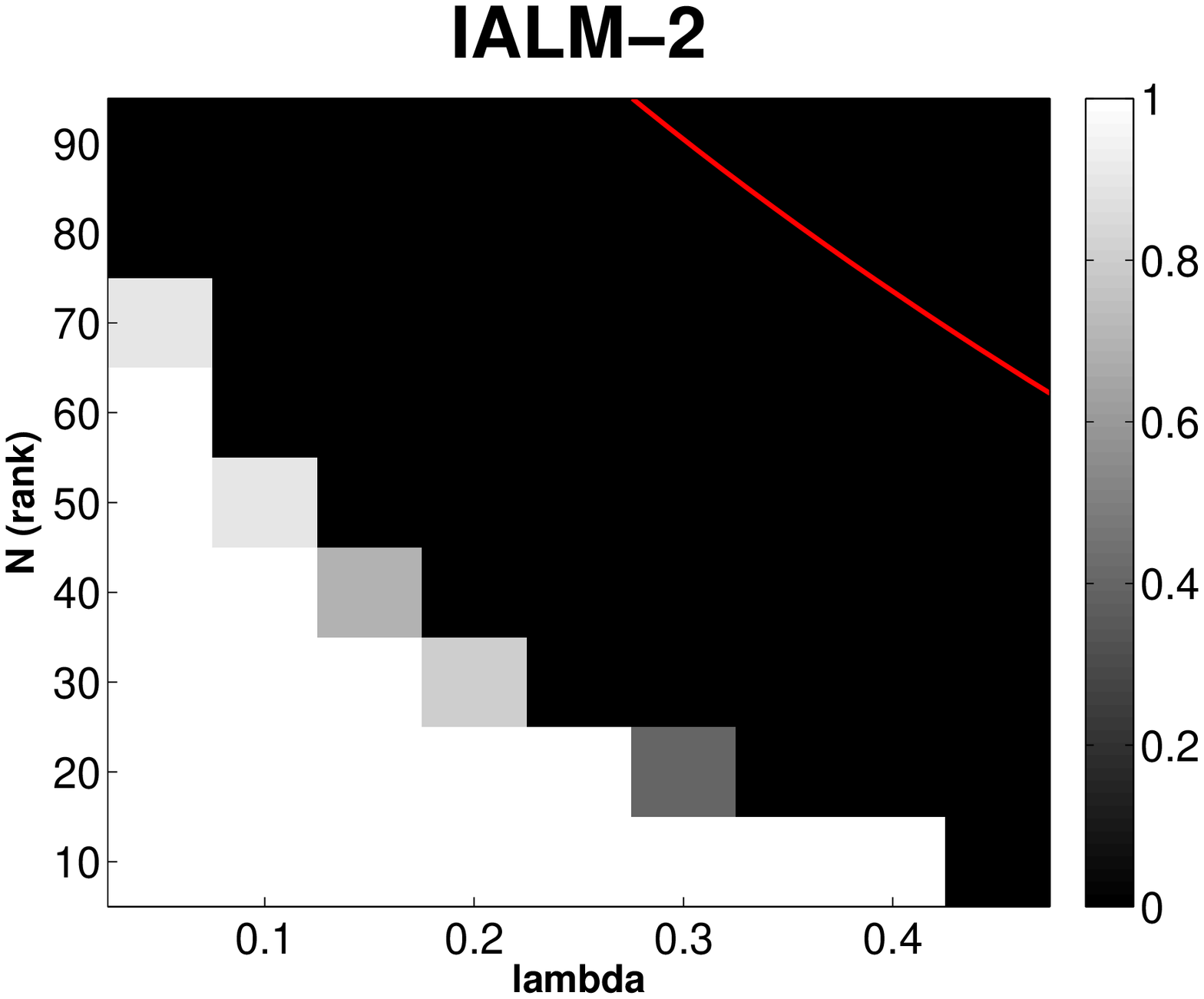}\\
\includegraphics[width=1.6in]{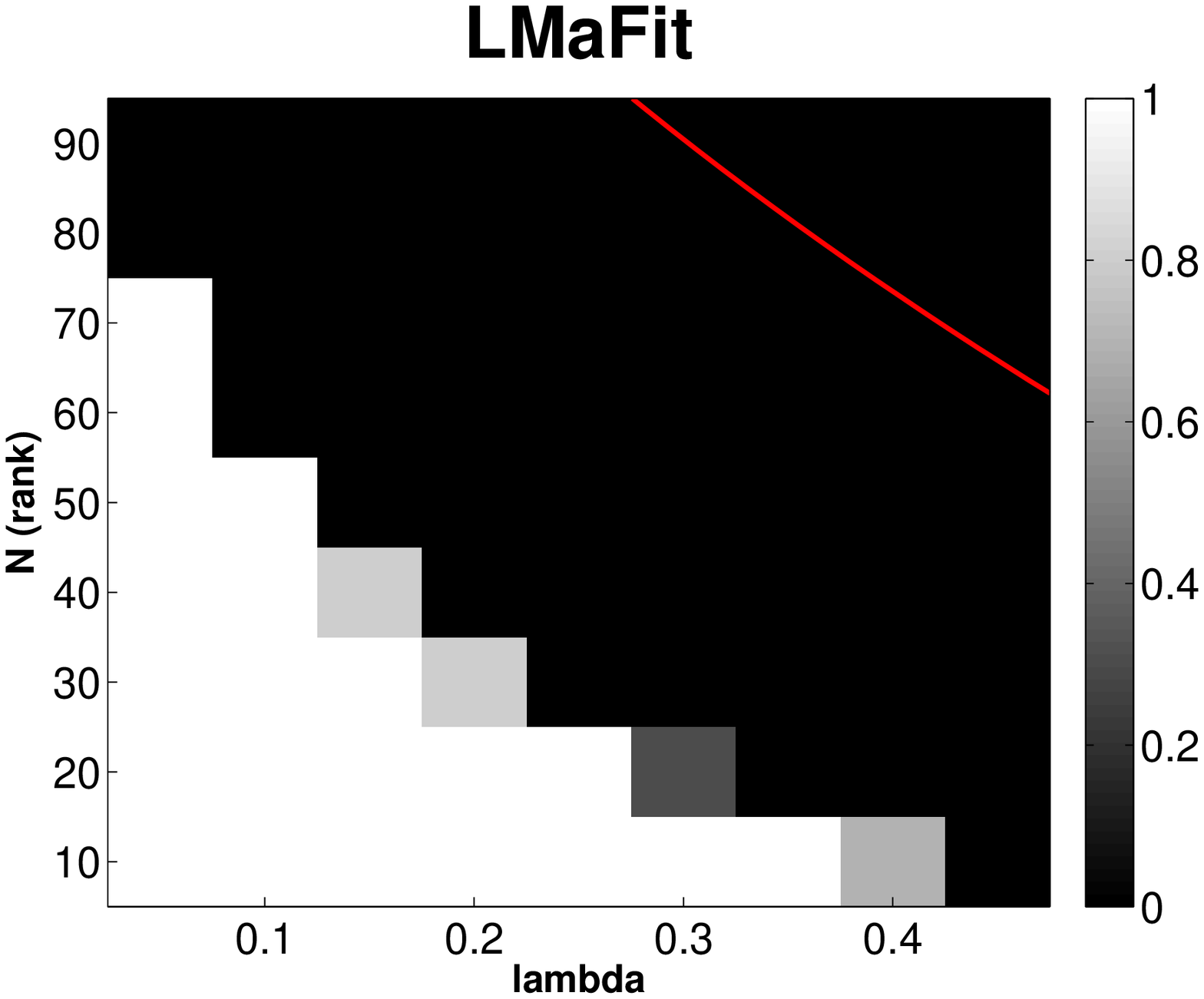}&
\includegraphics[width=1.6in]{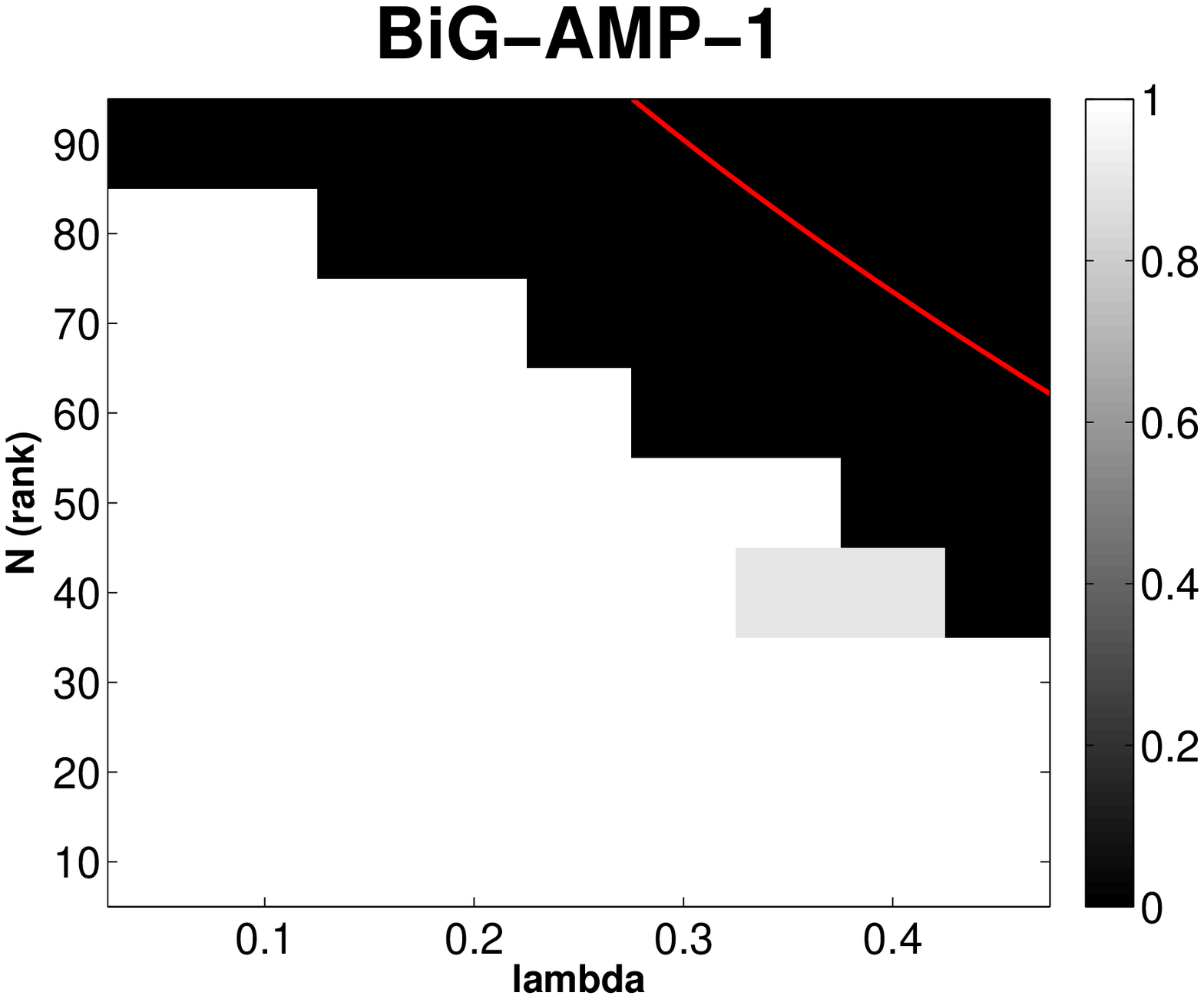}&
\includegraphics[width=1.6in]{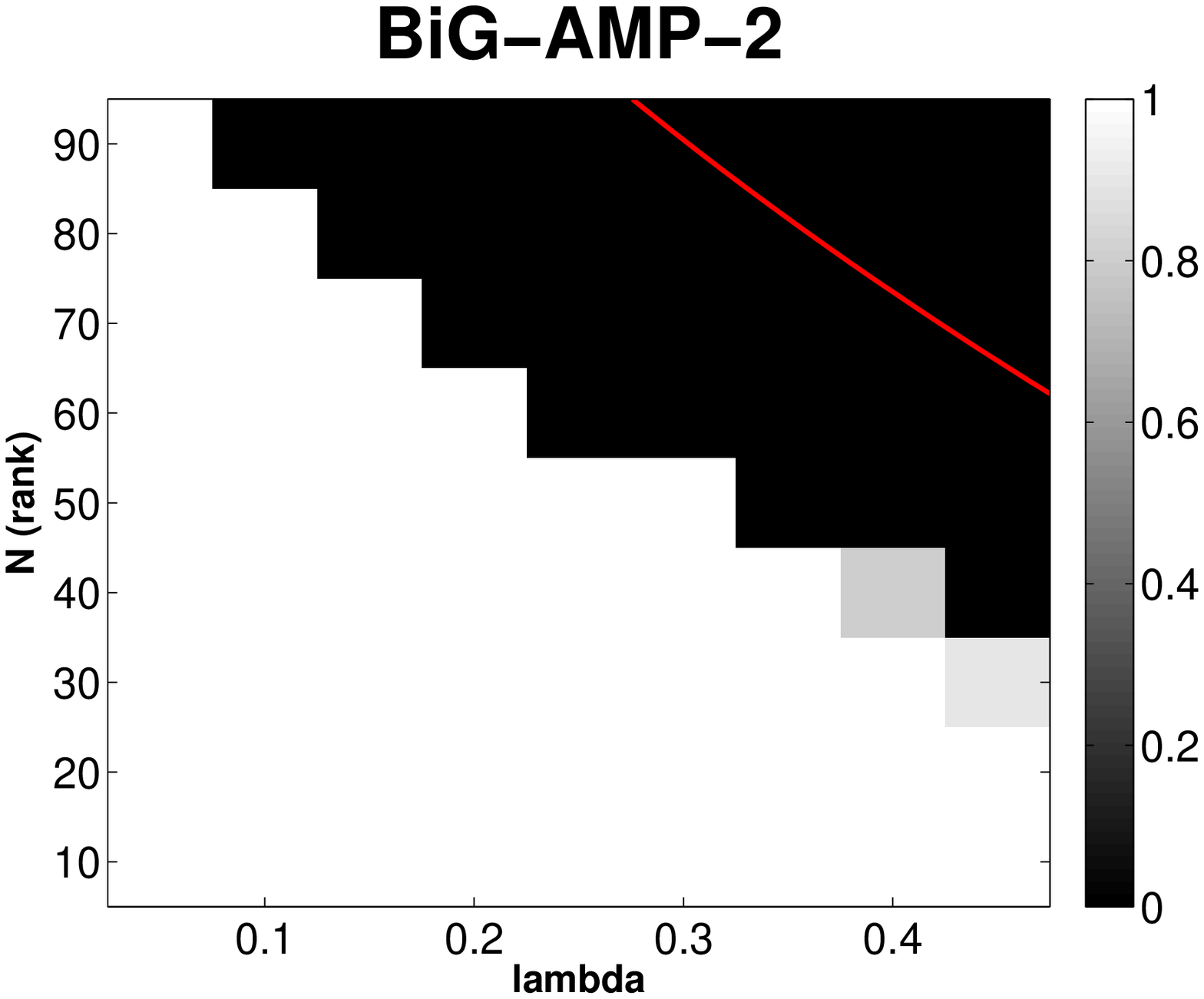}&
\includegraphics[width=1.6in]{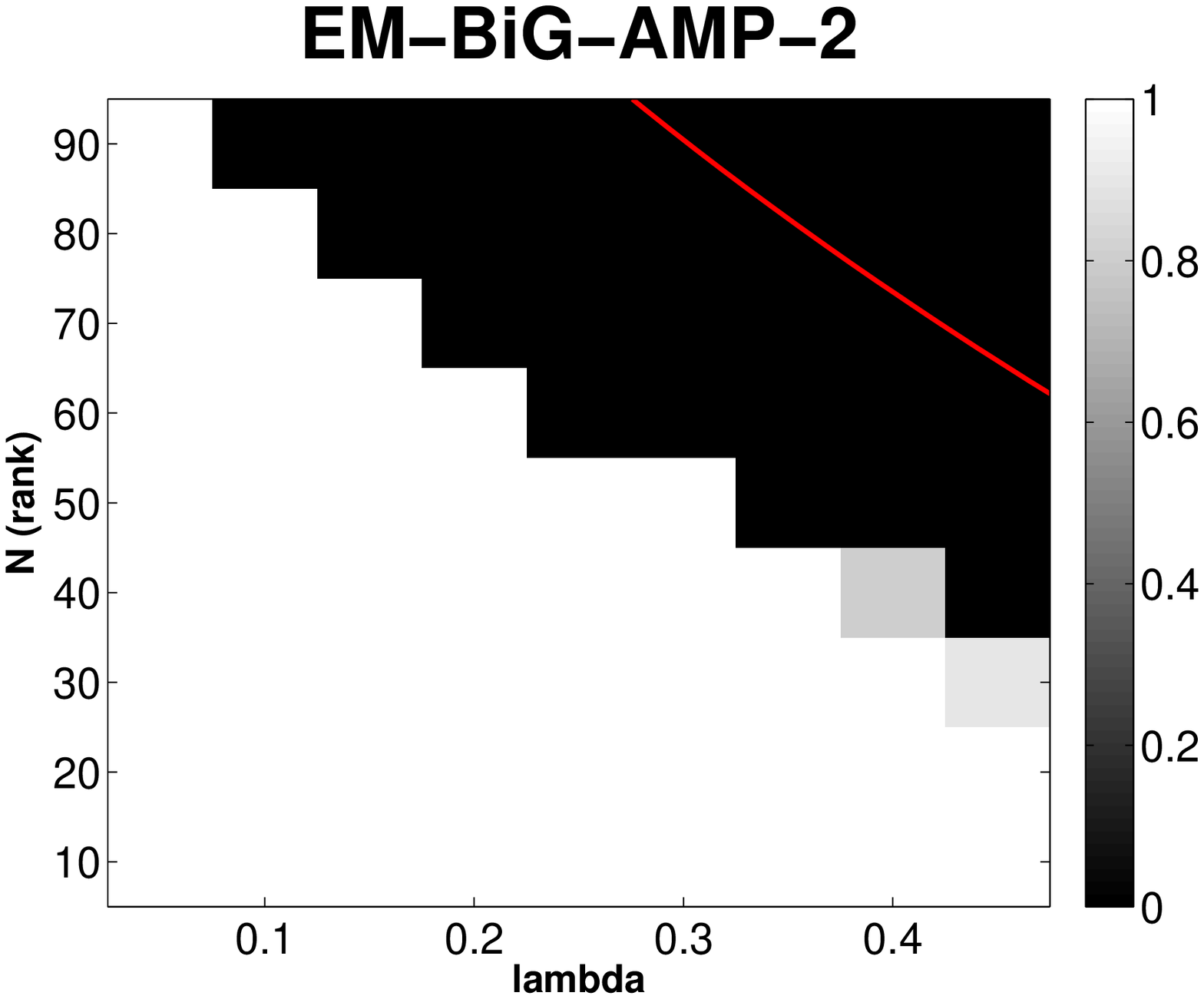}
\end{tabular}
\caption{Empirical success rates for RPCA with a $200\times 200$ matrix of rank $N$ corrupted by a fraction $\delta$ of outliers with amplitudes uniformly distributed on $[-10,10]$.  Here, ``success'' is defined as NMSE~$<-80$~dB, and success rates were averaged over $10$ problem realizations. Points above the red curve are infeasible, as described in the text.}
 \label{fig:phaseRPCAFull}
\end{figure*}

\Figref{phaseRPCAFull} shows the empirical success rate achieved by each algorithm as a function of corruption-rate $\delta$ and rank $N$, averaged over $10$ trials, where a ``success'' was defined as attaining an NMSE of $-80$~dB or better in the estimation of the low-rank component $\vec{Z}$.
The red curves in \figref{phaseRPCAFull} delineate the problem feasibility boundary: for points $(\delta,N)$ above the curve, $N(M+L-N)$, the degrees-of-freedom in $\vec{Z}$, exceeds $(1-\delta)ML$, the number of uncorrupted observations, making it impossible to recover $\vec{Z}$ without additional information.

\Figref{phaseRPCAFull} shows that all algorithms exhibit a relatively sharp phase-transition curve (PTC) separating success and failure regions, and that the BiG-AMP algorithms achieve substantially better PTCs than the other algorithms.
The PTCs of BiG-AMP-1 and BiG-AMP-2 are similar (but not identical), suggesting that both formulations are equally effective. 
Meanwhile, the PTCs of BiG-AMP-2 and EM-BiG-AMP-2 are nearly identical, demonstrating that the EM procedure was able to successfully learn the statistical model parameters used by BiG-AMP.
\Figref{phaseRPCAFull} also shows that all RPCA phase transitions remain relatively far from the feasibility boundary, unlike those for matrix completion (MC) shown in \figref{phaseMCFull}.
This behavior, also observed in \cite{jtp_Candes2011}, is explained by the relative difficulty of RPCA over MC: the locations of RPCA outliers (which in this case effectively render the corrupted observations as incomplete) are unknown, whereas in MC they are known.

\begin{figure*}[htb]
\centering
\psfrag{Z NMSE (dB)}[B][B][0.7]{\sf NMSE (dB)}
 \psfrag{N (rank)}[][][0.7]{$N$}
\psfrag{time (seconds)}[B][B][0.7]{\sf runtime (sec)}
\psfrag{lambda = 0.05}[][][0.7]{$\delta = 0.05$}
\psfrag{lambda = 0.2}[][][0.7]{$\delta = 0.2$}
\psfrag{lambda = 0.3}[][][0.7]{$\delta = 0.3$}
 \begin{tabular}{@{}c@{}c@{}c@{}}
\includegraphics[width=2.4in]{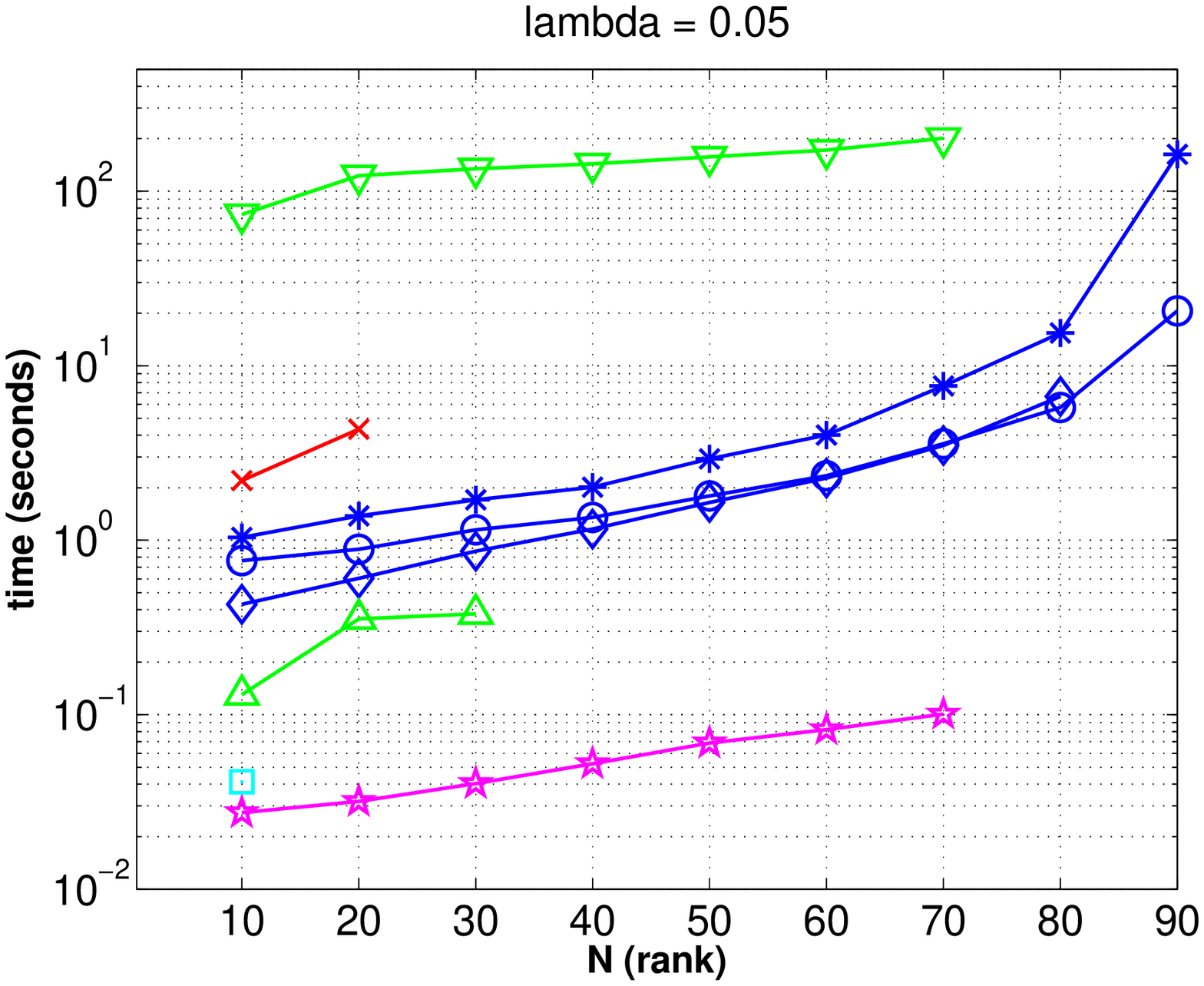}&
\includegraphics[width=2.4in]{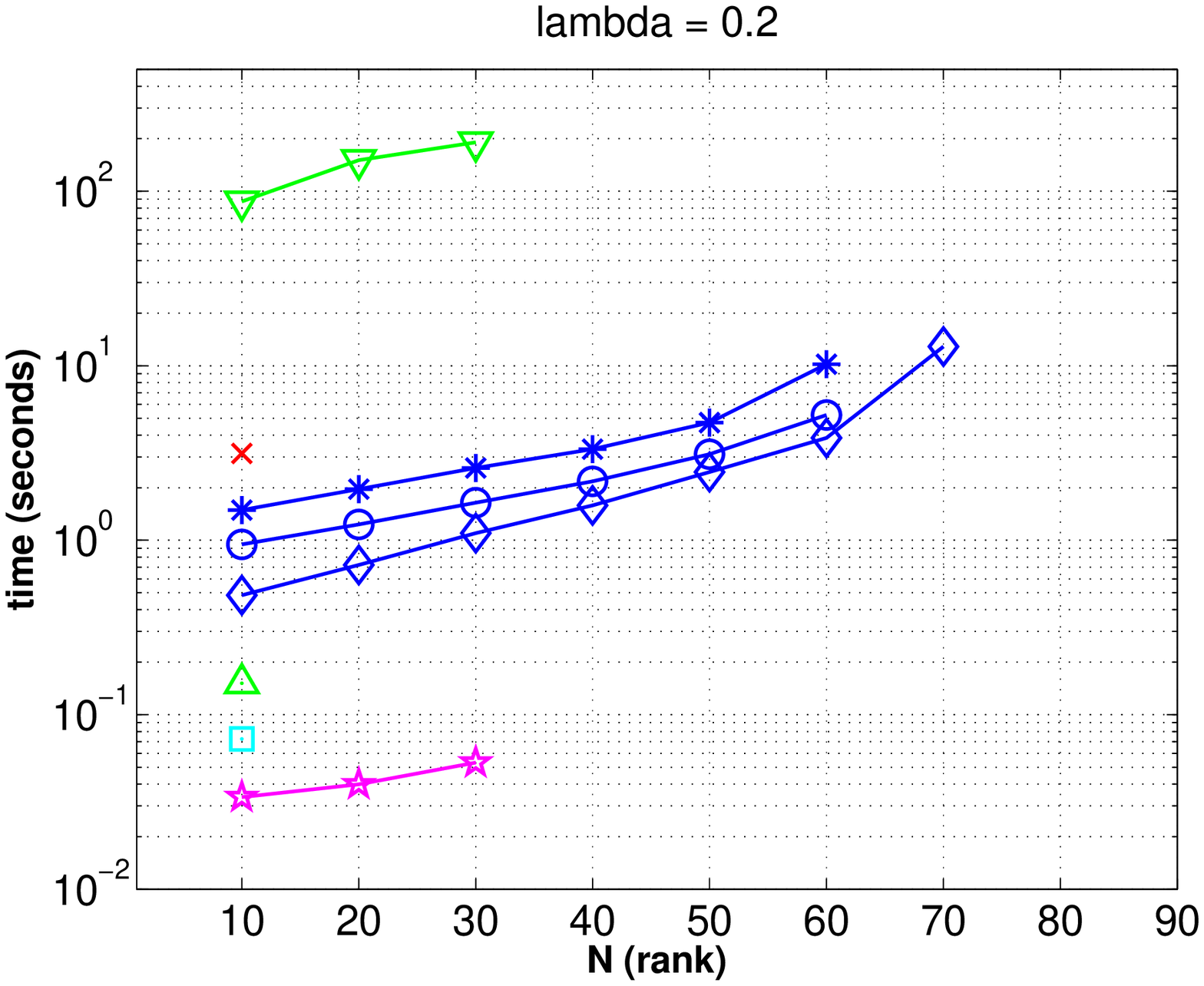}&
\includegraphics[width=2.4in]{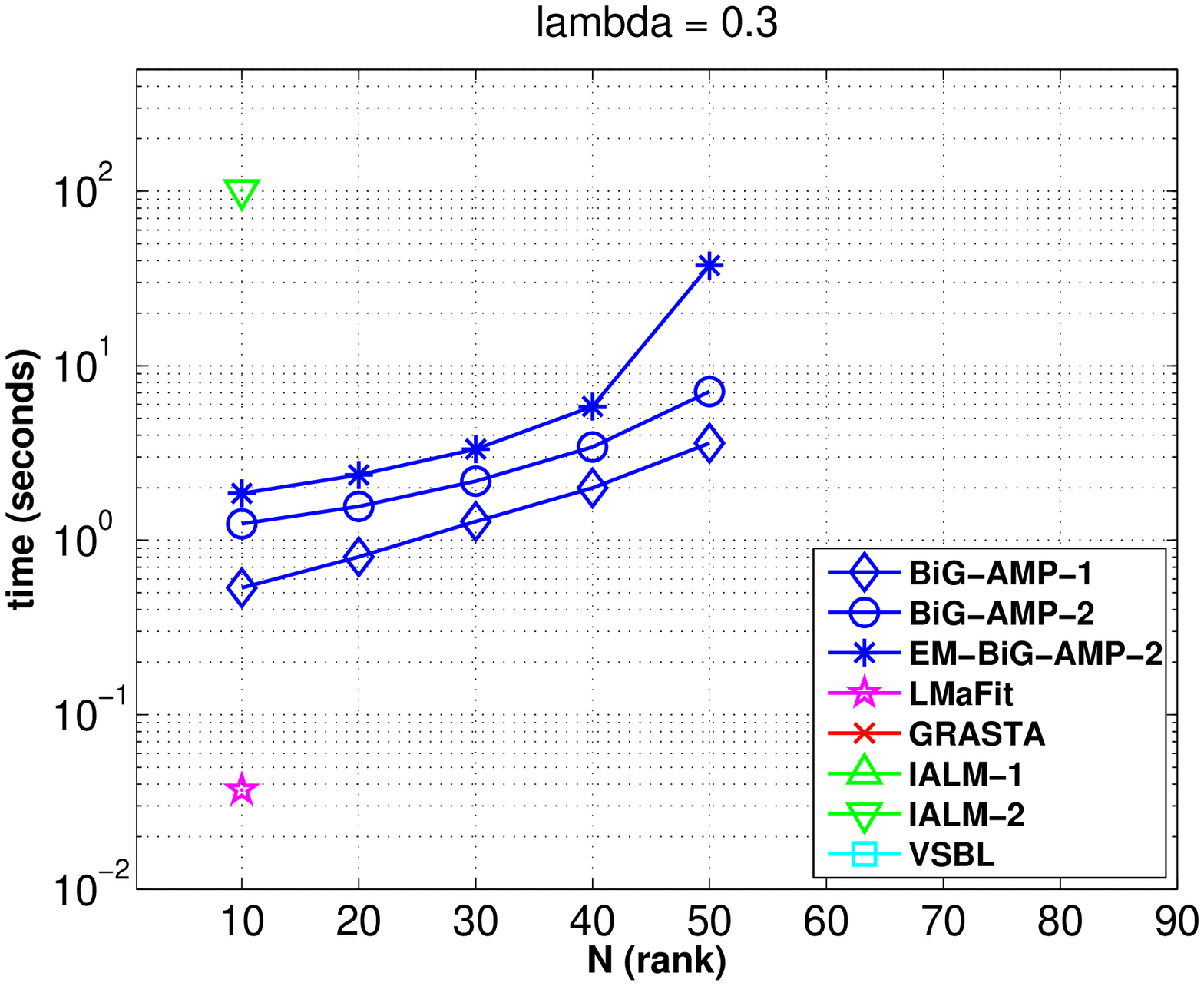}
\end{tabular}
 \caption{Runtime to NMSE$=-80$~dB for RPCA with a $200\times 200$ matrix of rank $N$ corrupted by a fraction $\delta \in\{0.05, 0.2, 0.3\}$ of outliers with amplitudes uniformly distributed on $[-10,10]$.
 All results represent median performance over $10$ trials.
 \textb{Missing values indicate that the algorithm did not achieve the required NMSE before termination and correspond to the black regions in \figref{phaseRPCAFull}.}}
 \label{fig:PhaseCutsRPCA}
\end{figure*}

\Figref{PhaseCutsRPCA} plots runtime to NMSE~$=-80$~dB as a function of rank $N$ for various outlier fractions.
The results suggest that the BiG-AMP algorithms are moderate in terms of speed, being faster than GRASTA\footnote{We note that, for this experiment, GRASTA was run as a Matlab M-file and not a MEX file, because the MEX file would not compile on the supercomputer used for the numerical results.  That said, since BiG-AMP was also run as an unoptimized M-file, the comparison could be considered ``fair.''}
and much faster than the grid-tuned IALM-2, but slower than IALM-1, VSBL, and LMaFit.
Notably, among the non-BiG-AMP algorithms, LMaFit offers both the fastest runtime and the best phase-transition curve on this synthetic test problem.

In summary, the results presented here suggest that BiG-AMP achieves state-of-the-art PTCs while maintaining runtimes that are competitive with existing approaches.

\subsubsection{Rank Estimation}	\label{sec:RPCArank}
We now investigate the ability to estimate the underlying rank, $N$, for 
EM-BiG-AMP-2 (using the rank-contraction strategy from \Xsecref{contraction}\footnote{The rank-selection rule \Xeqref{contract} was used with $\tau_{\text{MOS}} = 5$, up to $50$ EM iterations, and a minimum of $30$ and maximum of $500$ BiG-AMP iterations per EM iteration.}) IALM-1, IALM-2, LMaFit, and VSBL, all of which include either explicit or implicit rank-selection mechanisms.
For this, we generated problem realizations of the form $\vec{Y}=\vec{Z}+\vec{E}+\vec{W}$, where the $200\times 200$ rank-$N$ matrix $\vec{Z}$ and $\delta=0.1$-sparse outlier matrix $\vec{E}$ were generated as described in \secref{RPCAptc} and the noise matrix $\vec{W}$ was constructed with iid $\Nor(0,10^{-3})$ elements.
The algorithms under test were not provided with knowledge of the true rank $N$, which was varied between $5$ and $90$.
LMaFit, VSBL, and EM-BiG-AMP, were given an initial rank estimate of $\overline{N}=90$, which enforces an upper bound on the final estimates that they report.

\begin{figure}[htb]
\centering
 \begin{tabular}{c}
\includegraphics[width=\columnwidth]{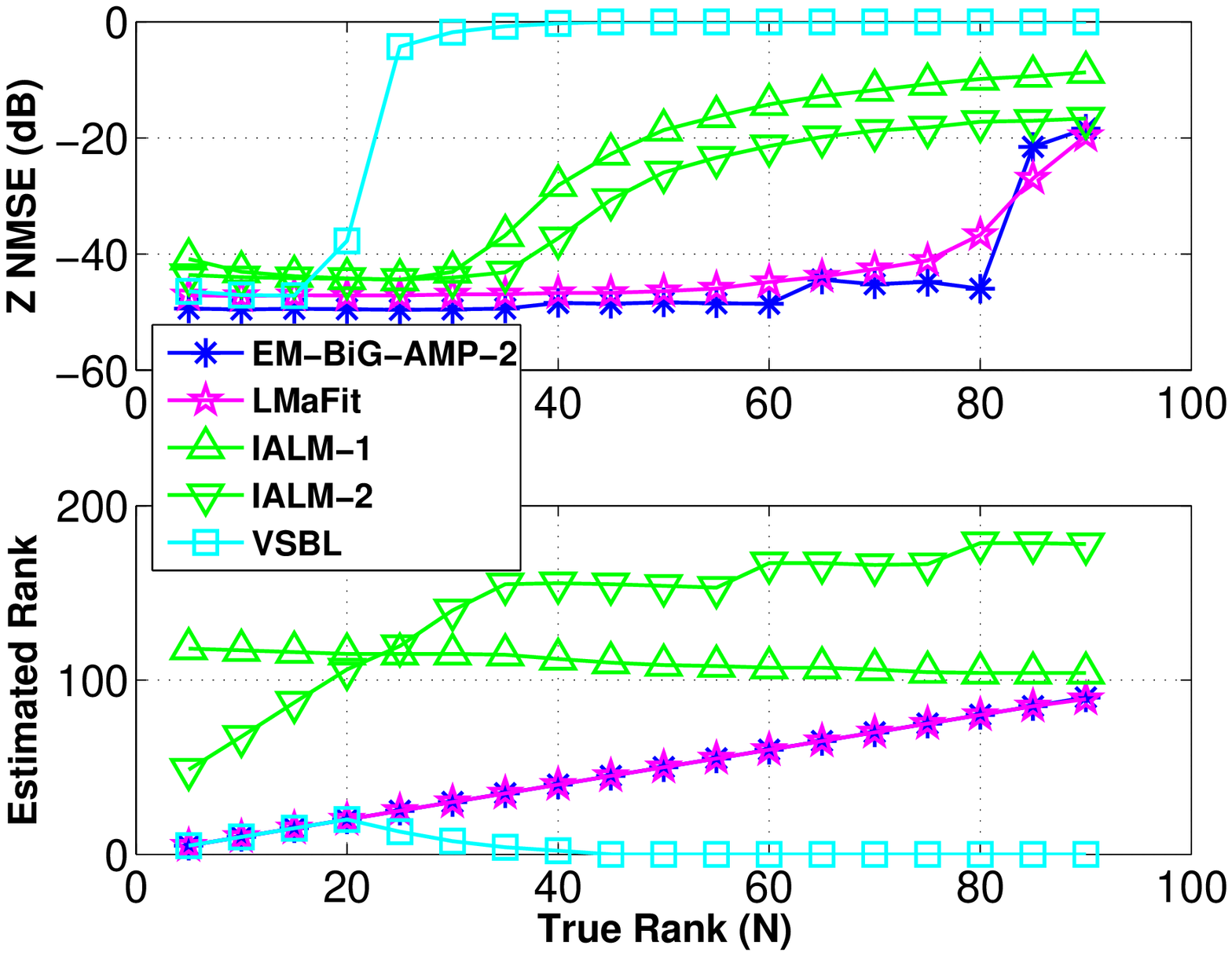}
\end{tabular}
 \caption{NMSE (top) and estimated rank $\hat{N}$ (bottom) versus true rank $N$ for several algorithms performing RPCA on a $200\times 200$ matrix in the presence of additive $\Nor(0,10^{-3})$ noise and a fraction $\delta = 0.1$ of outliers with amplitudes uniformly distributed on $[-10,10]$.  All results represent the median over $10$ trials.}
 \label{fig:rpcaRANK}
\end{figure}

\Figref{rpcaRANK} reports RPCA performance versus (unknown) true rank $N$ in terms of the estimated rank $\hat{N}$ and the NMSE on the estimate $\hvec{Z}$.
All results represent median performance over $10$ Monte-Carlo trials.
The figure shows that EM-BiG-AMP-2 and LMaFit returned accurate rank estimates $\hat{N}$ over the full range of true rank $N\in[5,90]$, whereas VSBL returned accurate rank estimates only for $N\leq 20$, and both IALM-1 and IALM-2 greatly overestimated the rank at all $N$.
Meanwhile, \figref{rpcaRANK} shows that EM-BiG-AMP-2 and LMaFit returned accurate estimates of $\hvec{Z}$ for all $N\leq 80$ (with EM-BiG-AMP-2 outperforming LMaFit by several dB throughout this range), whereas VSBL and IALM-1 and IALM-2 returned accurate estimates of $\hvec{Z}$ only for small values of $N$.
We note that the relatively poor MSE performance of LMaFit and EM-BiG-AMP-2 for true rank $N>80$ is not due to poor rank estimation but rather due to the fact that, at $\delta=0.1$, these operating points lie above the PTCs shown in \figref{phaseRPCAFull}.

\subsubsection{Application to Video Surveillance} \label{sec:RPCAvideo}
We now apply EM-BiG-AMP-2 to a video surveillance problem, where the goal is to separate a video sequence into a static ``background'' component and a dynamic ``foreground'' component.
To do this, we stack each frame of the video sequence into a single column of the matrix $\vec{Y}$, run EM-BiG-AMP-2 as described in \secref{RPCA}, extract the background frames from the estimate of the low-rank component $\vec{Z}=\vec{AX}$, and extract the foreground frames from the estimate of the (sparse) outlier component $\vec{E}$.
We note that a perfectly time-invariant background would correspond to a rank-one $\vec{Z}$ and that the noise term $\vec{W}$ in \eqref{RPCA} can be used to account for small perturbations that are neither low-rank nor sparse.

\begin{figure}[!htb]
\centering
 \begin{tabular}{ccc}
\includegraphics[width=1in]{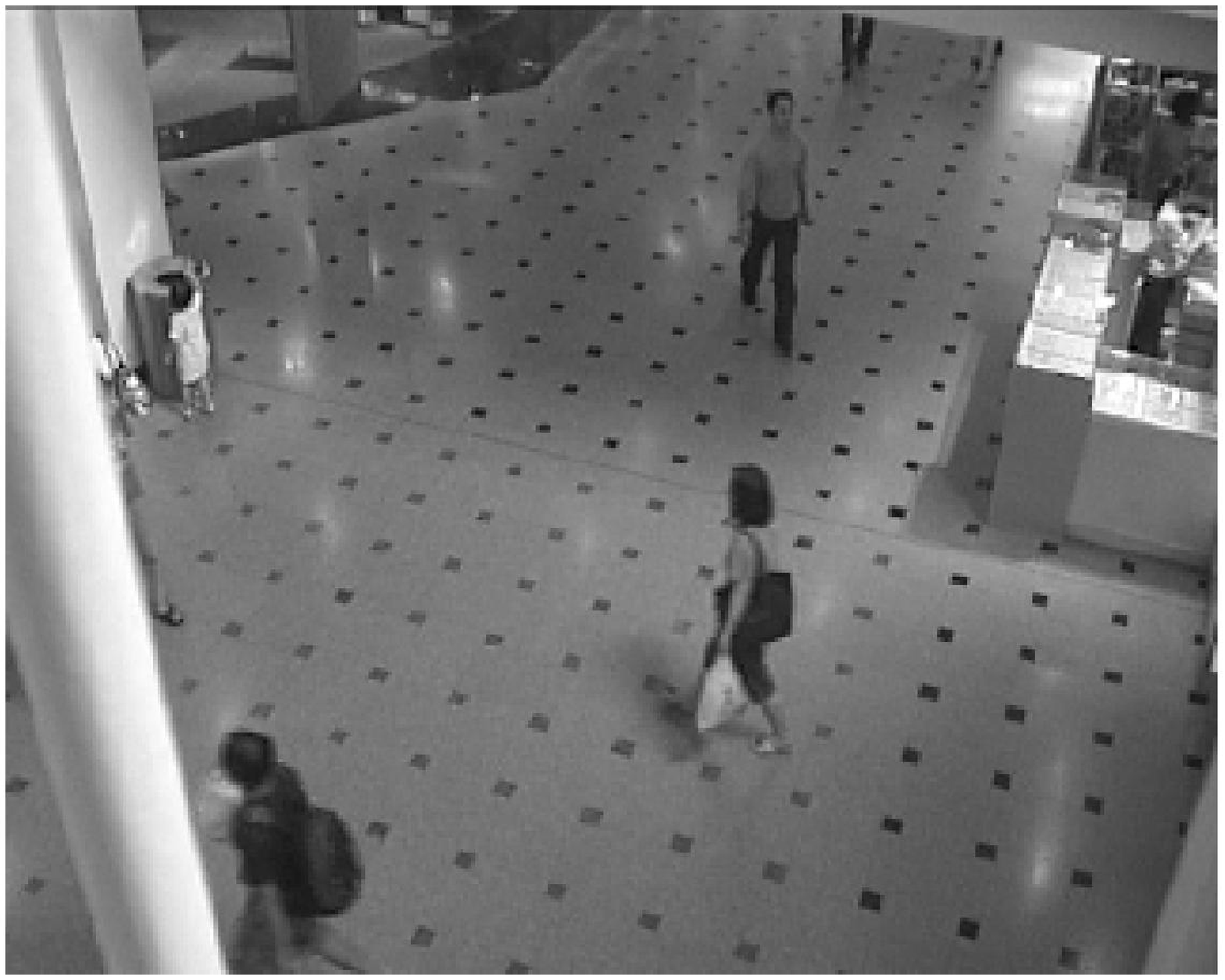}&
\includegraphics[width=1in]{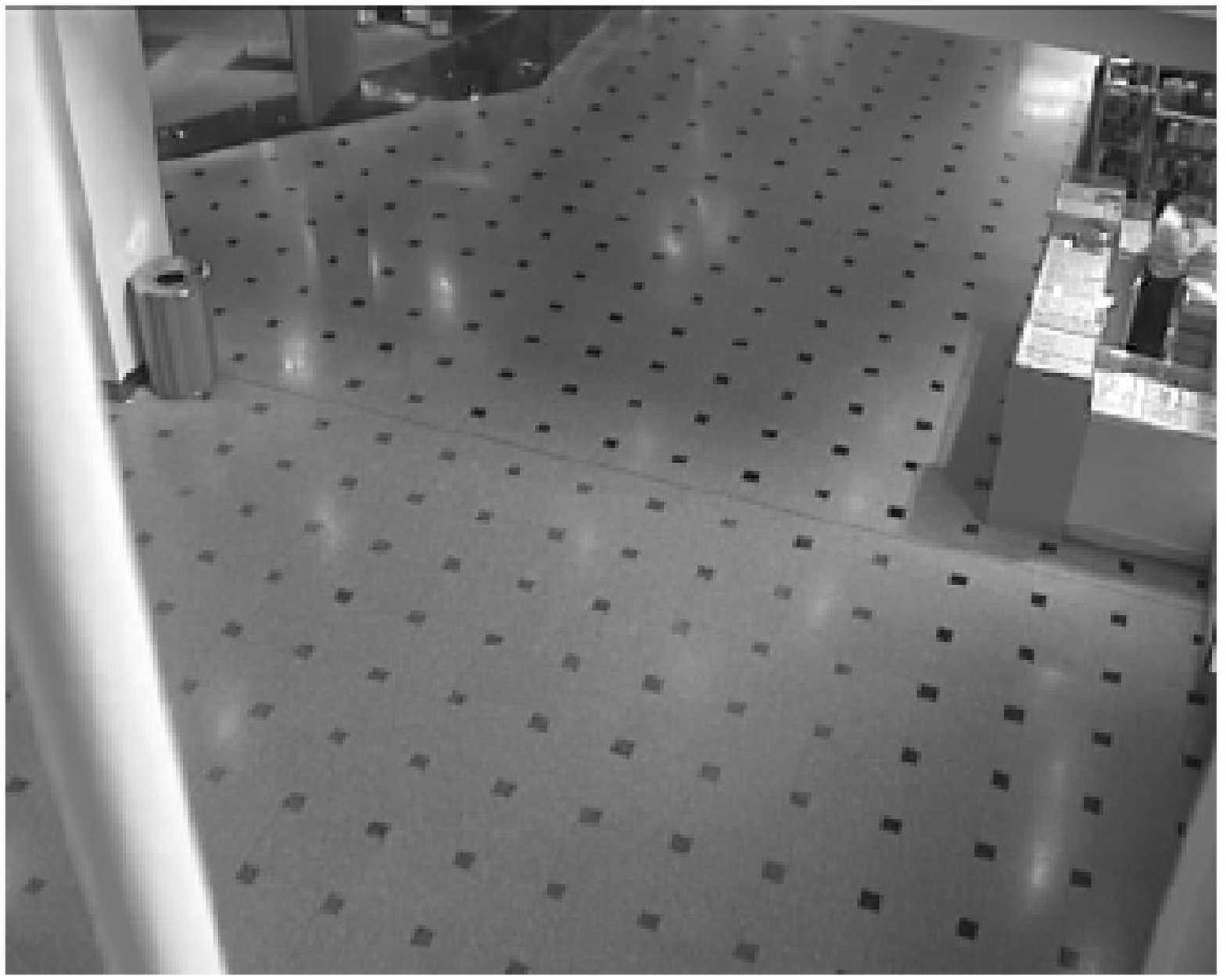}&
\includegraphics[width=1in]{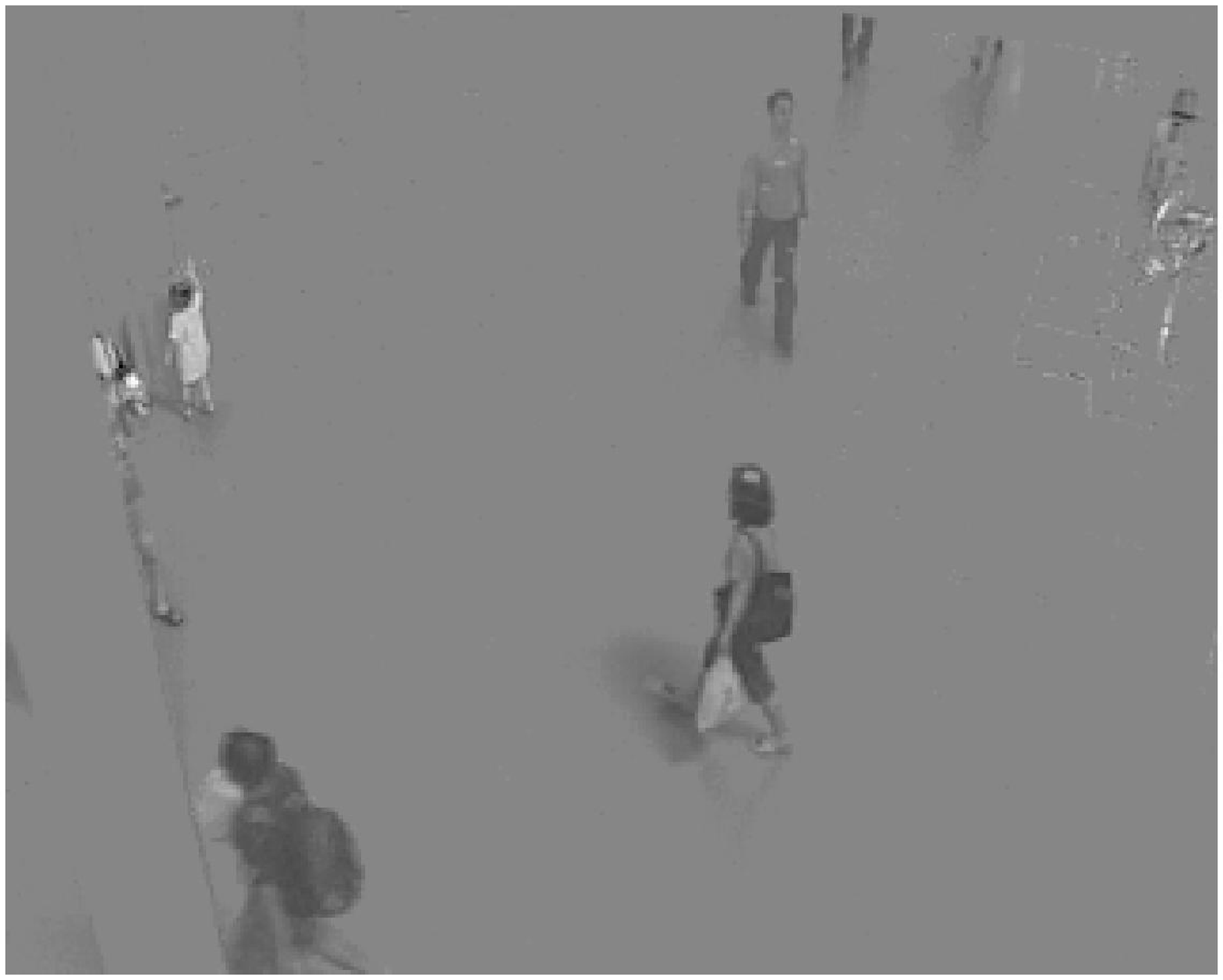}\\
\includegraphics[width=1in]{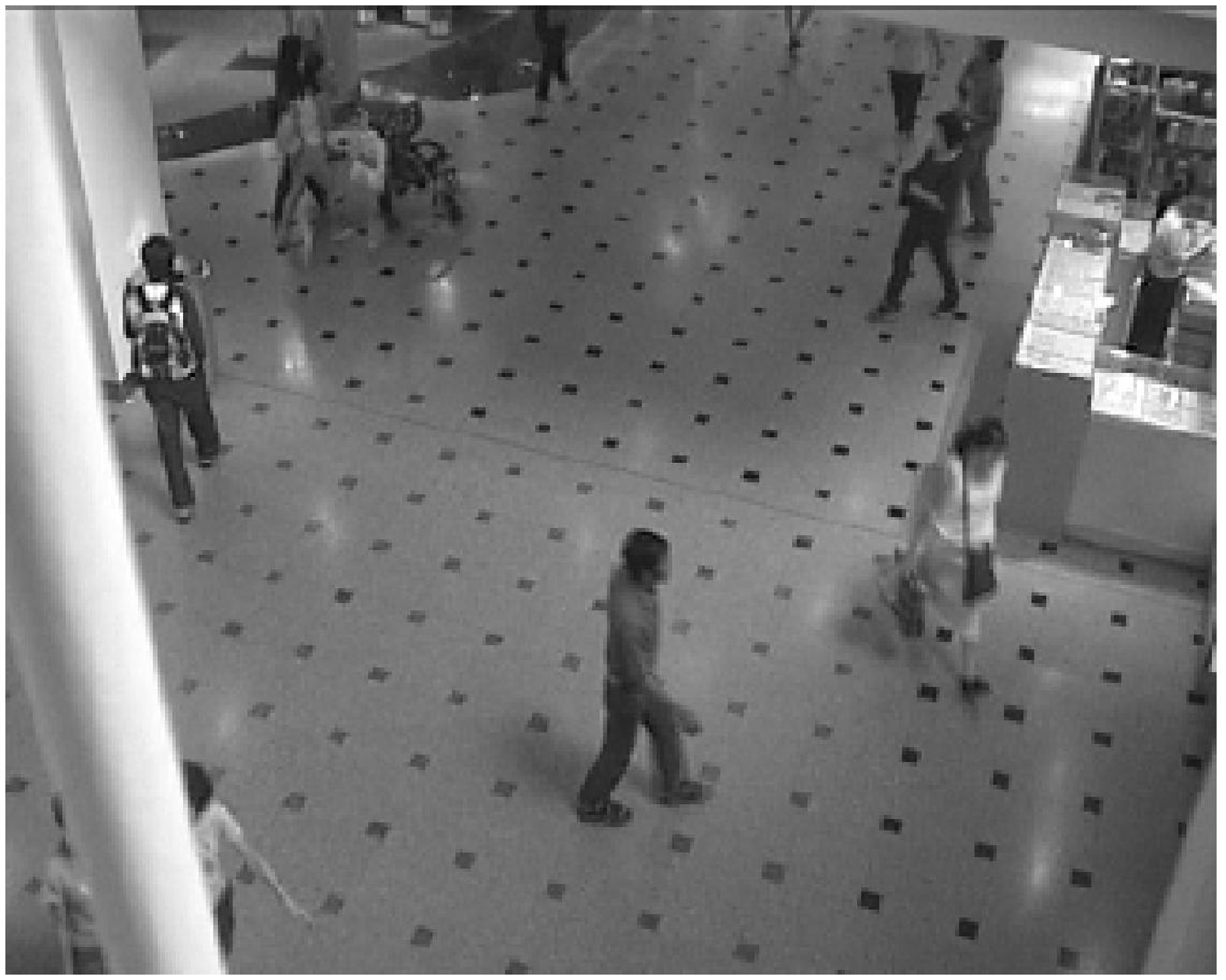}&
\includegraphics[width=1in]{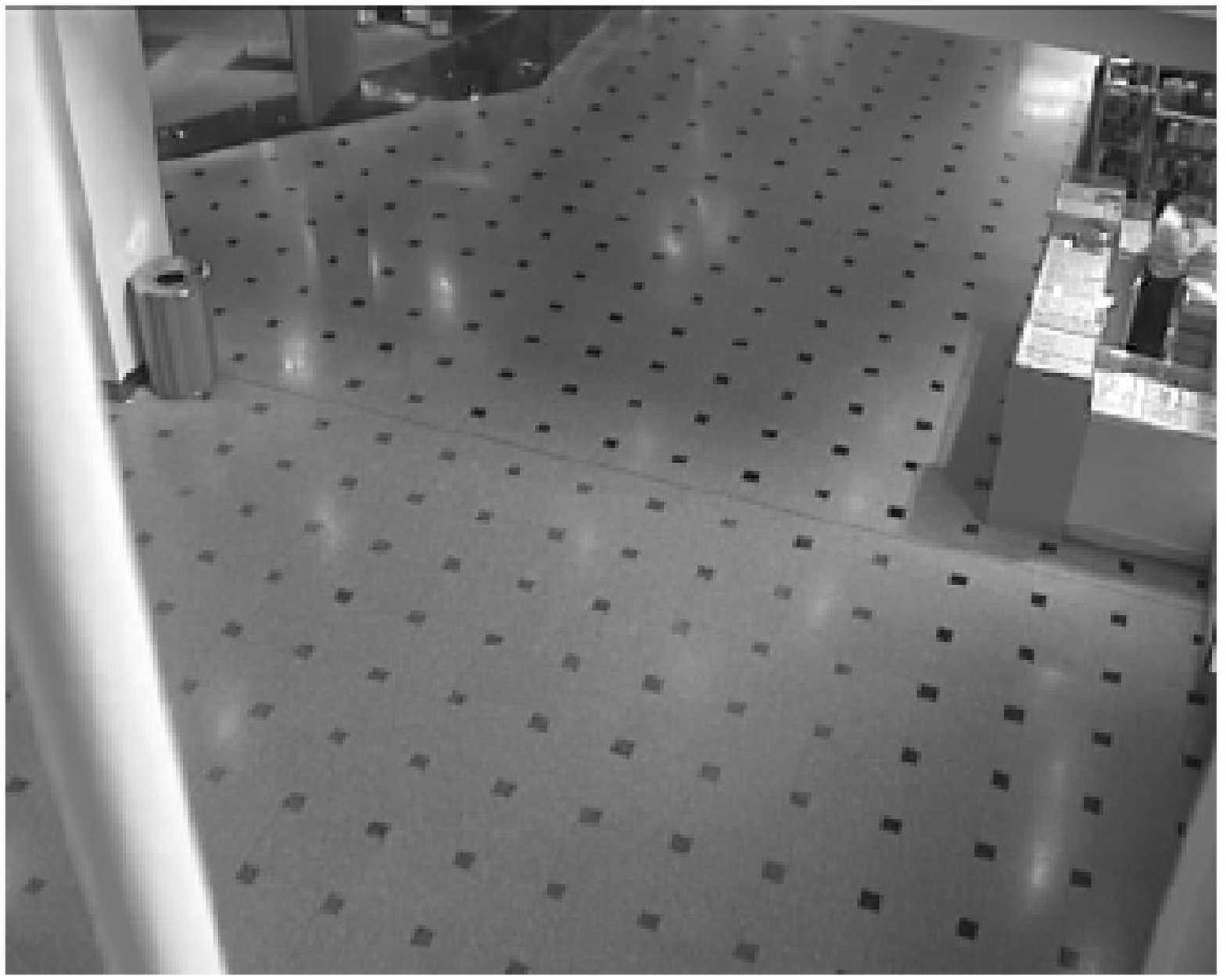}&
\includegraphics[width=1in]{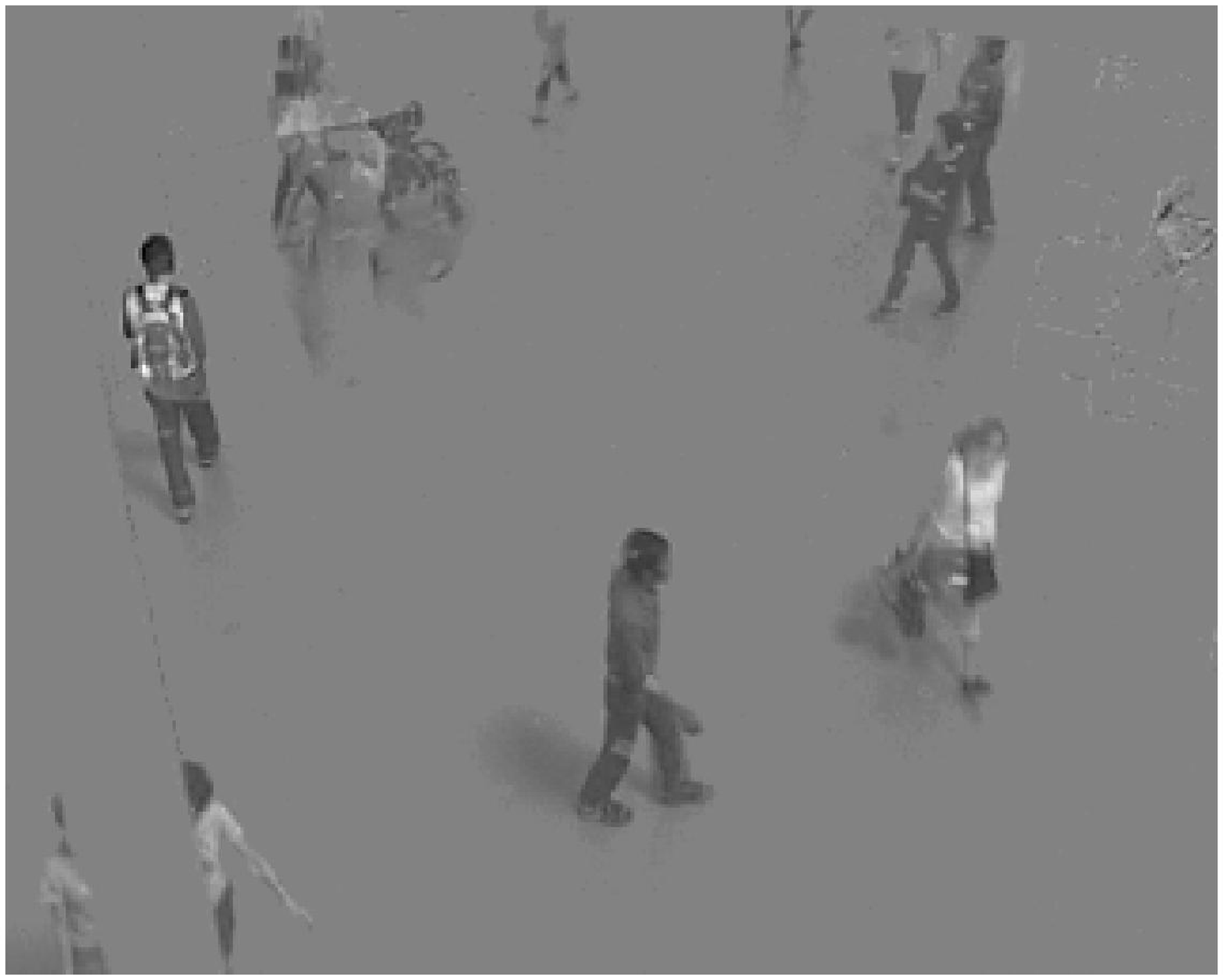}\\
\includegraphics[width=1in]{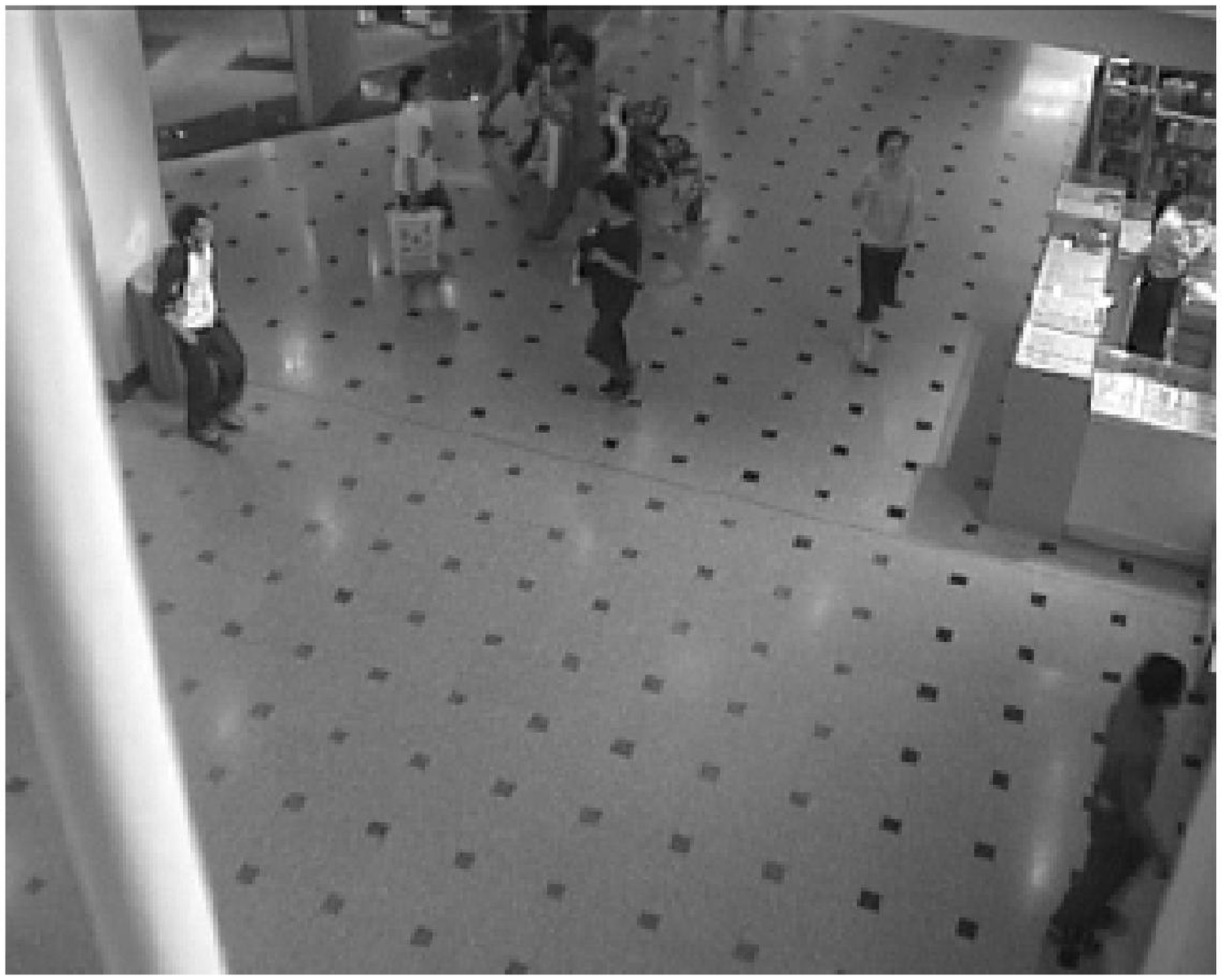}&
\includegraphics[width=1in]{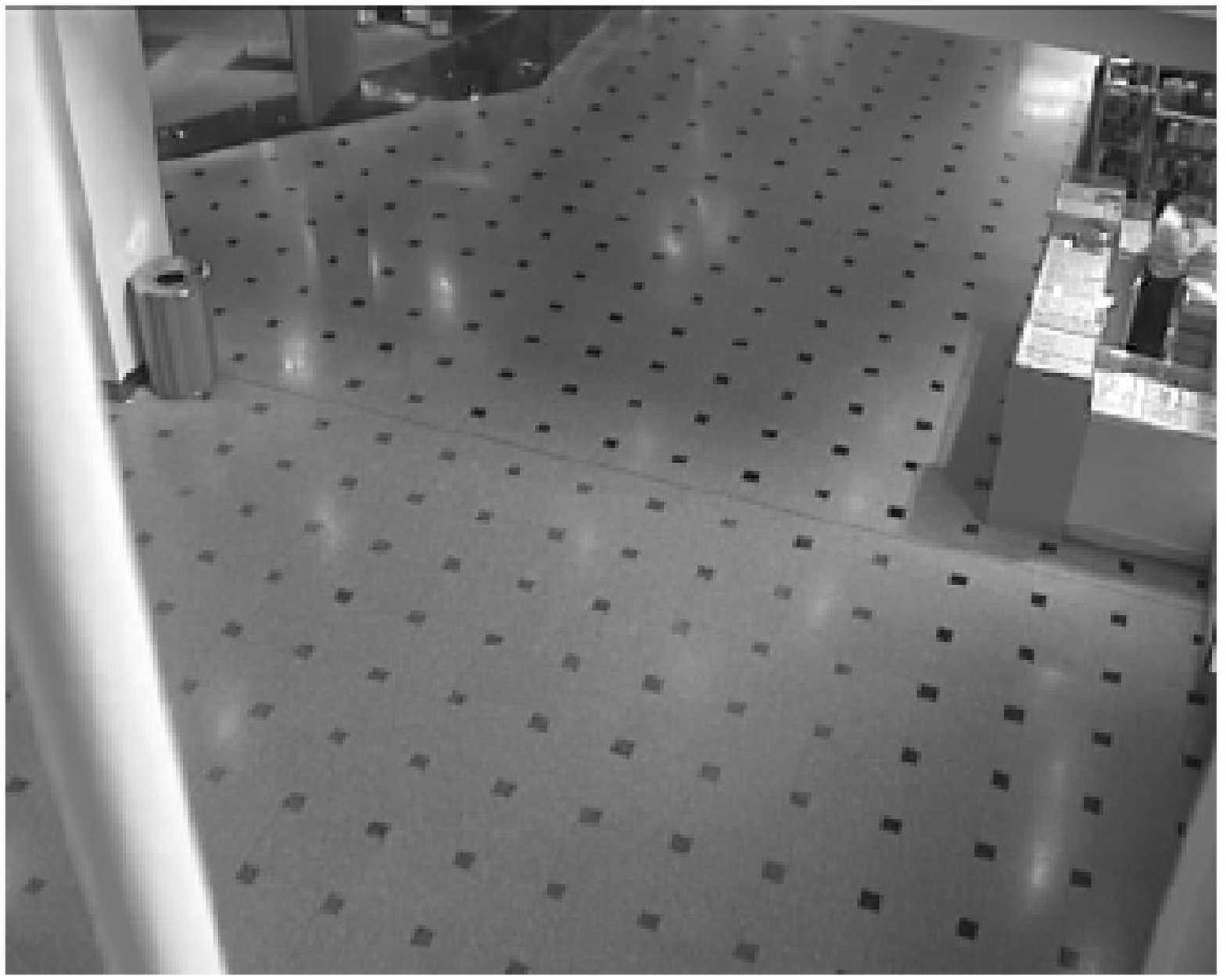}&
\includegraphics[width=1in]{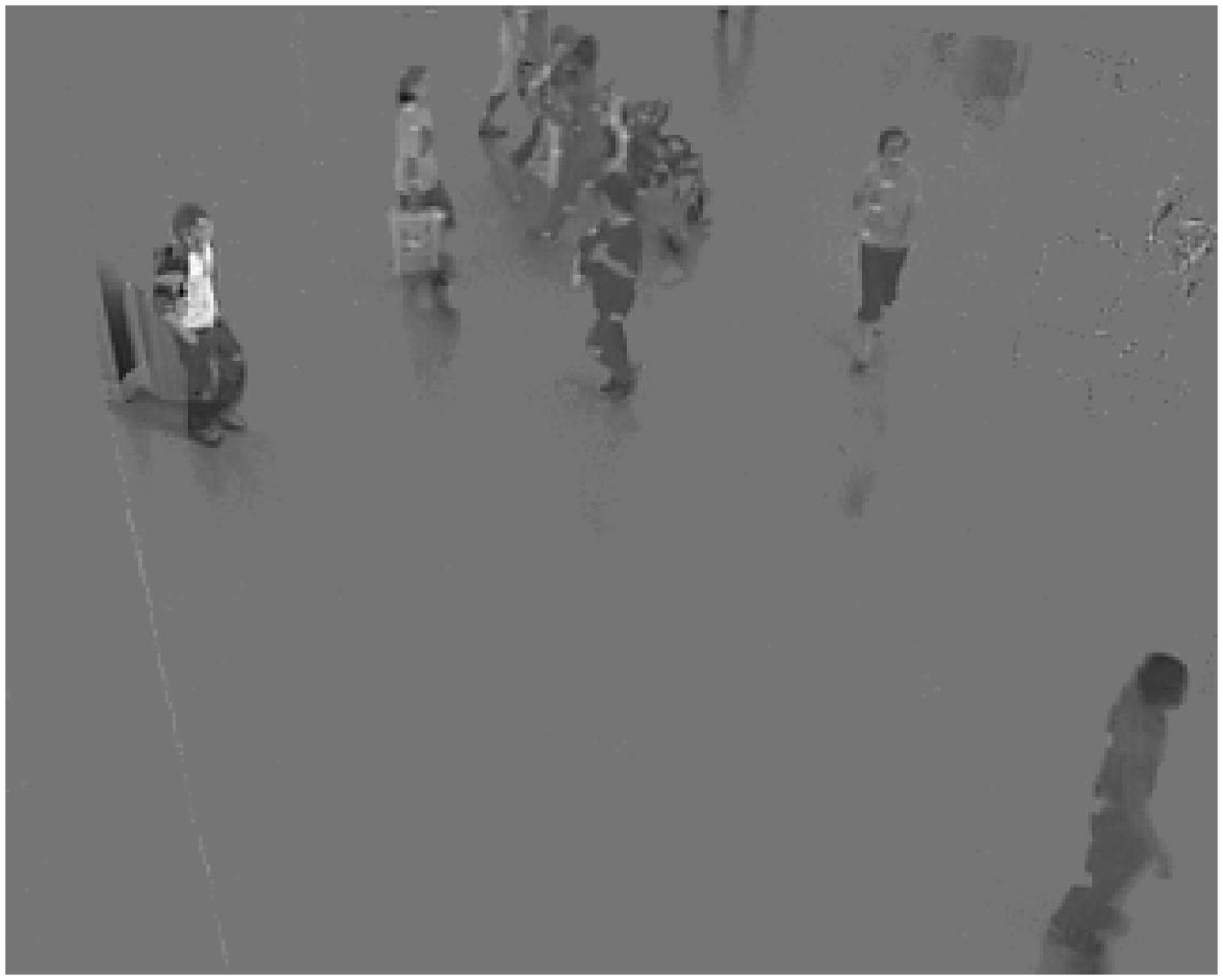}
\end{tabular}
 \caption{Three example frames from the ``mall'' video sequence. The left column shows original frames, the middle column EM-BiG-AMP-2 estimated background, and the right column EM-BiG-AMP-2 estimated foreground.}
  \label{fig:videoMall}
\end{figure}

We tested EM-BiG-AMP\footnote{The maximum allowed damping was reduced to $\texttt{stepMax} = 0.125$ for this experiment.  To reduce runtime, a relatively loose tolerance of $5 \times 10^{-4}$ was used to establish EM and BiG-AMP convergence.}  
on the popular ``mall'' video sequence,\footnote{See \url{http://perception.i2r.a-star.edu.sg/bk_model/bk_index.html}.} 
processing $200$ frames (of $256 \times 320$ pixels each) using an initial rank estimate of $\overline{N}=5$. 
\Figref{videoMall} shows the result, with original frames in the left column and EM-BiG-AMP-2 estimated background and foreground frames in the middle and right columns, respectively.
We note that, for this sequence, the rank-contraction strategy reduced the rank of the background component to $1$ after the first EM iteration.
Similar results (not shown here for reasons of space) were obtained with other video sequences.


\section{Dictionary Learning}
\label{sec:DL}

\subsection{Problem setup}
In dictionary learning (DL)~\cite{jtp_Rubinstein2010}, one seeks a dictionary $\vec{A} \in \Real^{M \times N}$ that allows the training samples $\vec{Y} \in \Real^{M \times L}$ to be encoded as $\vec{Y} = \vec{A}\vec{X}+\vec{W}$ for some sparse coefficient matrix $\vec{X}$ and small perturbation $\vec{W}$. 
One chooses $N=M$ to learn a square dictionary or $N>M$ (where $N$ is often a small multiple of $M$) to learn an overcomplete dictionary.
In general, one must have a sufficient number of training examples, $L \gg N$, to avoid over-fitting.\footnote{See \cite{jtp_Spielman2012} for a discussion of sample-size requirements for exact recovery of square dictionaries.} 

The BiG-AMP methodology is particularly well-suited to the DL problem, since both are inherently bilinear. 
In this work, for simplicity, we model the entries of $\vA$ using the iid standard normal prior \eqref{a_gauss2} and the entries of $\vX$ using the iid zero-mean Bernoulli-Gaussian (BG) prior 
\begin{align}
p_{\X_{nl}}(x) &= (1 - \xi) \delta(x) + \xi \Nor(x;0,\nu^x),
\end{align} 
where $\xi$ represents the activity rate and $\nu^x$ the active-component variance. 
However, other priors could be considered, such as truncated Gaussian mixtures \textb{with column-dependent prior parameters} in the case of non-negative matrix factorization \cite{Vila:SPIE:13}.
For the likelihood $p_{\vY|\vZ}$, we again select the PIAWGN model \textb{\eqref{PIAWGNPart2}}, but note that in most applications of DL the observations are completely observed.

\subsection{Initialization} \label{sec:DLinit}
In general, we advocate initializing $\hat{x}_{nl}(1)$ at the mean of the assumed prior on $\X_{nl}$, and initializing the variances $\nu^x_{nl}(1)$ and $\nu^a_{mn}(1)$ at $10$ times the variance of $\X_{nl}$ and $\A_{mn}$, respectively.
We now discuss several strategies for initializing the dictionary estimates $\hat{a}_{mn}(1)$.
One option is to draw $\hat{a}_{mn}(1)$ randomly from the assumed prior on $\A_{mn}$, as suggested for MC and RPCA. 
Although this approach works reasonably well, the prevalence of local minima in the DL problem motivates us to propose two alternative strategies.
The first alternative is to exploit prior knowledge of a ``good'' sparsifying dictionary $\vec{A}$, in the case that such knowledge exists.
With natural images, for example, the discrete cosine transform (DCT) and discrete wavelet transform (DWT) are known to yield reasonably sparse transform coefficients, and so the DCT and DWT matrices make appropriate initializations of $\hvec{A}(1)$. 

The second alternative is to initialize $\hvec{A}(1)$ using an appropriately chosen subset of the columns of $\vec{Y}$, which is well motivated in the case that there exists a very sparse representation $\vec{X}$.
For example, if there existed a decomposition $\vec{Y}=\vec{AX}$ in which $\vec{X}$ had $1$-sparse columns, then the columns of $\vec{A}$ would indeed match a subset of the columns of $\vec{Y}$ (up to a scale factor).
In the general case, however, it is not apriori obvious which columns of $\vec{Y}$ to choose, and so we suggest the following greedy heuristic, which aims for a well-conditioned $\hvec{A}(1)$: 
select (normalized) columns from $\vec{Y}$ sequentially, in random order, accepting each candidate if the mutual coherences with the previously selected columns and the condition number of the resulting submatrix are all sufficiently small.
If all columns of $\vec{Y}$ are examined before finding $N$ acceptable candidates, then the process is repeated using a different random order.
If repeated re-orderings fail, then $\hvec{A}(1)$ is initialized using a random draw from $\vA$.

\subsection{EM-BiG-AMP} \label{sec:EMdl}
To tune the distributional parameters $\vec{\theta} = [\nu^w, \nu_0^x, \xi   ]^T$, we can straightforwardly apply the EM approach from \Xsecref{EM}.
For this, we suggest initializing $\xi = 0.1$ (since \secref{DLexamples} shows that this works well over a wide range of problems) and initializing $\nu^x_0$ and $\nu^w$ using a variation on the procedure suggested for MC that accounts for the sparsity of $\X_{nl}$: 
\begin{align}
\nu^w &= \frac{\norm{P_\Omega(\vec{Y})}_F^2}{(\textrm{SNR}^0 + 1) |\Omega|}\\
\nu^x_0 &= \frac{1}{N\xi}\bigg[ \frac{\norm{P_\Omega(\vec{Y})}_F^2}{|\Omega|} - \nu^w\bigg].
\end{align}

\subsection{Avoiding Local Minima} \label{sec:DLreinit}
The DL problem is fraught with local minima (see, e.g., \cite{jtp_Spielman2012}), and so it is common to run iterative algorithms several times from different random initializations.
For BiG-AMP, we suggest keeping the result of one initialization over the previous if both\footnote{As an alternative, if both the previous and current solutions achieve sufficiently small residual error, then only the average sparsity is considered in the comparison.} the residual error $\norm{\vec{\hat{Z}} - \vec{Y}}_F$ and the average sparsity (as measured by $\frac{1}{NL}\sum_{nl} \Pr\{\X_{nl}\!\neq\! 0\,|\,\vY\!=\!\vec{Y}\}$) decrease.


\subsection{Dictionary Learning Experiments} \label{sec:DLexamples}
In this section, we numerically investigate the performance of EM-BiG-AMP for DL, as described in \secref{DL}. 
Comparisons are made with the greedy K-SVD algorithm~\cite{jtp_Aharon2006}, the SPAMS implementation of the online approach~\cite{jtp_Mairal2010}, and the ER-SpUD(proj) approach for square dictionaries~\cite{jtp_Spielman2012}. 

\subsubsection{Noiseless Square Dictionary Recovery} \label{sec:DLptc}
We first investigate recovery of square (i.e., $N=M$) dictionaries from noise-free observations, repeating the experiment from \cite{jtp_Spielman2012}. 
For this, realizations of the true dictionary $\vec{A}$ were created by drawing elements independently from the standard normal distribution and subsequently scaling the columns to unit $\ell_2$ norm. 
Meanwhile, realizations of the true $\vec{X}$ were created by selecting $K$ entries in each column uniformly at random and drawing their values from the standard normal distribution, while setting all other entries to zero. 
Finally, the observations were constructed as $\vec{Y} = \vec{A}\vec{X}$, from which the algorithms estimated $\vec{A}$ and $\vec{X}$ (up to a permutation and scale).
The accuracy of the dictionary estimate $\hvec{A}$ was quantified using the relative NMSE metric \cite{jtp_Spielman2012}
\begin{align}
\text{NMSE}(\hvec{A}) \defn
\min_{\vec{J}\in\mc{J}} \frac{\norm{\vec{\hat{A}J} - \vec{A}}^2_F}{\norm{\vec{A}}^2_F}, \label{eq:dlRecoveryError}
\end{align}
where $\vec{J}$ is a generalized permutation matrix used to resolve the permutation and scale ambiguities. 

\begin{figure}[htb]
\centering
\psfrag{K-SVD}[B][B][0.7]{\sf \textbf{K-SVD}}
\psfrag{SPAMS}[B][B][0.7]{\sf \textbf{SPAMS}}
\psfrag{EM-BiG-AMP}[B][B][0.7]{\sf \textbf{EM-BiG-AMP}}
\psfrag{ER-SpUD (proj)}[B][B][0.7]{\sf \textbf{ER-SpUD(proj)}}
\psfrag{Dictionary Size}[t][][0.7]{\sf \quad dictionary size $N$}
\psfrag{Sparsity}[][][0.65]{\sf training sparsity $K$}
\begin{tabular}{cccc}
\includegraphics[width=1.6in]{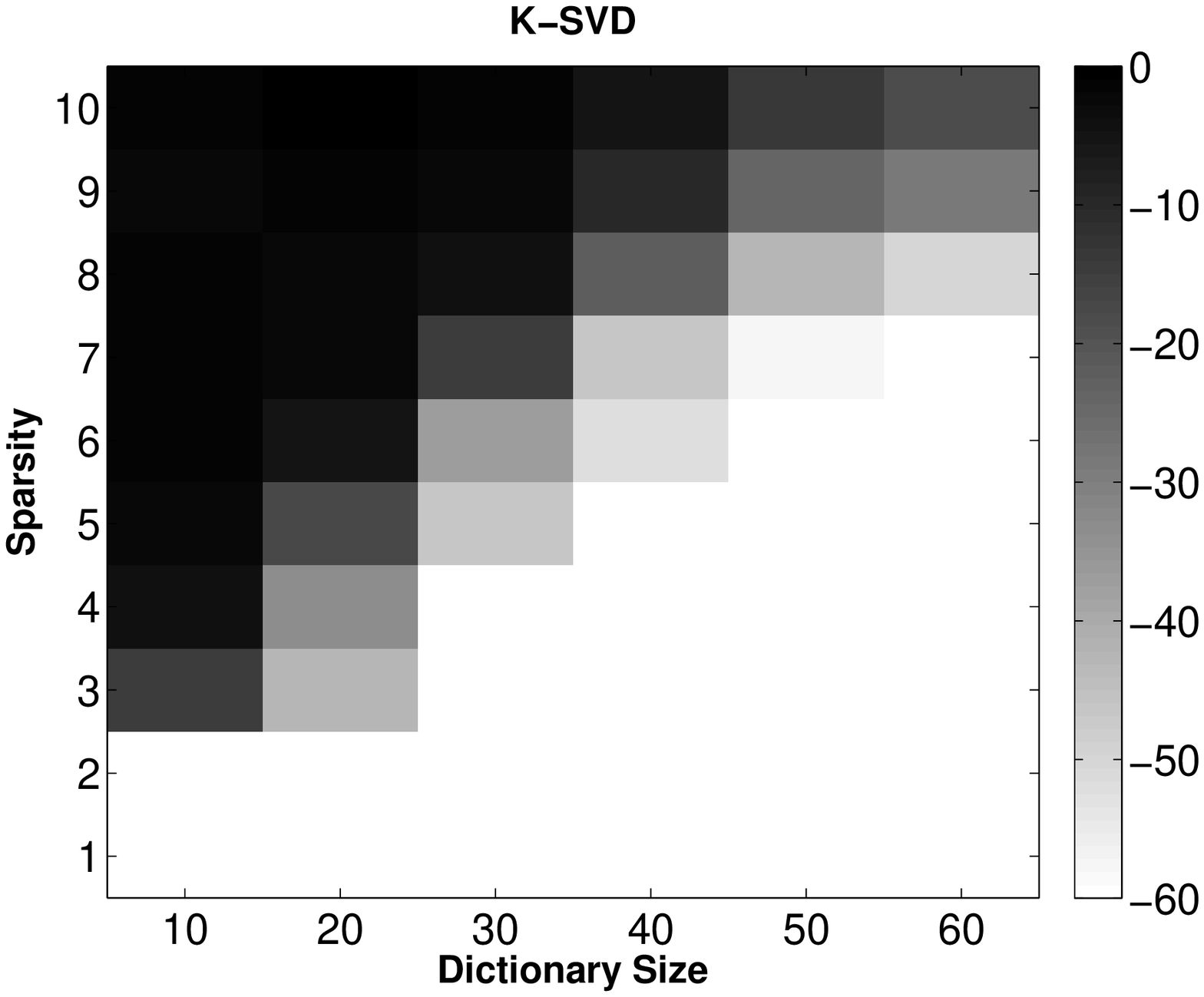}&
\includegraphics[width=1.6in]{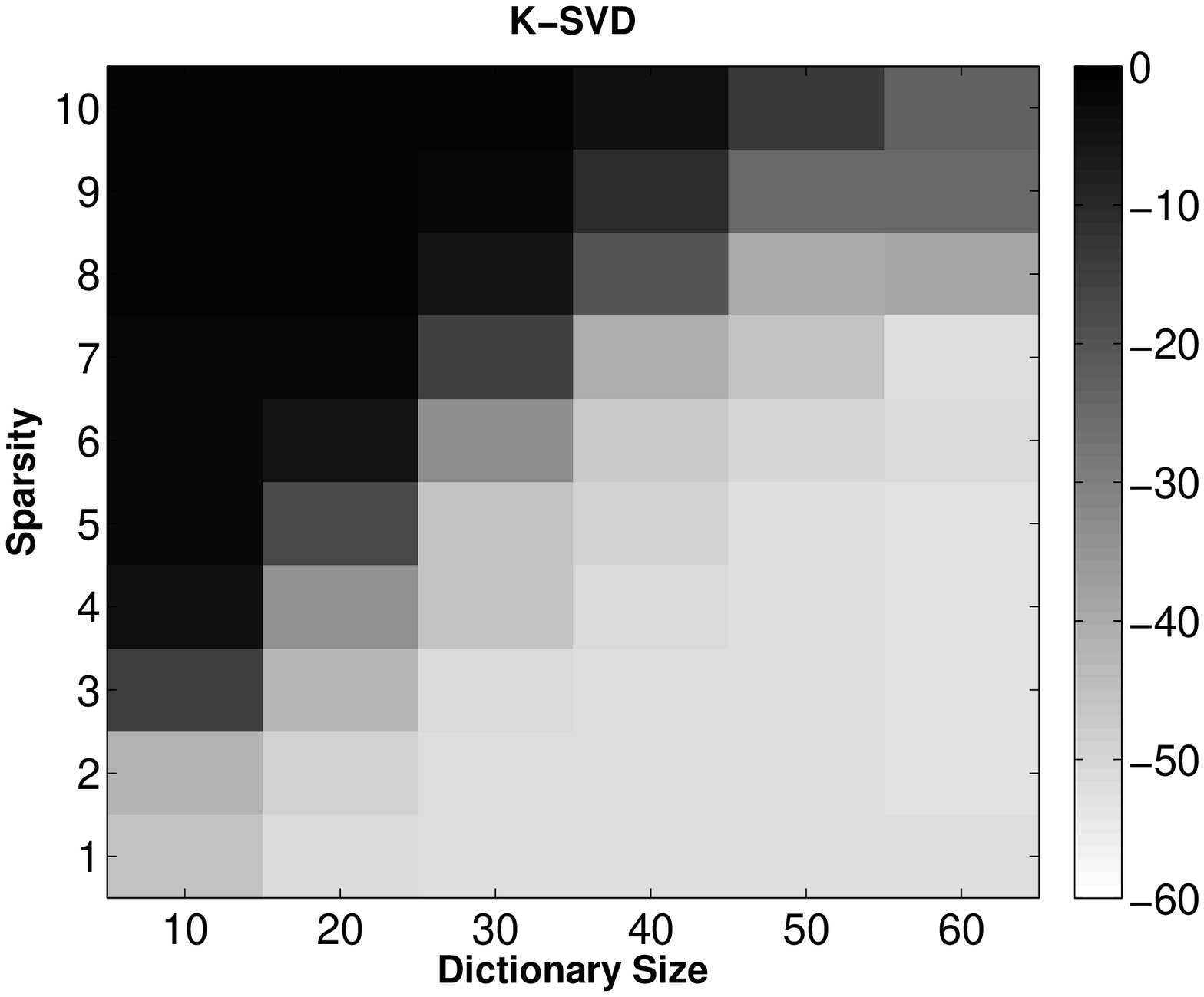}\\[2mm]
\includegraphics[width=1.6in]{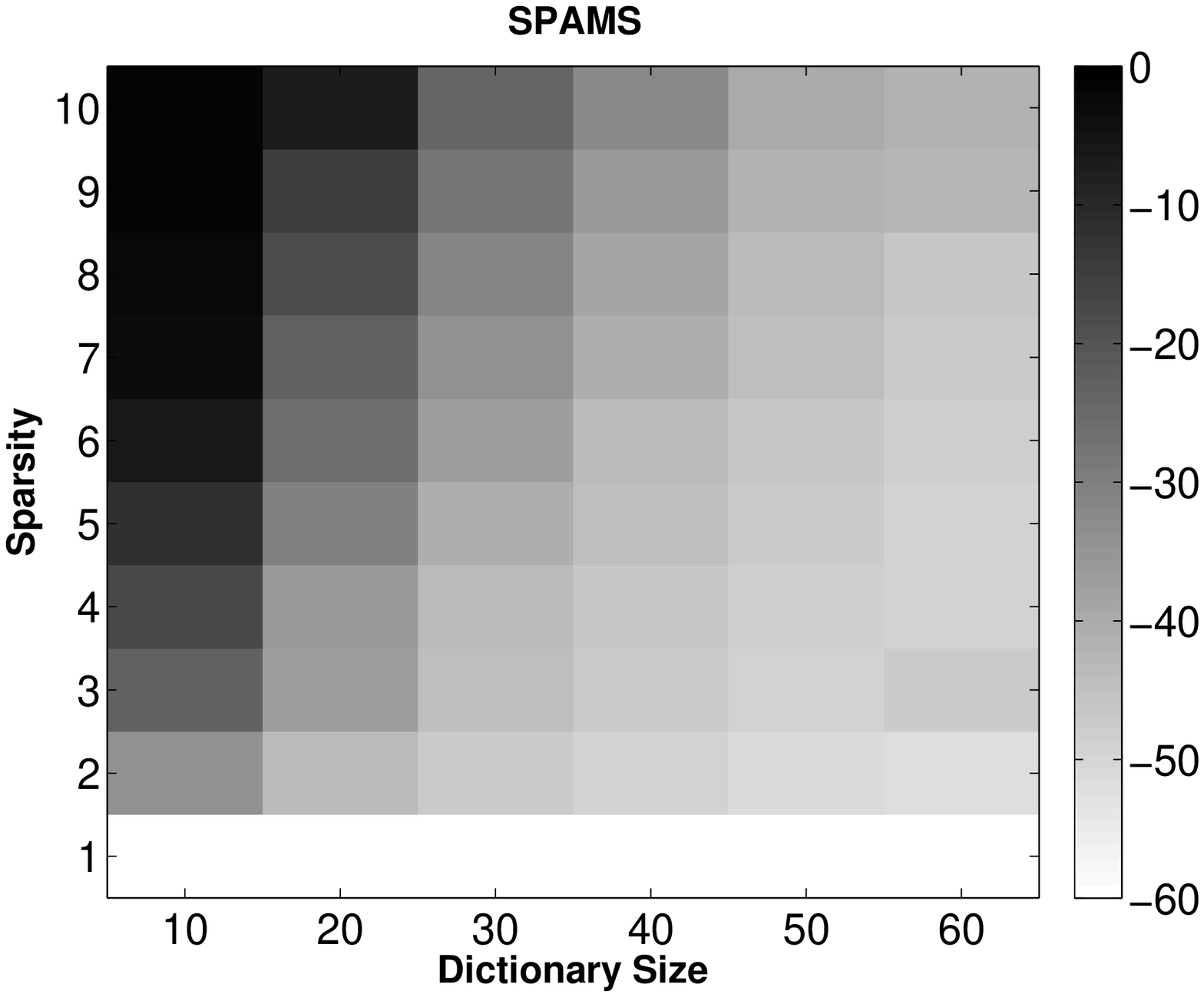}&
\includegraphics[width=1.6in]{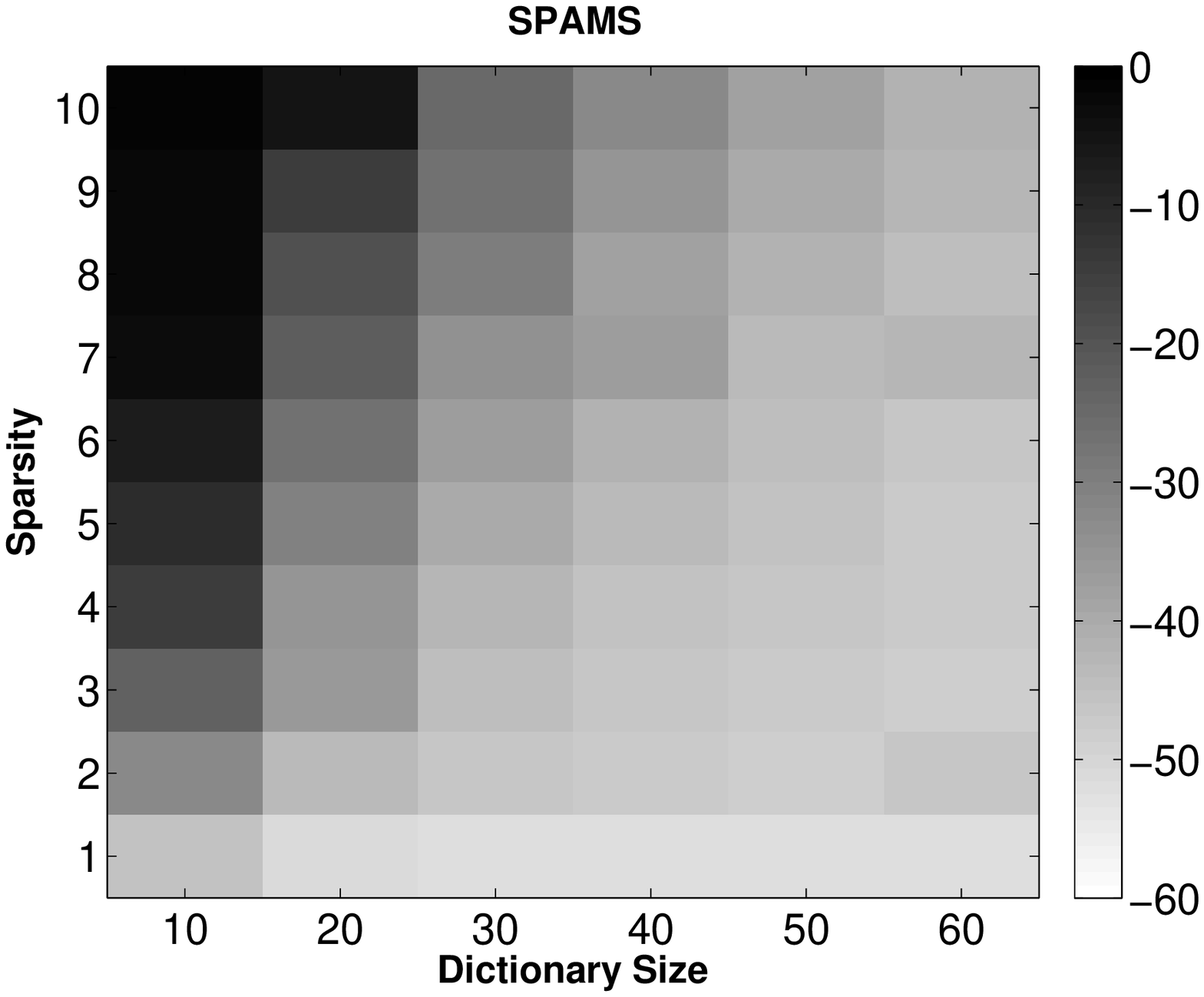}\\[2mm]
\includegraphics[width=1.6in]{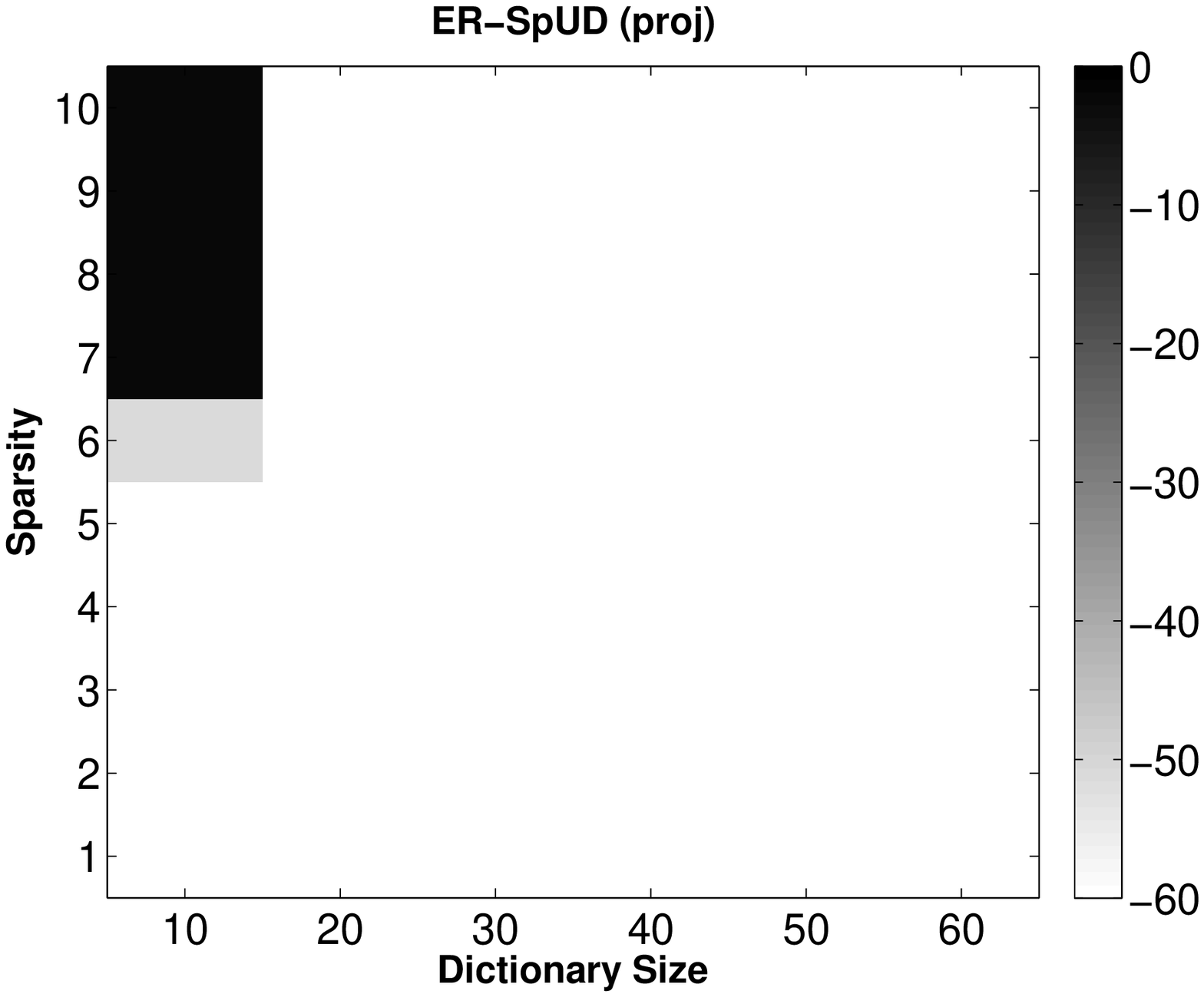}&
\includegraphics[width=1.6in]{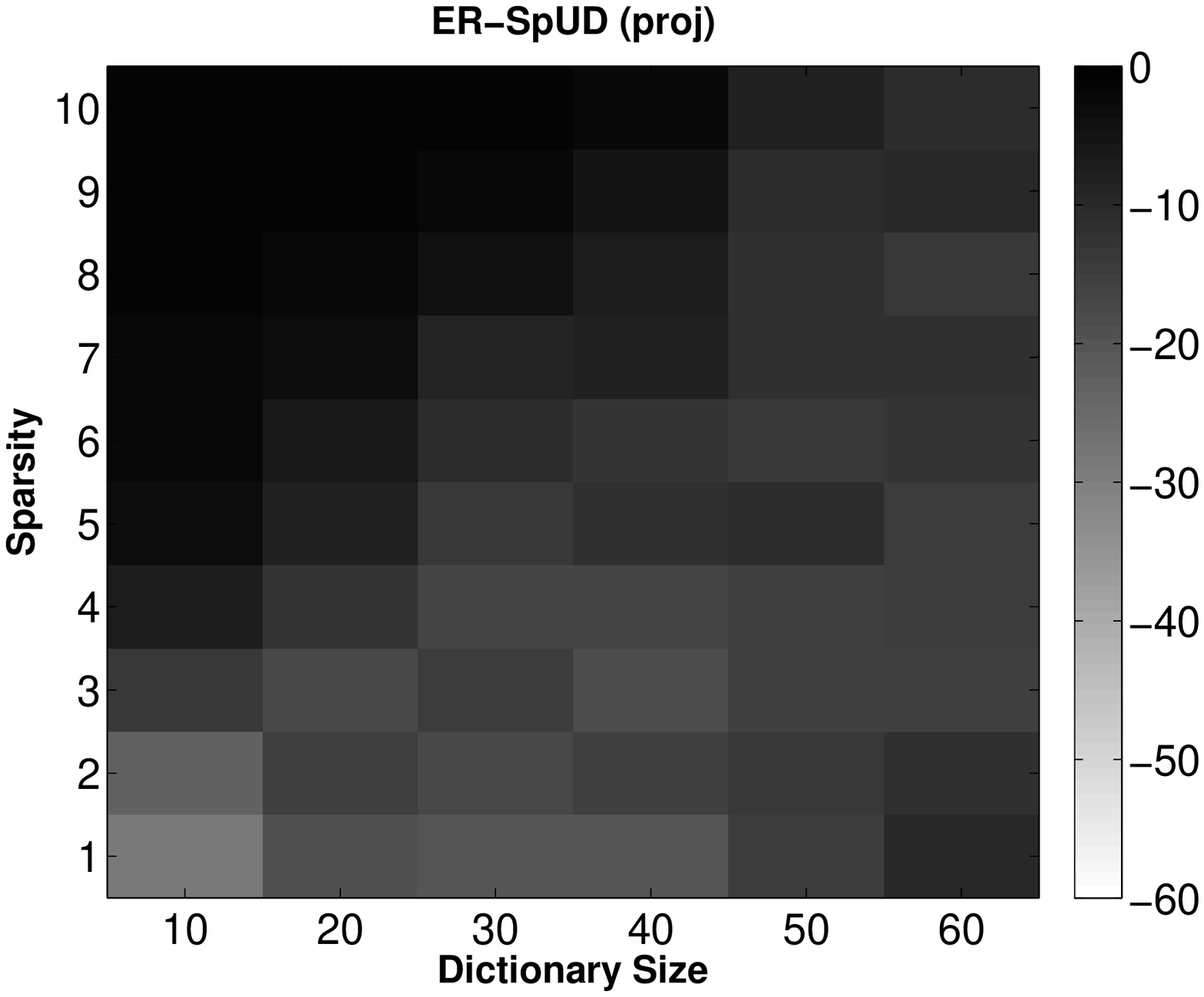}\\[2mm]
\includegraphics[width=1.6in]{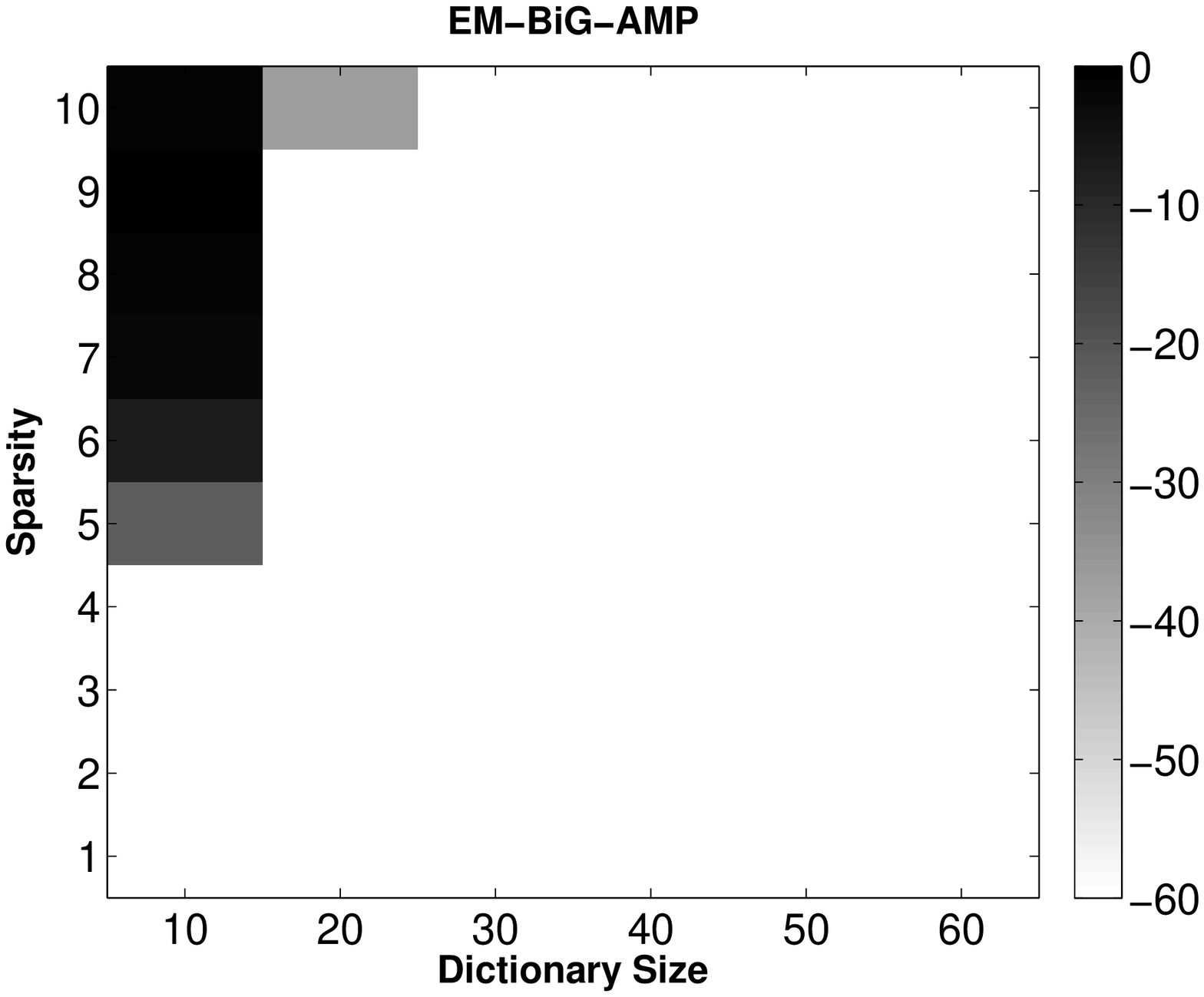}&
\includegraphics[width=1.6in]{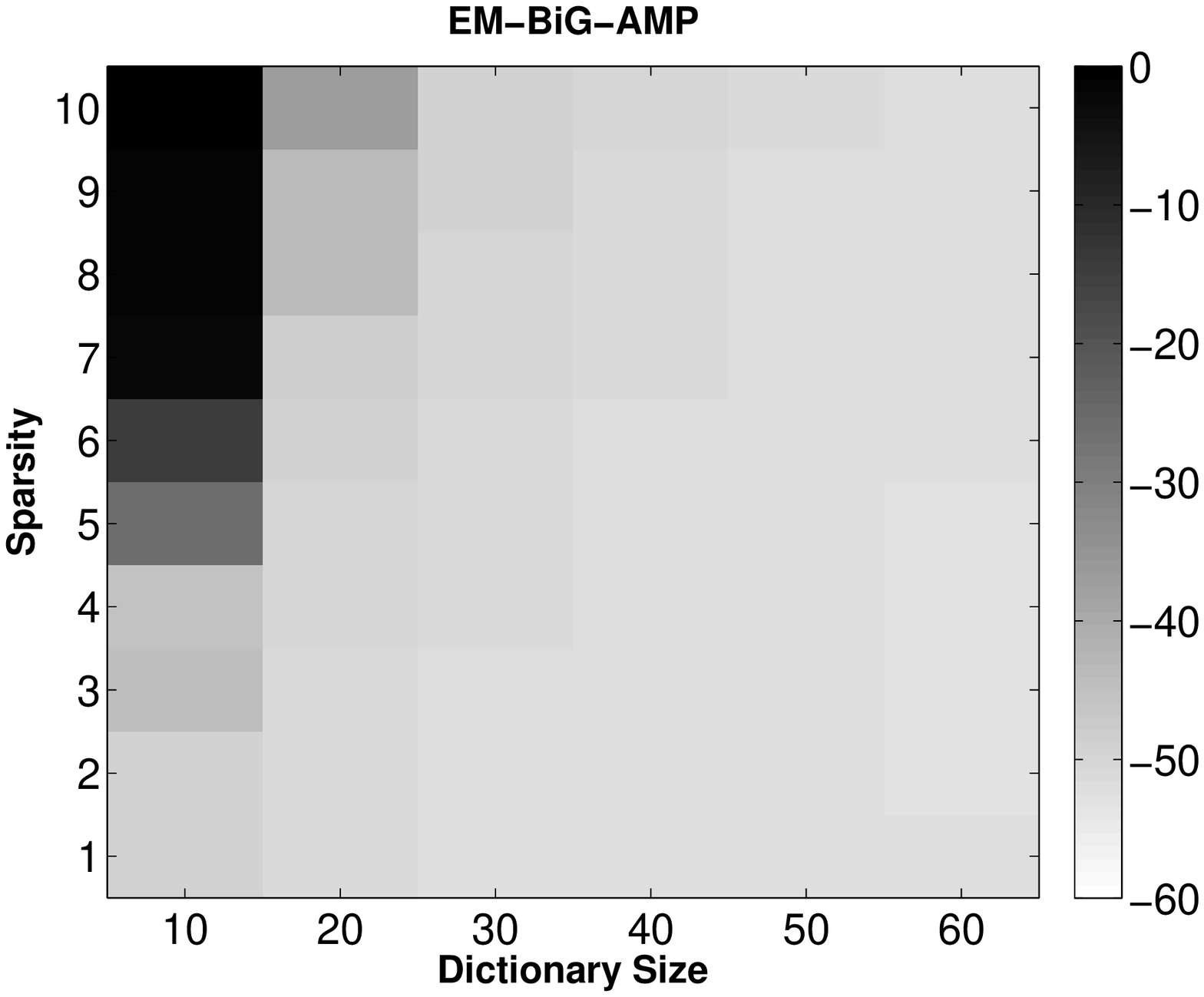}
\end{tabular}
\caption{Mean NMSE (over $10$ trials) for recovery of an $N\times N$ dictionary from $L = 5 N \log N$ training samples, each of sparsity $K$, in the noiseless case (left) and under AWGN of $40$~dB SNR (right), for several algorithms.}
\label{fig:PhaseDL}
\end{figure}

The subplots on the left of \figref{PhaseDL} show the mean NMSE achieved by K-SVD, SPAMS, ER-SpUD(proj), and EM-BiG-AMP,\footnote{%
EM-BiG-AMP was allowed up to $20$ EM iterations, with each EM iteration allowed a minimum of $30$ and a maximum of $1500$ BiG-AMP iterations. 
K-SVD was allowed up to $100$ iterations and provided with knowledge of the true sparsity $K$. 
SPAMS was allowed $1000$ iterations and run using the hand-tuned penalty $\lambda = 0.1/\sqrt{N}$. 
The non-iterative ER-SpUD(proj) was run using code provided by the authors without modification.}
respectively, over $50$ problem realizations, for various combinations of dictionary size $N\in\{10,\dots,60\}$ and data sparsity $K\in\{1,\dots,10\}$, 
using $L=5N\log N$ training examples.
K-SVD, SPAMS, and EM-BiG-AMP were run with $10$ different random initializations for each problem realization. 
To choose among these initializations, EM-BiG-AMP used the procedure from \secref{DLreinit}, while K-SVD and SPAMS used oracle knowledge to choose the NMSE-minimizing initialization. 

The left column in \figref{PhaseDL} shows that the K-SVD, ER-SpUD(proj), and EM-BiG-AMP algorithms all exhibit a relatively sharp phase-transition curve (PTC) separating success and failure regions, and that ER-SpUD(proj)'s PTC is the best, while EM-BiG-AMP's PTC is very similar.
Meanwhile, K-SVD's PTC is much worse and SPAMS performance is not good enough to yield a sharp phase transition,\footnote{Our results for SPAMS and ER-SPUD(proj) in the left column of \figref{PhaseDL} are nearly identical to those in \cite[Fig.~1]{jtp_Spielman2012}, while our results for K-SVD are noticeably better.}
despite the fact that both use oracle knowledge.
EM-BiG-AMP, by contrast, was not provided with any knowledge of the DL problem parameters, such as the true sparsity or noise variance (in this case, zero).

\begin{figure*}[htb]
\centering
\psfrag{Dictionary Size}[t][][0.7]{\sf \quad dictionary size $N$}
\psfrag{time (seconds)}[B][B][0.7]{\sf runtime (sec)}
\psfrag{K = 1}[][][0.6]{\sf training sparsity $K = 1$}
\psfrag{K = 5}[][][0.6]{\sf training sparsity $K = 5$}
\psfrag{K = 10}[][][0.6]{\sf training sparsity $K = 10$}
 \begin{tabular}{ccc}
\includegraphics[width=2.2in]{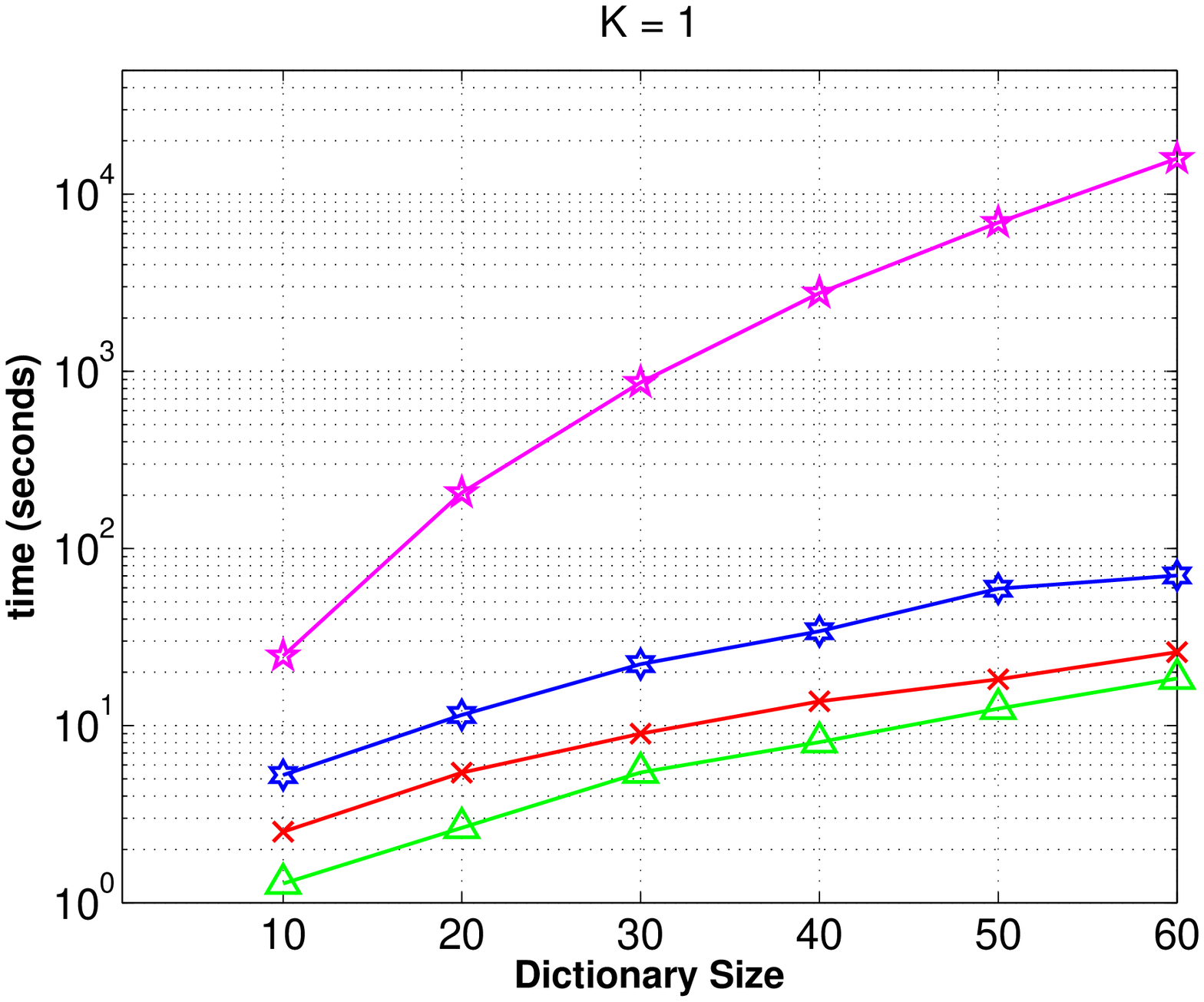}&
\includegraphics[width=2.2in]{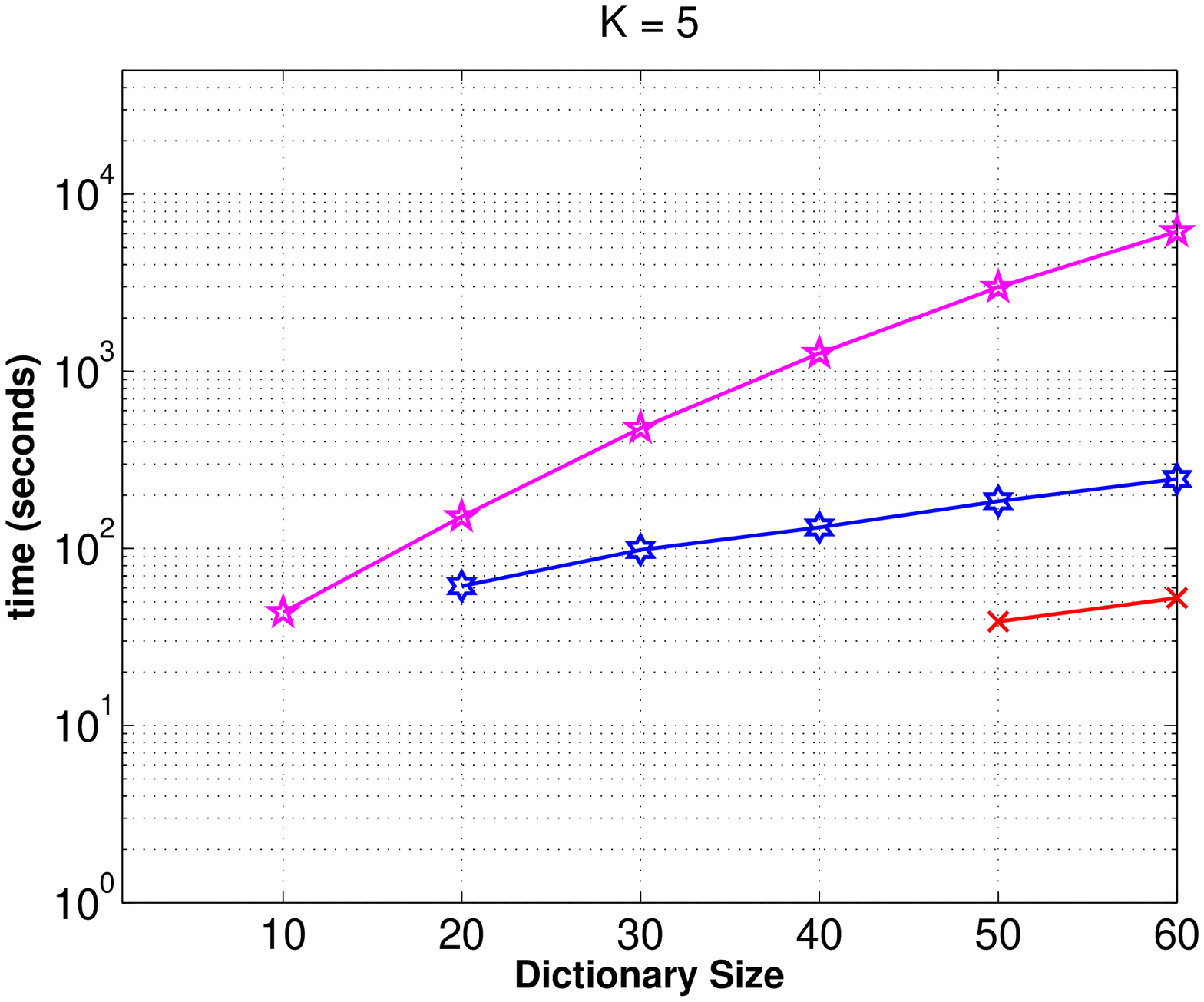}&
\includegraphics[width=2.2in]{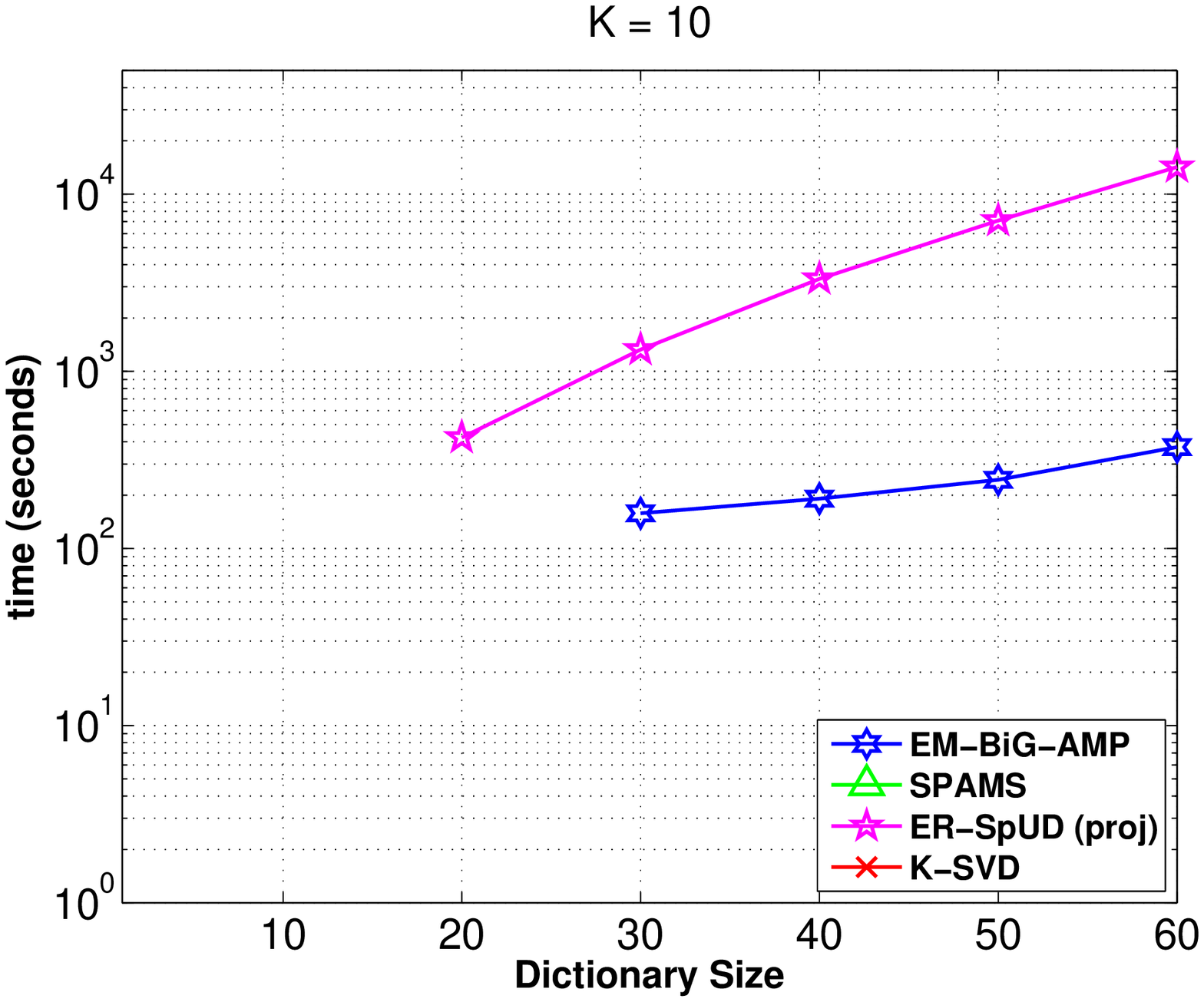}
\end{tabular}
 \caption{\textb{Median runtime until termination (over $10$ trials) versus dictionary size $N$, for noiseless recovery of a square dictionary from $L = 5 N \log N$ $K$-sparse samples, for several values of training sparsity $K$. Missing values indicate that the algorithm did not achieve the required NMSE before termination and correspond to the black regions in the panes on the left of \figref{PhaseDL}.}}
 \label{fig:PhaseCutsDL}
\end{figure*}

For the same problem realizations, \figref{PhaseCutsDL} shows the runtime to NMSE~$=-60$~dB (measured using \textb{MATLAB's} \texttt{tic} and \texttt{toc}) versus dictionary size $N$.
The results show that EM-BG-AMP runs within an order-of-magnitude of the fastest algorithm (SPAMS) and more than two orders-of-magnitude faster than ER-SpUD(proj)\footnote{%
The simpler ``SC'' variant of ER-SpUD reduces the computational cost relative to the ``proj'' variant, but results in a significantly worse PTC (see \cite[Fig.~1]{jtp_Spielman2012}) and remains slower than EM-BiG-AMP for larger problems.}
for larger dictionaries.

\subsubsection{Noisy Square Dictionary Recovery} \label{sec:DLnoisy}
Next we examined the recovery of square dictionaries from AWGN-corrupted observations. 
For this, we repeated the experiment from the previous section, but constructed the observations as $\vec{Y}=\vec{Z}+\vec{W}$, where $\vec{Z}=\vec{AX}$ and $\vec{W}$ contained AWGN samples with variance adjusted to achieve an SNR $=\E\{\sum_{m,l}|\Z_{ml}|^2\}/\E\{\sum_{m,l}|\Y_{ml}-\Z_{ml}|^2\}$ of $40$~dB.

The right subplots in \figref{PhaseDL} show the mean value (over $10$ trials) of the relative NMSE from \eqref{dlRecoveryError} when recovering an $N\times N$ dictionary from $L=5N\log N$ training samples of sparsity $K$, for various combinations of $N$ and $K$. 
These subplots show that ER-SpUD(proj) falls apart in the noisy case, which is perhaps not surprising given that it is intended only for noiseless problems. 
Meanwhile, the K-SVD, SPAMS, and EM-BiG-AMP algorithms appear to degrade gracefully in the presence of noise, yielding NMSE~$\approx-50$~dB at points below the noiseless PTCs.

\subsubsection{Recovery of Overcomplete Dictionaries} \label{sec:DLnonSquare}
Finally, we consider recovery of overcomplete $M\times N$ dictionaries, i.e., the case where $M<N$. 
In particular, we investigated the twice overcomplete case, $N=2M$. 
For this, random problem realizations were constructed in the same manner as described earlier, except for the dictionary dimensions.

\begin{figure}[htb]
\centering
\psfrag{K-SVD}[B][B][0.7]{\sf \textbf{K-SVD}}
\psfrag{SPAMS}[B][B][0.7]{\sf \textbf{SPAMS}}
\psfrag{EM-BiG-AMP}[B][B][0.7]{\sf \textbf{EM-BiG-AMP}}
\psfrag{Dictionary Size}[t][][0.7]{\sf \quad dictionary rows $M$}
\psfrag{Sparsity}[][][0.65]{\sf training sparsity $K$}
\begin{tabular}{cc}
\includegraphics[width=1.6in]{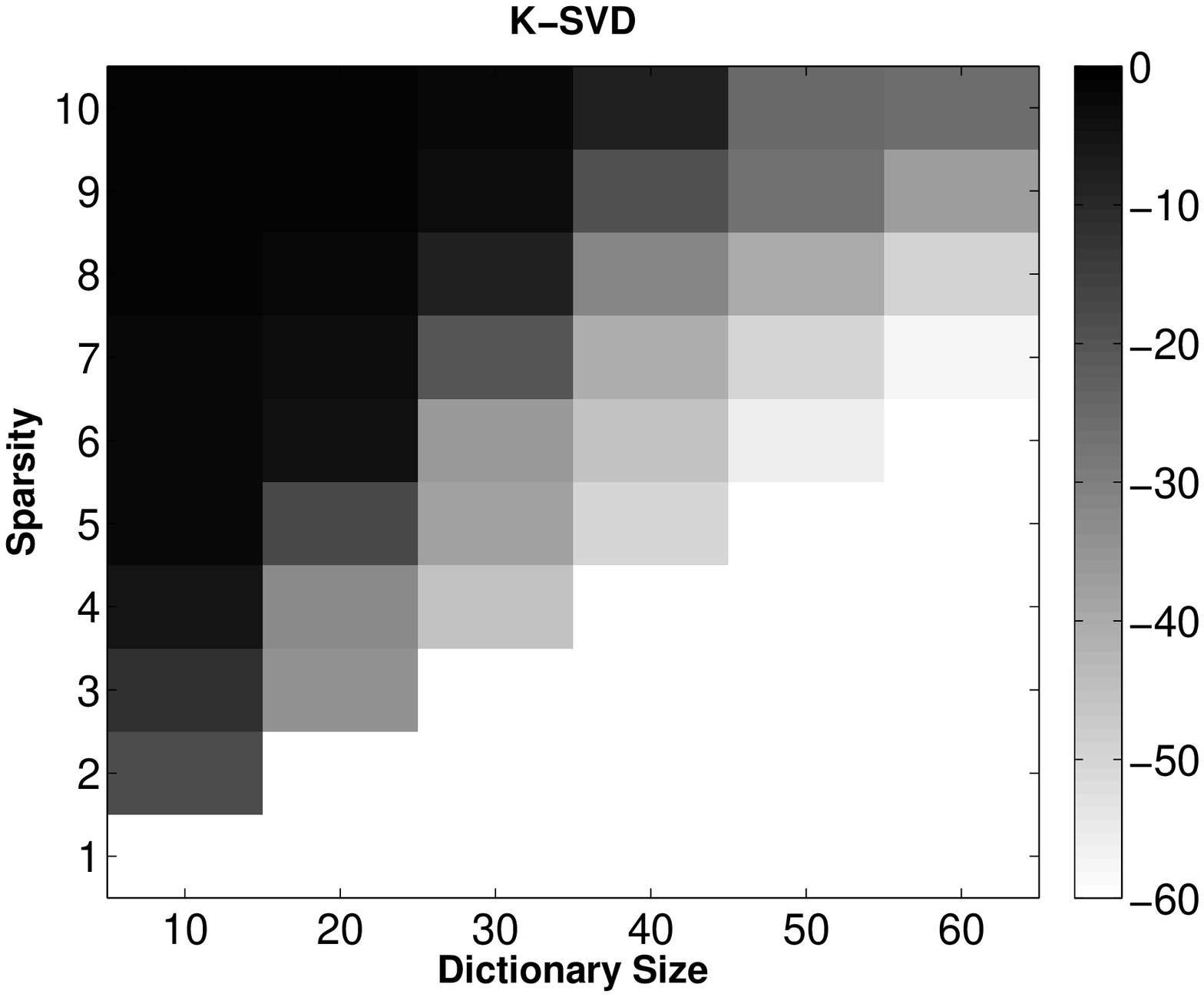}&
\includegraphics[width=1.6in]{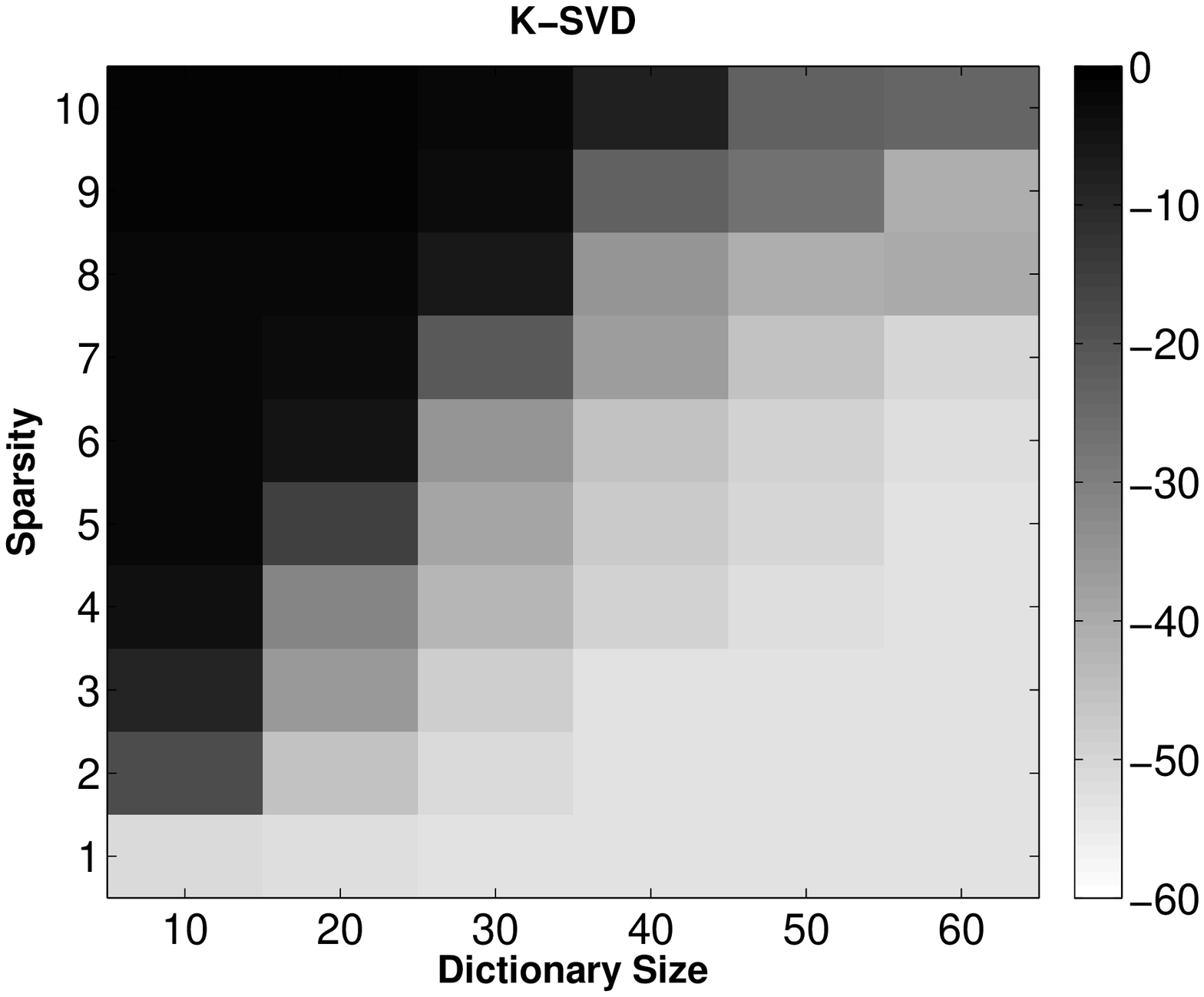}\\[2mm]
\includegraphics[width=1.6in]{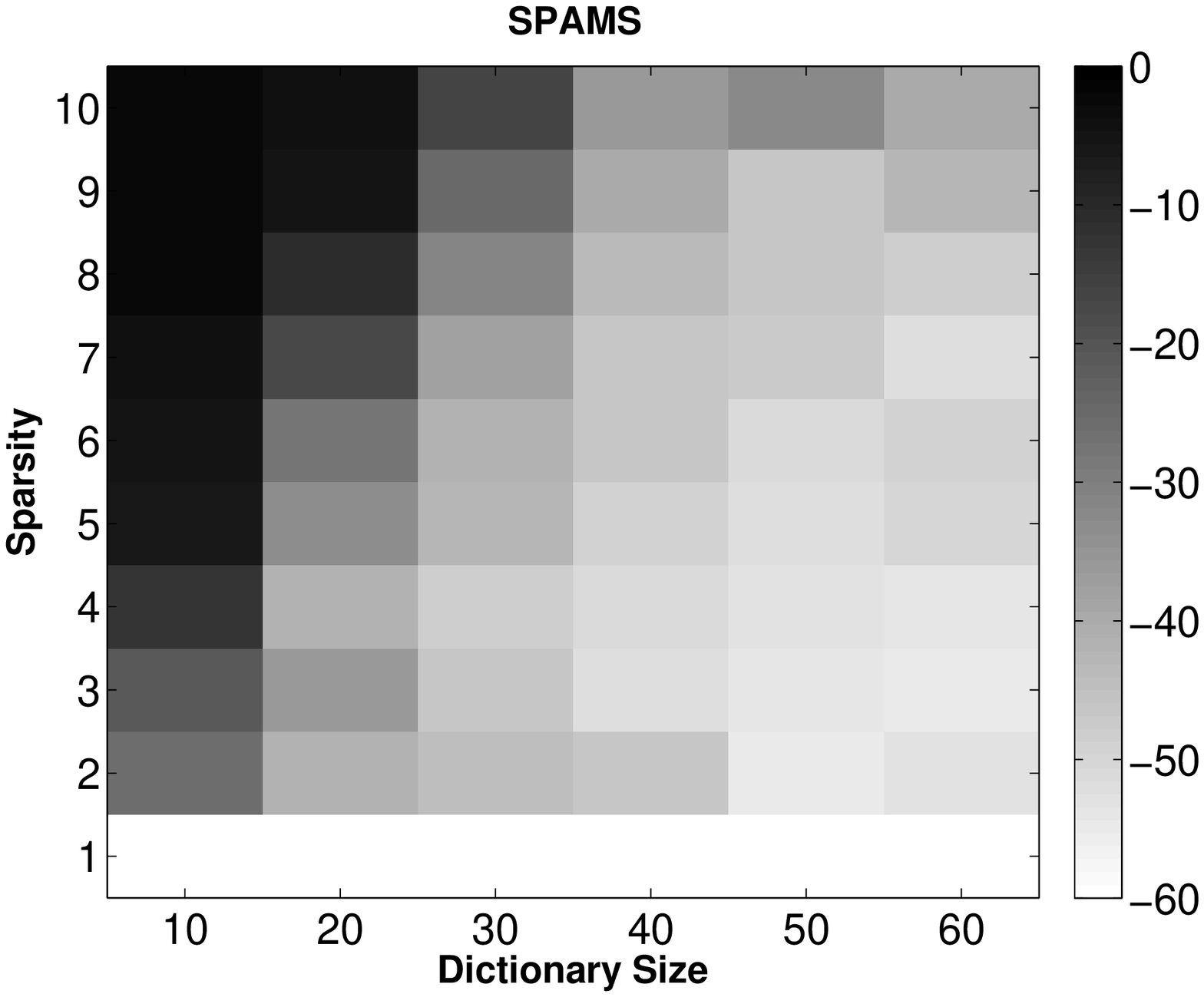}&
\includegraphics[width=1.6in]{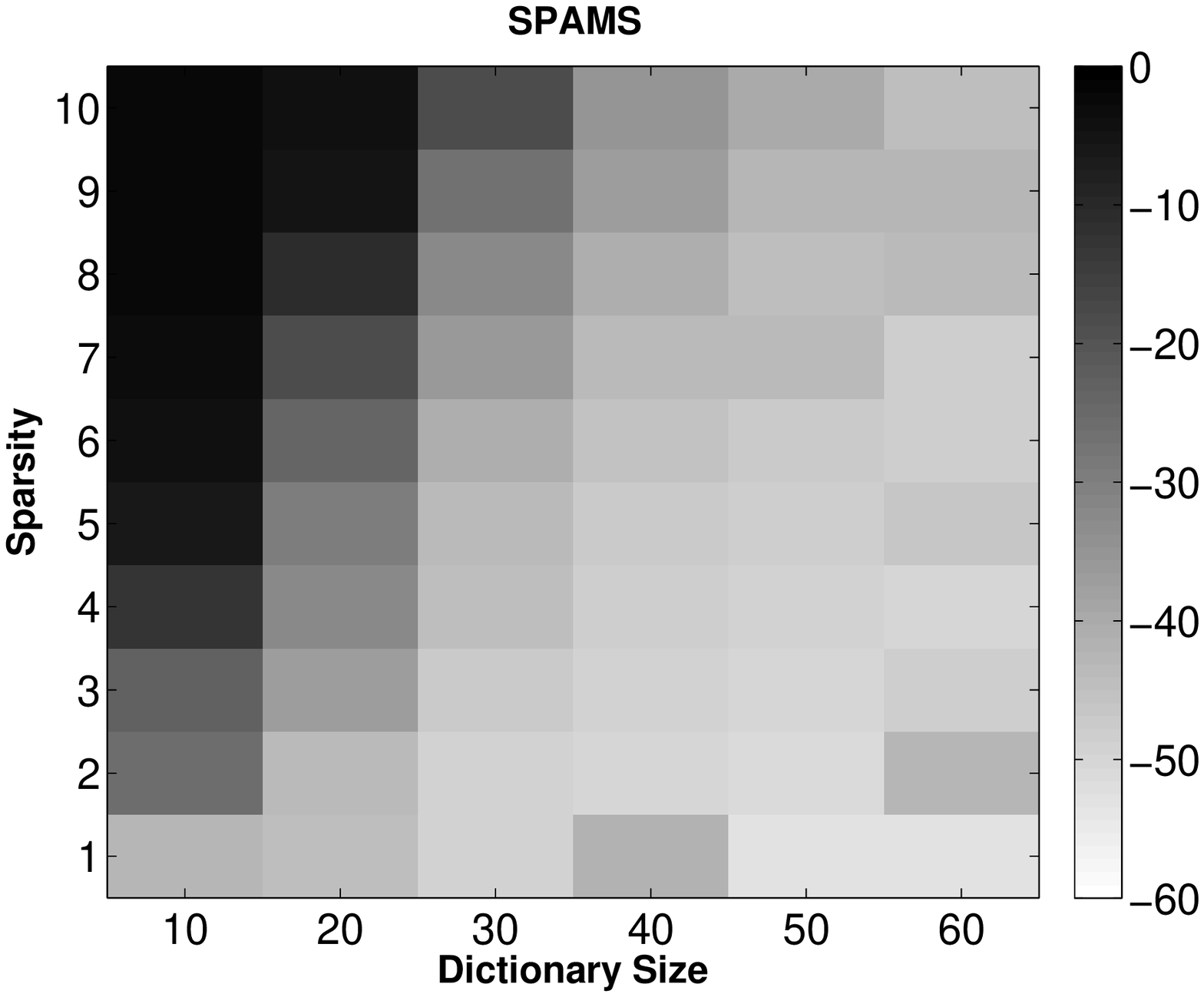}\\[2mm]
\includegraphics[width=1.6in]{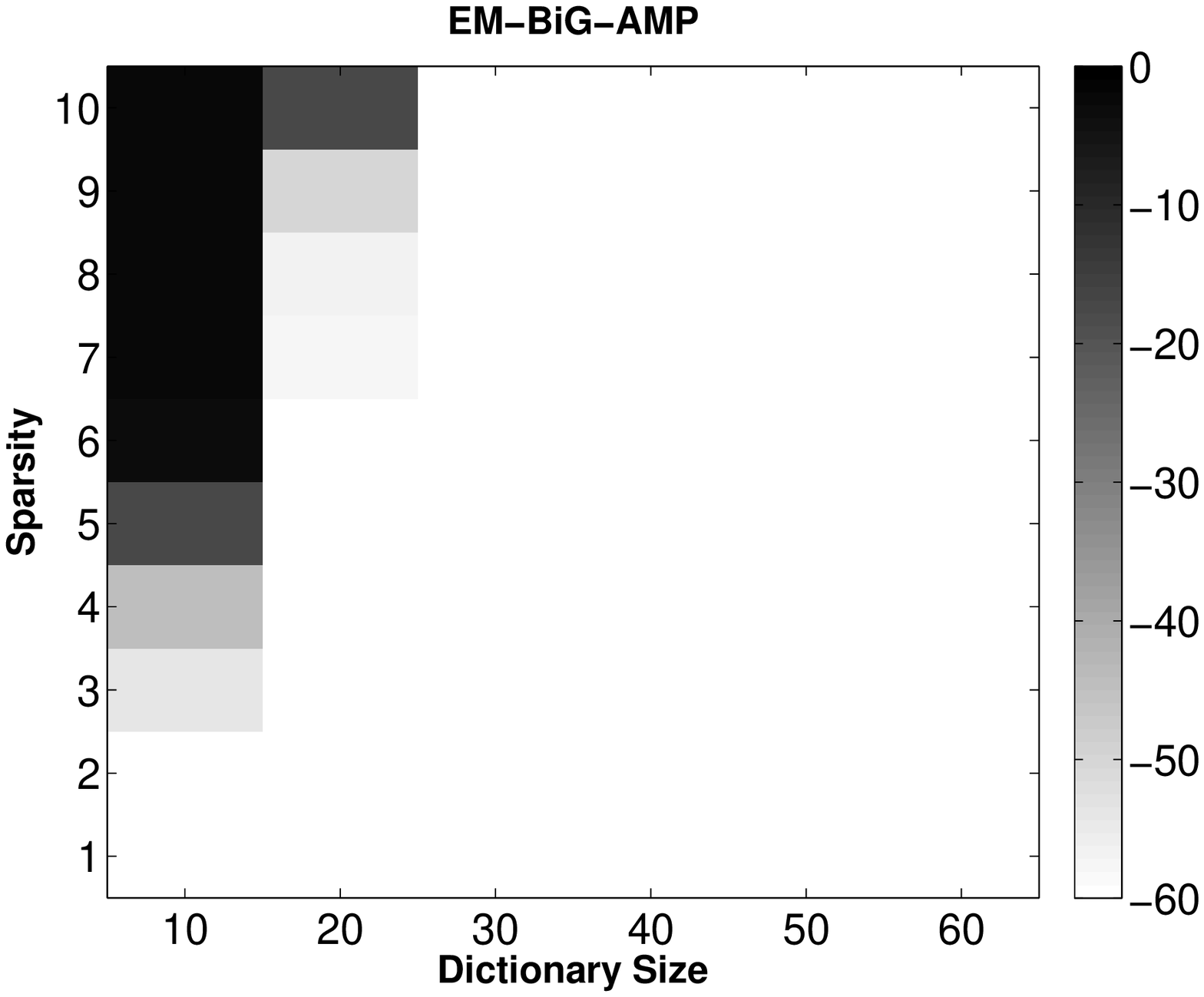}&
\includegraphics[width=1.6in]{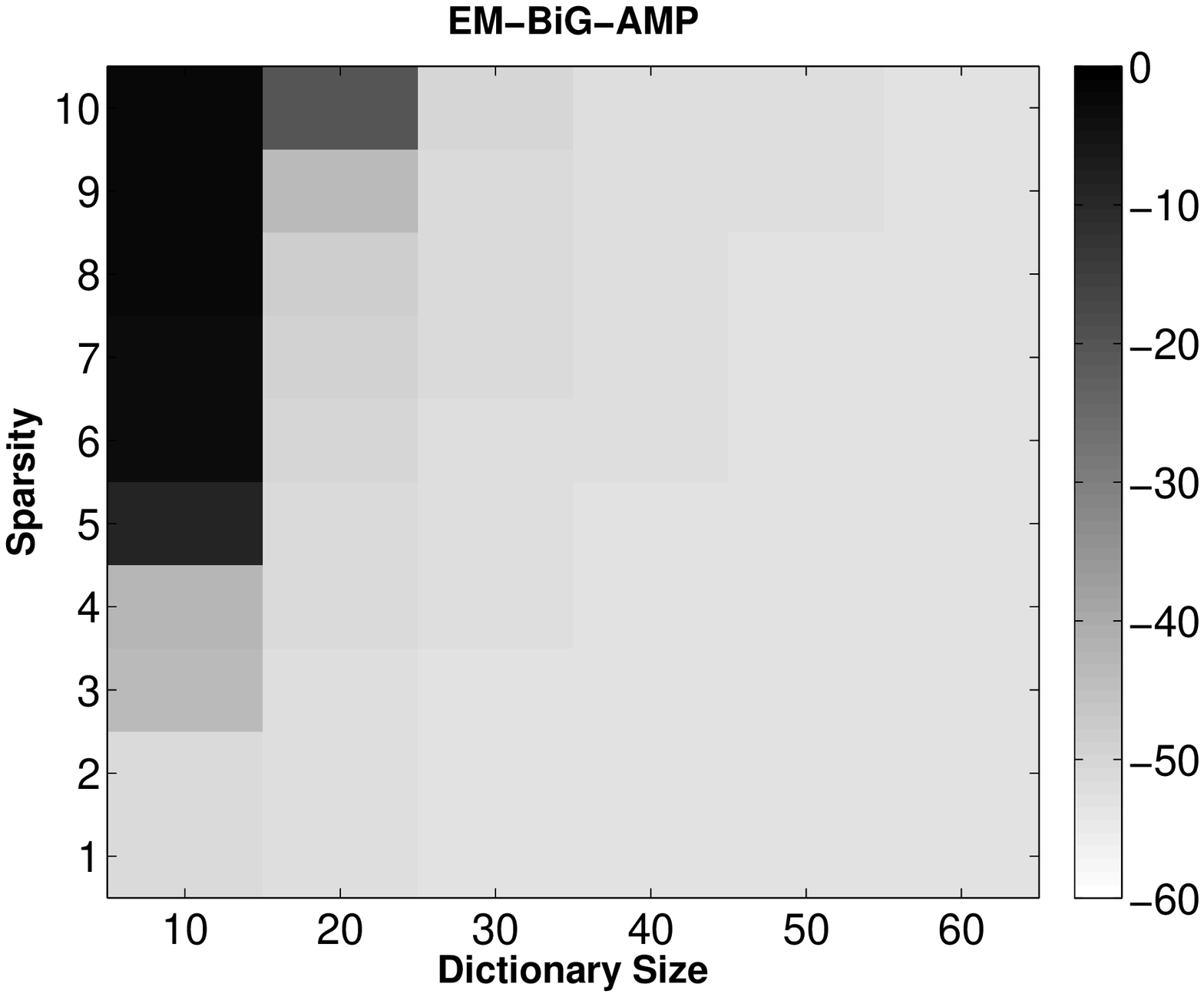}
\end{tabular}
\caption{Mean NMSE (over $10$ trials) for recovery of an $M\times (2M)$ dictionary from $L = 10M \log (2M)$ training samples, each of sparsity $K$, in the noiseless case (left) and under AWGN of $40$~dB SNR (right), for several algorithms.}
\label{fig:PhaseDLNonSquare}
\end{figure}

The left column of \figref{PhaseDLNonSquare} shows the mean value (over $10$ trials) of the relative NMSE for \emph{noiseless} recovery, while the right column shows the corresponding results for \emph{noisy} recovery. 
In all cases, $L=5N\log N = 10M\log(2M)$ training samples were provided. 
EM-BiG-AMP, K-SVD, and SPAMS all give very similar results to \figref{PhaseDL} for the square-dictionary case, verifying that these techniques are equally suited to the recovery of over-complete dictionaries. 
ER-SpUD(proj), however, only applies to square dictionaries and hence was not tested here.

\subsubsection{Summary}
In summary, Figs.~\ref{fig:PhaseDL}-\ref{fig:PhaseDLNonSquare} show that, for noiseless square dictionary learning, EM-BiG-AMP yields an empirical PTC that is nearly as good as the state-of-the-art ER-SpUD(proj) algorithm and much better than those of (genie-aided) K-SVD and SPAMS.
However, the figures show that, in comparison to ER-SpUD(proj), EM-BiG-AMP is fast for large (square) dictionaries, robust to AWGN, and applicable to non-square dictionaries. 

We recall that Krzakala, M{\'e}zard, and Zdeborov{\'a} recently proposed an AMP-based approach to blind calibration and dictionary learning \cite{jtp_Krzakala2013} that bears similarity to our scalar-variance BiG-AMP under AWGN-corrupted observations (recall \Xfoot{Krzakala}).
Although their approach gave good results for blind calibration, they report that it was ``not able to solve'' the DL problem \cite{jtp_Krzakala2013}.
We attribute EM-BiG-AMP's success with DL (as evidenced by Figs.~\ref{fig:PhaseDL}-\ref{fig:PhaseDLNonSquare}) to the adaptive damping procedure proposed in \Xsecref{adapt}, the initialization procedure proposed in \secref{DLinit}, the EM-learning procedure proposed in \secref{EMdl}, and the re-initialization procedure proposed in \secref{DLreinit}.


\section{Conclusion}    \label{sec:conc}

In this work, we presented BiG-AMP, an extension of the G-AMP algorithm proposed for high-dimensional generalized-\emph{linear} regression in the context of compressive sensing, to generalized-\emph{bilinear} regression, with applications in matrix completion, robust PCA, dictionary learning, and related matrix-factorization problems.
In addition, we proposed an adaptive damping mechanism to aid convergence under realistic problem sizes, an expectation-maximization (EM)-based method to automatically tune the parameters of the assumed priors, and two rank-selection strategies.
Extensive numerical results, conducted with synthetic and realistic datasets for matrix completion, robust PCA, and dictionary learning problems, demonstrated that BiG-AMP yields excellent reconstruction accuracy (often best in class) while maintaining competitive runtimes, and that the proposed EM and rank-selection strategies successfully avoid the need to tune algorithmic parameters.

The excellent empirical results reported here motivate future work on the analysis of EM-BiG-AMP, on the extension of EM-BiG-AMP to, e.g., structured-sparse or parametric models, and on the application of EM-BiG-AMP to practical problems in high-dimensional inference.
For example, preliminary results on the application of EM-BiG-AMP to hyperspectral unmixing have been reported in \cite{Vila:SPIE:13} and are very encouraging.

\appendices

\section{}
\label{app:Hderivatives}


Here we derive \eqref{s}-\eqref{nus}, which are stated without a detailed derivation in \cite{Rangan:ISIT:11}. 
Recalling \eqref{H} and omitting the $ml$ subscripts for brevity, it can be seen that
\begin{align}
\lefteqn{H' \big(\hat{q},\nu^q; y\big)}\non\\
&= \frac{\partial}{\partial \hat{q}}\log \int p_{\Y|\Z}(y \giv z) 
	\frac{1}{\sqrt{2\pi \nu^q}} \exp \Big(-\frac{1}{2\nu^q}(z - \hat{q})^2\Big) dz\non\\
&= \frac{\partial}{\partial \hat{q}} \Big\{
  	\log \int \exp \Big( \log p_{\Y|\Z}(y \giv z) 
	-\frac{z^2}{2\nu^q} +\frac{\hat{q}z}{\nu^q} \Big) dz 
	-\frac{\hat{q}^2}{2\nu^q} 
	\Big\}\non\\
  &= -\frac{\hat{q}}{\nu^q} +
  	\frac{\partial}{\partial \hat{q}} \log 
  	\int \exp \big( \phi(u) + \hat{q}u \big) du
	\text{~~via $u\defn\frac{z}{\nu^q}$} \label{eq:H1}
\end{align}
for an appropriately defined function $\phi(\cdot)$.
Now, defining
$p_{\textsf{u}|\textsf{q}}(u\giv \hat{q}) \defn Z(\hat{q})^{-1}\exp(\phi(u)+\hat{q}u)$ 
with normalization term $Z(\hat{q}) \defn \int \exp\big(\phi(u)+\hat{q}u\big) du$,
simple calculus yields
\begin{align}
  \frac{\partial}{\partial \hat{q}} \log Z(\hat{q})
  &= \E\{\textsf{u}\giv \textsf{q} = \hat{q}\} \label{eq:dlogZ}\\
  \frac{\partial^2}{\partial \hat{q}^2} \log Z(\hat{q})
  &=  \var\{ \textsf{u} \giv \textsf{q} = \hat{q}\}. \label{eq:d2logZ}
\end{align}
Thus, from \eqref{H1} and \eqref{dlogZ} it follows that
\begin{align}
\lefteqn{H' \big(\hat{q},\nu^q; y\big)}\non\\
  &= -\frac{\hat{q}}{\nu^q} +
  	\int u \frac{\exp \big( \phi(u) +\hat{q}u \big) }{Z(\hat{q})} du \non\\
  &= -\frac{\hat{q}}{\nu^q} +
  	\int \frac{z}{\nu^q} 
	\frac{\exp \big( \log p_{\Y|\Z}(y\giv z) -\frac{z^2}{2\nu^q} 
	+ \frac{\hat{q}z}{\nu^q}\big) }{Z(\hat{q})} \frac{dz}{\nu^q}  \non\\
  &= -\frac{\hat{q}}{\nu^q} +
  	\frac{1}{\nu^q}\int z 
	\frac{p_{\Y|\Z}(y\giv z) \mc{N}(z;\hat{q},\nu^q)}
	{\int p_{\Y|\Z}(y\giv \bar{z}) \mc{N}(\bar{z};\hat{q},\nu^q)d\bar{z}}
	dz. 
	\label{eq:H'}  
\end{align}
Equation \eqref{s} is then established by applying definitions \eqref{zhat} and \eqref{pZgivYP} to \eqref{H'}. 

Similarly, from \eqref{H1} and \eqref{d2logZ},
\begin{align}
\lefteqn{-H'' \big(\hat{q},\nu^q; y\big)}\non\\
&= \frac{\partial}{\partial \hat{q}} \Big\{
  	\frac{\hat{q}}{\nu^q} -
  	\frac{\partial}{\partial \hat{q}} \log Z(\hat{q}) \Big\}
  =  \frac{1}{\nu^q} - \var\{ \textsf{u} \giv \textsf{q} = \hat{q}\} \non \\
    &=  \frac{1}{\nu^q} - \int \big(u- \E\{\textsf{u}\giv \textsf{q} = \hat{q}\} \big)^2  \,
	\frac{\exp \big( \phi(u) +\hat{q}u \big) }{Z(\hat{q})} du\non \\
  &=  \frac{1}{\nu^q} - \frac{1}{(\nu^q)^2}
  	\int\big(z- \hat{z} \big)^2
	\frac{p_{\Y|\Z}(y\giv z) \mc{N}(z;\hat{q},\nu^q)}
	{\int p_{\Y|\Z}(y\giv \bar{z}) \mc{N}(\bar{z};\hat{q},\nu^q)d\bar{z}}
	dz, 
	\label{eq:H''}
\end{align}
where $\hat{z}$ is the expectation from \eqref{zhat}. 
Equation \eqref{nus} is then established by applying the definitions \eqref{zvar} and \eqref{pZgivYP} to \eqref{H''}.

\color{black}

\section{}
\label{app:s2nus}


Here we explain the approximations \eqref{nur2}-\eqref{nuq2}.
The term neglected in going from \eqref{nur} to \eqref{nur2} can be written using \eqref{s}-\eqref{nus} as
\begin{align}
\lefteqn{\sum_{m=1}^M \nu^a_{mn}(t) \big(\hat{s}^2_{ml}(t) - \nu^s_{ml}(t) \big)}\non\\
&= \sum_{m=1}^M \nu^a_{mn}(t) 
	\Bigg[ \frac{(\hat{z}_{ml}(t) - \hat{p}_{ml}(t))^2 + \nu^z_{ml}(t)}{\nu^p_{ml}(t)^2}
			- \frac{1}{\nu^p_{ml}(t)} \Bigg]\\
&= \sum_{m=1}^M \frac{\nu^a_{mn}(t)}{\nu^p_{ml}(t)} 
	\Bigg[\E\left\{\frac{(\Z_{ml} - \hat{p}_{ml}(t))^2}{\nu^p_{ml}(t)}\right\} - 1\Bigg] \label{eq:Sdiff}
\end{align}
where the expectations are taken over 
$\Z_{ml}\sim p_{\Z_{ml}|\p_{ml}}\big(\cdot\giv\hat{p}_{ml}(t);\nu^p_{ml}(t)\big)$
from \eqref{pZgivYP}.
For GAMP, \cite[Sec.~VI.D]{Rangan:ISIT:11} clarifies that, in the large system limit, under i.i.d priors and scalar variances, the true $z_m$ and the iterates $\hat{p}_m(t)$ converge empirically to a pair of random variables $(\textsf{z},\textsf{p})$ that satisfy $p_{\textsf{z}\giv \p}(z\giv\hat{p}(t)) = \mc{N}(z;\hat{p}(t),\nu^p(t))$. This result leads us to believe that the expectation in \eqref{Sdiff} is approximately unit-valued when averaged over $m$, and thus \eqref{Sdiff} is approximately zero-valued. Similar reasoning applies to \eqref{nuq2}.

\color{black}

\section*{Acknowledgment}
The authors would like to thank Sundeep Rangan, Florent Krzakala, and Lenka Zdeborov{\'a} for insightful discussions on various aspects of AMP-based inference.
We would also like to thank Subhojit Som, Jeremy Vila, and Justin Ziniel for helpful discussions about EM and turbo methods for AMP.

\bibliographystyle{IEEEtran}
\bibliography{parker,macros_abbrev,books,sparse,comm,misc,machine,multicarrier,hsi,phase}

\end{document}